\documentclass[12pt]{article}
\usepackage{graphicx} 

\usepackage{feynmp-auto}
\usepackage{slashed}
\usepackage{multicol}
\usepackage{tikz}
\usepackage{geometry}
\usepackage{amsmath}
\usepackage{hyperref}
\usepackage{fixltx2e}
\usepackage{amssymb}
\usetikzlibrary{arrows}
\usepackage{bbm}
\usepackage[utf8]{inputenc} 
\usepackage[T1]{fontenc} 

\usepackage{pstricks}
\usepackage{color}
\usepackage[english]{babel}
\usepackage[english]{layout}
\usepackage{amsmath}
\usepackage{amssymb}
\usepackage{enumerate}
\usepackage{tabularx}
\usepackage{calc}
\usepackage{indentfirst} 

\usepackage{fancyhdr} 
\usepackage{graphicx}
\usepackage{epstopdf}
\usepackage{lastpage}
\usepackage{ifthen}
\usepackage{float}
\usepackage{lineno} 
\usepackage{setspace}
\usepackage{enumitem}
\usepackage{booktabs}
\usepackage{titlesec}
\usepackage{etoolbox} 
\usepackage{tabto} 
\usepackage{colortbl}
\usepackage{soul}
\usepackage{microtype}
\usepackage{totcount}
\usepackage{changepage}
\usepackage{attrib}
\usepackage{upgreek}
\usepackage{array}
\usepackage{pbox}
\usepackage{ragged2e}
\usepackage{tocloft}
\usepackage{marginnote}
\usepackage{marginfix}
\usepackage{enotez}
\usepackage{amsthm}
\usepackage{cleveref}
\usepackage{scrextend}
\usepackage{newfloat}
\usepackage{seqsplit}

\usepackage{caption}
\usepackage{subcaption}
\usepackage{textcomp}
\usepackage{url}
\usepackage{epstopdf}
\usepackage{graphicx}
\usepackage{tikz}
\usepackage{tikz-feynman}
\usetikzlibrary{backgrounds}
\usepackage{feynmp-auto}
\usepackage{ulem}
\usepackage{multirow}
\usepackage{xcolor}
\usepackage{comment}
\usepackage{cancel}
\usepackage{slashed}
\DeclareMathOperator{\Tr}{Tr}
\usepackage{url}

\usepackage[backend=bibtex,sorting=none,style=numeric-comp,citestyle=phys]{biblatex} 

\allowdisplaybreaks

\hypersetup{colorlinks=true,
citecolor=magenta,
linkcolor=blue}

\usepackage[autostyle=true]{csquotes} 

\begin{document}

\captionsetup{width=0.8\linewidth}

\vspace*{5cm}
\begin{center}
    \Large \textbf{Hadronic Light by Light Corrections to the Muon Anomalous Magnetic Moment}\vspace{0.5cm}

\normalsize Daniel Melo $^{a,\,}$\footnote{Corresponding author: dgmelop@unal.edu.co}, Edilson Reyes  $^{b,\,}$\footnote{eareyesro@unal.edu.co} and Raffaele Fazio $^{a,\,}$\footnote{arfazio@unal.edu.co}

$^{a}$ \quad Departamento de F\'{i}sica, Universidad Nacional de Colombia, Bogot\'{a} D.C., Colombia.\\
$^{b}$ \quad Departamento de F\'{i}sica, Universidad de Pamplona, Pamplona, Colombia.


\end{center}

\abstract{We review the Hadronic Light-by-Light (HLbL) contribution to the muon anomalous magnetic moment. Upcoming measurements will reduce the experimental uncertainty of this precision observable by a factor of four, thus breaking the current balance with the theoretical prediction. A necessary step to restore it is to decrease the HLbL contribution error, which implies a study of the high-energy intermediate states that are neglected in dispersive estimates. We focus on the maximally symmetric high-energy regime and in quark loop approximation of perturbation theory we check the kinematic-singularity/zero-free tensor decomposition of the HLbL amplitude.}

\clearpage

\section*{Introduction}
\begin{figure}
    \centering
    \begin{fmffile}{aMu}
    \setlength{\unitlength}{1cm}\small
    \hspace{0cm}
    \begin{fmfgraph*}(5,5)
        \fmfleft{i1,i2} 
        \fmfright{o1,o2} 
        \fmf{phantom}{i2,v2,o2} 
        \fmf{fermion}{i1,v1,o1}
        \fmf{photon}{v2,v1}
        \fmfset{arrow_len}{3mm}
    \end{fmfgraph*}\hspace{2cm}
    \begin{fmfgraph*}(5,5)
        \fmfleft{i1,i2} 
        \fmfright{o1,o2} 
        \fmf{phantom}{i2,v2,o2} 
        \fmf{fermion}{i1,v1,o1}
        \fmf{photon}{v2,v1}
        \fmfblob{1cm}{v1}
        \fmfset{arrow_len}{3mm}
    \end{fmfgraph*}
\end{fmffile}    
    \caption{Interaction of a fermion with a classical electromagnetic field at tree level (left) vs. corrections due to virtual particles (right).}
    \label{fig:aMu}
\end{figure}
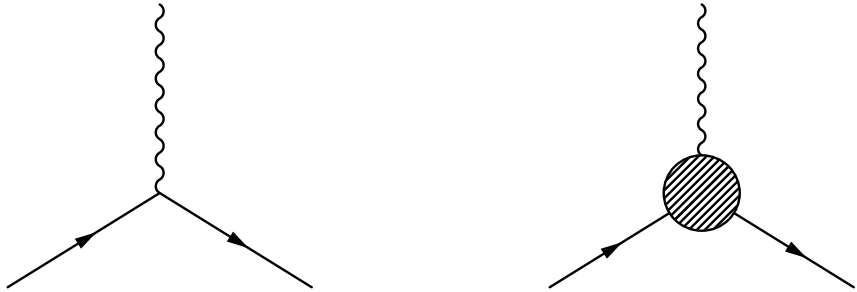

The Standard Model (SM) is the current theoretical paradigm for particle physics at its most fundamental level. This fact is rooted in the SM's mathematical consistency and specially in its highly accurate predictions for precision experiments. In fact, one of the most precisely verified theoretical predictions in the history of physics and the true triumph of quantum field theory is the SM magnetic moment of the electron \cite{eMagneticMomentMeasurement_1,eMagneticMomentMeasurement_2,eMagneticMomentComputation} $\vec{\mu}=g\left(\frac{e}{2m}\right)\vec{S}$, being $m$ the electron mass and $\vec{S}$ its spin operator. The so called anomalous part is expressed by the quantity $a=\frac{g-2}{2}$, quantifying the deviation of the Land$\acute{e}$ factor from the classical value $g=2$, and is entirely due to quantum-mechanical phenomena: the ``cloud'' of virtual particles with which the electron is constantly interacting, slightly changes the way it interacts with a classical magnetic field (see figure~\ref{fig:aMu}). Therefore, the measurement of the anomalous part of a particle's magnetic moment makes possible to test which kind of other particles it interacts with and what is the strength of the interaction. Consequently, this quantity is of the utmost interest for theoretical physicists when testing the SM itself and also theories so called Beyond the Standard Model (BSM). For the electron this anomalous part has been computed to $O(\alpha^{5})$ in QED, for weak contributions the uncertainty is $\sim10^{-16}$ and for hadronic contributions it is $\sim10^{-14}$~\cite{eMagneticMomentComputation}. The discrepancy with measurements is $8.76\cdot10^{-13}$~\cite{eMagneticMomentMeasurement_2} or $2.42$ times the standard uncertainty (often represented as $\sigma$), which is still far enough from the discovery threshold. 

For the muon, the tension between the SM theoretical prediction for the anomalous magnetic moment $a_{\mu}$ and its experimental measurement is bigger. Therefore, it has attracted very much attention since the Brookhaven National Laboratory (BNL) experiment results shed light on the issue in 2004~\cite{BNL}. Currently, the tension stands at $4.2\sigma$ if the 2021 results from Fermilab (FNAL)~\cite{Fermilab1,Fermilab2} are taken into account in addition to the BNL ones: 
\begin{align}
    a_{\mu}^{BNL} &=116\;592\;089(63)\cdot 10^{-11}\;,\\
    a_{\mu}^{Fermilab} &=116\;592\;055(24)\cdot 10^{-11}\;,\\
    a_{\mu}^{exp} &=116\;592\;059(22)\cdot 10^{-11}\;,
\end{align}
As usual the combination of the two measurements is obtained from the principle of maximum likelihood which, mathematically realized by the method of least squares due to the Gaussian probability distribution, provides the above weighted average. 

\begin{figure}
    \centering
    \begin{fmffile}{HVP} 
    \setlength{\unitlength}{1cm}\small
    \hspace{0cm} 
        \begin{fmfgraph*}(5,5)
            \fmfleft{i1,i2}
            \fmfright{o1,o2}
            \fmf{fermion}{i1,v4,v3,v1,o1}
            \fmf{phantom}{i2,v2,o2}
            \fmf{boson,photon}{v3,v2}
            \fmf{photon,tension=0.01,label.side=left}{v1,v5,v4}
            \fmfset{arrow_len}{3mm}
            \fmfblob{0.9cm}{v5}
        \end{fmfgraph*}
    \end{fmffile}
    \caption{Hadronic vacuum polarization contribution to the anomalous magnetic moment of the muon. The blob contains only strongly interacting virtual particles.}
    \label{fig:HVP}
\end{figure}
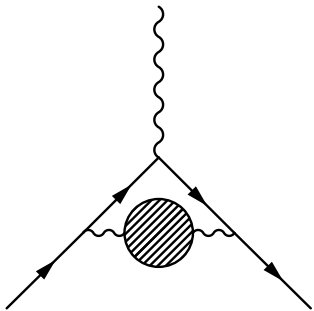

In contrast, the most recent consensus SM prediction $a_{\mu}^{\text{SM}}$ is:
\begin{equation}
     a_\mu^{\text{SM}}=116\;591\; 810(43)\cdot10^{-11}
\end{equation}
which has been obtained by the ``Muon $g-2$ Theory Initiative'' and is described in~\cite{WP}. Consequently, the tension between the SM value and the measurement is: 
\begin{align}
    \Delta a_{\mu} &= a_{\mu}^{\text{exp}} -a_{\mu}^{\text{SM}} = 249(48)\cdot 10^{-11} \text{ or } 5.2\;\sigma\;.
\end{align}
Although this discrepancy is already beyond the five sigma discovery threshold, it is still not considered to be a sign of New Physics because there are unresolved inconsistencies between the dispersive and lattice estimations of the Hadronic Vacuum Polarization (HVP) contribution (see~figure~\ref{fig:HVP}), which we will describe below.

From the three fundamental interactions considered in the SM, only the strong force contribution currently has an uncertainty that is relevant with respect to the tension's value \cite{WP}. Therefore these strong contributions to $a_{\mu}$ are the main focus of the theoretical work towards reducing uncertainty. Hadronic contributions affect $a_{\mu}$ in two ways: HVP and Hadronic Light by Light Scattering (HLbL).

The HVP contribution is much larger than the HLbL one and moreover it can be computed from experimental measurements in the well--known approach of dispersive integrals~\cite{WP}. More specifically, the contribution from HVP amounts to $6845\times 10^{-11}$ and from HLBL it is $92\times 10^{-11}$. HVP is essentially an hadronic correction to the photon propagator, then, because of analyticity of Green functions and the unitarity of the theory, it can be computed from the cross section of a virtual photon decaying into hadrons. This cross section can be extracted from $e^{+}e^{-} \longrightarrow Hadrons$ data from detectors such as BaBar~\cite{babar2013}, KLOE~\cite{kloe}, BESIII~\cite{BESIII}, SND~\cite{SND20} and CMD--3~\cite{cmd3}. By the nature of the dispersive method, it is necessary to know the hadron production cross section at different values of the center of mass energy. This can be achieved either by directly changing the energy of the $e^{-}$ and $e^{+}$ beams, called \textit{direct scan} \cite{DirectScanRussians,DirectScanChinese}, or by fixing it and letting the (measured) initial--state radiation do the work of varying the energy of the virtual photon which then decays into hadrons, called \textit{radiative return} \cite{RadiativeReturn_1,RadiativeReturn_2}. There are also alternative methods of measuring HVP by $\tau$ decay experiments~\cite{Tau} and by measuring the hadronic contribution to the running of the fine structure constant $\alpha=e^{2}/4\pi$ from $\mu^{-} e^{-}$ elastic cross sections, called the MUonE project \cite{Muone_1,Muone_2,Muone_3}. 

Before 2021, the main goals of the community for the HVP contribution were to improve the uncertainty on dispersive estimates by solving tensions between data sets, improving accountability of radiative corrections to the measured cross sections and considering contributions from further channels. Even though these goals still remain, the publication of the HVP contribution estimate by the BMW lattice collaboration~\cite{BMW2021} upended the priorities. This result, the first of its kind to have competitive uncertainties with respect to the dispersive estimates, reduced the tension with experiment to 1.5 $\sigma$ when considered alone. Although this prediction is in tension with dispersive estimates, it has been since confirmed by several other lattice collaborations~\cite{hpqcdHVP,ukqcdHVP,MainzHVP,etmcHVP,HVPrc}. Interesting discussions regarding comparison between dispersive and lattice estimates can be found in~\cite{recentG2review,windowColangelo2022,Colangelo2023discussion}. The most recent results from the CMD--3 experiment for $\pi\pi$ production show a significant deviation from all other previous experimental results, including CMD--2~\cite{cmd2}. Much like recent lattice HVP results, dispersive estimates of HVP with $\pi\pi$ contributions coming only from this new set of measurements also reduce the tension between the SM prediction and $a_\mu^{\text{exp}}$~\cite{Colangelo2023discussion}, which has further added to the confusion. Consistency checks from analyticity and unitarity constraints on the pion vector form factor have not shed much light on the discrepancies between experiments~\cite{Colangelo2023discussion}.

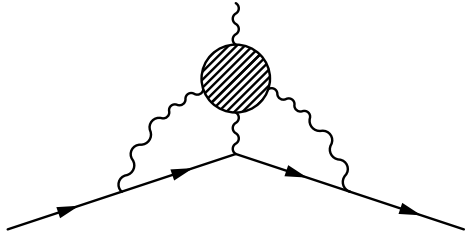
\begin{figure}
    \centering
    \begin{fmffile}{HLbL} 
    \setlength{\unitlength}{1.2cm}\small
        \begin{fmfgraph*}(5,5)
            \fmfleft{i}
            \fmfright{o}
            \fmftop{u}
            \fmf{fermion}{i,v1,v2,v3,o}
            \fmf{photon,tension=1}{v2,v4,u}
            \fmffreeze
            \fmf{photon,left=0.3,tension=1,label.side=left, label=$q_{1}$}{v1,v4}
            \fmf{photon,right=0.3,tension=1,label.side=right, label=$q_{2}$}{v3,v4}
            \fmfblob{0.9cm}{v4}          \fmflabel{$q_{4}\longrightarrow0$}{u}
            \fmfv{label=$q_{3}$,label.angle=135}{v2}
            \fmfset{arrow_len}{3mm}
        \end{fmfgraph*}
    \end{fmffile}
    \vspace{-2.5cm}
    \caption{Hadronic light--by--light scattering contribution to the anomalous magnetic moment of the muon. The blob contains only strongly interacting virtual particles.}
    \label{fig:HLbL}
\end{figure}

Now we go on to the HLbL scattering which gives a contribution (see figure~\ref{fig:HLbL}) of $92(18)\cdot10^{-11}$. In contrast with HVP, the theoretical side of the HLbL scattering computation had been much less understood until recently. The added complexity is due to the fact that four currents are involved, instead of only two. This introduces several difficulties. First of all, HLbL scattering cannot be as cleanly related to $e^{+}e^{-}$ annihilation or other experiments. Furthermore, the HLbL amplitude has a much more complex tensor decomposition: it is a linear combination of 43 tensors, even after gauge invariance constraints have been considered. Moreover, it is necessary to expand this set to a redundant one with 54 elements in order to avoid kinematic, meaning spurious or in general non-dynamical, singularities, that spoil the dispersive approach. In the end, for the purpose of computing $a_{\mu}^{\text{HLbL}}$, it is only necessary to know 7 of these scalar coefficients, since the rest are related to them by crossing symmetry of Mandelstam's variables. Meanwhile, for HVP one initially has two tensor structures which are then reduced to one due to gauge invariance. In fact, this dispersion--fit tensor decomposition for HLbL was only recently found for the first time~\cite{Colangelo2015,Colangelo2017}. This multiplicity of scalar coefficients makes the dispersive approach much more complex for HLbL than it is for HVP, because each coefficient requires its own dispersive integral. In spite of this, contribution of intermediate states including pseudoscalar poles, box topologies and rescattering diagrams~\cite{WP} have been successfully computed. Particular applications with pions can be found in~\cite{PionPole_1,Colangelo2017,ColangeloSWaves}. The dispersive treatment of scalar and axial contributions is not yet in satisfactory state, but significant progress has been made in that direction lately~\cite{holographicQCD,holographicQCD_2,holographicQCD_3,newScalar2021,Zanke2021,holographicQCD_4}. Before recent breakthroughs with the dispersive method, the low energy regime of the HLbL scattering was studied mostly with hadronic models, whose uncertainty was harder to assess. An advantage on the computation of the HLbL contribution with respect to the HVP one is that the first one appears at one further order of $\alpha$ than the second one and thus its computation requires slightly less accuracy. Finally, a common feature for both HLbL and HVP is that they are dominated by very different degrees of freedom at low and high energies, namely, hadrons and then quarks and gluons, respectively. The fact that HVP and HLbL amplitudes enter the muon vertex as an insertion of one and two loops, respectively, makes it necessary to properly ``sew'' the contributions from different approaches at different kinematic regions.

Much like with HVP, lattice estimations for HLbL have become competitive with dispersive ones on recent years~\cite{ETMChlbl2022,ETMChlbl2023,ukqcdHLbL,BMWhlbl2023,Mainz2021,MainzHLbL2023}. Even though the differences between the two are within uncertainty values, lattice estimates also push the SM prediction towards the experimental value.

As mentioned previously, the HLbL amplitude at low photon virtualities (see figure~\ref{fig:HLbL}) was obtained from low--energy QCD models (scalar QED for the pion, Nambu Jona-Lasinio model and vector meson dominance, for example)~\cite{HadronicModel1,HadronicModel2} or, more recently, from dispersive integrals on hadronic production from multiple virtual photons~\cite{Colangelo2014}. Of course, these methods have a certain high energy limit of validity, be it conceptual or practical. For the dispersive approach it is the latter case. Extension to heavier intermediate states has been hindered by a lack of data on the necessary subprocesses and the increasing complexity of unitarity diagrams with multiple particles. Fortunately, heavy intermediate states contributions are suppressed in dispersive integrals by a narrower phase space and thus one can consider states up to certain mass and still obtain a useful result. Nevertheless, to assess or reduce the uncertainty coming from the neglected heavier states it is necessary to resort to tools that complement, replace or evaluate the dispersive approach at high energies. These tools are called \textit{short distance constraints} (SDC). For example, in the high energy regions of dispersive integrals, data for an hadronic form factor can be replaced by the expression for its known asymptotic behaviour. One can also evaluate the asymptotic behaviour of HLbL scattering amplitude itself and use it to evaluate how well the set of intermediate states considered in the dispersive approach resembles such behaviour. For a finite number of intermediate states it is not possible to completely mimic such behaviour~\cite{BijnensSDC2003,KnechtNyffelerSDC}, but this fact can be used to measure how well a set of intermediate states represent high energy contributions.  Such studies are a key complement of dispersive computations and play a central role in uncertainty assessment~\cite{ColangeloSDC,ColangeloSDC_2,ColangeloSDC_3}. There are two loop momenta configurations that lead to a high energy regime in HLbL scattering: $|q_{1}^{2}|\sim |q_{2}^{2}|\sim |q_{3}^{2}|\gg \Lambda_{QCD}^{2}$ and $|q_{1}^{2}|\sim |q_{2}^{2}|\gg \Lambda_{QCD}^{2}$, where $q_{1}$, $q_{2}$ and $q_{3}$ are the virtual photon momenta, $q_{4}$ is the real soft photon momentum representing the electromagnetic external field (see figure~\ref{fig:HLbL}) and $\Lambda_{QCD}$ is the QCD hadronic regime threshold. The main purpose of this research is to compute the HLbL scattering amplitude by the methods of perturbative QCD in the regime in which the absolute values of the three virtualities, $q_1^2$, $q_2^2$, $q_3^2$, are much larger than the hadronic threshold. In particular, we perform an operator product expansion (OPE) in an electromagnetic background for the HLbL scattering amplitude following~\cite{Bijnens2019,Bijnens2020,Bijnens2021}, but we fully harness the background field method to provide an original and, in our view, more systematic framework, to include the hadronic contributions in the same spirit of QCD sum rules. The main result of the work is nevertheless the computation of the quark loop amplitude that constitutes the leading contribution of the HLbL scattering amplitude at high energies. Our computation can be considered an extension of the literature's result, because we obtain and study the full tensor structure of the amplitude and obtain a complete series expansion of ligth quark mass corrections up to arbitrary order. The computation is implemented using original \textit{Mathematica} scripts in combination with state--of--the--art packages, \textit{FeynCalc}~\cite{FeynCalc1,FeynCalc2,FeynCalc3} and \textit{MBConicHulls}~\cite{packageFriot}, for computations in high energy physics.

\section{\texorpdfstring{$a_\mu$}{Lg} in QFT}\label{Chapter1}
 In this section we review the basics of the computation of the HLbL contribution to $a_\mu$ with special focus on the dispersive approach and the Mandelstam representation on which it is based. At the end of the section we present a brief review of the low--energy contributions that are well--understood so far.

\subsection{Basics}
The magnetic moment of a particle is defined through its scattering amplitude on a classical magnetic field. More specifically, for a particle with spin $s$ and magnetic moment $\mu$ interacting with a classical magnetic field $\boldsymbol{B}$, the matrix element of the interaction hamiltonian $H_{int}$ between an initial state $\psi_{\boldsymbol{p}\sigma}$ with momentum $\boldsymbol{p}$ and spin projection $\sigma$ and final state $\psi_{\boldsymbol{p}'\sigma'}$ is:
\begin{align}\label{eq:defG}
    \langle\psi_{\boldsymbol{p}'\sigma'}|H_{int}|\psi_{\boldsymbol{p}\sigma}\rangle &= -\frac{\mu}{ s}(\boldsymbol{J}^{(s)})_{\sigma'\sigma}\cdot\boldsymbol{B}\;\delta^{3}(\boldsymbol{p'}-\boldsymbol{p}) \times 2m\;,
\end{align}
where $\delta^{3}$ represents the Dirac delta in three dimensions, $m$ stands for the particle's mass and $\boldsymbol{J}^{(s)}$ is the little group generator associated to a massive particle of spin $s$. The factor $2m$ appears only due to the relativistic normalization of the states.\\

For a relativistic charged particle, the corresponding matrix element is:
\begin{align}\label{eq:defQED}
    \langle\psi_{\boldsymbol{p}'\sigma'}|H_{int}|\psi_{\boldsymbol{p}\sigma}\rangle &= -e_{q}j_{\mu}A^{\mu}\;,
\end{align}
where $A_{\mu}$ is the classical electromagnetic potential, $j_{\mu}$ is the matrix element of the particle's current operator and $e_{q}$ represents its electric charge. For the muon we have $s=1/2$, $e_{q}=-e$\footnote{$e$ represents the absolute value of the electric charge of the electron.} and:
\begin{align}\label{eq:defCurrent}
    j_{\mu}(x) &=  e^{i(p-p')x}\langle\mu^{-}_{\boldsymbol{p}'\sigma'}|J^{\mu}(0)|\mu^{-}_{\boldsymbol{p}\sigma}\rangle =e^{i(p-p')x}\overline{u}_{\boldsymbol{p}'\sigma'}\Gamma_{\mu}(p',p)u_{\boldsymbol{p}\sigma}\;,
\end{align}
where $J^{\mu}$ represents the electromagnetic current Heisenberg operator of the muon,  $\Gamma_{\mu}$ (in QFT) is the amplitude of the full on--shell vertex diagram to the right of Figure~\ref{fig:aMu} and $u_{\boldsymbol{p}s}$ and $\overline{u}_{\boldsymbol{p}'s'}$ are the spinors associated to the incoming and outgoing muon, respectively. 

Considering the fact that $j_{\mu}$ has to behave as a Lorentz vector and it may contain Dirac matrices, then $\Gamma_{\mu}$ must be a linear combination of the four--momenta, the Levi--Civita symbol $\epsilon^{\mu\nu\lambda\rho}$ and Dirac bilinears, which are listed in the first column of table~\ref{tab:vertexDecomposition1}. The complete set of independent structures that can be built including index contractions is written in the second column table ~\ref{tab:vertexDecomposition1}, where the total momentum $P^{\mu}\equiv p^{\mu}+p^{'\mu}$ and the transferred momentum $q^{\mu}\equiv p^{'\mu}-p^{\mu}$ have been used. However, such set can be greatly reduced using the on--shell character of the spinors in~\eqref{eq:defCurrent} (i.e. Dirac' equation) and gauge invariance.

\begin{table}
    \centering
    \begin{tabular}{|c|c|c|c|c|c}\hline
        Dirac matrices' basis element & Available structures \\\hline\hline
        1 & $P^{\mu}$, $q^{\mu}$ \\
        $\gamma^{\mu}$ & $\gamma^{\mu}$, $q^{\mu}\slashed{q}$, $q^{\mu}\slashed{P}$, $P^{\mu}\slashed{q}$, $P^{\mu}\slashed{P}$, $\epsilon^{\mu\nu\lambda\rho}\gamma_{\nu}P_{\lambda}q_{\rho}$\\
        $\gamma^{5}$ & $P^{\mu}\gamma^{5}$, $q^{\mu}\gamma^{5}$\\
        $\gamma^{\mu}\gamma^{5}$ & $\gamma^{\mu}\gamma^{5}$, $q^{\mu}\slashed{q}\gamma^{5}$, $q^{\mu}\slashed{P}\gamma^{5}$, $P^{\mu}\slashed{q}\gamma^{5}$, $P^{\mu}\slashed{P}\gamma^{5}$, $\epsilon^{\mu\nu\lambda\rho}\gamma_{\nu}\gamma^{5}P_{\lambda}q_{\rho}$\\
        $\sigma_{\mu\nu}$ & $P^{\alpha}q^{\beta}\sigma_{\alpha\beta}P^{\mu}$, $P^{\alpha}q^{\beta}\sigma_{\alpha\beta}q^{\mu}$, $\sigma^{\mu\nu}P_{\nu}$, $\sigma^{\mu\nu}q_{\nu}$, $\epsilon^{\mu\nu\lambda\rho}\sigma_{\nu\lambda}P_{\rho}$, $\epsilon^{\mu\nu\lambda\rho}\sigma_{\nu\lambda}q_{\rho}$\\\hline
    \end{tabular}
    \caption{A priori available structures in the covariant decomposition of the muon electromagnetic on--shell vertex $\Gamma^{\mu}$ (see equation~\eqref{eq:defCurrent} and figure~\ref{fig:aMu}). We use $\sigma_{\mu\nu}=\frac{i}{2}[\gamma^{\mu},\gamma^{\nu}]$}
    \label{tab:vertexDecomposition1}
\end{table}
Indeed the most general tensor structure of $\Gamma_{\mu}(p',p)$ for on--shell spinors is:
\begin{adjustwidth}{-0.0cm}{0cm}
\begin{equation}
    \Gamma_{\mu}(p',p) = A_{1}(q^{2})\gamma_{\mu} +P^{\mu}A_{2}(q^{2}) +\left(\gamma^{\mu} -\frac{2mq^{\mu}}{q^{2}}\right)\gamma^{5}A_{3}(q^{2}) +P^{\mu}\gamma^{5}A_{4}(q^{2})\;.
\end{equation}
\end{adjustwidth}
By using the Gordon's identity: 
\begin{align}
\overline{u}_{\boldsymbol{p}'\sigma'}\gamma^{\mu}u_{\boldsymbol{p}\sigma} &=\frac{1}{2m}\overline{u}_{\boldsymbol{p}'\sigma'}\{P^{\mu} +i\sigma^{\mu\nu}q_{\nu}\}u_{\boldsymbol{p}\sigma}\label{eq:gordonIdentity}\;,\\\label{eq:tensorGamma}
\Gamma_{\mu}(p',p) &= F_{1}(q^{2})\gamma_{\mu} +i\sigma^{\mu\nu}\frac{q_{\nu}}{2m}F_{2}(q^{2}) +(\gamma^{\mu} -\frac{2mq^{\mu}}{q^{2}})\gamma^{5}F_{3}(q^{2}) +\sigma^{\mu\nu}\frac{q_{\nu}}{2m}\gamma^{5}F_{4}(q^{2})\;.
\end{align}
This convention is helpful because $\sigma_{\mu\nu}\equiv\frac{i}{2}[\gamma^{\mu},\gamma^{\nu}]$ is the generator of Lorentz transformations for covariant wave functions of Dirac fermions and therefore $\sigma_{ij}$ generates rotations and little group transformations. Thus the parallel with~\eqref{eq:defG} becomes straightforward. In~\eqref{eq:tensorGamma} $F_{1}$, $F_{2}$, $F_{3}$ and $F_{4}$ are Lorentz invariant coefficients, also called ``form factors''. The first two are associated to parity conserving contributions and are also known as electric and magnetic form factor, respectively. On the other hand, $F_{3}$ and $F_{4}$ are related to parity violating and CP violating contributions, respectively, and are also known as anapole moment and electric dipole moment.
For a pure on-shell derivation of the structure of the electromagnetic vertex including, after an appropriate analytic continuation of momenta, also the photon to be on-shell we refer to \cite{Barut}, \cite{Nima}.
Since the anomalous magnetic moment is related to a non--relativistic interaction, it is necessary to evaluate the muon vertex in the limit of zero exchanged momentum, that is $q\rightarrow0$. In such limit we have $F_{1}(0)=1$ in order to define $e$ as the physical electric charge measured in the interaction with a classical Coulomb field. We also have $F_{3}(0)=0$ in this limit. On the other hand, $F_{2}(0)$ and $F_{4}(0)$ are not constrained. Thus, in the limit of zero exchanged momentum and a slowly varying magnetic field~\eqref{eq:defQED} and~\eqref{eq:defCurrent} become: 
\begin{align}
    \langle\psi_{\boldsymbol{p}'\sigma'}|H_{int}|\psi_{\boldsymbol{p}\sigma}\rangle &= 2m\times \frac{e\,}{m}(1+F_{2}(0))(\boldsymbol{J}^{(1/2)})_{\sigma'\sigma}\cdot\boldsymbol{B}\;\delta^{3}(\boldsymbol{p'}-\boldsymbol{p})\;,
\end{align}
which implies:
\begin{align}
    \mu &= \frac{e}{2m}(1+F_{2}(0))\;.
\end{align}
At tree level we have $\Gamma_{\mu}=\gamma_{\mu}$ and thus $\mu=\frac{e_{q}}{2m}$, which agrees with Pauli's equation and Dirac's equation in the non--relativistic limit. Then, quantum corrections to this classical value can be singled out by the gyromagnetic factor $g$: 
\begin{align}\label{eq:defaMu}
    \mu\equiv g\frac{e_{q}}{2m}s \implies a\equiv\frac{1}{2}(g-2) =F_{2}(0)\;,
\end{align}
where $a$ is called the anomalous part of the magnetic moment of such particle. For the muon we will use the symbols $a_{\mu}$ and $g_{\mu}$.

Once it is clear which part of the on--shell vertex in~\eqref{eq:defCurrent} actually contributes to $\mu$, it is convenient to project it out. The projector needed for the magnetic form factor is~\cite{Brodsky1967Projector}: 
\begin{align}
    P_2^{\mu} &\equiv -\frac{m^{2}}{q^{2}(q^{2}-4m^{2})}\Big(\gamma^{\mu} +\frac{q^{2}+2m^{2}}{m(q^{2}-4m^{2})} P^{\mu}\Big)\;,
\end{align}
which is to be used in the following way:
\begin{align}\label{eq:aMuProjector1}
    F_{2}(q^{2}) &=\Tr\{(\slashed{p}+m)P_{2\mu}(\slashed{p}'+m)\Gamma^{\mu}\}\; , 
\end{align}
Note that there is a divergent $1/q^{2}$ factor in $P_{2\mu}$. Although expanding $\Gamma^{\mu}$ around $q=0$ is inevitable, we can truncate the expansion at first order:
\begin{align}
    \Gamma^{\mu}(q^{2}) &= \Gamma^{\mu}(0) +q_{\nu}\underbrace{\partial^{\nu}\Gamma^{\mu}(q^{2})|_{q=0}}_{\equiv \Gamma^{\mu\nu}}\;.
\label{GAMMAmu}
\end{align}
To prove that, we must first reorganize $P_{2}$ using a slightly different version of the Gordon identity:
\begin{align}\label{eq:cuasiGordon}
    (\slashed{p}'+m)\gamma^{\mu}(\slashed{p}+m) &=\frac{1}{2m}(\slashed{p}'+m)\{P^{\mu} +i\sigma^{\mu\nu}q_{\nu}\}(\slashed{p}+m)\;,\\\label{eq:reorganizeProjector}
     (\slashed{p}+m)P_{2}^{\mu}(\slashed{p}'+m) &= (\slashed{p}+m)\frac{-m}{2(q^{2}-4m^{2})}\Big(-i\sigma^{\mu\nu}\frac{q_{\nu}}{q^{2}} +\frac{3}{(q^{2}-4m^{2})} P^{\mu}\Big)(\slashed{p}'+m)\;.
\end{align}
We see that the divergent term in the projector is of order $1/q$, so the conclusion follows.

By inserting (\ref{GAMMAmu}) as well as (\ref{eq:reorganizeProjector}) into (\ref{eq:aMuProjector1}) with the substitutions $p'=\frac{1}{2}(P+q)$ and $p=\frac{1}{2}(P-q)$ we evaluate the corresponding expression at $q=0$ obtaining after some algebra:
\begin{adjustwidth}{-0cm}{0cm}
\begin{equation}\label{eq:aMuBeforeAverage}
\begin{split}
    a_{\mu} 
    &= \lim_{q\rightarrow0}\frac{im}{2(q^{2}-4m^{2})}\times\\
    &\hspace{0.0cm}\Tr\Big\{\Big(-\frac{1}{2}\slashed{q}\sigma_{\mu\nu}\frac{q^{\nu}}{q^{2}} \Big[\frac{\slashed{P}}{2}+m\Big] +\Big[\frac{\slashed{P}}{2}+m\Big]\sigma_{\mu\nu}\frac{q^{\nu}}{q^{2}} \Big[\frac{\slashed{P}}{2}+m\Big] +\frac{1}{2}\Big[\frac{\slashed{P}}{2}+m\Big]\sigma_{\mu\nu}\frac{q^{\nu}}{q^{2}} \slashed{q}\}\Big)\Gamma^{\mu}(0)\Big\}\\
    &+\lim_{q\rightarrow0}\frac{im}{2(q^{2}-4m^{2})} \Tr\Big\{\Big[\frac{\slashed{P}}{2}+m\Big]\sigma_{\mu\nu}\frac{q^{\nu}q_{\beta}}{q^{2}} \Big[\frac{\slashed{P}}{2}+m\Big]\Gamma^{\mu\beta}\Big\} -\frac{3}{8m^{2}} \Tr\Big\{p_{\mu}(m+\slashed{p})\Gamma^{\mu}(0)\Big\}\;,
\end{split}
\end{equation}
\end{adjustwidth}
where we have explicitly evaluated the limit when possible. We still have divergent terms together with tensor dependence on $q^{\mu}$. To get rid of the latter we will take advantage of the scalar character of $F_{2}(q^{2})$ to perform Lorentz transformations on $q^{\mu}$ before taking the $q\rightarrow0$ limit.

 In particular, we can perform spatial rotations, thus, we carry out an angular average over the spatial components of $q^{\mu}$ taking $P$ as reference. The results are: 
\begin{equation}\label{averages}
    \int\frac{d\Omega}{4\pi}q^{\mu} =0\;, \hspace{1cm}
    \int\frac{d\Omega}{4\pi}q^{\mu}q^{\nu} =\frac{q^{2}}{3}\Big(g^{\mu\nu}-\frac{P^{\mu}P^{\nu}}{P^{2}}\Big)\;.
\end{equation}
The first result is obvious by the oddness of the integrand. The tensor structure of the second result is evident from Lorentz covariance and the scalar factors can be obtained straightforwardly by computing the Lorentz trace on both sides. 

After inserting the angular averages (\ref{averages}) inside~\eqref{eq:aMuBeforeAverage} we obtain: 
\begin{equation}\label{eq:aMuHLbL}
    a_{\mu}= \Tr\Big\{\Big(\frac{1}{12}\gamma_{\mu}-\frac{1}{4}\frac{p_{\mu}}{m} -\frac{1}{3}\frac{1}{m^{2}}p_{\mu}\slashed{p} \Big)\Gamma^{\mu}(0)+\frac{1}{48m}(\slashed{p}+m)[\gamma_{\mu},\gamma_{\beta}] (\slashed{p}+m)\Gamma^{\mu\beta}\Big\}\;,
\end{equation}
which is the direct relation between $a_\mu$ and Feynman amplitudes ($\Gamma^\mu$) that we were looking for and opens the path to specialize the result to particular topologies of diagrams and, in particular, to the HLbL one. 

\subsection{Specializing \texorpdfstring{$a_{\mu}$}{Lg} to HLbL scattering amplitudes}\label{sec:aMuHLbL}

In this section we specialize the result obtained in previous one to compute $a_{\mu}$ from HLbL scattering amplitudes, whose diagrams are shown in Figure~\ref{fig:HLbL}. The term ``light--by--light'' makes reference to the subdiagram appearing in figure~\ref{fig:HLbL}, which has four external photons (three virtual and attached to the muon line and one representing and external field). The term ``hadronic'' is due to the fact that only strongly interacting particles (quarks and gluons) or hadrons (mesons and various resonances) are allowed to appear in the blob of figure~\ref{fig:HLbL}, either as virtual exchanged particles or as poles of the amplitude, for the HLbL contributions.

The first step to specialize $a_\mu$ to HLbL is to isolate the appropriate subdiagram amplitudes from the muon electromagnetic vertex ones. Making use of the Feynman rules for QED it is possible to read the result off the Feynman diagram in figure~\ref{fig:HLbL}:
\begin{align}\nonumber
    -e\overline{u}_{\boldsymbol{p}'\sigma'}\Gamma^{\mu_{4}}u_{\boldsymbol{p}\sigma}&=    \overline{u}_{\boldsymbol{p}'\sigma'}\int \frac{d^{4}q_{1}}{(2\pi)^{4}}\int \frac{d^{4}q_{2}}{(2\pi)^{4}}\frac{-i}{q_{1}^{2}}\frac{-i}{q_{2}^{2}}\frac{-i}{q_{3}^{2}}(-ie\gamma_{\mu_{1}})i\frac{\slashed{p}'+\slashed{q}_{1}+m}{(p'+q_{1})^{2}-m^{2}}(-ie\gamma_{\mu_{3}})\\\label{eq:vertexFeynAmplitude1}
    &\hspace{1cm}\times i\frac{\slashed{p}-\slashed{q}_{2}+m}{(p-q_{2})^{2}-m^{2}}(-ie\gamma_{\mu_{2}}) \times e^4\,\Pi^{\mu_{1}\mu_{2}\mu_{3}\mu_{4}}(q_{1},q_{2},q_{3})u_{\boldsymbol{p}\sigma}\;,
\end{align}
where $e^4\,\Pi^{\mu_{1}\mu_{2}\mu_{3}\mu_{4}}$ represents the amplitude of the hadronic blob inside figure~\ref{fig:HLbL}:
\begin{align}\label{eq:HLbLTensor}
    \Pi^{\mu_{1}\mu_{2}\mu_{3}\mu_{4}} &= -i\int d^{4}x\int d^{4}y \int d^{4}z e^{-i(q_{1}x+q_{2}y+q_{3}z)}\langle\Omega|J^{\mu_{1}}_{s}(x)J^{\mu_{2}}_{s}(y)J^{\mu_{3}}_{s}(z)J^{\mu_{4}}_s(0)|\Omega\rangle\;.
\end{align}
$|\Omega\rangle$ represents the QCD vacuum and $J_{s}$ stands for the electromagnetic current of strongly interacting particles. In the literature, $\Pi^{\mu_{1}\mu_{2}\mu_{3}\mu_{4}}$ is referred to as ``fourth rank vacuum polarization tensor''~\cite{Aldins1970,Knecht2002,HLbL_LO9} or ``HLbL tensor''~\cite{Colangelo2014,Bijnens2019}. Introducing this new convention into~\eqref{eq:vertexFeynAmplitude1} we obtain:
\begin{equation}\label{eq:GammaHLbL}
    \Gamma^{\mu_{4}}_{\text{HLbL}}=-e^{6}\int \frac{d^{4}q_{1}}{(2\pi)^{4}}\int \frac{d^{4}q_{2}}{(2\pi)^{4}}\frac{\gamma_{\mu_{1}}}{q_{1}^{2}}\frac{\slashed{p}'+\slashed{q}_{1}+m}{(p'+q_{1})^{2}-m^{2}}\frac{\gamma_{\mu_{3}}}{q_{3}^{2}}\frac{\slashed{p}-\slashed{q}_{2}+m}{(p-q_{2})^{2}-m^{2}}\frac{\gamma_{\mu_{2}}}{q_{2}^{2}}\, \Pi^{\mu_{1}\mu_{2}\mu_{3}\mu_{4}}\;.
\end{equation}
The next step towards finding the HLbL contribution to $a_{\mu}$, labelled $a_{\mu}^{\text{HLbL}}$, is to insert~\eqref{eq:GammaHLbL} into~\eqref{eq:aMuHLbL}.

The $\Gamma^\mu(0)$ term vanishes for HLbL, as can be deduced from the analysis presented in~\cite{low1958_SoftPhotonZeros}, which concludes that the cross section of a process in the limit in which an external photon becomes soft is equal to a sum of terms proportional to the amplitude of the process without the soft photon and its derivative plus vanishing contributions proportional to the soft photon momentum. In the context of HLbL scattering, the previous statement reads:
\begin{align}
    \Pi^{\mu_{1}\mu_{2}\mu_{3}\mu_{4}}(q_{1},q_{2},q_{3}) &\sim \Pi^{\mu_{1}\mu_{2}\mu_{3}}A^{\mu_{4}} + \partial\Pi^{\mu_{1}\mu_{2}\mu_{3}}B^{\mu_{4}} +O(q_{4})\;,
\end{align}
where $\Pi^{\mu_{1}\mu_{2}\mu_{3}}$ represents the three--photon scattering amplitude, $A_{\mu_{4}}$ and $B_{\mu_{4}}$ are two vectors of order $O(1/q_{4})$ and $O(q_{4}^{0})$, respectively, and $\partial$ represents derivative with respect to some kinematic variable of the problem. $\Pi^{\mu_{1}\mu_{2}\mu_{3}}$ vanishes due to Furry's theorem and therefore $\Pi^{\mu_{1}\mu_{2}\mu_{3}\mu_{4}}$ vanishes (at least) linearly in the static field limit\footnote{It is not possible to arrive at this conclusion using Weinberg's eikonal factor since the external soft photon may be emitted by an internal line in the strongly--interacting blob.}.

With respect to $\Gamma^{\mu\alpha}_{\text{HLbL}}$, from~\eqref{eq:GammaHLbL} we find:
\begin{equation}
    \Gamma^{\mu_4\nu_4}=e^{6}\int \frac{d^{4}q_{1}}{(2\pi)^{4}}\int \frac{d^{4}q_{2}}{(2\pi)^{4}}\frac{\gamma_{\mu_{1}}}{q_{1}^{2}}\frac{\slashed{p}+\slashed{q}_{1}+m}{(p'+q_{1})^{2}-m^{2}}\frac{\gamma_{\mu_{3}}}{q_{3}^{2}}\frac{\slashed{p}-\slashed{q}_{2}+m}{(p-q_{2})^{2}-m^{2}}\frac{\gamma_{\mu_{2}}}{q_{2}^{2}}\, \partial^{\mu_4}\Pi^{\mu_{1}\mu_{2}\mu_{3}\nu_{4}}\Big|_{q_{4}\rightarrow0}\;,
\end{equation}
where we have used the antisymmetry between $\mu_{4}$ and $\nu_{4}$ of $\partial^{\mu_{4}}\Pi^{\mu_{1}\mu_{2}\mu_{3}\nu_{4}}|_{q_{4}\rightarrow0}$, which can be deduced by differentiating the Ward identity. Terms proportional to $q_4^{\nu_4}\Pi^{\mu_{1}\mu_{2}\mu_{3}\mu_{4}}$ were disregarded in the static limit. Finally, turning back to $a_{\mu}^{\text{HLbL}}$ one obtains: 
\begin{align}\nonumber
    a_{\mu}^{\text{HLbL}} &=\frac{e^{6}}{48m} \int \frac{d^{4}q_{1}}{(2\pi)^{4}}\int \frac{d^{4}q_{2}}{(2\pi)^{4}}\frac{1}{q_{1}^{2}}\frac{1}{q_{2}^{2}}\frac{1}{q_{3}^{2}}\frac{1}{(p+q_{1})^{2}-m^{2}}\frac{1}{(p-q_{2})^{2}-m^{2}}\,\frac{\partial}{\partial q_{4\mu
    _{4}}}\Pi^{\mu_{1}\mu_{2}\mu_{3}\nu_{4}}\Big|_{q_{4}\rightarrow0}\\
    &\hspace{1cm}\times \Tr\Big\{(\slashed{p}+m)[\gamma_{\mu_{4}},\gamma_{\nu_{4}}] (\slashed{p}+m)\gamma_{\mu_{1}}(\slashed{p}+\slashed{q}_{1}+m)\gamma_{\mu_{3}}(\slashed{p}-\slashed{q}_{2}+m)\gamma_{\mu_{2}}\Big\}\;.\label{eq:aMuHLbLIntermediate}
\end{align}
There are only three steps left to obtain $a_{\mu}^{\text{HLbL}}$: (i) compute the Dirac trace, (ii) compute $\partial^{\mu_{4}}\Pi^{\mu_{1}\mu_{2}\mu_{3}\nu_{4}}|_{q_{4}\rightarrow0}$ and (iii) compute the two--loop integral. The trace can be performed straightforwardly. On the other hand, the computation of the HLbL amplitude is very complex and it is therefore necessary to study it in depth before advancing further.

\subsection{Dispersive computation of the HLbL amplitude}\label{sec:mandelstamRepresentation}

In the previous section we expressed $a_{\mu}$ in terms of the HLbL scattering amplitude $\Pi^{\mu_{1}\mu_{2}\mu_{3}\mu_{4}}$ and in this section we will present the dispersive approach for computing it.

Since $\Pi^{\mu_{1}\mu_{2}\mu_{3}\mu_{4}}$ appears inside a two loop integral on $q_{1}$ and $q_{2}$, it is necessary to compute it at different energy regions involving perturbative and non--perturbative regimes. This is due to asymptotic freedom, which invalidates perturbation theory at energy scales below $\Lambda_{QCD}\sim$1 GeV because the coupling $\alpha_s$ approaches $1$. Non--perturbative contributions give the bulk of $a_\mu^{\text{HLbL}}$, but high energy studies are important for error estimation, so it is convenient to perform computations in both regimes in a unified framework.

QFT in the lattice and dispersive integrals are two of very few tools that allow for the computation of amplitudes in non--perturbative regimes. The first one tries to solve the QFT equations in a finite spacetime cube of side length $L$ with discrete Euclidean spacetime coordinates of spacing $a$. The observables of interest are then computed for different values of large $L$ and $1/a$ and these results are then extrapolated to $L,1/a\rightarrow\infty$ in order to recover the standard QCD results. Using a very different perspective, the dispersive approach~\cite{Colangelo2014,Colangelo2015,Colangelo2017,PionPole_1} relies on the analyticity of scattering amplitudes and the unitarity of the S--matrix (probability conservation) to relate the amplitudes of a process with the cross sections of its sub--processes, which are fitted to data. Lattice computations have high numerical complexity due to the very large number of degrees of freedom that appear when both the size $L$ of the system and its resolution $1/a$ become large, as needed to reduce the uncertainty produced by the extrapolation to $L,1/a\rightarrow\infty$. Therefore, even for the simpler case of HVP only till very recently have its results become competitive with the dispersive ones in terms of uncertainty~\cite{HLbLLattice,BMW2021,Mainz2021}. Furthermore, for HLbL the first lattice computations are relatively new and they still are not competitive with dispersive ones. Although the dispersive approach for the computation of the HLbL contribution to $a_{\mu}$ also has its drawbacks, the main one relating to the Mandelstam representation of $\Pi^{\mu_{1}\mu_{2}\mu_{3}\mu_{4}}$ have been recently overcome. This has allowed to obtain the most reliable accounts of $a_{\mu}^{\text{HLbL}}$ in recent years~\cite{WP}\footnote{Although the approach proposed in~\cite{Colangelo2014,Colangelo2015,Colangelo2017} has been the standard scheme in recent estimations of $a_{\mu}^{\text{HLbL}}$, there exist alternate dispersive frameworks. For example, in \cite{AlternateDispersive1,AlternateDispersive2,AlternateDispersive3,AlternateDispersive4,AlternateDispersive5} a dispersive equation is applied directly to the magnetic form factor $F_{2}$ instead of the HLbL Feynman amplitude.}.

As mentioned at the beginning of this subsection, in this work we focus on the dispersive approach. It is based on four fundamental pillars: unitarity of the $S$--matrix, the Sugawara--Kanazawa theorem for functions of a complex variable and the Schwarz reflection theorem. We will review such pillars in order. 

\subsubsection{Unitarity of the S--matrix}
A key concept in relativistic quantum theories of fundamental interactions\footnote{Not any quantum theory conserves probability, non--relativistic systems with absorptive potentials are an example.} is conservation of probability. For transition rates, it is equivalent to the unitarity of the S--matrix. Let us explore the consequences of such feature for the transition matrix $T$:
\begin{align}
    S^{\dagger} &=S^{-1} \implies (1+iT)(1-iT^{\dag}) =1 \implies TT^{\dag} = i(T^{\dag}-T)\;.
\end{align}
If we evaluate a certain matrix element of $S$ and insert a complete set of momentum eigenstates in the last equation we obtain~\cite{BookPeskin}:
\begin{align}\label{eq:opticalTheorem}
    2\text{Im}\mathcal{M}(i\rightarrow f) &=\displaystyle \sum_{n}\Big(\Pi^{n}_{i=1}\int\frac{d^{3}\boldsymbol{q}_{i}}{(2\pi)^{3}}\frac{1}{2E_{i}}\Big)\mathcal{M}^{*}(f\rightarrow \{q_{i}\})\mathcal{M}(i\rightarrow \{q_{i}\})\, (2\pi)^{4}\delta^{4}(P_{i} -\sum_{i}q_{i})\;,
\end{align}
where $\mathcal{M}$ stands for a Feynman amplitude, $i$ represents the initial state, $f$ the final one and ${q_{i}}$ the on--shell intermediate--states, which come from the insertion of the identity resolved in terms of the momentum eigenstates. If the initial and final states are the same, like for HVP (see figure~\ref{fig:HVP}), then we obtain the optical theorem.

At this point~\eqref{eq:opticalTheorem} does not seem to be of much help. First, it can only be applied in principle to the on--shell electromagnetic vertex amplitude. Secondly, we only have the imaginary part of $\Pi^{\mu_{1}\mu_{2}\mu_{3}\mu_{4}}$ and we need it in full to compute observables such as $a_{\mu}^{\text{HLbL}}$ in~\eqref{eq:aMuHLbL}. Additionally, we are tasked to compute one amplitude in terms of infinitely many and infinitely complex different subamplitudes.  The first issue can be fixed by simplifying terms in both sides of the equation such that it applies just as well to the HLbL subdiagrams. To deal with the second one we need to make use of the Sugawara--Kanazawa theorem, which reconstructs a complex variable function based on its poles and branch cuts. It will also show us why we can consider only intermediate states up to certain energy scale in~\eqref{eq:opticalTheorem} and still get meaningful predictions.

\subsubsection{Sugawara--Kanazawa theorem and Schwarz reflection identity}
Consider a function of a complex variable $z$ with (possibly) two branch cuts along the real axis: one to the right starting at $c_{1}$ and extending (possibly) to positive infinity and one to the left starting at $-c_{2}$ and extending (possibly) to negative infinity. Based on the following three requirements:

\begin{itemize}
    \item $f(z)$ has finite limits in the positive real infinity direction above and below the right--hand cut.
    \item The limit of $f(z)$ in the negative real infinity direction above and below the left--hand cut exists.
    \item If $f(z)$ is divergent in certain infinite direction, such a divergence is weaker than a polynomial with finite power $N$ such that $N\geq1$.
\end{itemize}

then the Sugawara--Kanazawa theorem~\cite{SugawaraKanazawa} claims that $f(z)$ may be represented as\footnote{There is also another result to this theorem that essentially claims that $f(z)$ has the same limit at infinity in any direction with positive (negative) imaginary part as it has along and above (under) the right (left) cut.}:
\begin{align}\label{eq:Sugawara}
    &f(z) =\displaystyle\sum_{i}\frac{R_{i}}{z-x_{i}} +\frac{1}{\pi}\Big(\int_{c_{1}}^{\infty} +\int_{-\infty}^{-c_{2}}\Big)\frac{\Delta_{x} f(x)}{x-z} \;dx +\lim_{x\rightarrow\infty}\overline{f}(x)\;,\\\nonumber
    &\Delta_{x} f(x) =\frac{1}{2i}\{f(x+i\epsilon) -f(x-i\epsilon)\}\;,\hspace{1cm}
    \overline{f}(x) = \frac{1}{2}\{f(x+i\epsilon) +f(x-i\epsilon)\}\;,
\end{align}
where $R_{i}$ represents the residue of $f(z)$ in $x_{i}$ which lies on the real interval $[-c_{2},c_{1}]$. The two integrals in~\eqref{eq:Sugawara} are performed along the real axis. This representation of $f(z)$ is usually called ``dispersion relation''. The last term is referred to as the ``subtraction constant'' and accounts for possible divergences of $f(z)$ in the infinity, which enter the equation as the contribution of the circumference of a Cauchy integration path at infinity.

The HLbL process may be regarded as a function of two of the usual Mandelstam variables $s$, $t$ and $u$ for two--two scattering. Therefore, the Sugawara--Kanazawa theorem has to be applied twice; one for each independent variable, leaving the rest constant. However, the procedure is to be applied to a function that unifies all three channels' amplitudes. The resulting double dispersive integral is known as Mandelstam representation~\cite{Mandelstam1958}. The first step to build it is to write a single dispersive representation for, say, $s$, which we will consider to be unsubtracted for simplicity. Since we are expecting $\mathcal{M}$ to ``contain'' the amplitudes for the three channels, it must be invariant under crossing. As such, given that $t$ has a fixed value, we expect to have: 
\begin{align}
    \mathcal{M}(s,t) &=\displaystyle\sum_{i}\frac{R_{i}^{s}(t)}{s-x^{s}_{i}} +\frac{1}{\pi}\int_{c_{1}}^{\infty}\frac{\Delta_{s'} \mathcal{M}(s',t)}{s'-s} \;ds' +\frac{1}{\pi}\int_{c_{1}}^{\infty}\frac{\Delta_{u'} \mathcal{M}(u',t)}{u'-u} \;du' \;.
\end{align}
Now we perform an analytic continuation on $t$ to whichever value we require. Via interaction form factors the $t$ dependence of the residues is usually well--known, but not for $\Delta\mathcal{M}$. To deal with this we apply~\eqref{eq:Sugawara} once again, but this time for $\Delta_{s}\mathcal{M}$ with fixed $s'$:
\begin{equation}
    \Delta_{s'}\mathcal{M}(s',t) =\frac{1}{\pi}\int_{c_3(s')}^{\infty}\frac{\Delta_{t'}\Delta_{s'} \mathcal{M}(s',t')}{t'-t} \;dt' +\frac{1}{\pi}\int_{c_3(s')}^{\infty}\frac{\Delta_{u'}\Delta_{s'} \mathcal{M}(s',u')}{u'-\overline{u}} \;du'
\end{equation}
where $\overline{s}$ is the Mandelstam variable associated to $u'$ and $t$, while $\overline{u}$ is the Mandelstam variable associated to $s'$ and $t$. An similar equation applies for $\Delta_{u}\mathcal{M}$. Inserting this into the single dispersion relation we obtain:
\begin{align}\nonumber
    \mathcal{M}(s,t)
    &=\frac{1}{\pi^{2}}\int_{c_{1}}^{\infty}ds' \int_{c_3(s')}^{\infty}dt'\frac{\Delta_{t'}\Delta_{s'} \mathcal{M}(s',t')}{(s'-s)(t'-t)} +\frac{1}{\pi^{2}}\int_{c_{1}}^{\infty}ds' \int_{c_3(s')}^{\infty}du'\frac{\Delta_{u'}\Delta_{s'} \mathcal{M}(s',u')}{(s'-s)(u'-u)}\\\label{eq:MandelstamRepresentation}
    &+\frac{1}{\pi^{2}}\int_{c_{1}}^{\infty}du' \int_{c_3(u')}^{\infty}dt'\frac{\Delta_{t'}\Delta_{s'} \mathcal{M}(t',u')}{(u'-u)(t'-t)} +\displaystyle\sum_{i}\frac{R_{i}^{s}(t)}{s-x^{s}_{i}}\;.
\end{align}
Some terms were simplified by noting that $s'-\overline{s} =s'-s+u'-u =u'-\overline{u}$. 

The $\Delta_{i}\Delta_{j} \mathcal{M}$ are called double spectral functions and can be obtained from the optical theorem with help from the Schwarz reflection principle. It states that if a function $f(z)$ of a complex variable $z$ is real along certain finite segment $\Gamma$ of the real axis, then $f^{*}(z) = f(z^{*})$ in a domain $D$ of the $z$ complex plane that contains $\Gamma$ and in which $f(z)$ is analytic. This is always the case for any amplitude because  at a sufficiently low center--of--mass (CM) energy no intermediate state is allowed and thus the amplitude becomes real.

On the other hand, this theorem implies the existence of a discontinuity across the physical region of the real axis for each kinematic variable because there is always at least one possible intermediate state for any process: the initial one. Let $z$ be a kinematic variable of an amplitude $\mathcal{M}$ in a physical region, then:  
\begin{equation}
    2i\text{Im}\mathcal{M}(z+i\epsilon) = \mathcal{M}(z+i\epsilon) -\mathcal{M}^{*}(z+i\epsilon) =\mathcal{M}(z +i\epsilon) -\mathcal{M}(z -i\epsilon)=\Delta\mathcal{M}(z)\;.
\end{equation}
This result relates the Mandelstam representation with~\eqref{eq:opticalTheorem}. We can see that the constant $c_{1}$ of the dispersive integral is in fact the CM--frame energy of the lightest multiparticle intermediate state. One--particle intermediate states correspond to poles. Analogously, this helps us understand the meaning of $c_{3}$. For a physical value of $s$ and $t$, $\Delta\mathcal{M}$ is just $\text{Im}\mathcal{M}$ and is hence real. However, if we analytically continue $\Delta\mathcal{M}$ beyond the physically permissible boundaries of $t$, it may (and does) become complex, betraying the existence of a discontinuity. $c_{2}(s)$ is the point where that happens. Note that this is true even if $s$ takes a physically allowed value, as is the case in the initial fixed--$t$ single dispersive integral of $\mathcal{M}$. Note that contributions from heavy intermediate states are suppressed by: 1) the $1/s'$ integral kernel and 2) a reduced integration region.

In summary, from this theoretical framework it is possible to compute scattering amplitudes starting from~\eqref{eq:Sugawara} and then obtaining the spectral functions from experimental data for subprocesses via the optical theorem and the Schwarz reflection principle\footnote{This sentence implicitly claims that the analytic properties of an amplitude can be obtained entirely from its dynamics, that is, from the intermediate states it allows. This claim is known as the ``Mandelstam hypothesis'' and it is related to the causality requirement of the theory.}.

\subsubsection{Tensor decomposition of \texorpdfstring{$\Pi^{\mu_{1}\mu_{2}\mu_{3}\mu_{4}}$}{Lg}}\label{subsubsec:tensorDecompositionPi}

Dispersive integrals are suitable for scalar functions and $\Pi^{\mu_{1}\mu_{2}\mu_{3}\mu_{4}}$  is clearly not one. Hence, it is necessary to decompose it, like we did for the electromagnetic vertex, into form factors, which can then be dispersively represented.  Nevertheless, the Mandelstam hypothesis cannot be considered to apply in general for these scalar coefficients. The key point behind this is that the tensor structures of the decomposition may have kinematic singularities and/or zeroes, which have to be cancelled by corresponding terms in the form factors, because the amplitude does not have such terms according to the Mandelstam hypothesis. In such cases, zeroes (singularities) change the asymptotic (analytic) behaviour of the coefficients and this has an impact on the dispersion relation in the form of subtraction constants in~\eqref{eq:Sugawara} and spurious residues. Such input has to be determined experimentally and its presence in~\eqref{eq:Sugawara} further hinders the computations\footnote{It may even introduce ambiguities in the soft photon limit of $\text{Im}\Pi^{\mu_{1}\mu_{2}\mu_{3}\mu_{4}}$ and its derivatives. See~\cite{Colangelo2014}.}. In order to avoid these issues it is necessary to find a tensor decomposition of the amplitude free of kinematic singularities and/or zeroes, which is called a Bardeen--Tarrach--Tung (BTT) decomposition and was found recently~\cite{Colangelo2015} for $\Pi^{\mu_{1}\mu_{2}\mu_{3}\mu_{4}}$. Let us review the main steps followed in~\cite{Colangelo2015}. 

The only covariant objects which we can work with are the metric and the momenta of the four photons. There are 138 possible combinations of said objects. The most general structure is therefore:
\begin{equation}\label{eq:generalDecomposition}
\begin{split}
    \Pi^{\mu_{1}\mu_{2}\mu_{3}\mu_{4}} &=g^{\mu_{1}\mu_{2}}g^{\mu_{3}\mu_{4}}\Pi^{1} +g^{\mu_{1}\mu_{3}}g^{\mu_{2}\mu_{4}}\Pi^{2} +g^{\mu_{1}\mu_{4}}g^{\mu_{3}\mu_{2}}\Pi^{3}\\
    &+\sum_{k,l}g^{\mu_{1}\mu_{2}}q_{k}^{\mu_{3}}q_{l}^{\mu_{4}}\Pi^{4}_{kl} +\sum_{j,l}g^{\mu_{1}\mu_{3}}q_{j}^{\mu_{2}}q_{l}^{\mu_{4}}\Pi^{5}_{jl} +\sum_{j,k}g^{\mu_{1}\mu_{4}}q_{k}^{\mu_{3}}q_{j}^{\mu_{2}}\Pi^{6}_{jk}\\
    &+\sum_{i,l}g^{\mu_{3}\mu_{2}}q_{i}^{\mu_{1}}q_{l}^{\mu_{4}}\Pi^{7}_{il} +\sum_{i,k}g^{\mu_{4}\mu_{2}}q_{k}^{\mu_{3}}q_{i}^{\mu_{1}}\Pi^{8}_{ik} +\sum_{i,j}g^{\mu_{3}\mu_{4}}q_{i}^{\mu_{1}}q_{j}^{\mu_{2}}\Pi^{9}_{ij}\\
    &+\sum_{i,j,k,l}q_{i}^{\mu_{1}}q_{j}^{\mu_{2}}q_{k}^{\mu_{3}}q_{l}^{\mu_{4}}\Pi^{10}_{ijkl}\;,
\end{split}
\end{equation}
where the indices are $i\in\{2,3,4\}$, $j\in\{1,3,4\}$, $k\in\{1,2,4\}$ and $l\in\{1,2,3\}$\footnote{Each index may be any three--element subset of the four photons momenta. This choice in particular is due to~\cite{KarplusNeuman1950} and is useful to determine constraints from crossing symmetry.}. There are no kinematic singularities in the scalar coefficients untill this point. However there are kinematic zeroes coming from two constraints that we have not yet explicitly accounted for: gauge invariance and crossing symmetry.

Gauge invariance may be explicitly imposed by projecting each one of the Lorentz indices of the amplitude onto the orthogonal space of the associated virtual photon momentum. To this end the following projectors may be used~\cite{BardeenTung}:
\begin{equation}\label{eq:gaugeProjectors}
    I_{12}^{\mu\nu} = g^{\mu\nu} -\frac{q_{1}^{\mu}q_{2}^{\nu}}{q_{1}\cdot q_{2}}, \hspace{1cm}
    I_{34}^{\mu\nu} = g^{\mu\nu} -\frac{q_{3}^{\mu}q_{4}^{\nu}}{q_{3}\cdot q_{4}}\;,
\end{equation}
\begin{align}\label{eq:gaugeInvariantAmplitude}
    \Pi^{\mu_{1}\mu_{2}\mu_{3}\mu_{4}} &= I_{12}^{\mu_{1}'\mu_{1}}I_{12}^{\mu_{2}\mu'_{2}}I_{34}^{\mu_{3}'\mu_{3}}I_{34}^{\mu_{4}\mu'_{4}} \Pi_{\mu'_{1}\mu'_{2}\mu'_{3}\mu'_{4}}\;,
\end{align}
and only 43 tensor structures remain, which means that we have taken into account 95 constraints coming from gauge invariance.

Of course these are not the only projectors suitable for the job. We could have given each momentum its own projector. For example: 
\begin{align}\nonumber
    g^{\mu_{1}\mu_{2}} -\frac{q_{1}^{\mu_{1}}q_{1}^{\mu_{2}}}{q_{1}^{2}} \hspace{1cm} \text{or} \hspace{1cm} g^{\mu_{1}\mu_{2}} -\frac{q_{2}^{\mu_{1}}q_{2}^{\mu_{2}}}{q_{2}^{2}}\;.
\end{align}
However, by applying any transverse projectors to $\Pi^{\mu_{1}\mu_{2}\mu_{3}\mu_{4}}$ we are introducing kinematic singularities of the type $1/q_{i}\cdot q_{j}$. Thus, the less projectors we introduce and the simpler they are, the less types of kinematic singularities will be introduced. 

In our case, two appearances of $I_{12}$ and two of $I_{34}$ in the tensor structures lead to poles associated to all the possible combinations of $q_{1}\cdot q_{2}$ and $q_{3}\cdot q_{4}$ with up to two repetitions of each. This can be solved by building linear combinations of these singular tensor structures such that the poles cancel. The precise procedure proposed by~\cite{BardeenTung} deals with the poles by decreasing singularity order. In the first step poles of the form $(q_{1}\cdot q_{2})^2(q_{3}\cdot q_{4})^2$ are eliminated, first by linear combinations and, once this is not possible, by multiplying them by $q_{1}\cdot q_{2}$ or $q_{3}\cdot q_{4}$. The next step is to deal with the single--double poles in the same way, that is, $(q_{1}\cdot q_{2})(q_{3}\cdot q_{4})^{2}$ and $(q_{1}\cdot q_{2})^{2}(q_{3}\cdot q_{4})$. Single--single poles then come and so forth until there are no kinematic singularities left. The decomposition obtained after this procedure is completed is not yet suitable for a dispersive representation because the tensor ``basis'' found is actually not linearly independent in $q_{1}\cdot q_{2}=0$ or $q_{3}\cdot q_{4}=0$ nor it actually spans the complete space of possible gauge invariant tensors for the HLbL amplitude in those cases. This is due to the existence of 11 linear combinations of the basis elements that are proportional to $q_{1}\cdot q_{2}$ and/or $q_{3}\cdot q_{4}$ and some new tensor structure, thus introducing the linear dependence and span issues just mentioned.\footnote{This phenomenon was first described by Tarrach~\cite{Tarrach} for the BTT decomposition of the $\gamma\gamma\rightarrow\pi\pi$ process.} These new 11 structures have to be added to the previous 43 to build a set that spans all relevant gauge--invariant tensors even at $q_{1}\cdot q_{2}=0$ and $q_{3}\cdot q_{4}=0$~\cite{Colangelo2015}: 
\begin{equation}\label{eq:HLbLdecomposition}
    \Pi^{\mu_1\mu_2\mu_3\mu_4} =\sum_i \hat{T}^{\mu_1\mu_2\mu_3\mu_4}_i\hat{\Pi}_i \, .
\end{equation}
Since the set is linearly dependent, the definition of the scalars $\hat{\Pi}_i$ is obviously redundant.

It is not surprising that the 43--element basis does not work as expected in the singular points of the projectors~\eqref{eq:gaugeProjectors}, but where do the new tensors come from? To shed light on that issue let us consider, for example, $\Pi^{7}_{21}$, which is one of the form factors that disappears after using the gauge--invariance projectors. That means that it has been replaced by a sum of other form factors using a constraint equation. It can be explicitly obtained by studying the coefficient of $g^{\mu_{3}\mu_{2}}q_{1}^{\mu_{4}}$ in the equation $q_{1\mu'_1}\Pi^{\mu'_1\mu_2\mu_3\mu_4}=0$, which yields the constraint:     
\begin{equation}\label{eq:constraintExample}
    0= \Pi^{3} +(q_{1}\cdot q_{2})\Pi^{7}_{21} +(q_{1}\cdot q_{3})\Pi^{7}_{31} +(q_{1}\cdot q_{4})\Pi^{7}_{41}\;.
\end{equation}
It is clear that $\Pi^{7}_{21}$ cannot be simplified using this constraint when $q_1\cdot q_2=0$, thus the simplified tensor structures obtained using this constraint no longer span the amplitude fully. Furthermore, the remaining three form factors in~\eqref{eq:constraintExample} cease to be linearly independent when $q_1\cdot q_2=0$, thus hinting at the other issue we found. One might think that this is just an unfortunate result due to a poor solution of the constraint system, i.e. bad choice of projectors, and that we should look for one that solves for $\Pi^3$ instead of $\Pi_{21}^7$, for example. Unfortunately, this is not possible: it is not hard to see that for all four gauge invariance constraint equations only $\Pi^1$, $\Pi^2$ and $\Pi^3$ can be replaced without compromising the kinematics of the problem and there are 95 different constraint equations, so there will always be at least 92 replacements that will have problematic limits.
    
Finally, it is worth noting that, thanks to the choice of the $i,j,k,l$ in~\eqref{eq:generalDecomposition}, the 54 tensor ``basis''\footnote{It is a basis in the sense that it spans the tensor structures of the amplitude, but it is not linearly independent.} is closed under crossing. In fact, only 7 of the 54 tensors are actually independent in terms of crossing transformations\footnote{There are some elements that are actually crossing antisymmetric, but the corresponding kinematic zero does not affect the computation of $a_{\mu}$~\cite{Colangelo2015}.}.

\subsection{Master formula for the HLbL contribution to \texorpdfstring{$a_{\mu}$}{Lg}}\label{sec:MasterFormula}

In this section we obtain $a_{\mu}^{\text{HLbL}}$ from the scalar coefficients of~\eqref{eq:HLbLdecomposition}, which is key to connect it to experimental data in the low energy regime via dispersion relations.

Since the tensors $\hat{T}_i^{\mu_1\mu_2\mu_3\mu_4}$ from~\eqref{eq:HLbLdecomposition} have all the kinematic constrains and symmetries incorporated, they must vanish in the soft photon limit like $\Pi^{\mu_1\mu_2\mu_3\mu_4}$ does, as argued previously. Besides, it is possible to find $\hat{T}_{i}^{\mu_{1}\mu_{2}\mu_{3}\mu_{4}}$ such that in the limit $q_4\rightarrow0$ the derivative of 35 of this tensors vanish,  which leads us from~\eqref{eq:aMuHLbLIntermediate} to:
\begin{align}\nonumber
    a_{\mu}^{\text{HLbL}} &=e^{6}\int \frac{d^{4}q_{1}}{(2\pi)^{4}}\int \frac{d^{4}q_{2}}{(2\pi)^{4}}\frac{1}{q_{1}^{2}}\frac{1}{q_{2}^{2}}\frac{1}{(q_{1}+q_{2})^{2}}\frac{1}{(p+q_{1})^{2}-m^{2}}\frac{1}{(p-q_{2})^{2}-m^{2}}\,\sum_{i}^{19}\hat{T}_{i}\hat{\Pi}_{i}\;,\\\nonumber
    \hat{T}_{i} &\equiv \frac{1}{48m} \Tr\Big\{(\slashed{p}+m)[\gamma_{\nu_{4}},\gamma_{\mu_{4}}] (\slashed{p}+m)\gamma_{\mu_{1}}(\slashed{p}+\slashed{q}_{1}+m)\gamma_{\mu_{2}}(\slashed{p}-\slashed{q}_{2}+m)\gamma_{\mu_{2}}\Big\}\frac{\partial}{\partial q_{4\nu
    _{4}}}\hat{T}_{i}^{\mu_{1}\mu_{2}\mu_{3}\mu_{4}}\Big|_{q_{4}\rightarrow0}\;.
\end{align}
The objects $\hat{T}_{i}$ act as kernels for the two loop integral. Their number can be further reduced to 12 by harnessing the symmetry of the integral and some of the kernels under the $q_{1}\leftrightarrow-q_{2}$ exchange, which implies that some pairs of kernels actually give the same result and can be absorbed into one.

At this point $a_{\mu}^{\text{HLbL}}$ seems to depend on $p$, but we know from momentum conservation that $F_2=F_2(q_4^2)$ and hence $a_{\mu}^{\text{HLbL}}$ is just a number, not a function of any momentum. We can remove the spurious dependence on $p$ through angular averages. Let us start by performing a Wick rotation, in which we essentially render all four--vectors' time components imaginary. This causes all scalar products to acquire a minus sign and renders the corresponding metric euclidean, transforming the two loops accordingly. This has non--trivial consequences, because it amounts to a rotation of the real axis of time--component integration into the imaginary one. In case that the region swept by such rotation contains singularities, the resulting contributions have to be taken into account accordingly. However, for $a_{\mu}$ there no such issues~\cite{Colangelo2015} and the Wick rotation may be performed without problems. The Wick--rotated version of the momenta is represented by $Q_{1}$, $Q_{2}$ and $P$\footnote{This capital ``P'' still represents the initial momentum of the muon. It should not be confused with $p+p'$, as was represented in previous sections.}, $\hat{P}$ and $\hat{Q}_{i}$ represent their unit vectors, while $|P|$ and $|Q_{i}|$ represent their norm. Then it is possible to remove the spurious dependence on $P$ by averaging $a_\mu^{\text{HLbL}}$ over all possible orientations of $P$:
\begin{equation}
    a_\mu^{\text{HLbL}} =\int\frac{d^4\Omega(P)}{2\pi^2} a_\mu^{\text{HLbL}}
\end{equation}
Wick rotated propagators strongly resemble the generating function of Gegenbauer polynomials~\cite{BookBateman}, which allows to represent each one of the former as a linear combination of the latter. Finally, the integrals are found using the polynomials' orthogonality~\cite{Bjorken1964_GegenbauerPolynomials}. After performing the corresponding average, it is possible to perform five of the six four--dimensional angular integrals on $Q_{1}$ and $Q_{2}$ and the final result is:
\begin{align}\label{eq:masterFormula}
     a_{\mu}^{\text{HLbL}} &=\frac{2\alpha^{3}}{3\pi^{2}}\int_{0}^{\infty} dQ_{1}\int_{0}^{\infty} dQ_{2}\int_{-1}^{1} d\tau \sqrt{1-\tau^{2}}|Q_{1}|^{3}|Q_{2}|^{3}\times\sum_{i}^{12}T_{i}\overline{\Pi}_{i}\;. 
\end{align}
There are three aspects of this last step that are worth noting: 
\begin{itemize}
    \item The integral over $Q_{2}$ in spherical coordinates is considered in the first place. It is possible to take any four--momentum as a reference for the angular integral; it does not matter because the integrals will go over all the possible values anyway. We take $Q_{1}$ as a reference.
    
    \item The integrand is only dependent on one angle (in $\tau=\hat{Q}_{1}\cdot \hat{Q}_{2}$) and it is therefore convenient to assign $\tau$ as one of the three euclidean angles over which the angular integrals of the four momentum $Q_{2}$ is performed. It is relevant which of these three angle we are referring to because it will determine what the Jacobian will look like. In the master formula there is a term $1-\tau^{2}$, a sine squared, which means $\tau$ does not represent neither polar nor the azimuthal angle of the three dimensional sphere embedded in the four dimensional space. Thus, the angular integral on $Q_{2}$ yields:     
    \begin{align}
        \int d\tau \sqrt{1-\tau^{2}}\int d\theta d\phi\sin{\theta} =4\pi\int d\tau \sqrt{1-\tau^{2}}\;,
    \end{align}    
    where $\theta$ and $\phi$ represent the three--dimensional polar and azimuthal angles of the four--dimensional $Q_{2}$ space.
    \item Once the angular integrals on $Q_{2}$ have been performed there is no dependence on $\tau$ or another angle left on the integrand. This means that we can perform the four--dimensional solid angle integral on $Q_{1}$ which yields $2\pi^{2}$.
\end{itemize}

\section{Review of low energy contributions to \texorpdfstring{$a_{\mu}^{\text{HLbL}}$}{Lg}}

Up to this point we have presented an approach to compute low--energy contributions to $a_{\mu}^{\text{HLbL}}$ based on a dispersive description of the HLbL scattering amplitude. Now let us review the results obtained when such approach is put to use.

As we stated in section~\ref{sec:mandelstamRepresentation}, the dispersive calculation of an amplitude offers a way to establish a hierarchy of contributions for the intermediate states that enter the computation via the unitarity relation of~\eqref{eq:opticalTheorem}. Heavier intermediate states induce cuts that appear further to the right of the dispersive integration region compared to their lighter counterparts and therefore these former's contributions are expected to be smaller. The $1/(s-s')$, $1/(t-t')$ or $1/(u-u')$ weights further suppress such contributions. In the context of HLbL low energy scattering, the most relevant intermediate states are expected to be the lightest one-- or two--hadron intermediate states.

\subsection{One--particle intermediate states contribution to \texorpdfstring{$a_{\mu}$}{Lg}}

Baryons are not suitable single--particle intermediate states of the HLbL scattering amplitude due to baryonic number conservation. Thus, they can only contribute from two--particle intermediate states onwards. However, since the proton, the lightest baryon, has a mass of $\sim938.3$ MeV~\cite{protonMass}, its corresponding two--particle state has a mass of about $1.9$ GeV, which is already in the perturbative regime of QCD. This means that baryon contributions can and should be accounted for perturbatively.

The lightest mesons with masses up to $\sim 1~\text{GeV}$ are $\pi^{0}$, $\pi^{\pm}$, $K^{\pm}$, $K^{0}$, $\eta$ and $\eta'$\footnote{$\eta$ and $\eta'$ are actually not stable in QCD and therefore they do not appear in the unitarity relation~\eqref{eq:opticalTheorem} as a one-particle intermediate state, however, its decay width is small enough to be considered as a one.}. Charge conservation obviously prohibits the appearance of charged pions and kaons as one--particle intermediate states of HLbL. Furthermore, a single neutral kaon cannot be produced since strong interactions conserve strangeness. The contributions from the $\pi^{0}$, $\eta$ and $\eta'$ intermediate states to $a_{\mu}^{\text{HLbL}}$ are of the form:
\begin{align}\label{eq:aMuPole}
    a_{\mu}^{P-\text{Pole}} &=\frac{2\alpha^{3}}{3\pi^{2}}\int_{0}^{\infty} dQ_{1}\int_{0}^{\infty} dQ_{2}\int_{-1}^{1} d\tau \sqrt{1-\tau^{2}}|Q_{1}|^{3}|Q_{2}|^{3}\Big(T_{1}\overline{\Pi}_{1}^{P} +T_{2}\overline{\Pi}_{2}^{P}\Big)\;,\\\nonumber
    \overline{\Pi}_{1}^{P} &=-\frac{\mathcal{F}_{P\gamma^{*}\gamma^{*}}(-Q_{1}^{2},-Q_{2}^{2})\mathcal{F}_{P\gamma^{*}\gamma^{*}}(-Q_{3}^{2},0)}{Q_{3}^{2}+M_{P}^{2}}\;,\\\nonumber
    \overline{\Pi}_{2}^{P} &=-\frac{\mathcal{F}_{P\gamma^{*}\gamma^{*}}(-Q_{1}^{2},-Q_{3}^{2})\mathcal{F}_{P\gamma^{*}\gamma^{*}}(-Q_{2}^{2},0)}{Q_{2}^{2}+M_{P}^{2}}\;,
\end{align}
where $P$ states for the corresponding meson in the one--particle intermediate state, $M_{P}$ is the meson's mass and $\mathcal{F}_{P\gamma^{*}\gamma^{*}}$ represents its doubly virtual transition form factor, which is defined by:
\begin{align}\label{eq:TFF}
    i\int d^{4}x\;e^{-iq_{1}x}\langle0|T\{J_{\mu}(x)J_{\nu}(0)\}|P(q_{1}+q_{2})\rangle &=\epsilon_{\mu\nu\lambda\rho}q_{1}^{\lambda}q_{2}^{\rho}\mathcal{F}_{P\gamma^{*}\gamma^{*}}(q_{1}^{2},q_{2}^{2})\;.
\end{align}
The main hurdle for the computation of $a_{\mu}^{P-\text{pole}}$ is to obtain reliable values for the transition form factor, specially for the double virtual sector, for which there is usually little experimental data available~\cite{CA_pion}. Furthermore, data available below $1$ GeV is particularly scarce, which means that a fit extrapolation is required in order to obtain the transition form factors at that range. This is specially troublesome because the main contribution to~\eqref{eq:aMuPole} actually comes from the low energy region. For the neutral pion, however, the negative effects of these issues have been addressed by a dispersive reconstruction of the transition form factor~\cite{PionPole_1,dispersiveTFF_pion,HLbL_LO4} starting from~\eqref{eq:TFF}. When $\mathcal{F}_{\pi^{0}\gamma^{*}\gamma^{*}}$ is computed in that fashion, the corresponding contribution to $a_{\mu}^{\text{HLbL}}$ is: 
\begin{align}
    a_{\mu}^{\pi^{0}-\text{pole}}(\text{dispersive}) &=63.0^{+2.7}_{-2.1}\times10^{-11}\;.
\end{align}
There are of course alternate approaches for evaluating $\mathcal{F}_{\pi^{0}\gamma^{*}\gamma^{*}}$ such as the Canterbury Approximants (CA)~\cite{bookCA,CA_pion,CA_pion2}, which reproduces the low energy behaviour of the transition form factor approximating it via rational functions of polynomials in $Q_{1}^{2}$ and $Q_{2}^{2}\,$\footnote{For the singly--virtual transition form factor the Padé Approximants approach is used, which are essentially the univariate version of the CA.}. The CA computation of the pion transition form factor yields the result~\cite{WP} $a_{\mu}^{\pi^{0}-\text{pole}}(\text{CA}) =63.2(2.7)\times10^{-11}$, which is very much compatible with the dispersive one. 

For $\eta$ and $\eta'$ a dispersive computation of the corresponding form factor in its doubly virtual region is not yet available, although the framework has been established for the singly--virtual one~\cite{dispersiveTFF_eta}. Furthermore, progress has been made in the doubly--virtual case as well~\cite{dispersiveTFF_eta2,dispersiveTFF_eta3}. In any case, until a full dispersive computation of the transition form factor is obtained the CA approach offers an option whose reliability is supported by the excellent compatibility with the dispersive results in the case of $\pi^0$. For the $\eta$ and $\eta'$ pole contributions the CA approach concludes~\cite{CA_pion}: 
\begin{equation}
    a_{\mu}^{\eta-\text{pole}}(\text{CA}) = 16.3(1.4)\times10^{-11}\;, \hspace{1cm}   a_{\mu}^{\eta'-\text{pole}}(\text{CA}) =14.5(1.9)\times10^{-11}\; .
\end{equation}
It is clear that the neutral pion contribution is by far the largest from the one--particle intermediate states. Since the $\pi^{0}$ is lighter than the $\eta$ and it in turn is lighter than the $\eta'$, these results for the contribution of each of these intermediate states agree with the expected hierarchy mentioned at the beginning of this section. In summary, the one--particle intermediate states contribution to $a_{\mu}$ is~\cite{WP}: 
\begin{align}
    a_{\mu}^{\pi^{0}+\eta+\eta'} &=93.8^{+4.0}_{-3.6}\times10^{-11}\;.
\end{align}
\subsection{Two--particle intermediate states contribution to \texorpdfstring{$a_{\mu}$}{Lg}}

The lightest two--meson intermediate states with mass up to $\sim1$ GeV from lightest to heaviest are $\pi^{0}\pi^{0}$, $\pi^{+}\pi^{-}$, $\eta\pi^{0}$, $K^{+}K^{-}$ and $K^{0}\overline{K}^{0}$. Note that, again due to strangeness conservation in QCD, no intermediate states with a single kaon is allowed.

The dispersive frame work for the computation of the two--pions contribution has been studied in~\cite{Colangelo2015}. In there, by performing the unitarity cut across the s--channel, this amplitude is split into two $\gamma\gamma\rightarrow\pi\pi$; one with two virtual photons and one with one virtual and one real soft photon. Then, this two amplitudes are split again by performing the unitarity cut across the t--channel, which is the application of a second dispersion relation that leads to the Mandelstam representation that was described in section~\ref{sec:mandelstamRepresentation}. Although in this second cut there is the possibility for multi--pion intermediate states to appear, the biggest contribution is expected to come from the pion pole (lightest intermediate state) as well. This introduces the pion--box topology as an intermediate state.\footnote{It is worth noting that due to angular momentum conservation and Bose symmetry, a photon does not couple to two neutral pions and therefore the $\gamma^{*}\gamma^{*}\rightarrow\pi^{0}\pi^{0}$ does not allow a neutral pion intermediate state across the t--channel.} Since the four pieces in which the original HLbL is split are essentially pion electromagnetic vertices, it is reasonable to expect the pion--box topology contribution to $a_{\mu}$ to be proportional to four pion electromagnetic vector form factors $\mathcal{F}^{V}_{\pi}(q_{i}^{2})$. Furthermore in~\cite{Colangelo2015} this relation is taken further and via explicit computation of the double spectral functions associated to the pion--box contribution it was shown that the pion--box contribution to $\Pi^{\mu\nu\lambda\rho}$ is in fact equal to the scalar QED pion loop amplitude multiplied by four pion vector form factors due to each of the four off-shell photons. The mathematical expression of these statements is: 
\begin{align}\nonumber
    \Pi_{i}^{\pi-\text{box}} &=\mathcal{F}^{V}_{\pi}(q_{1}^{2})\mathcal{F}^{V}_{\pi}(q_{2}^{2})\mathcal{F}^{V}_{\pi}(q_{3}^{2})\mathcal{F}^{V}_{\pi}(q_{4}^{2})\times\frac{1}{\pi^{2}}\Bigg(\int_{c_{1}}^{\infty}ds' \int_{c_{2}(s')}^{\infty}dt'\frac{\Delta_{t'}\Delta_{s'} \mathcal{M}(s',t')}{(s'-s)(t'-t)}\\
    &\hspace{-1cm}+\int_{c_{1}}^{\infty}ds' \int_{c_{2}(s')}^{\infty}du'\frac{\Delta_{u'}\Delta_{s'} \mathcal{M}(s',u')}{(s'-s)(u'-u)} +\int_{c_{1}}^{\infty}du' \int_{c_{2}(u')}^{\infty}dt'\frac{\Delta_{t'}\Delta_{s'} \mathcal{M}(t',u')}{(u'-u)(t'-t)}\Bigg)\;,
\end{align}
where $\mathcal{M}$ is the scalar QED one loop amplitude with pions. The corresponding double spectral functions can of course be obtained used Cutkosky's rules to perform unitarity cuts to loops, however, it is much more efficient to compute the light by light scattering amplitude directly in perturbative scalar QED and then insert into the master formula~\eqref{eq:masterFormula} with the appropriate vector form factors. This allows for the contribution of the pion box to be written in terms of compact and well known Feynman parameters integrals. In conclusion, the pion box contribution to $a_{\mu}^{\text{HLbL}}$ reads:
\begin{align}\nonumber
     a_{\mu}^{\pi-\text{box}} &=\frac{2\alpha^{3}}{3\pi^{2}}\int_{0}^{\infty} dQ_{1}\int_{0}^{\infty} dQ_{2}\int_{-1}^{1} d\tau \sqrt{1-\tau^{2}}|Q_{1}|^{3}|Q_{2}|^{3}\mathcal{F}^{V}_{\pi}(-Q_{1}^{2})\mathcal{F}^{V}_{\pi}(-Q_{2}^{2})\mathcal{F}^{V}_{\pi}(-Q_{3}^{2})\sum_{i}^{12}T_{i}\overline{\Pi}_{i}^{sQED}\\
     &=-15.9(2)\times10^{-11}\;,
\end{align}
where $\overline{\Pi}_{i}^{sQED}$'s are the scalar coefficients of the BTT decomposition of the scalar QED one--loop light--by--light scattering amplitude, which obviously is a subset of the HLbL BTT structures. In the previous equation only three vector form factors appear, instead of four, because one of the photons of this HLbL process is actually on--shell.

Regarding intermediate states heavier than one pion in the t--channel cut of the $\gamma\gamma\rightarrow\pi\pi$ subprocess, they can be classified into two categories: one with one pion--pole contribution in one subamplitude and a multiparticle cut in the other, and one where both subprocesses contain multiparticle cuts. The computation of this contributions can be performed through a partial wave expansion of the amplitude. One of the advantages of this approach is that the unitarity relation~\eqref{eq:opticalTheorem} is diagonal in the helicity partial wave basis, that is, helicity partial waves of HLbL $h^{J}_{\lambda_{1}\lambda_{2}\lambda_{3}\lambda_{4}}$ are only connected by unitarity with the $\gamma^{*}\gamma^{*}\rightarrow\pi\pi$ partial waves $h^{J}_{\lambda_{i}\lambda_{j}}$ with the same total angular momentum $J$ and photon helicities $\lambda_{i}$:
\begin{equation}
\begin{split}
    \text{Im}^{\pi\pi}\;h^{J}_{\lambda_{1},\lambda_{2},\lambda_{3},\lambda_{4}} =\frac{\sigma_{\pi}}{16\pi S}h^{J}_{\lambda_{1}\lambda_{2}}h^{*J}_{\lambda_{3}\lambda_{4}}\;,&\hspace{0.5cm}    H_{\lambda_{1}\lambda_{2}\lambda_{3}\lambda_{4}} =\sum_{J}(2J+1)d_{m_{1}m_{2}}^{J}(z)h^{J}_{\lambda_{1}\lambda_{2}\lambda_{3}\lambda_{4}}\;,\\
    H_{\lambda_{i}\lambda_{j}} =\sum_{J}(2J+&1)d_{m0}^{J}(z)h^{J}_{\lambda_{i}\lambda_{j}}\;,
\end{split}
\end{equation}
\begin{equation*}
\begin{split}
    m=|\lambda_{i}-\lambda_{j}|\;,& \hspace{1cm} m_{1} = |\lambda_{1}-\lambda_{2}|\;,\\\nonumber
    m_{2} = |\lambda_{3}-\lambda_{4}|\;,&\hspace{1cm}
    \sigma_{\pi} =\sqrt{1-\frac{4M^{2}_{\pi}}{s}}\;,
\end{split}
\end{equation*}
where $H_{\lambda_{1}\lambda_{2}\lambda_{3}\lambda_{4}}$ and $H_{\lambda_{i}\lambda_{j}}$ are helicity amplitudes for the HLbL and the $\gamma^{*}\gamma^{*}\rightarrow\pi\pi$ processes, respectively, which are obtained by contracting polarization vectors of a given helicity with the corresponding process amplitude. Furthermore, $d_{m_{1}m_{2}}^{J}$ are Wigner's matrices, $z$ is the scattering angle and $M_{\pi}$ represents the pion's mass. Helicity partial waves are, however, not suitable for a double Mandelstam representation, because they contain kinematic singularities (see section 3 of chapter 7 in~\cite{BookSpearman}). To retain the advantages of the partial wave expansion, but solve the kinematic singularities issues it is necessary to relate them to the BTT basis described in section~\ref{sec:mandelstamRepresentation}. This is done by first inverting the linear relation that is obtained between helicity amplitudes and BTT scalar functions via its definition from contraction with photon polarization vectors. Then helicity partial waves are projected out of the corresponding helicity amplitude by using orthogonality properties of Wigner's matrices. As we mentioned in section~\ref{sec:mandelstamRepresentation}~\cite{Mandelstam1958}, the BTT decomposition of $\Pi^{\mu_{1}\mu_{2}\mu_{3}\mu_{4}}$ actually trades linear independence in order to span the whole amplitude, therefore introducing redundancies. This redundancies are constrained in the soft limit ($q_{4}\rightarrow0$) by using dispersive sum rules derived from the asymptotic properties of the scalar coefficients~\cite{Colangelo2017}. Finally, the S--wave contribution to the two pion HLbL intermediate states with one pion as and intermediate state in one $\gamma^{*}\gamma^{*}\rightarrow\pi\pi$ subprocess and a multiparticle cut in the other is~\cite{Colangelo2017}: 
\begin{align}\label{eq:twopionSwave}
    a_{\mu}^{\pi\pi,\;\pi-\text{pole LHC}} =-8(1)\times10^{-11}\;,
\end{align}
where ``LHC'' stands for ``left hand cut''. Contributions from higher angular momentum partial waves is difficult due to the lack of experimental input for the doubly virtual subprocess $\gamma^{*}\gamma^{*}\rightarrow\pi\pi$. Furthermore, heavier singularities in the left hand cut have to be taken into account for the computation of higher angular momentum partial waves, but in such case some assumptions that simplified the computation of~\eqref{eq:twopionSwave} are no longer valid. Therefore, the dispersive framework developed in~\cite{Colangelo2017} would have to be studied from a more general perspective. 

In~\cite{WP} it is proposed that the comparison between the size of the $\gamma\gamma\rightarrow\pi\pi$ cross section and $\gamma\gamma\rightarrow MM$ cross section for some given processes gives input to determine which two--particle intermediate state contribution $MM$ may be relevant beyond the $\pi\pi$ contribution. Following this approach, the relevant heavier intermediate states are $\pi^{0}\eta$ and $K^{+}K^{-}$, which are actually the next two--particle intermediate states in terms of mass. For the two kaon contribution, the same arguments of the pion box apply and therefore, via the scalar QED kaon loop and using vector meson dominance to obtain the electromagnetic form factor, the corresponding contribution has bee computed~\cite{WP}: 
\begin{align}
    a_{\mu}^{KK\text{--loop}} &=-0.5(1)\times10^{-11}\;.
\end{align}
For the $\pi^{0}\eta$ the real version of the process a dispersive framework has been developed~\cite{pionEta2} and the results performed well against data of the $\eta\rightarrow\pi^{0}\gamma\gamma$ crossed process. Work on the singly and doubly virtual processes is still under way~\cite{pionEta1}.

\subsection{Intermediate states with more than two particles and heavier than \texorpdfstring{$KK$}{Lg}}

The dispersive computation of contributions heavier than the ones we presented previously, that is, with masses between $1$ GeV and $2$ GeV, is more difficult due to the more complex unitarity diagrams than come into play. Perhaps the most straightforward upgrade is the contribution from D--\footnote{Each pion has isospin equal to one, therefore the possible values for total isospin are $|1-1|=0$ and $1+1=2$. This means that only even partial waves are allowed.} and higher waves from the two pion contribution, for which there is a pretty much developed framework~\cite{ColangeloSWaves}. Contributions from states heavier that two pions are expected to be small. For example, the $K^{+}K^{-}$ intermediate state has mass $\sim1$ GeV and its contribution to $a_{\mu}$ is already below $10^{-11}$. Consequently, heavier intermediate states contributions are expected to be non--negligible only if they are associated to resonant helicity partial waves. To estimate such contributions a narrow--width resonance approximation is followed in~\cite{HLbL_LO7} (updated in~\cite{HLbL_LO8}) and~\cite{HLbL_L010}, which essentially means that narrow resonances are treated as dispersive poles. In summary, the contribution considering the $f_{0}(980 \text{ MeV})$, $a_{0}(980 \text{ MeV})$, $f_{0}(1370 \text{ MeV})$, $a_{0}(1450 \text{ MeV})$, $f_{0}(1500 \text{ MeV})$, $f_{2}(1270 \text{ MeV})$, $f_{2}(1565 \text{ MeV})$, $a_{2}(1320 \text{ MeV})$ and $a_{2}(1700 \text{ MeV})$ \footnote{In~\cite{HLbL_LO7} and its updated version~\cite{HLbL_LO8} the first three (scalar) resonances and the last four (tensor) resonances are considered. In~\cite{HLbL_L010} considers all the scalar resonances mentioned.} is: 
\begin{align}
    a_{\mu}^{\text{scalars}+\text{tensor}} &=-1(3)\times10^{-11}\;,
\end{align}
which introduces a significant uncertainty due to the unknown error introduced by the narrow--width resonance approximation. In principle, such approximation may be tested in the lightest scalar and tensor resonances against the $\pi\pi$ and $KK$ contributions described previously. However, only a full tower of resonant poles satisfy the sum rules found in~\cite{Colangelo2017}, which is a necessary condition for a tensor--basis independent contribution.

Axial vector mesons (meson bound states with total angular momentum equal to 1, but even parity) have not been considered in the previous paragraphs, because\footnote{Note that the Landau-Yang theorem only forbids the coupling of two real photons to axial vector mesons, therefore in virtual HLbL it is allowed.} the lightest associated resonances are actually heavy compared with other states: $a_{1}(1260 \text{ MeV})$, $f_{1}(1285 \text{ MeV})$ and $f_{1}(1420 \text{ MeV})$. They contribute to $a_{\mu}$ as resonant poles and their respective transition form factors are obtained from experimental data using a non--relativistic quark model computation~\cite{AxialFormFactor1,AxialFormFactor2,AxialFormFactor3,AxialFormFactor4} which simplifies the tensor structure of the $\gamma^{*}\gamma^{*}\rightarrow A$ amplitude, as described in~\cite{HLbL_LO7}. The current agreed contribution (see~\cite{WP}) is obtained from the mean of three results~\cite{HLbL_LO7,HLbL_L010,AxialResult1} that consider the three lightest axial vector mesons:
\begin{align}
    a_{\mu}^{\text{axial vectors}} &=6(6)\times10^{-11}\;.
\end{align}

In summary, we have presented the basics of the computation of $a_{\mu}$ and in particular $a_{\mu}^{\text{HLbL}}$, which led to the~\eqref{eq:masterFormula} master formula, which is based on a BTT tensor decomposition of the HLbL amplitude with 54 elements such that its scalar coefficients are free of kinematic singularities and zeroes. The tensor ``basis'' is actually not linearly independent in the whole phase space due to 11 structures that need to be added by hand in order to span the amplitude at some kinematic points $q_{1}\cdot q_{2}=0$ or $q_{3}\cdot q_{4}=0$. Furthermore, in four space--time dimensions there are two additional linear relations among the 138 tensors of~\eqref{eq:generalDecomposition}~\cite{FourDimensionsRelations} which the 54 BTT structures inherite. Of course, the linear dependence of the set of structures introduces redundancies in the definition of the associated scalar coefficients. However, in~\cite{ColangeloSWaves} it was shown that these redundancies do not affect the observables, as expected, due to a series of dispersive sum rules obeyed by the form factors.

We also discussed contributions of intermediate states with masses up to $\sim 2$ GeV, which are expected to give the man contribution to $a_{\mu}^{\text{HLbL}}$. However, it is as well necessary to understand the behaviour of the HLbL tensor beyond that threshold. The high energy regime of the HLbL contribution has been already reached from the light intermediate states when the high energy part of the integral in~\eqref{eq:masterFormula} has been carried out for high virtuality values of the transition and vector form factors. Although these penetrations into the high energy regime are expected to take into account most of the contribution, in this regime heavier states with more complex topologies that are not taken into account dispersively can also contribute. It is those contributions and other uses for computations in the high energy regime of the QCD in the context of $a_{\mu}$ with which the next sections will deal.

\section{Operator product expansion of \texorpdfstring{$\Pi^{\mu_1\mu_2\mu_3\mu_4}$}{Lg}}\label{sec:OPEfirst}

Computations in the high energy regime of QCD play several important roles in the determination of $a_{\mu}$. First, for data--driven computations they provide the asymptotic behaviour of transition and electromagnetic form factors for the high energy ``tail'' of the integral in the master formula~\eqref{eq:masterFormula} and dispersive integrals. Examples of these short distance constraints (SDC) can be found in~\cite{BL1,BL2,TFFAsymptotic1,TFFAsymptotic2,TFFAsymptotic3,TFFAsymptotic4} and in~\cite{pionTFFAsymptotic1,pionTFFAsymptotic2} for the specific case of the pion. Second, SDC can be found for the HLbL amplitude itself, which are used to evaluate how much of this asymptotic behaviour is recovered by dispersive computations and thus assess the uncertainty produced by the missing high energy contributions. These come from the heavy intermediate states with topologically more complex unitarity diagrams that are ignored in the dispersive approach.

In this work we are interested in the SDC for the HLbL tensor. There are two high energy regimes for the HLbL process with one real soft photon: one where all three Euclidean virtualities are similarly high ($Q^2_{1}\sim Q^2_{2}\sim Q^2_{3}\gg \Lambda^2_{QCD}$), (we refer to the Euclidean virtualities $q_i^2=-Q_i^2$) and one where two are similarly high and much greater than the third Euclidean virtuality ($Q^2_{1}\sim Q^2_{2} \gg Q^2_{3}\sim \Lambda^2_{QCD}$ and crossed versions). It is worth to remind that we refer to the Euclidean virtualities, because the considered photons are far off-shell since when the space-time separation goes to zero the corresponding momenta $q_i$ become space-like. Each of these regimes impose asymptotic behaviour constraints on different subsets of BTT scalar coefficients and therefore they allow for the independent evaluation of the different sets of intermediate states. In this section we will focus on the $Q^2_{1}\sim Q^2_{2}\sim Q^2_{3}\gg \Lambda^2_{QCD}$ regime of the HLbL tensor, which we will study by performing an Operator Product Expansion (OPE) where the soft photon is introduced as a background electromagnetic field, as was done in~\cite{Bijnens2020}.

\subsection{OPE of \texorpdfstring{$\Pi^{\mu_{1}\mu_{2}\mu_{3}\mu_{4}}$}{Lg}}

A product of operators carrying very high momenta can be expressed in terms of a linear combination of local operators carrying zero momentum with singular coefficients that carry the momentum dependence of the original operator. The mathematical expression of this statement for the product of two operators $O_{1}$ and $O_{2}$ is: 
\begin{align}\label{eq:OPE}
    \int d^{4}xe^{-iqx} O_{1}(x)O_{2}(0) &=\sum_{n}C_{n}^{12}(q^{2})O_{n}(0)\;,
\end{align}
where $C_{n}^{12}$ is a c--number function that decreases for large $Q^{2}$ and is known as ``Wilson coefficient''. The operator basis $\{O_{n}\}$ of the expansion only admits elements with the same quantum numbers and symmetries of the original operator product. The generalization for higher numbers of operators with arbitrary tensor structures is straightforward. Such identity is called operator product expansion and was originally introduced by Kenneth Wilson~\cite{OPE1}. By engineering dimensional analysis it can be noticed that the Wilson coefficients vanish increasingly quicker for $O_{n}$ of higher mass dimensions, which implies that the OPE has an implicit and useful hierarchy of contributions in which simpler operators (with less derivatives and fields) give bigger contributions than more complex operators. Since the operators on both sides of~\eqref{eq:OPE} are renormalized at some scale $\mu$, a more careful study finds that dimensional analysis must be done considering the anomalous dimensions of the operators $O_{n}$, but for asymptotically free theories such as QCD, at high energy the simple power counting suffices. It is also worth noting that the relation in~\eqref{eq:OPE} is given in terms of operators, that is, it does not depend on specific matrix elements an therefore Wilson coefficients do not either. For further details see chapter 20 of volume II in~\cite{BookWeinberg} and chapter 18 in~\cite{BookPeskin}.

To perform the OPE in~\eqref{eq:OPE} it is first necessary to fix the maximum number of mass dimensions of the operators that are going to be considered and then the task is to find all compatible operators with the same quantum numbers and symmetries of the original operator product. Once the complete set of relevant operators $O_{n}$ is known, each Wilson coefficient is found by choosing an appropriate matrix element of the original operator product $O_{1}O_{2}$ and then transforming it into the corresponding element of the operator basis $O_{n}$. Such transformation is done by expanding the correlation function and leaving the elements that form $O_{n}$ uncontracted. The result of the contraction of the rest of the parts of the original product and the fields in the correlation function vertices constitutes the Wilson coefficient of the element $O_{n}$ for the operator basis.

From the definition of the OPE it is evident that it constitutes a very well suited framework for the evaluation of the HLbL tensor in the $Q_{1}^2\sim Q_{2}^2\sim Q_{3}^2\gg \Lambda_{QCD}^2$ regime. However, the limit $q_{4}\rightarrow 0$ does not allow to include the fourth current of~\eqref{eq:HLbLTensor} in the OPE. One could in principle start the construction of such OPE for the four currents of $\Pi_{\text{HLbL}}^{\mu_{1}\mu_{2}\mu_{3}\mu_{4}}$, but it does not work for our particular problem. For example, the Wilson coefficient associated to the identity operator is just the HLbL tensor in perturbative QCD, which involves an expansion in terms of the strong coupling constant and the usual large logarithms for increasing powers: $\alpha^{n}_{s}(\mu)\ln^{n}{\big\{Q_{4}^2/\mu^2\big\}}$, where $\mu$ represents the renormalization subtraction point in the $\overline{MS}$ scheme. In order for the logarithms not to blow up it is necessary to have $\mu\sim Q_{4}$, but in such case $\alpha_{s}$ would enter the non--perturbative domain of QCD and the expansion would be spoiled anyway. Wilson coefficients for higher dimensional operators also suffer from infrared singularities since they depend upon the $1/q_{4}^{2}$ propagator. Including the fourth current $J^{\mu_{4}}$ into the OPE even though its momentum is not large is not the only problem. Even if the OPE were performed only for the three currents with high momenta, there would still be matrix elements of the type $\langle0|O_{n}J^{\mu_{4}}(q_{4})|0\rangle$, which cannot be perturbatively computed in QCD. In general, these issues are different consequences of the fact that perturbative QCD is not the correct framework to describe the soft interaction that is required by $a_{\mu}$. It is therefore necessary to perform the OPE of the three high--momentum currents only and also take the fourth one into account properly. This can be done by letting the soft photon be introduced by an external electromagnetic field instead of a quark current and we will present such procedure in the next section of this review\footnote{This approach was first used in~\cite{OPEBackgroundViejo} in the context of the computation of the magnetic moment of nucleons, then it was used in~\cite{OPEBackgroundReciente} for the hadronic corrections to the electroweak contribution to $a_{\mu}$ and finally it was again picked up in~\cite{Bijnens2019,Bijnens2020,Bijnens2021} for the HLbL tensor. A pedagogical review of the framework is presented in~\cite{OPEBackgroundReview}.}.

\subsection{OPE of \texorpdfstring{$\Pi^{\mu_{1}\mu_{2}\mu_{3}\mu_{4}}$}{Lg} in an electromagnetic background field: A first look}

A new object suitable for the OPE described at the end of the previous part is: 
\begin{align}\nonumber
    \Pi^{\mu_{1}\mu_{2}\mu_{3}} &=\frac{1}{e}\int d^{4}x\int d^{4}y\;e^{-i(q_{1}x+q_{2}y)}\langle0|T\,J^{\mu_{1}}(x)J^{\mu_{2}}(y)J^{\mu_{3}}(0)|\gamma(q_{4})\rangle=-\epsilon_{\mu_{4}}(q_{4})\Pi^{\mu_{1}\mu_{2}\mu_{3}\mu_{4}}\;,
\end{align}
where the soft photon $q_4\rightarrow 0$ is included implicitly in the initial state. In addition, this time $J^{\mu}$ makes reference to the electromagnetic current of the three lightest quarks, namely, up, down and strange or $u$, $d$ and $s$ and thus: 
\begin{equation}
\begin{split}
    J^{\mu} &= \overline{\Psi}\hat{Q}\gamma^{\mu}\Psi \hspace{2cm} \hat{Q} = \text{diag}\Big(\frac{2}{3},-\frac{1}{3},-\frac{1}{3}\Big)\;,
\end{split}
\end{equation}
where $\hat{Q}$ is the charge matrix and now $\Psi$ is a vector of bispinors with quark flavor and color indices, which are summed upon and suppressed in the current.

Since $\langle0|...|\gamma(q_{4})\rangle$ is an on--shell matrix element only gauge invariant operators contribute to the OPE. From these only $F_{\mu\nu}$, the field--strength tensor, contributes to first order in $q_{4}$, hence only operators that have the same quantum numbers and symmetries of $F_{\mu\nu}$ are relevant. In summary, this means that at the first order in the external electromagnetic field we have for the regime of high virtualities at hand:
\begin{equation}\label{eq:OPEandHLbL1}
\begin{split}\Pi^{\mu_{1}\mu_{2}\mu_{3}} &\equiv i\Pi^{\mu_{1}\mu_{2}\mu_{3}\mu_{4}\mu_{5}}_{F}(q_{1},q_{2})\langle0|F_{\mu_{4}\mu_{5}}(0)|\gamma(q_{4})\rangle=q_{4\mu_{4}}\epsilon_{\mu_{5}}(q_{4})\Pi^{\mu_{1}\mu_{2}\mu_{3}[\mu_{4}\mu_{5}]}_{F}\;.\\
\frac{\partial \Pi^{\mu_{1}\mu_{2}\mu_{3}\mu_{4}}}{\partial q_{4\mu_{5}}}\Bigg|_{q_{4}\rightarrow0} &=\Pi^{\mu_{1}\mu_{2}\mu_{3}[\mu_{4}\mu_{5}]}_{F}\;.
\end{split}
\end{equation}
We see here a confirmation of the antisymmetric nature of $\partial\Pi^{\mu_1\mu_2\mu_3\mu_4}$ arising from gauge invariance. We can also see that the real object of interest for $a_{\mu}^{\text{HLbL}}$ is actually $\Pi^{\mu_{1}\mu_{2}\mu_{3}[\mu_{4}\mu_{5}]}_{F}$, which is made from the Wilson coefficients of the OPE. Note that t is explicitly free of any $q_{4}$ dependence and therefore does no suffer from singularities at $q_4^2\rightarrow 0$.

We already discussed that the OPE elements for $\Pi^{\mu_1\mu_2\mu_3}$ have the same Lorentz structure and symmetries of $F_{\mu_{1}\mu_{2}}$, that is:
\begin{itemize}
    \item second rank antisymmetric tensor; 
    \item odd charge--conjugation parity (remember in this regard the famous Furry's theorem).
\end{itemize}

In~\cite{Bijnens2020} operators with these features and mass dimension up to $6$ are taken into account and the rest are neglected. This choice is ultimately supported by the fact the contribution of higher dimensional operators turns out to be at least two orders of magnitude smaller than the leading order. It is however also true that the non perturbative matrix elements of the dimension seven operators are less known. We focus here, for simplicity of reading at this step of our analysis, to the case of only one flavour and therefore on the following list of operators 
\begin{equation}\label{eq:OPEelements}
\begin{split}
    S_{1,\mu\nu} &\equiv ee_{f}F_{\mu\nu}\;, \hspace{1.625cm}
    S_{2,\mu\nu} \equiv \overline{\Psi}\sigma_{\mu\nu}\Psi\;,\\
    S_{3,\mu\nu} &\equiv ig_{S}\overline{\Psi}G_{\mu\nu}\Psi\;,\hspace{1cm}
    S_{4,\mu\nu} \equiv ig_{S}\overline{\Psi}\,\overline{G}_{\mu\nu}\gamma_{5}\Psi\;,\\
    S_{5,\mu\nu} &\equiv \overline{\Psi}\Psi\;ee_{f}F_{\mu\nu}\;,\hspace{1cm}
    S_{6,\mu\nu} \equiv \frac{\alpha_{s}}{\pi}G^{\alpha\beta}_{a}G^{a}_{\alpha\beta}\; ee_{f}F_{\mu\nu}\;,\\
    S_{7,\mu\nu} &\equiv g_{S}\overline{\Psi}(G_{\mu\lambda}D_{\nu} +D_{\nu}G_{\mu\lambda})\gamma^{\lambda}\Psi +g_{S}\overline{\Psi}(G_{\nu\lambda}D_{\mu} +D_{\mu}G_{\nu\lambda})\gamma^{\lambda}\Psi\;,\\
    S_{\{8\},\mu\nu} &\equiv \alpha_{s}(\overline{\Psi}\Gamma \Psi\overline{\Psi}\Gamma 
    \Psi)_{\mu\nu}\;,
\end{split}
\end{equation}
where $\Psi$ represents again a quark field in a given flavour of electric charge $ee_f$, the colour indices indices are implicitly summed upon, $\Gamma$ represents a combination of Dirac gamma matrices, $D_{\nu}$ represents the gauge--covariant derivative, $G_{\mu\nu}^{a}$ represents the gluon field strength tensor, $G_{\mu\nu}\equiv it^{a}G_{\mu\nu}^{a}$ and $\overline{G}_{\mu\nu}\equiv\frac{i}{2}\epsilon^{\mu\nu\alpha\beta}G_{\alpha\beta}$. Since the largest non--perturbative QCD energy scale is the perturbative threshold $\Lambda_{QCD}$, then the contributions from operators with mass dimension $d$ are expected to be suppressed like $\Big(\frac{\Lambda_{QCD}}{Q_{i}}\Big)^{d-1}$\footnote{Note that the mass dimensions of the matrix element $\langle0|...|\gamma\rangle$ of an operator with mass dimension $d$ actually has mass dimension $d-1$.}, which means that an $O\Big(\frac{\Lambda_{QCD}^{d}}{Q_{i}^{d}}\Big)$ error is introduced when a cut--off dimension $d$ is imposed. When included in the high--energy integration region  of the master formula~\eqref{eq:masterFormula}, the integration domain should be used coherently in agreement with the mass dimension cut--off of the OPE.

Due to the quantum numbers of the $S_{i,\mu\nu}$ operators we will be able to factorize the plane wave of the external electromagnetic field times the non--zero expectation value in the true QCD vacuum of a Lorentz, gauge invariant and charge-conjugation even operator, so called a condensate $X_i^S$: 
\begin{align}
    \langle0| S_{i,\mu\nu}|\gamma\rangle \equiv X_{i}^{S}\langle0|F_{\mu\nu}|\gamma\rangle\;,
\end{align}
which render the connection between the $
S_{i\mu\nu}$ operators and $\Pi_F$ more evident.

Now that the operator basis is known up to dimension six, the next step is to obtain the Wilson coefficients of its elements. Since these are local operators, evaluated at $x=0$, the obvious step is to perform a Taylor expansion of the field variables. However, such an expansion hinders the computation of the Wilson coefficients because it involves terms that are not even gauge covariant while the $S_{i,\mu\nu}$ are gauge invariant. Hence the computation of the Wilson coefficients becomes very complex, particularly for the gluon operators, as can be seen in~\cite{OPEBackgroundObsoleto_1,OPEBackgroundObsoleto_2}. 

Instead, if one chooses the Fock--Schwinger gauge~\cite{Fock1937} for the photon and gluon fields, defined by the constraint $A_\mu (x-x_0)^\mu=0$ for some constant $x_0$, the following expansions hold~\cite{GaugeInvariantExpansion}:
\begin{equation}
\begin{split}
    \Psi(x) &= \Psi(0) +x^{\mu_{1}}D_{\mu_{1}}\Psi(0) +\frac{1}{2!}x^{\mu_{1}}x^{\mu_{2}}D_{\mu_{1}}D_{\mu_{2}}\Psi(0) +...\;,\\
    A^{a}_{\alpha}(x) &= \frac{1}{2\times 0!}x^{\mu_{1}}G^{a}_{\mu_{1}\alpha}(0)
+\frac{1}{3\times 1!}x^{\mu_{1}}x^{\mu_{2}}D_{\mu_{1}}G_{\mu_{2}\alpha}^{a}(0) +...\;,\\
A_{\alpha}(x) &= \frac{1}{2\times 0!}x^{\mu_{1}}F_{\mu_{1}\alpha}(0)
+\frac{1}{3\times 1!}x^{\mu_{1}}x^{\mu_{2}}D_{\mu_{1}}F_{\mu_{2}\alpha}(0) +...\;,\\
\end{split}
\end{equation}
where $A^a$ represents the gluon field and we have chosen $x_0=0$ for simplicity. The gauge--covariant form of these expansions offers a clear advantage for the computation of Wilson coefficients associated to gauge--invariant operators. On the other hand, the gauge--fixing constraint evidently breaks traslational invariance and the propagator, being gauge dependent, inherits this feature, which results in a rather complex expression~\cite{FockSchwingerPropagator}. As a result the gain in simplicity due to explicit gauge covariance may be lost in perturbative corrections due to the complexity of the propagators.

A way around this issue is to split the gauge field into a background part that acts as a classical field that parameterizes the effects of the non--perturbative QCD vacuum on perturbative dynamics dictated by the asymptotic freedom and a fluctuation part that represents the perturbative oscillations around the vacuum solution. The former leads to the non--zero values $\langle S_{i,\mu\nu}\rangle$ and can be fixed by the radial gauge constraint, but it is not a dynamical field variable and hence does not have an associated propagator, and viceversa for the fluctuation part, which is fixed by a different gauge constraint that we will explore in the next section. Such a split can be represented as:
\begin{equation}
    A_{\mu}(x) = a_{\mu}(x) +A^{'}_{\mu}(x)\;,\hspace{0.7cm}
    A^{a}_{\mu}(x) = a^{a}_{\mu}(x) +A^{'a}_{\mu}(x)\;,
\end{equation}
where the unprimed variables are the classical fields an the primed ones represent the fluctuations around this classical value. It is not needed to include the specific form of the QCD vacuum fields but just to parameterize them as external fields, as it is for instance done in the Coleman-Weinberg approach. In the next section we present in detail the theoretical framework that supports the background field method for these split fields.

\subsection{OPE of $\Pi^{\mu_{1}\mu_{2}\mu_{3}\mu_{4}}$ in an electromagnetic background field: Theoretical framework}

First, we will relate the results obtained with the split fields to the ones obtained with the conventional approach. Let $Z[J_n]$ represent the generating functional for the HLbL interaction:
\begin{equation}\label{eq:radialGaugeExpansions}
    Z[J_n] =\int\prod_{n}\mathcal{D}\Phi_{n}\exp{i\Big(S[\Phi_{n}]+J_{n}\Phi_{n}\Big)}\;,
\end{equation}
where $\Phi$ represents different types of fields (fermion, vector, etc.) and $n$ is a collective index that is meant to label all (color, flavor and Lorentz) degrees of freedom. Complementary one can define a background generating functional:
\begin{equation}
    \mathcal{Z}[J_n,\phi_{n}] =\int\prod_{n}\mathcal{D}\phi'_{n}\exp{i\Big(S[\phi_{n}+\phi'_{n}]+J_{n}\phi'_{n}\Big)}\;,
\end{equation}
where primed and unprimed variables represent quantum fluctuations and background fixed fields, respectively, just as in the previous section. From these objects one can obtain the corresponding quantum effective action by performing a Legendre transform on the generator of connected diagrams:
\begin{align}
    W[J_{n}] &= -i\ln{Z[J_n]}\;, & \Gamma[\langle\Phi_{n}\rangle_{J}] &=W[J_{n}] -\int d^{4}x\,J_{n}(x)\langle\Phi_{n}\rangle_{J}(x)\;,\\
    \mathcal{W}[J_n,\phi_n] &= -i\ln{\mathcal{Z}[J_n,\phi_n]}\;, & H[\langle\phi'_{n}\rangle_{J},\phi_{n}] &=\mathcal{W}[J_n] -\int d^{4}x\,J_{n}(x)\langle\phi'_{n}\rangle_{J}(x)\;,
\end{align}
where $\langle\Phi_{n}\rangle_{J}$ and $\langle\phi'_{n}\rangle_{J}$ represent the VEVs of $\Phi_{n}$ and $\phi'_{n}$ in the presence of sources $J_{n}$:
\begin{align}
    \langle\Phi_{n}\rangle_{J} &\equiv \frac{\delta Z}{\delta J_{n}}\;,   &\langle\phi'_{n}\rangle_{J} &\equiv \frac{\delta \mathcal{Z}}{\delta J_{n}}\;.
\end{align}
Furthermore, by shifting  the integration variable in $\mathcal{Z}$ it is possible to conclude that these objects, although defined for different generating functionals, are related to each other:
\begin{align}
    \langle\phi'_{n}\rangle_{J} &= \langle\Phi_{n}\rangle_{J} -\phi_{n} &\implies&&
    H[\langle\phi'_{n}\rangle_{J},\phi_{n}] &= \Gamma[\langle\phi'_{n}\rangle_{J}+\phi_{n}]\;,\\
     \phi_{n}&= \langle\Phi_{n}\rangle_{J} & \implies&&  H[0,\phi_{n}] &= \Gamma[\phi_{n}]\;.
\end{align}
The last equation states that the quantum effective action of a theory can be computed from its corresponding background effective action by turning off the VEV of the fluctuations. Moreover, since one--particle irreducible (1PI) diagrams are obtained by functional differentiation of the quantum effective action one can conclude that: 1) $H[0,\phi_{n}]$ diagrams contain no external lines of quantum fluctuations, 2) matrix elements in the original theory (represented by $Z[J_n]$) can be computed by functionally differentiating vacuum--to--vacuum diagrams in the background theory with respect to the background fields, which is the result we required at the start of this section. This last result is what justifies the fact that a product of background fields, say $\overline{\psi}\sigma_{\mu\nu}\psi$, is related to the composite operator $\overline{\Psi}\sigma_{\mu\nu}\Psi$.

There is an important caveat however: the gauge fixing procedure requires one to be careful not to break the symmetries of the ``original'' $Z[J_n]$ when building $\mathcal{Z}[J_n,\phi_{n}]$ in order for these conclusions to hold. In particular, we want to fix the gauge redundancy of the fluctuation:
\begin{equation}
    \delta'A^{'a}_{\mu} =\partial_{\mu}\epsilon^{a}(x) +g_{S}f^{abc}\epsilon^{c}(a_{\mu}^{b} +A^{'b}_{\mu})\;,
\end{equation}
with matter fields transforming as usual and $\delta' a_\mu^a=0$, while keeping the lagrangian's invariance under the background gauge transformations:
\begin{equation}
    \delta a_{\mu}^{a} =\langle0|\partial_{\mu}\epsilon^{a} +g_{S}f^{abc}A^{b}_{\mu}\epsilon^{c}|0\rangle = \partial_{\mu}\epsilon^{a} +g_{S}f^{abc}a_{\mu}^{b}\epsilon^{c}\;, \hspace{0.6cm}\delta A^{'a}_{\mu} = \delta A^{a}_{\mu} -\delta a_{\mu}^{a} =g_{S}f^{abc}A_{\mu}^{'b}\epsilon^{c}\;.
\end{equation}
There is in fact only one gauge fixing term for the background generating functional that fulfills such a requirement: $-\frac{1}{2\xi}(\overline{D}^{\mu}A^{'a}_{\mu})^2$, where $\overline{D}^{\mu}A^{'a}_{\mu} \equiv \partial^{\mu}A^{'a}_{\mu} +g_{S}f^{abc}a^{b}_{\mu}A^{'c\mu}$ acts as a background covariant derivative and $\xi$ is the arbitrary gauge--fixing parameter.

Now that we have presented the theoretical framework of the separation of fields that we will use, we are ready to obtain the computational tools that we will need to build the OPE. Let us read in detail the Feynman rules from the lagrangian $\mathcal{L}^{\text{HLbL}}$ after the separation of the fields in classical background and quantum parts:
\begin{equation}
\begin{split}
    \mathcal{L}^{\text{HLbL}} &=-\frac{1}{4}(f_{\mu\nu}^{a} +\overline{D}_{\mu}A^{'a}_{\nu} -\overline{D}_{\nu}A^{'a}_{\mu} +g_{S}f^{abc}A^{'b}_{\mu}A^{'c}_{\nu})^{2} -\frac{1}{4}(f_{\mu\nu} +\partial_{\mu}A^{'a}_{\nu} -\partial_{\nu}A^{'}_{\mu})^{2}\\ 
    &+(\overline{\psi}_{l} +\overline{\psi}'_{l}) (\{i\slashed{\partial} -m\}\delta_{lk} +e\hat{Q}\delta_{lk}\slashed{a} +g_{S}t^{a}_{lk}\slashed{a}^{a})(\psi_{l} +\psi'_{l})\\
    &+(\overline{\psi}_{l} +\overline{\psi}'_{l}) (e\hat{Q}\delta_{lk}\slashed{A}' +g_{S}t^{a}_{lk}\slashed{A}^{'a})(\psi_{k} +\psi'_{k}) -\frac{1}{2\xi}\overline{D}^{\mu}A^{'a}_{\mu}\overline{D}^{\mu}A^{'a}_{\mu} -\mathcal{L}_{\text{ghosts}}\;,
\end{split}
\end{equation}
where we have split the quark fields following the same (un)primed convention and we have included colour indices $l$ and $k$. Besides, $f^{a}_{\mu\nu}$ and $f_{\mu\nu}$ are background counterparts of the usual field strength tensors.

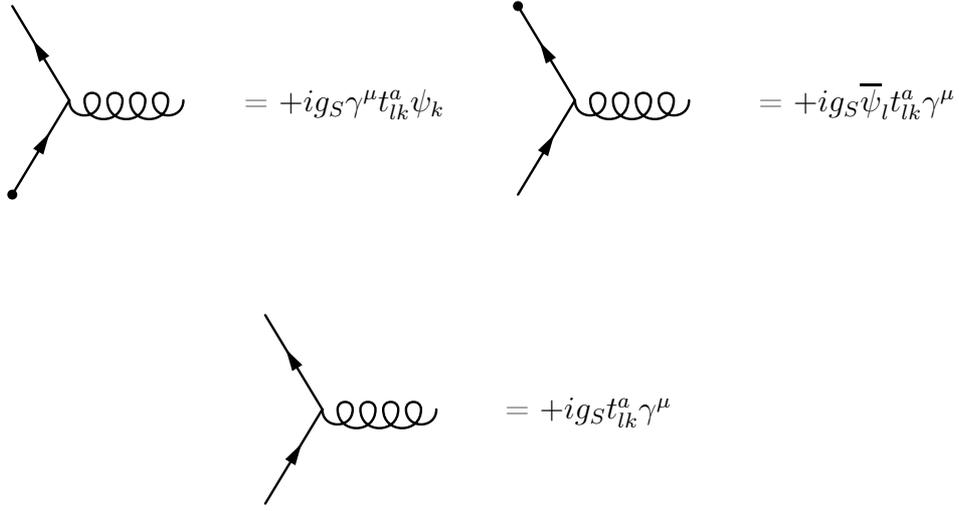
\begin{figure}
    \centering
    \begin{tabular}{cccc}
        \multirow{3}{*}{\begin{fmffile}{psiPsiPrimeAaPrime}
        \setlength{\unitlength}{0.5cm}\small
        \begin{fmfgraph*}(5,5)
            \fmfleft{i1,i2}
            \fmflabel{$l$}{i2}
            \fmfright{o}
            \fmflabel{$a$}{o}
            \fmf{fermion}{i1,v,i2}
            \fmf{gluon}{v,o}
            \fmfset{arrow_len}{3mm}
            \fmfv{decor.shape=circle,decor.size=1mm}{i1}
        \end{fmfgraph*}
    \end{fmffile}} &  &
    \multirow{3}{*}{\begin{fmffile}{psiPrimePsiAaPrime}
        \setlength{\unitlength}{0.5cm}\small
        \begin{fmfgraph*}(5,5)
            \fmfleft{i1,i2}
            \fmflabel{$k$}{i1}
            \fmfright{o}
            \fmflabel{$a$}{o}
            \fmf{fermion}{i1,v,i2}
            \fmf{gluon}{v,o}
            \fmfset{arrow_len}{3mm}
            \fmfv{decor.shape=circle,decor.size=1mm}{i2}
        \end{fmfgraph*}
    \end{fmffile}} & 
    \\
    &&&\\ & \;\;= $+ig_S\gamma^{\mu}t^{a}_{lk}\psi_{k}$ & &\;\; = $+ig_S\overline{\psi}_{l}t^{a}_{lk}\gamma^{\mu}$\\ 
    &&&\\&&&\\&&&\\&&&\\&&&\\
    &\multirow{3}{*}{\begin{fmffile}{psiPrimePsiPrimeAaPrime}
        \setlength{\unitlength}{0.5cm}\small
        \begin{fmfgraph*}(5,5)
            \fmfleft{i1,i2}
            \fmflabel{$l$}{i2}
            \fmflabel{$k$}{i1}
            \fmfright{o}
            \fmflabel{$a$}{o}
            \fmf{fermion}{i1,v,i2}
            \fmf{gluon}{v,o}
            \fmfset{arrow_len}{3mm}
        \end{fmfgraph*}
    \end{fmffile}} &&\\ &&&\\
    &&= $+ig_St^{a}_{lk}\gamma^{\mu}$&\vspace{1.5cm}
    \end{tabular}
    \caption{These figures show the three different types of quark interactions with a quantum gluon. In the two diagrams at the top one quark line is non--pertubatively annihilated by the vacuum. $l$ and $k$ represent the colour of the quarks, $a$ represents the colour of the gluon and trivial quark flavor indices are suppressed.}
    \label{fig:quarkVertices}
\end{figure}

The previous lagrangian can be simplified very much. First, the parts with only background field variables are constant from the point of view of the path integral and therefore they may be adsorbed by its normalization constant, as is usual for amplitudes of disconnected diagrams. In addition, since HLbL scattering only involves vertices from strong interactions, the quantum fluctuation part of the photon field is not relevant. In second place, it is necessary to recall that the classical background field minimize the action\footnote{The classical background fields actually minimize the \emph{quantum} effective action, which in absence of external sources and to leading order is equivalent to the same statement on the classical action. For quark and gluon fields the vacuum expectation values of course do not receive perturbative contributions since tadpole perturbative diagrams are zero at all orders.}, therefore the first functional derivative of the action with respect to each field fluctuation vanishes, which means that terms of the lagrangian that are linear in the field fluctuations do not contribute to the dynamics. Finally for the gauge fixing parameter we choose $\xi=1$. In summary, the parts of the lagrangian that are relevant to our case is: 
\begin{equation}\label{eq:backgroundLagrangian}
\begin{split}
    \mathcal{L}^{\text{HLbL}} &=-\frac{g_{S}}{2}f^{a\mu\nu}f^{abc}A^{'b}_{\mu}A^{'c}_{\nu} -\frac{1}{2}\Big(\overline{D}^{\mu}A^{'a\nu}\overline{D}_{\mu}A^{'a}_{\nu} +\overline{D}^{\mu}A^{'a}_{\mu}\overline{D}^{\nu}A^{'a}_{\nu} -\overline{D}^{\mu}A^{'a\nu}\overline{D}_{\nu}A^{'a}_{\mu}\Big)\\
    &-g_{S}f^{a\overline{b} \overline{c}}A^{'\overline{b}\mu}A^{'\overline{c}\nu}\overline{D}_{\mu}A^{'a}_{\nu} -\frac{1}{4}g_{S}^{2}f^{abc}f^{a\overline{b} \overline{c}}A^{'b}_{\mu}A^{'c}_{\nu}A^{'\overline{b}\mu}A^{'\overline{c}\nu}\\
    &+\overline{\psi}'_{l} (\{i\slashed{\partial} -m\}\delta_{lk} +e\hat{Q}\delta_{lk}\slashed{a} +g_{S}t^{a}_{lk}\slashed{a}^{a})\psi'_{k} +\overline{\psi}_{l} (e\hat{Q}\delta_{lk}\slashed{A}' +g_{S}t^{a}_{lk}\slashed{A}^{'a})\psi'_{k}\\
    & +\overline{\psi}'_{l} (e\hat{Q}\delta_{lk}\slashed{A}' +g_{S}t^{a}_{lk}\slashed{A}^{'a})\psi_{k} +\overline{\psi}'_{l} (e\hat{Q}\delta_{lk}\slashed{A}'+g_{S}t^{a}_{lk}\slashed{A}^{'a})\psi'_{k}\\
    &-w^{*}_{a}\overline{D}_{\mu}\{\overline{D}^{\mu}w_{a} +f_{abc}w_{c}A^{'\mu}_{b}\}\;,
\end{split}
\end{equation}
where ghost fields are represented by $w$ and $w^*$ and the last line corresponds to $\mathcal{L}_{\text{ghosts}}$, whose invariance under background gauge transformations is guaranteed by the fact that ghost fields transform like matter. As usual, the kernel of quadratic terms in a specific path--integral variable is the inverse of corresponding free propagator, while the rest are interaction vertices. These are very similar to the ones in a theory with no background fields, the only difference being that any one line can now be created or annihilated by the vacuum. Feynman rules for quark--gluon vertices are summarized in figure~\ref{fig:quarkVertices}. The vertices for gluon self interactions can be found straightforwardly, but we do not quote them here because they are not relevant at the perturbative level we have working at.

The quark free propagator is modified by the background fields in the usual way:
\begin{align}
    (\{i\slashed{\partial} -m\}\delta_{l'l}\delta^{f'f} +e\hat{Q}^{f'f}\delta_{l'l}\slashed{a} +g_{S}\delta^{f'f}t^{a}_{l'l}\slashed{a}^{a}) S_{lk}^{fs}(x,y) &=i\delta^{4}(x-y)\delta^{f's}\delta_{l'k}\;,
\end{align}
therefore it can be computed recursively by supposing that the strength of the background gauge fields is much smaller than the characteristic momentum of the process of interest, which is a reasonable hypothesis in our context. Note that the new indices $f'f$ represent quark flavor and had been suppressed previously. In the approximation of weak external electromagnetic field, an expansion to $O(e)$ is enough. For the background gluon field it is necessary to go to $O(g_{S}^{2})$ in agreement with the highest mass dimension of the local operators that were chosen for the OPE.

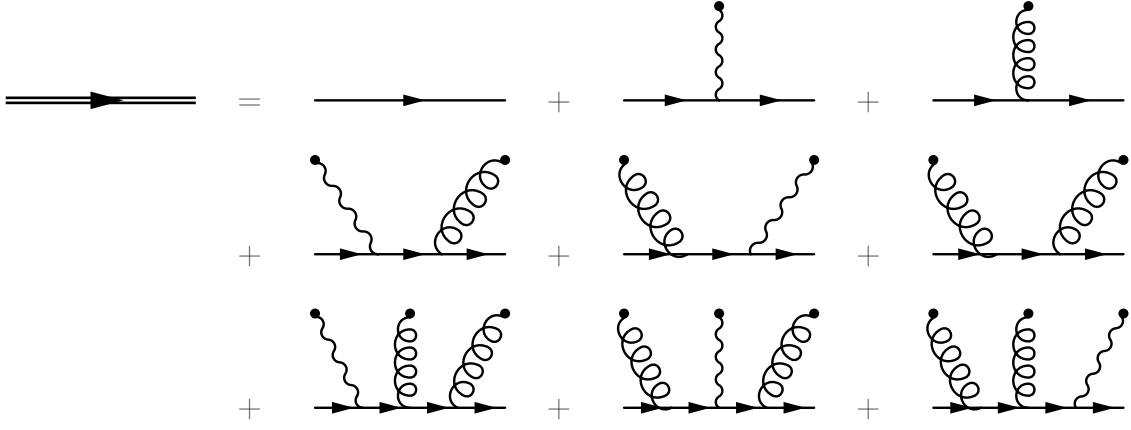
\begin{figure}
    \centering
    \hspace{-1cm}
    \begin{tabular}{ccccccc}
        \multirow{3}{*}{\begin{fmffile}{quarkProp}
        \setlength{\unitlength}{0.5cm}\small
        \begin{fmfgraph*}(5,5)
            \fmfleft{i}
            \fmfright{o}
            \fmf{heavy}{i,o}
            \fmfset{arrow_len}{3mm}
        \end{fmfgraph*}
    \end{fmffile}} &  & 
    \multirow{3}{*}{\begin{fmffile}{quarkProp0}
        \setlength{\unitlength}{0.5cm}\small
        \begin{fmfgraph*}(5,5)
            \fmfleft{i}
            \fmfright{o}
            \fmf{fermion}{i,o}
            \fmfset{arrow_len}{3mm}
        \end{fmfgraph*}
    \end{fmffile}} &  &
    \multirow{3}{*}{\begin{fmffile}{quarkPropE}
        \setlength{\unitlength}{0.5cm}\small
        \begin{fmfgraph*}(5,5)
            \fmfstraight
            \fmfleft{i}
            \fmfright{o}
            \fmftop{t}
            \fmf{fermion}{i,v,o}
            \fmffreeze
            \fmf{photon}{t,v}
            \fmfv{decor.shape=circle,decor.size=1mm}{t}
            \fmfset{arrow_len}{3mm}
        \end{fmfgraph*} 
    \end{fmffile}} &  & 
    \multirow{3}{*}{\begin{fmffile}{quarkPropG}
        \setlength{\unitlength}{0.5cm}\small
        \begin{fmfgraph*}(5,5)
            \fmfstraight
            \fmfleft{i}
            \fmfright{o}
            \fmftop{t}
            \fmf{fermion}{i,v,o}
            \fmffreeze
            \fmf{gluon}{t,v}
            \fmfv{decor.shape=circle,decor.size=1mm}{t}
            \fmfset{arrow_len}{3mm}
        \end{fmfgraph*}
    \end{fmffile}}\\&&&&&&\\ & = &  & + &  & + &\\&&&&&&\\
    &  & \multirow{3}{*}{\begin{fmffile}{quarkPropEG}
        \setlength{\unitlength}{0.5cm}\small
        \begin{fmfgraph*}(5,5)
            \fmfstraight
            \fmfleft{i}
            \fmfright{o}
            \fmftop{t1,t2}
            \fmf{fermion}{i,v1,v2,o}
            \fmffreeze
            \fmf{photon}{t1,v1}
            \fmf{gluon}{t2,v2}
            \fmfv{decor.shape=circle,decor.size=1mm}{t1,t2}
            \fmfset{arrow_len}{3mm}
        \end{fmfgraph*} 
    \end{fmffile}} &  &
    \multirow{3}{*}{\begin{fmffile}{quarkPropGE}
        \setlength{\unitlength}{0.5cm}\small
        \begin{fmfgraph*}(5,5)
            \fmfstraight
            \fmfleft{i}
            \fmfright{o}
            \fmftop{t1,t2}
            \fmf{fermion}{i,v1,v2,o}
            \fmffreeze
            \fmf{gluon,wiggly_len=4mm}{t1,v1}
            \fmf{photon}{t2,v2}
            \fmfv{decor.shape=circle,decor.size=1mm}{t1,t2}
            \fmfset{arrow_len}{3mm}
        \end{fmfgraph*}
    \end{fmffile}} &  &
    \multirow{3}{*}{\begin{fmffile}{quarkPropGG}
        \setlength{\unitlength}{0.5cm}\small
        \begin{fmfgraph*}(5,5)
            \fmfstraight
            \fmfleft{i}
            \fmfright{o}
            \fmftop{t1,t2}
            \fmf{fermion}{i,v1,v2,o}
            \fmffreeze
            \fmf{gluon}{t1,v1}
            \fmf{gluon}{t2,v2}
            \fmfv{decor.shape=circle,decor.size=1mm}{t1,t2}
            \fmfset{arrow_len}{3mm}
        \end{fmfgraph*}
    \end{fmffile}}\\&&&&&&\\ & + &  & + &  & + &\\&&&&&&\\
    &  & \multirow{3}{*}{\begin{fmffile}{quarkPropEGG}
        \setlength{\unitlength}{0.5cm}\small
        \begin{fmfgraph*}(5,5)
            \fmfstraight
            \fmfleft{i}
            \fmfright{o}
            \fmftop{t1,t2,t3}
            \fmf{fermion}{i,v1,v2,v3,o}
            \fmffreeze
            \fmf{photon}{t1,v1}
            \fmf{gluon}{t2,v2}
            \fmf{gluon}{t3,v3}
            \fmfv{decor.shape=circle,decor.size=1mm}{t1,t2,t3}
            \fmfset{arrow_len}{3mm}
        \end{fmfgraph*}
    \end{fmffile}} &  &
    \multirow{3}{*}{\begin{fmffile}{quarkPropGEG}
        \setlength{\unitlength}{0.5cm}\small
        \begin{fmfgraph*}(5,5)
            \fmfstraight
            \fmfleft{i}
            \fmfright{o}
            \fmftop{t1,t2,t3}
            \fmf{fermion}{i,v1,v2,v3,o}
            \fmffreeze
            \fmf{gluon,wiggly_len=4mm}{t1,v1}
            \fmf{photon}{t2,v2}
            \fmf{gluon}{t3,v3}
            \fmfv{decor.shape=circle,decor.size=1mm}{t1,t2,t3}
            \fmfset{arrow_len}{3mm}
        \end{fmfgraph*}
    \end{fmffile}} &  &
    \multirow{3}{*}{\begin{fmffile}{quarkPropGGE}
        \setlength{\unitlength}{0.5cm}\small
        \begin{fmfgraph*}(5,5)
            \fmfstraight
            \fmfleft{i}
            \fmfright{o}
            \fmftop{t1,t2,t3}
            \fmf{fermion}{i,v1,v2,v3,o}
            \fmffreeze
            \fmf{gluon,wiggly_len=4mm}{t1,v1}
            \fmf{gluon}{t2,v2}
            \fmf{photon}{t3,v3}
            \fmfv{decor.shape=circle,decor.size=1mm}{t1,t2,t3}
            \fmfset{arrow_len}{3mm}
        \end{fmfgraph*}
    \end{fmffile}}\\&&&&&&\\ & + &  & + &  & + &\\&&&&&&
    \end{tabular}
    \caption{This figure shows the expansion of the free quark propagator in a background of gauge fields in terms of diagrams with interactions with gluons and photons that are created/annihilated in the vacuum. The order in which diagrams appear in the sum corresponds to the order of terms in equation~\eqref{eq:momentumQuarkPropagator1}.}
    \label{fig:quarkPropagator}
\end{figure}

In addition, we will see that due to the presence of a background the propagator is not traslationally invariant. This of course also breaks momentum conservation along the propagator in its Fourier transformed version. Therefore there are in general three different types of momentum--space quark propagators: 
\begin{align}
    S_{lk}^{fs}(p_{1},p_{2}) &\equiv \int d^{4}x \int d^{4}y \;e^{ip_{1}x}e^{-ip_{2}y} S_{lk}^{fs}(x,y)\;,\\
    S_{lk}^{fs}(p_{1}) &\equiv \int d^{4}x\;e^{ip_{1}x}S_{lk}^{fs}(x,0) =\int \frac{d^{4}p_{2}}{(2\pi)^{4}}S_{lk}^{fs}(p_{1},p_{2})\;,\\
    \tilde{S}_{lk}^{fs}(p_{2}) &\equiv \int d^{4}x\;e^{-ip_{2}x} S_{lk}^{fs}(0,x) =\int \frac{d^{4}p_{1}}{(2\pi)^{4}}S_{lk}^{fs}(p_{1},p_{2})\;,
\end{align}
By Taylor expanding the gauge fields we have intrinsically assumed that they are soft, but not even that restores translation invariance of the propagator. In summary, the free quark propagator in the momentum space (and in the presence of background gluon and photon fields) at $O(eg^{2}_{S})$ is: 
\begin{equation}
\begin{split}
    S_{lk}^{fs}(p_{1},p_{2}) &= (2\pi)^{4}\delta^{4}(p_{1}-p_{2})S^{0}_{p_{1}}\delta^{fs}\delta_{lk}\\ &+iS^{0}_{p_{1}}\int_{q_{1}}\{e\hat{Q}^{fs}\delta_{lk}\slashed{a}_{q_{1}} +g_{S}t^{a}_{lk}\delta^{fs}\slashed{a}^{a}_{q_{1}}\}S^{0}_{p_{1}+q_{1}}(2\pi)^{4}\delta^{4}(p_{1}-p_{2}+q_{1})\\
    &-e\hat{Q}^{fs}g_{S}t^{a}_{lk}S^{0}_{p_{1}}\int_{q_{1}}\slashed{a}_{q_{1}}S^{0}_{p_{1}+q_{1}}\int_{q_{2}}\slashed{a}^{a}_{q_{2}}S^{0}_{p_{1}+q_{1}+q_{2}}
    (2\pi)^{4}\delta^{4}(p_{1}-p_{2}+q_{1}+q_{2})\\
    &-e\hat{Q}^{fs}g_{S}t^{a}_{lk}S^{0}_{p_{1}}\int_{q_{1}}\slashed{a}^{a}_{q_{1}}S^{0}_{p_{1}+q_{1}}\int_{q_{2}}\slashed{a}_{q_{2}}S^{0}_{p_{1}+q_{1}+q_{2}}(2\pi)^{4}\delta^{4}(p_{1}-p_{2}+q_{1}+q_{2})\\ &-g_{S}^{2}t^{a}_{ll'}t^{b}_{l'k}\delta^{fs}S^{0}_{p_{1}}\int_{q_{1}}\slashed{a}^{a}_{q_{1}}S^{0}_{p_{1}+q_{1}}\int_{q_{2}}\slashed{a}^{b}_{q_{2}}S^{0}_{p_{1}+q_{1}+q_{2}}(2\pi)^{4}\delta^{4}(p_{1}-p_{2}+q_{1}+q_{2})\\
    &\hspace{0cm}-ie\hat{Q}^{fs}g^{2}_{S}t^{a}_{ll'}t^{b}_{l'k}\Big(S^{0}_{p_{1}}\int_{q_{1}}\slashed{a}_{q_{1}} S^{0}_{p_{1}+q_{1}}\int_{q_{2}}\slashed{a}^{a}_{q_{2}}S^{0}_{p_{1}+q_{1}+q_{2}}\int_{q_{3}}\slashed{a}^{b}_{q_{3}}S^{0}_{p_{1}+q_{1}+q_{2}+q_{3}}\\
    &\hspace{2.4cm}+S^{0}_{p_{1}}\int_{q_{1}}\slashed{a}^{a}_{q_{1}} S^{0}_{p_{1}+q_{1}}\int_{q_{2}}\slashed{a}_{q_{2}}S^{0}_{p_{1}+q_{1}+q_{2}}\int_{q_{3}}\slashed{a}^{b}_{q_{3}}S^{0}_{p_{1}+q_{1}+q_{2}+q_{3}}\\
    &\hspace{2.4cm}+S^{0}_{p_{1}}\int_{q_{1}}\slashed{a}_{q_{1}}^{a} S^{0}_{p_{1}+q_{1}}\int_{q_{2}}\slashed{a}^{b}_{q_{2}}S^{0}_{p_{1}+q_{1}+q_{2}}\int_{q_{3}}\slashed{a}_{q_{3}}S^{0}_{p_{1}+q_{1}+q_{2}+q_{3}}\Big) (2\pi)^{4}\delta^{4}(p_{1}-p_{2}+\sum_{i=1,2,3}q_{i})\;.
\end{split}
\end{equation}
Note that the Dirac delta is always under the effect of the integrals to its left. Additionally, we use the convention: 
\begin{equation}
    S^{0}(p) = i\frac{\slashed{p} +m}{p^{2}-m^{2} +i\epsilon} \equiv S^{0}_{p}\;.
\end{equation}
Note that we have kept the momentum--conservation delta of the $y$ vertex explicitly for $S(p_{1},p_{2})$ in order to make the relation with $S(p)$ and $\tilde{S}(p)$ more evident. At this point, we can use the expansions of the gauge fields in~\eqref{eq:radialGaugeExpansions}. Since the local operators considered for the OPE contain no derivatives of the photon fields and contain only up to one derivative of the gluon field, it is enough to retain the first term of the expansion for the photon field and the first two terms for the gluon. Therefore in the momentum representation we can use the replacement\footnote{Note that these expressions were derived for soft insertions where momentum is leaving the diagram. From the distributional point of view the derivatives are supposed to act on test-functions~\cite{BookStrichartz}.}:
\begin{equation}
    \begin{split}
        a^{\mu}(q) &=\frac{i}{2}(2\pi)^{4}f^{\nu\mu}(0)\frac{\partial}{\partial q^{\nu}}\delta^{4}(q) =-\frac{i}{2}(2\pi)^{4}\delta^{4}(q)f^{\nu\mu}(0)\frac{\partial}{\partial q^{\nu}}\\
        a^{a\mu}(q) &=\frac{i}{2}f^{a\nu\mu}(0)\frac{\partial}{\partial q^{\nu}}(2\pi)^{4}\delta^{4}(q) -\frac{1}{3}D^{\tau}f^{a\nu\mu}(0)\frac{\partial}{\partial q^{\nu}}\frac{\partial}{\partial q^{\tau}}(2\pi)^{4}\delta^{4}(q)\\
        &=(2\pi)^{4}\delta^{4}(q)\Big(-\frac{i}{2}f^{a\nu\mu}(0)\frac{\partial}{\partial q^{\nu}} -\frac{1}{3}D^{\tau}f^{a\nu\mu}(0)\frac{\partial}{\partial q^{\nu}}\frac{\partial}{\partial q^{\tau}}\Big)\;,
    \end{split}
\end{equation}
which yields the following result:
\begin{equation}\label{eq:momentumQuarkPropagator1}
\begin{split}
    S_{lk}^{fs}(p_{1},p_{2}) &= (2\pi)^{4}\Bigg(S^{0}_{p_{1}}\delta^{fs}\delta_{lk}\delta^4(p_1-p_2)\\
    &+\frac{1}{2}\{e\hat{Q}^{fs}\delta_{lk}f^{\mu_{1}\nu_{1}}(0) +g_{S}t^{a}_{lk}\delta^{fs}f^{a\mu_{1}\nu_{1}}(0)\}\frac{\partial}{\partial q^{\mu_{1}}_{1}}S^{0}_{p_{1}}\gamma_{\nu_{1}}S^{0}_{p_{1}+q_{1}}\delta^{4}(p_{1}-p_{2}+q_{1})\\
    &-\frac{i}{3}g_{S}t^{a}_{lk}\delta^{fs}\overline{D}^{\tau}f^{a\mu_{1}\nu_{1}}\frac{\partial}{\partial q^{\tau}_{1}}\frac{\partial}{\partial q^{\mu_{1}}_{1}}S^{0}_{p_{1}+q_{1}}S^{0}_{p_{1}}\gamma_{\nu_{1}}\delta^{4}(p_{1}-p_{2}+q_{1})\\
    &+\frac{1}{4}e\hat{Q}^{fs}g_{S}t^{a}_{lk}\Big(f^{\mu_{1}\nu_{1}}f^{a\mu_{2}\nu_{2}} +f^{a\mu_{1}\nu_{1}}f^{\mu_{2}\nu_{2}}\Big)\\
    &\hspace{1cm}\times\frac{\partial}{\partial q^{\mu_{1}}_{1}}\frac{\partial}{\partial q^{\mu_{2}}_{2}}\Big(S^{0}_{p_{1}}\gamma_{\nu_{1}}S^{0}_{p_{1}+q_{1}}\gamma_{\nu_{2}}S^{0}_{p_{1}+q_{1}+q_{2}}\delta^{4}(p_{1}-p_{2}+q_{1}+q_{2})\Big)\Bigg|_{q_{1,2}=0}\\
    &+\frac{1}{4}g_{S}^{2}t^{a}_{ll'}t^{b}_{l'k}\delta^{fs}\Big(f^{a\mu_{1}\nu_{1}}f^{b\mu_{2}\nu_{2}}\Big)\\
    &\hspace{1cm}\times\frac{\partial}{\partial q^{\mu_{1}}_{1}}\frac{\partial}{\partial q^{\mu_{2}}_{2}}\Big(S^{0}_{p_{1}}\gamma_{\nu_{1}}S^{0}_{p_{1}+q_{1}}\gamma_{\nu_{2}}S^{0}_{p_{1}+q_{1}+q_{2}}\delta^{4}(p_{1}-p_{2}+q_{1}+q_{2})\Big)\Bigg|_{q_{1,2}=0}\\
    &+\frac{1}{8}e\hat{Q}^{fs}g^{2}_{S}t^{a}_{lk'}t^{b}_{k'k}\Big(f^{\mu_{1}\nu_{1}}f^{a\mu_{2}\nu_{2}}f^{b\mu_{3}\nu_{3}} +f^{a\mu_{1}\nu_{1}}f^{\mu_{2}\nu_{2}}f^{b\mu_{3}\nu_{3}} +f^{a\mu_{1}\nu_{1}}f^{b\mu_{2}\nu_{2}}f^{\mu_{3}\nu_{3}}\Big)\\
    &\hspace{1cm}\times\frac{\partial}{\partial q^{\mu_{1}}_{1}}\frac{\partial}{\partial q^{\mu_{2}}_{2}}\frac{\partial}{\partial q^{\mu_{3}}_{3}}\Big(S^{0}_{p_{1}}\gamma_{\nu_{1}} S^{0}_{p_{1}+q_{1}}\gamma_{\nu_{2}}S^{0}_{p_{1}+q_{1}+q_{2}}\gamma_{\nu_{3}}S^{0}_{p_{1}+q_{1}+q_{2}+q_{3}}\delta^{4}(p_{1}-p_{2}+\sum_{i=1,2,3}q_{i})\Big)\Big|_{q_{1,2,3}=0}\Bigg)\;.
\end{split}
\end{equation}
This expression for the free quark propagator can be understood as an expansion in terms of diagrams (see figure~\ref{fig:quarkPropagator}) with increasing number of (background) gauge bosons. 

The ``free'' gluon propagator contains interactions with the background gluon fields in a similar but more involved way than the quark one. However, at the order that we are interested in there appear no such propagators, so we refer the interested reader to see the details of the computation in section 2 of~\cite{FockSchwingerPropagatorInversion}.

\section{Computation of un--renormalized Wilson coefficients}\label{sec:unrenormalizedWilson}

In the previous section we obtained expressions for the quark and gluon fluctuations propagators which contained background insertions of vacuum expectation values (VEV) such as $f^{a\mu\nu}$ and $f^{\mu\nu}$. Furthermore we saw that vertices from the Dyson series also introduce VEVs of quark operators, thus giving us all the tools required to build the OPE of $\Pi^{\mu_{1}\mu_{2}\mu_{3}}$ with background fields and find the Wilson coefficients that require to compute $\Pi_{F}^{\mu_{1}\mu_{2}\mu_{3}\mu_{4}\mu_{5}}$. Concerning the actual computation of Wilson coefficients, let us start by considering the one related to $S_{1,\mu\nu}=ee_{f}F_{\mu\nu}$. This term represents the configuration in which hard momenta travels through all internal lines of the diagrams, thus, there are no cut lines. The leading order contribution for this configuration is given by the quark loop (see figure~\ref{fig:quarkLoop}), where different contributions are obtained by inserting the soft photon in different sides of the triangle and/or inverting the orientation of the loop. Since $S_{1,\mu\nu}$ is the operator with the lowest dimension in the OPE, its Wilson coefficient is expected to give the most relevant contribution to $\Pi^{\mu_{1}\mu_{2}\mu_{3}\mu_{4}\mu_{5}}_{F}$ and therefore to $a_{\mu}$. When splitting the quark fields in the currents of $\Pi^{\mu_1\mu_2\mu_3}$, this diagram comes from the term that contains only quantum fluctuations. In the end, the contribution from the Wilson coefficient of $S_{1,\mu\nu}$ to $\Pi^{\mu_{1}\mu_{2}\mu_{3}\mu_{4}\mu_{5}}_{F}$:
\begin{equation}\label{eq:quarkLoop}
\begin{split}
    \Pi^{\mu_{1}\mu_{2}\mu_{3}\mu_{4}\nu_{4}}_{F(S_{1})} &= i\frac{N_c}{2}\int\frac{d^{4}p}{(2\pi)^{4}}\sum_{f} e_{f}^{4}\frac{\partial}{\partial q_{4\nu_{4}}}\sum_{\sigma(1,2,4)}\Tr\Big\{\gamma^{\mu_{3}}S^{0}(p+q_{1}+q_{2}+q_{4})\gamma^{\mu_{4}}\\
    &\hspace{1cm}\times S^{0}(p+q_{1}+q_{2})\gamma^{\mu_{1}}S^{0}(p+q_{2})\gamma^{\mu_{2}}S^{0}(p)\Big\}\Bigg|_{q_{4}=0}\;,
\end{split}
\end{equation}
where $N_c$ is the number of quark colors, $f$ represents quark flavor and $\sigma(1,2,4)$ represents a permutation over the set $\{(q_{i},\mu_{i})|\,i\in\{1,2,4\}\}$. The derivative with respect to the soft photon momentum can be traced back to~\eqref{eq:momentumQuarkPropagator1}. Note that propagator depends implicitly on the quark flavor through the masses, assumed to be all equal for the light flavours considered. Besides, the effect of the derivative on the propagators is to duplicate them:
\begin{equation}\label{eq:propagatorDifferentiation}
    \lim_{q_{4}\rightarrow0}\frac{\partial}{\partial q_{\nu_{4}}} S(p+q_{4}) =i\lim_{q_{4}\rightarrow0} S^{0}(p+q_{4})\gamma^{\nu_{4}}S(p+q_{4}) =iS^{0}(p)\gamma^{\nu_{4}}S^{0}(p)\;.
\end{equation}
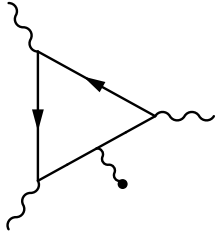
\begin{figure}
    \centering
    \begin{fmffile}{quarkLoop}
        \setlength{\unitlength}{0.6cm}\small
        \begin{fmfgraph*}(5,5) 
            \fmfleft{i1,i2}
            \fmfright{o1,o2,o3}         
            \fmf{phantom}{v1,v2,v3,v4,v5}
            \fmf{phantom}{v5,v6,v7,v8,v9}
            \fmf{phantom}{v9,v10,v11,v12,v1}
            \fmf{photon}{i1,v1}
            \fmf{photon}{v9,i2}
            \fmf{photon}{v5,o2}
            \fmffreeze
            \fmf{photon}{v3,o1}
            \fmf{plain}{v1,v2,v3,v4,v5}
            \fmf{fermion}{v5,v9,v1}
            \fmfset{arrow_len}{3mm}
            \fmfv{decor.shape=circle,decor.size=1mm}{o1}
            \fmffreeze
            \fmfshift{(-0.3w,0.2h)}{o1}
        \end{fmfgraph*}
    \end{fmffile}
    \caption{Representative diagram of the leading order contribution to the Wilson coefficient of $S_{1,\mu\nu}$ in the OPE of $\Pi^{\mu_{1}\mu_{2}\mu_{3}}$. The black dot represents creation/annihilation of a line by the background fields in the vacuum.}
    \label{fig:quarkLoop}
\end{figure}

The focus of our work is on the contribution from $S_{1,\mu\nu}$, which gives a much larger contribution to $a_\mu$ than any other operator, so we will not discuss the contributions from other operators in detail. Instead, we give a brief overview of the computation. We refer the interested reader to~\cite{Bijnens2020}. 

\begin{figure}
    \centering
    \begin{tabular}{ccc}
        \begin{fmffile}{cutQuarkLine}
        \setlength{\unitlength}{0.6cm}\small
        \begin{fmfgraph*}(5,5) 
            \fmfleft{i1,i2}
            \fmfright{o1,o2,o3}         
            \fmf{phantom}{v1,v2,v3,v4,v5}
            \fmf{phantom}{v5,v6,v7,v8,v9}
            \fmf{phantom}{v9,v10,v11,v12,v1}
            \fmf{photon}{i1,v1}
            \fmf{photon}{v9,i2}
            \fmf{photon}{v5,o2}
            \fmffreeze
            \fmf{photon}{v11,o1}
            \fmf{fermion}{v1,v5,v9}
            \fmf{plain}{v9,v10}
            \fmf{plain}{v12,v1}
            \fmfset{arrow_len}{3mm}
            \fmfv{decor.shape=circle,decor.size=1mm}{o1,v10,v11,v12}
            \fmffreeze
            \fmfshift{(-0.9w,0.5h)}{o1}
            \fmfshift{(0,-0.06h)}{v10}
            \fmfshift{(0,0.06h)}{v12}
        \end{fmfgraph*}
        \end{fmffile} & 
        \begin{fmffile}{cutQuarkLineGluon}
        \setlength{\unitlength}{0.6cm}\small
        \begin{fmfgraph*}(5,5) 
            \fmfleft{i1,i2}
            \fmfright{o1,o2,o3}         
            \fmf{phantom}{v1,v2,v3,v4,v5}
            \fmf{phantom}{v5,v6,v7,v8,v9}
            \fmf{phantom}{v9,v10,v11,v12,v1}
            \fmf{photon}{i1,v1}
            \fmf{photon}{v9,i2}
            \fmf{photon}{v5,o2}
            \fmffreeze
            \fmf{photon}{v11,o3}
            \fmf{gluon}{v3,o1}
            \fmf{plain}{v1,v2,v3,v4,v5}
            \fmf{fermion}{v5,v9}
            \fmf{plain}{v9,v10}
            \fmf{plain}{v12,v1}
            \fmfset{arrow_len}{3mm}
            \fmfv{decor.shape=circle,decor.size=1mm}{o1,v10,v11,v12,o3}
            \fmffreeze
            \fmfshift{(-0.3w,0.2h)}{o1}
            \fmfshift{(-0.9w,-0.5h)}{o3}
            \fmfshift{(0,-0.06h)}{v10}
            \fmfshift{(0,0.06h)}{v12}
        \end{fmfgraph*}
        \end{fmffile} & 
        \begin{fmffile}{cutQuarkLinePhoton}
        \setlength{\unitlength}{0.6cm}\small
        \begin{fmfgraph*}(5,5) 
            \fmfleft{i1,i2}
            \fmfright{o1,o2,o3}         
            \fmf{phantom}{v1,v2,v3,v4,v5}
            \fmf{phantom}{v5,v6,v7,v8,v9}
            \fmf{phantom}{v9,v10,v11,v12,v1}
            \fmf{photon}{i1,v1}
            \fmf{photon}{v9,i2}
            \fmf{photon}{v5,o2}
            \fmffreeze
            \fmf{photon}{v3,o1}
            \fmf{plain}{v1,v2,v3,v4,v5}
            \fmf{fermion}{v5,v9}
            \fmf{plain}{v9,v10}
            \fmf{plain}{v12,v1}
            \fmfset{arrow_len}{3mm}
            \fmfv{decor.shape=circle,decor.size=1mm}{o1,v10,v12}
            \fmffreeze
            \fmfshift{(-0.3w,0.2h)}{o1}
            \fmfshift{(0,-0.075h)}{v10}
            \fmfshift{(0,0.075h)}{v12}
        \end{fmfgraph*}
        \end{fmffile}
    \end{tabular}
    
    \caption{Representative diagrams of the leading order contribution to the Wilson coefficient of $S_{2,\mu\nu}$ (first diagram), $S_{3,4,7,\mu\nu}$ (second diagram) and $S_{5,\mu\nu}$ (third diagram) in the OPE of $\Pi^{\mu_{1}\mu_{2}\mu_{3}}$. The black dot represents creation/annihilation of a line by the background fields in the vacuum.}
    \label{fig:cutQuarkLine}
\end{figure}
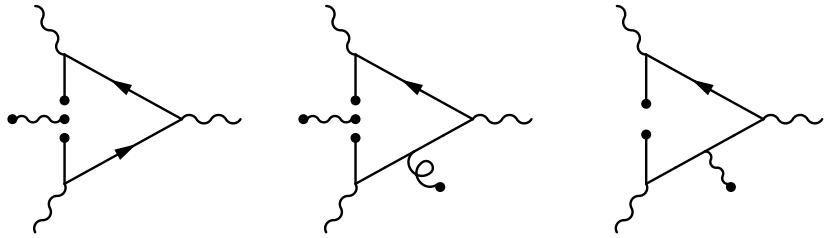

Contributions with one cut quark line and at most one soft gauge boson insertion ($S_{2-5,\mu\nu}$ and $S_{7,\mu\nu}$) are obtained at leading order from the diagrams in figure~\ref{fig:cutQuarkLine}. Their corresponding amplitudes are computed from terms in $\Pi^{\mu_1\mu_2\mu_3}$ that contain two soft quark fields and require no vertices from the Dyson series expansion. Soft gluon or photon insertions on quark hard lines, if necessary, come from propagators of quark fluctuations as seen in the previous section.

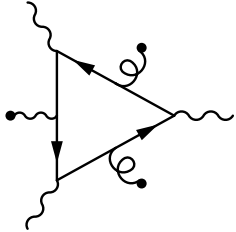
\begin{figure}
    \centering
    \begin{fmffile}{twoCutGluonQuarkLoop}
        \setlength{\unitlength}{0.6cm}\small
        \begin{fmfgraph*}(5,5) 
            \fmfleft{i1,i2,i3}
            \fmfright{o1,o2,o3}         
            \fmf{phantom}{v1,v2,v3,v4,v5}
            \fmf{phantom}{v5,v6,v7,v8,v9}
            \fmf{phantom}{v9,v10,v11,v12,v1}
            \fmf{photon}{i1,v1}
            \fmf{photon}{v9,i3}
            \fmf{photon}{v5,o2}
            \fmffreeze
            \fmf{gluon}{v3,o1}
            \fmf{gluon}{v7,o3}
            \fmf{photon}{v11,i2}
            \fmf{plain}{v1,v3}
            \fmf{fermion}{v3,v5}
            \fmf{plain}{v5,v7}
            \fmf{fermion}{v7,v9}
            \fmf{plain}{v9,v11}
            \fmf{fermion}{v11,v1}
            \fmfset{arrow_len}{3mm}
            \fmfv{decor.shape=circle,decor.size=1mm}{o1,o3,i2}
            \fmffreeze
            \fmfshift{(0.025w,0h)}{i2}
            \fmfshift{(-0.3w,-0.2h)}{o3}
            \fmfshift{(-0.3w,0.2h)}{o1}
        \end{fmfgraph*}
    \end{fmffile}
    \caption{Representative diagram of the leading order contribution to the Wilson coefficient of $S_{6,\mu\nu}$ in the OPE of $\Pi^{\mu_{1}\mu_{2}\mu_{3}}$. The black dot represents creation/annihilation of a line by the background fields in the vacuum.}
    \label{fig:twoCutGluonQuarkLoop}
\end{figure}

Let us now consider the operator with two cut gluon lines, that is, $S_{6,\mu\nu}$. Diagrams contributing to this operator are very similar to the quark loop of $S_{1,\mu\nu}$, but they have two soft gluon insertions (see figure~\ref{fig:twoCutGluonQuarkLoop}). As with the first quark loop, these insertions must be permuted in all possible ways to obtain the full contribution.

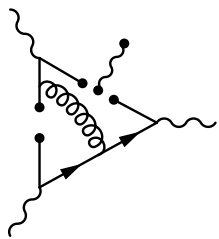
\begin{figure}
    \centering
    \begin{tabular}{ccc}
        \begin{fmffile}{twoCutQuarkLines}
        \setlength{\unitlength}{0.6cm}\small
        \begin{fmfgraph*}(5,5) 
            \fmfleft{i1,i2}
            \fmfright{o1,o2,o3}         
            \fmf{phantom}{v1,v2,v3,v4,v5}
            \fmf{phantom}{v5,v6,v7,v8,v9}
            \fmf{phantom}{v9,v10,v11,v12,v1}
            \fmf{photon}{i1,v1}
            \fmf{photon}{v9,i2}
            \fmf{photon}{v5,o2}
            \fmffreeze
            \fmf{gluon,wiggly_len=7mm,right=0.2,tension=0.8}{v3,v11}
            \fmf{photon}{v7,o3}
            \fmf{fermion}{v1,v3}
            \fmf{fermion}{v3,v5}
            \fmf{plain}{v5,v6}
            \fmf{plain}{v8,v9,v10}
            \fmf{plain}{v12,v1}
            \fmfset{arrow_len}{3mm}
            \fmfv{decor.shape=circle,decor.size=1mm}{v6,v7,v8,v10,v12,o3}
            \fmffreeze
            \fmfshift{(0,-0.075h)}{v10}
            \fmfshift{(0,0.075h)}{v12}
            \fmfshift{(-0.3w,-0.15h)}{o3}
            \fmfshift{(0.060w,-0.040h)}{v8}
            \fmfshift{(-0.060w,0.030h)}{v6}
            \fmfshift{(0,0.160h)}{v11}
            \fmfset{curly_len}{2mm}
        \end{fmfgraph*}
        \end{fmffile}    
    \end{tabular}
    \caption{Representative diagram of the leading order contribution to the Wilson coefficient of $S_{8,\mu\nu}$
    in the OPE of $\Pi^{\mu_{1}\mu_{2}\mu_{3}}$. The black dot represents creation/annihilation of a line by the background fields in the vacuum.}
    \label{fig:twoCutQuarkLines}
\end{figure}

Let us now consider operators with four quark background insertions ($S_{8,\mu\nu}$). Diagrams that contribute to the Wilson coefficients of this operator correspond to the quark loop with two cut quark lines, therefore the diagram is divided in two parts, which have to be connected by a gluon (see figure~\ref{fig:twoCutQuarkLines}). There are six different ways in which the virtual gluon line can connect the two parts of the diagram and all have to be accounted for. The corresponding two gluon--quark vertices are responsible for the $\alpha_{S}$ coefficient of $S_{8,\mu\nu}$.

Except for $S_{1,\mu\nu}$, contributions from all operators to $\Pi_{F}^{\mu_{1}\mu_{2}\mu_{3}\mu_{4}\mu_{5}}$ depend on the susceptibilities $X_{i}^{S}$. By definition these are non--perturbative quantities which are usually computed either by lattice, models and/or educated guesses. The most well--known one is $X_{5}$,\footnote{Note that we have suppressed the $S$ index. This was done because, as we will see in the next section, the elements of the OPE need renormalization and therefore a new set of susceptibilities $X_{i}$ is defined in terms of the renormalized operators.} because it is related to the quark condensate which is a common subject of study in lattice computations. A more recent version of the review cited in~\cite{Bijnens2020} can be found in~\cite{LatticeReview2021}, where figure~14, table~22 and references therein represent a thorough compilation of results for the quark condensate. The rest of the susceptibilities are not so well--known and their numerical values are estimated in~\cite{Bijnens2020} by a combination of models and educated guesses.

Up to this point we have presented all unrenormalized Wilson coefficients associated with operators in~\eqref{eq:OPEelements} and, more importantly, their contribution to $\partial^{\mu_{5}}\Pi^{\mu_{1}\mu_{2}\mu_{3}\mu_{4}}$. Computation of the Wilson coefficients is however not yet complete, for renormalization of the OPE elements has not been taken into account. In contrast to the usual situation in perturbative computations, we have not encountered ultraviolet divergences in the Wilson coefficients of this section. In fact, except for $S_{1,\mu\nu}$ and $S_{6,\mu\nu}$ all of their leading order contributions are at tree level. As we will see in the next sections for the quark loop, the Wilson coefficients of these two operators, although finite, have infrared contributions that are renormalized by the quark masses. Such singularities scale as logarithms and negative powers of $m_{f}$. These singular terms are problematic in a twofold way. From a computational perspective these singular factors may spoil convergence of the perturbative computation when the momenta of the process, namely $Q_{i}$, get much bigger that the mass scale of the quarks, which is actually our situation. From a conceptual point of view it is also questionable to have Wilson coefficients with infrared contributions: in the OPE framework they are meant to represent the contribution from the parts of the diagram through which the external very high momenta travel. In the next section we will present how renormalization of the the product of background fields ``cures'' these infrared divergences and thus completes the separation of low and high energy contributions of the OPE.

\section{OPE of $\Pi^{\mu_{1}\mu_{2}\mu_{3}\mu_{4}}$ in an electromagnetic background field: Renormalization}

In this section we will present the renormalization program for the operators that form the OPE for $\Pi^{\mu_{1}\mu_{2}\mu_{3}}$ in the $\overline{MS}$ scheme.

\begin{figure}
    \centering
    \begin{fmffile}{infraredDivergence}
        \setlength{\unitlength}{0.75cm}\small
        \begin{fmfgraph*}(5,5) 
            \fmfleft{i1,i2}
            \fmfright{o1,o2,o3}         
            \fmf{phantom,tension=0.5}{v1,v2,v3,v4,v5}
            \fmf{phantom,tension=0.5}{v5,v6,v7,v8,v9}
            \fmf{phantom,tension=0.5}{v9,v10,v11,v12,v1}
            \fmf{photon}{i1,v1}
            \fmf{photon}{v9,i2}
            \fmf{photon}{v5,o2}
            \fmffreeze
            \fmf{fermion}{v1,v5,v9}
            \fmf{plain}{v9,v10,v13}
            \fmf{plain}{v14,v12,v1}
            \fmf{photon}{v11,o3}
            \fmfset{arrow_len}{3mm}
            \fmfv{decor.shape=circle,decor.size=1mm}{v10,v11,v12,o3}
            \fmfv{decor.shape=circle,decor.filled=shaded,decor.size=5mm}{v13,v14}
            \fmffreeze
            \fmfshift{(0,-0.12h)}{v10}
            \fmfshift{(0,0.12h)}{v12}
            \fmfshift{(0,0.03h)}{v13}
            \fmfshift{(0,-0.03h)}{v14}
            \fmfshift{(-w,-0.5h)}{o3}
        \end{fmfgraph*}
    \end{fmffile}
    
    \caption{Diagram with infrared divergences affecting the Wilson coefficient of $S_{2,\mu\nu}$ in the OPE of $\Pi^{\mu_{1}\mu_{2}\mu_{3}}$. The black dot represents creation/annihilation of a line by the background fields in the vacuum. The shaded blob represents self--energy corrections to the soft quark line.}
    \label{fig:infraredDivergence}
\end{figure}
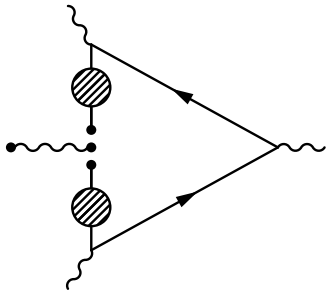

The Wilson coefficients that were presented in the previous section are UV finite at the computed order, but they do have infrared divergent terms such as $1/m_{f}^{2}$ and $\ln\{Q^{2}_{i}/m_{f}^{2}\}$ that are regularized by the quarks masses. These terms may spoil the convergence of the perturbative expansion. Moreover, the Wilson coefficients should not have infrared contributions in the first place, therefore it should be possible to safely to compute them in the massless quarks limit. There is an additional kind of low energy effects that may affect Wilson coefficients: the ones arising from diagrams where soft quark and gluon\footnote{This does not apply for photon soft lines since we do not consider photon fluctuations.} lines receive self--energy corrections. For example, figure~\ref{fig:infraredDivergence} shows how such divergences can arise in diagrams that contribute to the Wilson coefficient of $S_{2,\mu\nu}$. The gray blob of figure~\ref{fig:infraredDivergence} involves a perturbative series in $\alpha_{S}$ at zero momentum, which of course does not converge since the processes it is trying to describe belong to the non--perturbative domain. These diagrams however do not appear at the order we are considering and therefore we will not discuss them in detail this section.

The prescription of a renormalization program in this context is not very surprising considering that the OPE is built from composite operators which are known to require counterterms of their own to be renormalized. Therefore, one could expect that after carrying out the renormalization of these composite operators the infrared divergences of the Wilson coefficients are cancelled. In~\cite{Bijnens2020}, renormalization was done in the full--field framework by dressing operators $S_{i.\mu\nu}$, that is, by inserting them into the Dyson series expansion. 

However, from the point of view of the background field method that we have followed in the previous sections, these operators are simply products of classical background fields, so at first it may seem rather odd to assert that they require renormalization. Nevertheless, this is not at all surprising if we trace the step back to the analysis at the start of the previous section. There it was argued that Green functions in the ``original'' theory could be obtained from the background theory by functionally differentiating vacuum--to--vacuum amplitudes with respect to background fields. Thus, products of background fields were converted into Green functions. Insertions of composite operators however work differently, because they have their own source terms in the generating functional of the background theory. In contrast to the fields, whose source terms only involved the fluctuation part of the field, namely $J_n\phi'_n$, the insertion of a composite operator $O_i$ involves the full fields, that is, $L_iO_i(\{\phi+\phi'\})$, where $L_i$ is the source. This is required in order for the relation $H[0,\phi_n] =\Gamma[\phi_n]$ between the effective action and its background counterpart to remain valid. For example, the insertion of $S_{2,\mu\nu}$ in the background framework actually involves:
\begin{equation}\label{eq:fullCompositeOperator}
    (\overline{\psi} +\overline{\psi}')\sigma_{\mu\nu}(\psi+\psi')
\end{equation}
instead of just:
\begin{equation}\label{eq:backgroundCompositeOperator}
    \overline{\psi}\sigma_{\mu\nu}\psi\;.
\end{equation}
This does not mean that computation of Wilson coefficients of the previous section is wrong, for the operator in~\eqref{eq:backgroundCompositeOperator} is the one that is related to the matrix element we are interested in. We will later see that they can be related to the renormalized composite operator in a  straightforward manner. Instead, this means that operator mixing is naturally ingrained in the background field formalism. This also means that it is the operator in~\eqref{eq:fullCompositeOperator} the one that needs renormalization regardless of whether it ends up curing divergences in the Wilson coefficients of the operator in~\eqref{eq:backgroundCompositeOperator} as well. 

After justifying the need for renormalization in the background theory, now we can proceed to apply to the operators in~\eqref{eq:OPEelements}. In our renormalization program composite operators are inserted in Green's functions, which then are computed using dimensional regularization to preserve gauge invariance. Finally, the relation between singular terms and counterterms is defined by modified minimal subtraction $\overline{MS}$. As was mentioned earlier, counterterms required for renormalization of composite operators are a linear combination that includes other composite operators with singular coefficients. This is referred to as ``operator mixing''. For simplicity, we will compute Green's functions with an insertion of each composite operator and no other fields involved, for otherwise additional singularities renormalized by the lagrangian's counterterms would appear. However, not \emph{any} operators can mix under renormalization. Only operators with the same quantum numbers can. Furthermore, since the background field method does not break background gauge invariance, then this means that mixing also respects gauge invariance\footnote{Only if the Green's function in which it is inserted has only background quark and gauge fields.}~\cite{ZuberCompositeOperatorRenormalization1,ZuberCompositeOperatorRenormalization2,ZuberCompositeOperatorRenormalization3}. In the end, this means that the operators that form the OPE of $\Pi^{\mu_{1}\mu_{2}\mu_{3}}$ mix among themselves under renormalization 

In our context this means that the renormalization of the elements of our OPE will have the following shape:
\begin{equation}
    \boldsymbol{Q}^{0}_{\mu\nu} = \hat{Z}\boldsymbol{Q}_{\mu\nu} \hspace{1cm}
    Q^{0}_{i,\mu\nu} = Q^{0}_{i,\mu\nu}(\psi+\psi',a^{a\mu}+A^{'a\mu},a^{\mu})\;,
\end{equation}
where $\boldsymbol{Q}^{0}_{\mu\nu}$ represents the vector whose components are the bare elements of the OPE of~\eqref{eq:OPEelements} and it is a function of the full fields, that is, the sum of the background and fluctuation parts. $\boldsymbol{Q}_{\mu\nu}$ contains its renormalized versions. Consequently, $\hat{Z}$ is a $8\times8$ matrix containing constants with regularized ultraviolet divergences. As we will see in the following, the vector of operators $\boldsymbol{Q}_{\mu\nu}$ does not coincide with the $\boldsymbol{S}_{\mu\nu}$ that we defined earlier, but they are related by a constant matrix whose elements contain regularized infrared divergences. Consequently one can define:
\begin{equation}\label{eq:mixingMatrices}
\begin{split}
    \boldsymbol{Q}_{\mu\nu} &=\hat{U}\boldsymbol{S}_{\mu\nu} \implies
    \boldsymbol{Q}^{0}_{\mu\nu} = \hat{Z}\hat{U}\boldsymbol{S}_{\mu\nu}\;.
\end{split}
\end{equation}

Renormalization is used to separate contributions coming from different energy scales and in this case such objective is achieved since the elements in $\hat{U}$ are just the required ones to cancel the infrared contributions of the Wilson coefficients. Furthermore, it is important to note that we could not have avoided singular terms in the Wilson coefficients by using $\boldsymbol{Q}^{0}_{\mu\nu}$ instead of $\boldsymbol{S}_{\mu\nu}$, since in such case we would have traded infrared for ultraviolet contributions. Instead it is necessary to use renormalization to successfully separate low and high energy contributions and find $\boldsymbol{Q}_{\mu\nu}$. The renormalized Wilson coefficients $\boldsymbol{C}$ are free of infrared contributions and are defined in terms of the bare ones $\boldsymbol{C}_{S}$ as:
\begin{align}
    \Pi^{\mu_{1}\mu_{2}\mu_{3}} &=\boldsymbol{C}_{S}^{\mu_{1}\mu_{2}\mu_{3}\mu_{4}\mu_{5}}\cdot\langle0|\boldsymbol{S}_{\mu\nu}|\gamma\rangle \equiv\boldsymbol{C}^{\mu_{1}\mu_{2}\mu_{3}\mu_{4}\mu_{5}}\cdot\langle0|\boldsymbol{Q}_{\mu\nu}|\gamma\rangle\\
    \implies \boldsymbol{C}^{\mu_{1}\mu_{2}\mu_{3}\mu_{4}\mu_{5}} &= (\hat{U}^{-1})^{T}\boldsymbol{C}_{S}^{\mu_{1}\mu_{2}\mu_{3}\mu_{4}\mu_{5}}\;.
\end{align}
Note that renormalized susceptibilities $\boldsymbol{X}$ can also be defined for $\boldsymbol{Q}_{\mu\nu}$ and can be related to the unrenormalized ones $\boldsymbol{X}_{i}^{S}$ in a straightforward way: 
\begin{equation}
    \boldsymbol{Q}_{\mu\nu} \equiv \boldsymbol{X}F_{\mu\nu} \implies \boldsymbol{X} =\hat{U}\boldsymbol{X}^{S}\;.
\end{equation}

As always, it is of course necessary to specify an order at which renormalization constants will be truncated. The appearance of non--perturbative matrix elements in the OPE introduces non--perturbative expansion parameters ($\Lambda_{QCD}/Q$) besides the perturbative ones ($g_{S}$ and $e$). In terms of the latter, the cut--off is placed at $O(e^{-1}g^{2}_{S})$. With respect to the former we have $O(\Lambda_{QCD}^{6}/Q^{6})$, which, in addition to gauge invariance conservation of the background field theory, essentially means that operators $S_{i,\mu\nu}$ only mix among themselves. Since we are considering the three lightest quarks, its masses' effects can be regarded as perturbations as well therefore introducing another expansion parameter $m_{f}/\Lambda_{QCD}$. Nevertheless, we can obtain the full dependence of the mixing coefficients on the quarks' masses. 

It is important to note that perturbative and non--perturbative parameters must not be regarded independently: the mixing matrix $\hat{U}$ is meant to modify the Wilson coefficients as shown in the previous equation, therefore each element must be expanded up to the order of the Wilson coefficients which it modifies. This introduces an interplay between the dimension of the operators that are mixing and the order of their Wilson coefficients. The precise implications of this assertion should become clearer throughout the rest of this section. 

Now we are ready to put the renormalization program we just described to use. For $S_{1,\mu\nu}$ renormalization is at its simplest. Since the photon field does not have quantum fluctuations, then $Q_{1,\mu\nu}^{0}$ is just equal to $S_{1,\mu\nu}$ and hence it cannot mix with any other operator.

\subsection{Mixing of the $Q_{2,\mu\nu}^{0}$ operator}

The first and most non--trivial case is $Q_{2,\mu\nu}$. A Green's function with a full--field insertion of this composite operator is given by:
\begin{equation}\label{eq:Q2Decomposition}
    \langle0|Q_{2,\mu\nu}^{0}|\gamma\rangle = \overline{\psi}\sigma_{\mu\nu}\psi +\langle0|\overline{\psi}\sigma_{\mu\nu}\psi'|\gamma\rangle +\langle0|\overline{\psi}'\sigma_{\mu\nu}\psi|\gamma\rangle +\langle0|\overline{\psi}'\sigma_{\mu\nu}\psi'|\gamma\rangle\;,
\end{equation}
where we are evaluating the matrix element of a Heisenberg operator and therefore the Dyson series of interaction vertices has to be inserted. Mixing with $S_{1,\mu\nu}$ can only come from the fourth term and it requires the contraction of both quark fluctuations and a soft insertion of the photon field in the resulting propagator. As we will see later in this section, further soft insertions lead to mixing with other operators. Since the Wilson coefficient of $S_{2,\mu\nu}$ is $O(e^{-1}g_{S}^{0})$ and the mixing coefficient is $O(e^{0}g_{S}^{0})$, then the net mixing contribution is of order $O(e^{-1}g_{S}^{0})$, already the same as the Wilson coefficient of $S_{1,\mu\nu}$. Therefore we can cut off the mixing coefficient at this point. The result is: 
\begin{align}\nonumber
    \langle0|\overline{\psi}'\sigma_{\mu\nu}\psi'|\gamma\rangle &= -\Tr\{S^{ff}_{ll}(0,0)\sigma_{\mu\nu}\} =\frac{e\mu^{2\epsilon}}{2}f^{\mu_{1}\nu_{1}}\int\frac{d^{d}p_{1}}{(2\pi)^{d}}\frac{\partial}{\partial q^{\mu_{1}}_{1}}\Tr\{S^{0}_{p_{1}}\gamma_{\nu_{1}}S^{0}_{p_{1}+q_{1}}\sigma_{\mu\nu}\}\Bigg|_{q_{1}=0}\\\nonumber
    &=4iN_{c}e\mu^{2\epsilon}e_fm_{f}f_{\mu\nu}\int\frac{d^{d}p_{1}}{(2\pi)^{d}}\frac{1}{[p_{1}^{2}-m^{2}_{f}]^{2}}\\\label{eq:preliminarS2S1Mixing}
    &=-\frac{N_{c}ee_f}{4\pi^{2}}m_{f}f_{\mu\nu}\Gamma(\epsilon)\Big(\frac{4\pi\mu^{2}}{m^{2}_{f}}\Big)^{\epsilon}\;,
\end{align}
where $d\equiv 4-2\epsilon$ is the shifted dimension, $\mu$ is the mass parameter that carries the mass dimension of $e$ in the regularized theory and we have used the well--known formula: 
\begin{equation}\label{eq:scalarLoopIntegralIdentity}
    \int\frac{d^{d}p}{(2\pi)^{d}}\frac{1}{[p^{2}-\Delta]^{n}} =\frac{(-1)^{n}}{(4\pi)^{d/2}}i\frac{\Gamma(n-\frac{d}{2})}{\Gamma(n)}\Big(\frac{1}{\Delta}\Big)^{n-\frac{d}{2}}\;.
\end{equation}
Note that the derivation of this formula involves a Wick rotation of the integration variable, therefore it is necessary to ensure that $\Delta$ is positive, as it is of course in~\eqref{eq:preliminarS2S1Mixing}. Otherwise the integrand acquires a discontinuity when the norm of the spatial momentum $\boldsymbol{p}^{2}$ becomes smaller than the absolute value $|\Delta|$. This can be accounted for by giving the pole a vanishing imaginary part: $\Delta\longrightarrow\Delta-i0^{+}$, which plays a role analogous to the Feynman prescription for free propagators. 

Turning back to~\eqref{eq:preliminarS2S1Mixing}, it is necessary to expand the result around $\epsilon=0$ to expose the singular terms. The result  when one discards terms that vanish when $\epsilon\rightarrow0$ is:
\begin{equation}
    \langle0|\overline{\psi}'\sigma_{\mu\nu}\psi'|\gamma\rangle =-\frac{N_{c}}{4\pi^{2}}m_{f}\Big(\frac{1}{\hat{\epsilon}}+\ln{\Big\{\frac{\mu^{2}}{m^{2}_{f}}\Big\}}\Big)S_{1,\mu\nu}\;.
\end{equation}
where $\gamma_{E}\approx0.5772$ is the Euler--Mascheroni constant and we have used the expansion of the gamma function around $\epsilon\rightarrow0$:
\begin{equation}\label{eq:gammaExpansion}
    \Gamma(\epsilon) = \frac{1}{\epsilon} -\gamma_{E} +O(\epsilon)\;.
\end{equation}
As was mentioned previously, we define the singular term to be subtracted in the $\overline{MS}$ scheme:
\begin{equation}
    \frac{1}{\hat{\epsilon}} \equiv \frac{1}{\epsilon}+\ln{\{4\pi\}} -\gamma_{E}.
\end{equation}
This mixing coefficient states two facts about the insertions of the operator $Q_{2,\mu\nu}$: 1) a part of their singular ultraviolet behaviour can be effectively renormalized by mixing with $S_{1,\mu\nu}$ and 2) they involve a low--energy contribution from $S_{1,\mu\nu}$, which is represented by the mass logarithm. This is an explicit example of the appearance of the $\hat{Z}$ and $\hat{U}$ matrices that were defined in~\eqref{eq:mixingMatrices}.

The leading order contribution to the mixing of $Q_{2,\mu\nu}$ with $S_{3,4,5,\mu\nu}$ and $S_{7,\mu\nu}$ is given by the second and third terms at the right hand side of~\eqref{eq:Q2Decomposition}. They require the introduction two quark--gluon vertices. However, Wilson coefficients of these operators are all $O(e^{-1})$, so the mixing coefficient cannot receive perturbative corrections from interaction vertices at the relevant order. The same analysis of course applies the other way around, thus, $S_{2-5,\mu\nu}$ and $S_{7,\mu\nu}$ do not mix with each other at the order of our computation.

The mixing of $Q_{2,\mu\nu}$ with $S_{6,\mu\nu}$ can of course only come from the fourth term in~\eqref{eq:Q2Decomposition}. Just as with the mixing with $S_{1,\mu\nu}$ we can contract both fluctuations without introducing interaction vertices and then three soft insertions can be performed on the propagator. The corresponding result is:
\begin{align}\nonumber
    \langle0|\overline{\psi}'\sigma_{\mu\nu}\psi'|\gamma\rangle &=-\Tr\{S^{ff}_{ll}(0,0)\sigma_{\mu\nu}\}\\\nonumber
    &=-\frac{ee_{f}g^{2}_{S}}{8}\Tr\{t^{a}t^{b}\}\Big(f^{\mu_{1}\nu_{1}}f^{a\mu_{2}\nu_{2}}f^{b\mu_{3}\nu_{3}} +f^{a\mu_{1}\nu_{1}}f^{\mu_{2}\nu_{2}}f^{b\mu_{3}\nu_{3}} +f^{a\mu_{1}\nu_{1}}f^{b\mu_{2}\nu_{2}}f^{\mu_{3}\nu_{3}}\Big)\\\nonumber
    &\hspace{-0.5cm}\times\int\frac{d^{4}p}{(2\pi)^{4}}\frac{\partial}{\partial q^{\mu_{1}}_{1}}\frac{\partial}{\partial q^{\mu_{2}}_{2}}\frac{\partial}{\partial q^{\mu_{3}}_{3}}\Tr\{S^{0}_{p}\gamma_{\nu_{1}} S^{0}_{p+q_{1}}\gamma_{\nu_{2}}S^{0}_{p+q_{1}+q_{2}}\gamma_{\nu_{3}}S^{0}_{p+q_{1}+q_{2}+q_{3}}\sigma_{\mu\nu}\}\Big|_{q_{1,2,3}=0}\Bigg)\\
    &=-\frac{1}{72 m_{f}^{3}}S_{6,\mu\nu}\;.
\end{align}
Finally, the mixing coefficient of $Q_{2,\mu\nu}$ with $S_{8,\mu\nu}$ is evidently beyond the perturbative order that we are interested in.

\subsection{Mixing of the $Q_{3,\mu\nu}^{0}$ operator}

For this operator we have: 
\begin{equation}\label{eq:Q3Mixing}
\begin{split}
    \frac{1}{g_S}\langle
    0|Q_{3,\mu\nu}^{0}|\gamma\rangle &= -\langle0|\overline{\psi}t^{a}G^{a}_{\mu\nu}(a+A')\psi|\gamma\rangle-\langle
    0|\overline{\psi}t^{a}G^{a}_{\mu\nu}(a+A')\psi'|\gamma\rangle\\
    &-\langle
    0|\overline{\psi}'t^{a}G^{a}_{\mu\nu}(a+A')\psi|\gamma\rangle -\langle
    0|\overline{\psi}'t^{a}G^{a}_{\mu\nu}(a+A')\psi'|\gamma\rangle\;,
\end{split}
\end{equation}
where $G^{a}_{\mu\nu}(a+A')$ is the gluon field strength tensor separated between fluctuation and background parts, namely: 
\begin{equation}
    G^{a}_{\mu\nu}(a+A') =f_{\mu\nu}^{a} +\overline{D}_{\mu}A^{'a}_{\nu} -\overline{D}_{\nu}A^{'a}_{\mu} +g_{S}f^{abc}A^{'b}_{\mu}A^{'c}_{\nu}\;,
\end{equation}
as it was introduced in the previous sections. All the Wilson coefficients of the previous section were computed up to $O(e^{-1}g_{S}^{0})$, therefore no terms with gluon quantum fluctuations give relevant contributions to the mixing and we can replace $G^{a}_{\mu\nu}\longrightarrow f^{a}_{\mu\nu}$ in $\langle Q_{3,\mu\nu}^{0}\rangle$. This means that at the order that is relevant for us $Q_{3,\mu\nu}^{0}$ can only mix with operators that have at least one soft insertion of $f^{a\mu\nu}$. From~\eqref{eq:OPEelements} the only compatible one is $S_{6,\mu\nu}$.\footnote{In principle $S_{7,\mu\nu}$ is compatible as well, but to obtain terms with covariant derivatives it is necessary to introduce additional $g_{S}^{2}$ factors that take the mixing beyond the established cutoff.} The leading order contribution to that mixing coefficient is given by the term in~\eqref{eq:Q3Mixing} with two quark fluctuations when they are contracted with each other and one soft gluon and one soft photon insertion are performed on the resulting quark propagator:
\begin{equation}
\begin{split}
    -\langle
    0|\overline{\psi}'t^{a}f^{a}_{\mu\nu}\psi'|\gamma\rangle &=g_{S}\Tr\{S^{ff}_{kl}(0,0)\}t_{lk}^{a}f^{a}_{\mu\nu}\\
    &=\frac{1}{4}ee_{f}f^{a}_{\mu\nu}g_{S}^{2}\Tr\{t^{a}t^{b}\}\Big(f^{\mu_{1}\nu_{1}}f^{b\mu_{2}\nu_{2}} +f^{b\mu_{1}\nu_{1}}f^{\mu_{2}\nu_{2}}\Big)\\
    &\hspace{1cm}\times\int\frac{d^{4}p}{(2\pi)^{4}}\frac{\partial}{\partial q^{\mu_{1}}_{1}}\frac{\partial}{\partial q^{\mu_{2}}_{2}}\Tr\Big\{S^{0}_{p}\gamma_{\nu_{1}}S^{0}_{p+q_{1}}\gamma_{\nu_{2}}S^{0}_{p+q_{1}+q_{2}}\Big\}\Bigg|_{q_{1,2}=0}\\
    &=\frac{1}{36m_{f}}S_{6,\mu\nu}\;.
\end{split}
\end{equation}

\subsection{Mixing of the $Q_{4,\mu\nu}^{0}$ operator}

For $Q_{4,\mu\nu}^{0}$ we have: 
\begin{equation}\label{eq:Q4Mixing}
\begin{split}
    \frac{1}{g_S}\langle
    0|Q_{4,\mu\nu}^{0}|\gamma\rangle &=- \langle0|\overline{\psi}t^{a}\overline{G}^{a}_{\mu\nu}(a+A')\gamma_{5}\psi|\gamma\rangle -\langle
    0|\overline{\psi}t^{a}\overline{G}^{a}_{\mu\nu}(a+A')\gamma_{5}\psi'|\gamma\rangle\\
    &-\langle
    0|\overline{\psi}'t^{a}\overline{G}^{a}_{\mu\nu}(a+A')\gamma_{5}\psi|\gamma\rangle -\langle
    0|\overline{\psi}'t^{a}\overline{G}^{a}_{\mu\nu}(a+A')\gamma_{5}\psi'|\gamma\rangle\;,
\end{split}
\end{equation}
where $\overline{G}^{a\mu\nu}(a+A')$ is the dual of $G^{a\mu\nu}(a+A')$, that is: $\overline{G}^{a\mu\nu} =\frac{i}{2}\epsilon^{\mu\nu\alpha\beta}G^{a}_{\alpha\beta}$. The analysis of the relevant mixing coefficients at the order of interest is essentially the same as for $Q_{3,\mu\nu}^{0}$, therefore there is mixing only with $S_{6,\mu\nu}^{0}$ and the corresponding coefficient is:
\begin{align}\nonumber
    -\langle0|\overline{\psi}'t^{a}\overline{f}^{a}_{\mu\nu}\gamma_{5}\psi'|\gamma\rangle &=g_{S}\Tr\{S^{ff}_{kl}(0,0)\gamma_{5}\}t_{lk}^{a}\overline{f}^{a}_{\mu\nu}\\\nonumber
    &=\frac{1}{4}ee_{f}\overline{f}^{a}_{\mu\nu}g_{S}^{2}\Tr\{t^{a}t^{b}\}\Big(f^{\mu_{1}\nu_{1}}f^{b\mu_{2}\nu_{2}} +f^{b\mu_{1}\nu_{1}}f^{\mu_{2}\nu_{2}}\Big)\\\nonumber
    &\hspace{1cm}\times\int\frac{d^{4}p}{(2\pi)^{4}}\frac{\partial}{\partial q^{\mu_{1}}_{1}}\frac{\partial}{\partial q^{\mu_{2}}_{2}}\Tr\Big\{S^{0}_{p}\gamma_{\nu_{1}}S^{0}_{p+q_{1}}\gamma_{\nu_{2}}S^{0}_{p+q_{1}+q_{2}}\gamma_{5}\Big\}\Bigg|_{q_{1,2}=0}\\
    &=\frac{1}{24m_{f}}S_{6,\mu\nu}\;.
\end{align}

\subsection{Mixing of the $Q_{5,\mu\nu}^{0}$ operator}

For $Q_{5,\mu\nu}^{0}$ we have: 
\begin{equation}\label{eq:Q5Mixing}
\begin{split}
    \frac{1}{ee_f}\langle
    0|Q_{5,\mu\nu}^{0}|\gamma\rangle &= \langle
    0|\overline{\psi}\psi F_{\mu\nu}|\gamma\rangle +\langle
    0|\overline{\psi}\psi'F_{\mu\nu}|\gamma\rangle +\langle
    0|\overline{\psi}' F_{\mu\nu}\psi|\gamma\rangle +\langle
    0|\overline{\psi}'\psi'F_{\mu\nu}|\gamma\rangle\;.
\end{split}
\end{equation}
In a similar fashion as with $Q_{3,\mu\nu}$ and $Q_{4,\mu\nu}$, the soft photon insertion allows only for mixing with $S_{1,\mu\nu}$ and $S_{6,\mu\nu}$. The leading order contribution to both mixing coefficients is once again given by the term in~\eqref{eq:Q5Mixing} with two quark fluctuations:
\begin{equation}
\begin{split}
    \langle
    0|\overline{\psi}'\psi'ee_{f}F_{\mu\nu}|\gamma\rangle &=-\Tr\{S_{ll}^{ff}(0,0)\}ee_{f}f_{\mu\nu}\;.
\end{split}
\end{equation}
Mixing with $S_{1,\mu\nu}$ is obtained by performing no soft insertions in the quark propagator:
\begin{equation}
\begin{split}
    \langle
    0|\overline{\psi}'\psi'ee_{f}F_{\mu\nu}|\gamma\rangle &=-\frac{m_{f}^{3}}{4\pi^{2}}\Big(\frac{1}{\hat{\epsilon}}+\ln\Big\{\frac{\mu^{2}}{m_{f}^{2}}\Big\}+1\Big)S_{1,\mu\nu}\;,
\end{split}
\end{equation}
As mentioned in~\cite{Bijnens2020}, this mixing coefficient can be used to subtract the low--energy contributions to the $O(m^{4}_{f})$ correction to the massless part of the quark loop. On the other hand, mixing with $S_{6,\mu\nu}$ is obtained by inserting two soft gluons on the quark propagator:
\begin{equation}
\begin{split}
    \langle
    0|\overline{\psi}'\psi'ee_{f}F_{\mu\nu}|\gamma\rangle &=-\frac{1}{4}ee_{f}f_{\mu\nu}g_{S}^{2}\Tr\{t^{a}t^{b}\}f^{a\mu_{1}\nu_{1}}f^{b\mu_{2}\nu_{2}}\int\frac{d^{4}p}{(2\pi)^{4}}\frac{\partial}{\partial q^{\mu_{1}}_{1}}\frac{\partial}{\partial q^{\mu_{2}}_{2}}\Tr\Big\{S^{0}_{p}\gamma_{\nu_{1}}S^{0}_{p+q_{1}}\gamma_{\nu_{2}}S^{0}_{p+q_{1}+q_{2}}\Big\}\Bigg|_{q_{1,2}=0}\\
    &=-\frac{1}{12m_{f}}S_{6,\mu\nu}\;.
\end{split}
\end{equation}

\subsection{Mixing of the $Q_{6,\mu\nu}^{0}$ operator}

For this operator it can be performed the same separation of the gluon field strength tensor that was described for $Q_{3,\mu\nu}$ and $Q_{4,\mu\nu}$. However, since this operator is $O(eg_{S}^{2})$ and all other Wilson coefficients are $O(e^{-1}g_{S}^{0})$, then its mixing coefficients are beyond the perturbative order of the computation.

\subsection{Mixing of the $Q_{7,\mu\nu}^{0}$ operator}

With respect to $Q_{7,\mu\nu}^{0}$ we have: 
\begin{equation}\label{eq:Q7Mixing}
\begin{split}
     \frac{1}{ig_{S}}\langle
    0|Q_{7,\mu\nu}^{0}|\gamma\rangle &=\langle
    0|\overline{\psi}(t^{a}G^{a}_{\mu\lambda}D_{\nu} +D_{\nu}t^{a}G^{a}_{\mu\lambda})\gamma^{\lambda}\psi |\gamma\rangle +\langle
    0|\overline{\psi}(t^{a}G^{a}_{\mu\lambda}D_{\nu} +D_{\nu}t^{a}G^{a}_{\mu\lambda})\gamma^{\lambda}\psi'|\gamma\rangle -(\mu\longleftrightarrow\nu)\\
    &\hspace{-0.25cm}+\langle
    0|\overline{\psi}'(t^{a}G^{a}_{\mu\lambda}D_{\nu} +D_{\nu}t^{a}G^{a}_{\mu\lambda})\gamma^{\lambda}\psi'|\gamma\rangle +\langle
    0|\overline{\psi}'(t^{a}G^{a}_{\mu\lambda}D_{\nu} +D_{\nu}t^{a}G^{a}_{\mu\lambda})\gamma^{\lambda}\psi'|\gamma\rangle -(\mu\longleftrightarrow\nu)\;,
\end{split}
\end{equation}
where this time both the field strength tensor and the covariant derivative are implicitly divided into background and fluctuation parts. Since this operator is already $O(e^{0}g_{S})$, it can only mix with $S_{3,\mu\nu}$, $S_{4,\mu\nu}$ and $S_{6,\mu\nu}$. Mixing with the first two is relevant only up to terms that do not introduce higher orders of $g_{S}$, therefore only the first term in~\eqref{eq:Q7Mixing} may contribute. However, this would give just the background version $S_{7,\mu\nu}$ of $Q_{7,\mu\nu}$. With respect to the mixing with $S_{6,\mu\nu}$, the leading contribution comes from the last term in~\eqref{eq:Q7Mixing} when both quark fluctuations are contracted and their corresponding propagator has a soft gluon insertion. For this mixing only terms that introduce at most another order of $g_{S}$ are relevant, therefore only the fully background part of the $t^{a}G^{a}_{\mu\lambda}D_{\nu}$ terms is relevant. In the end, the mixing coefficient between $Q_{7,\mu\nu}$ and $S_{6,\mu\nu}$ is: 
\begin{align}\nonumber
    ig_{S}\langle
    0|\overline{\psi}'(t^{a}G^{a}_{\mu\lambda}D_{\nu} +D_{\nu}t^{a}G^{a}_{\mu\lambda})\gamma^{\lambda}\psi'|\gamma\rangle &=-2ig_{S}t^{a}_{lk}f^{a}_{\mu\lambda}\overline{D}_{\nu}\Tr\{S^{ff}_{kl}\gamma^{\lambda}\}\\\nonumber
    &\hspace{-3cm}=\frac{i}{2}\mu^{2\epsilon}ee_{f}g^{2}_{S}\Tr\{t^{a}t^{b}\}f^{a}_{\mu\lambda}\Big(f^{\mu_{1}\nu_{1}}f^{b\mu_{2}\nu_{2}} +f^{b\mu_{1}\nu_{1}}f^{\mu_{2}\nu_{2}}\Big)\\\nonumber
    &\hspace{-2cm}\times\int\frac{d^{d}p}{(2\pi)^{d}}ip_{\nu}\frac{\partial}{\partial q^{\mu_{1}}_{1}}\frac{\partial}{\partial q^{\mu_{2}}_{2}}\Tr\Big\{S^{0}_{p}\gamma_{\nu_{1}}S^{0}_{p+q_{1}}\gamma_{\nu_{2}}S^{0}_{p+q_{1}+q_{2}}\gamma^{\lambda}\Big\}\Bigg|_{q_{1,2}=0}\\
    &\hspace{-3cm}=-\frac{1}{6}\Big(\frac{1}{\hat{\epsilon}}+\ln\Big\{\frac{\mu^{2}}{m_{f}^{2}}\Big\} +\frac{7}{6}\Big)S_{6,\mu\nu}\;,
\end{align}
where we have implicitly included the ($\mu\longleftrightarrow\nu$) permutation.

\subsection{Mixing of the $Q_{8,\mu\nu}^{0}$ operator}

Finally, it is worth mentioning that operator $Q_{8,\mu\nu}^{0}$ is already $O(e^{0}g_{S}^{2})$ and therefore its mixing coefficients are not relevant for the Wilson coefficients at the computed order. Note that the order $O(e^{0}g_{S}^{2})$ of the operator $Q_{8,\mu\nu}^{0}$ is not arbitrary, but rather is due to the fact that we need to introduce two interaction vertices from the Dyson series to complete the two cut quark lines (see figure~\ref{fig:twoCutQuarkLines}).

Up to this point we have followed~\cite{Bijnens2020} to present the computation of the HLbL tensor $\Pi^{\mu_{1}\mu_{2}\mu_{3}\mu_{4}}$ in the high energy regime via an OPE in the presence of an electromagnetic background field. We have generalized such approach to include gluon and quark background fields as well. In the OPE there is a separation of perturbative contributions (which are bigger) and non--perturbative ones coming matrix elements of strongly interacting operators. All contributions are computed at leading order up to dimension six operators. Infrared contributions to the Wilson coefficients of the OPE represented both conceptual and computational problems, but they were dealt with by performing renormalization of the composite operators of the OPE. The need for a renormalization program in the background field method context, when composite operators are represented by products of background classical quark, gluon and photon fields, is not evident and it must be justified. We presented the rationale behind the renormalization scheme and performed all necessary computations within the background field method framework.

\section{Computation of the quark loop by the method of Bijnens}

The main contribution to $\Pi^{\mu_{1}\mu_{2}\mu_{3}\mu_{4}}$ (and, thus, to $a_{\mu}$) comes from the Wilson coefficient of the electromagnetic field--strength tensor $F_{\mu\nu}$,\footnote{After renormalization there are actually contributions from non--perturbative operators. However, such mixing contributions only affect the mass corrections of the quark loop.} which is the quark loop (see figure~\ref{fig:quarkLoop}) and its expression is given in~\eqref{eq:quarkLoop}. Consequently, in the remaining sections we focus on the computation of the quark loop contribution to $a_{\mu}$ from the high energy integration regions of the master formula~\eqref{eq:masterFormula}. From \eqref{eq:aMuHLbLIntermediate} we know exactly how $\partial_{\nu_{4}}\Pi^{\mu_{1}\mu_{2}\mu_{3}\mu_{4}}_{HLbL}$ and thus $\Pi_{F}^{\mu_{1}\mu_{2}\mu_{3}\mu_{4}\nu_{4}}$ contributes to $a_{\mu}$ without recurring to a specific tensor basis. However, it is convenient to express the result in the tensor basis used for the master formula in order to benefit from the Gegenbauer polynomials framework that allowed to simplify a full two loop integral, containing eight integrals, into a threefold one.

In~\cite{Bijnens2020} the computation was performed by applying projectors which extract the relevant contributions to the $\overline{\Pi}_{i}$ of the master formula~\eqref{eq:masterFormula} out of the amplitude
\begin{align}
    \hat{
    \Pi}_{i} &=P_{i}^{\mu'_{1}\mu'_{2}\mu'_{3}\mu'_{4}\nu'_{4}}\Pi_{F\;\mu'_{1}\mu'_{2}\mu'_{3}\mu'_{4}\nu'_{4}}\;.
\end{align}

Some denominator cancellations can be performed on the resulting scalar loop integrals such that they are written in terms of scalar tadpole, self-energy and triangle integrals. Note that no scalar box integrals arise due to the soft $q_{4}\rightarrow$ limit, which guarantees that, after applying the soft derivative, only three different propagators appear in the quark loop. These three scalar master integrals are then expanded as a function of the squared of the infinitesimal (in the considered regime) quark mass $m^{2}_{f}$. Finally, the infrared divergences that appear as $\ln{(Q_{3}^{2}/m_{f}^{2}})$ in the mass--suppressed corrections are cancelled via mixing with $S_{2,\mu\nu}$ as discussed in previous sections. The final result can be written as: 
\begin{align}
    \hat{\Pi}_{m}^{\overline{MS}} &= \hat{\Pi}_{m}^{0} +m_{f}^{2}\hat{\Pi}_{\overline{MS},m}^{m_{f}^{2}} +O(m_{f}^{4})\;,\\\nonumber
    \hat{\Pi}_{m}^{0} &=\frac{N_{c}e_{q}^{4}}{\pi^{2}}\sum_{i,j,k,n}\Big[c^{(m,n)}_{ijk} +f^{(m,n)}_{ijk} F +g^{(m,n)}_{ijk}\ln{\Big(\frac{Q_{2}^{2}}{Q_{3}^{2}}\Big)} +h^{(m,n)}_{ijk}\ln{\Big(\frac{Q_{1}^{2}}{Q_{2}^{2}}\Big)}\Big]\lambda^{-n}Q_{1}^{2i}Q_{2}^{2j}Q_{3}^{2k}\;,\\
    \hat{\Pi}_{\overline{MS},m}^{m_{f}^{2}} &=\frac{N_{c}e_{q}^{4}}{\pi^{2}}\sum_{i,j,k,n}\lambda^{-n}Q_{1}^{2i}Q_{2}^{2j}Q_{3}^{2k}\\\nonumber
    &\times\Big[d^{(m,n)}_{ijk} +p^{(m,n)}_{ijk} F +q^{(m,n)}_{ijk}\ln{\Big(\frac{Q_{2}^{2}}{Q_{3}^{2}}\Big)} +r^{(m,n)}_{ijk}\ln{\Big(\frac{Q_{1}^{2}}{Q_{2}^{2}}\Big)} +s^{(m,n)}_{ijk}\ln{\Big(\frac{Q_{3}^{2}}{\mu^{2}}\Big)}\Big]\;,
\end{align}
where $c^{(m,n)}_{ijk}$, $f^{(m,n)}_{ijk}$, $g^{(m,n)}_{ijk}$, $h^{(m,n)}_{ijk}$, $d^{(m,n)}_{ijk}$, $p^{(m,n)}_{ijk}$, $q^{(m,n)}_{ijk}$, $r^{(m,n)}_{ijk}$ and $s^{(m,n)}_{ijk}$ are constant coefficients and their values are given in appendix C.1 of~\cite{Bijnens2020}. $\lambda$ is the Källen function of the three virtual photon momenta:
\begin{equation}
    \lambda(q_{1}^{2},q_{2}^{2},q_{3}^{2}) \equiv q_{1}^{4} +q_{2}^{4} +q_{3}^{4} -2q_{1}^{2}q_{2}^{2} -2q_{1}^{2}q_{3}^{2} -2q_{2}^{2}q_{3}^{2}\;,
\end{equation}
where we have used the standard notation $q^{2n}\equiv (q^{2})^{n}$. In addition, $\mu$ represents the subtraction point of the $\overline{MS}$ renormalization scheme, which we introduced in the previous section. Finally, $F=F(Q_{1}^{2},Q_{2}^{2},Q_{3}^{2})$ is the massless triangle integral: 
\begin{equation}\label{eq:masslessTriangle}
    F(Q_{1}^{2},Q_{2}^{2},Q_{3}^{2}) \equiv (4\pi)^{2}i\int\frac{d^{4}p}{(2\pi)^{4}}\frac{1}{p^{2}}\frac{1}{(p-q_{1})^{2}}\frac{1}{(p-q_{1}-q_{2})^{2}}\;.
\end{equation}

Note that the expressions of the form factors $\hat{\Pi}_{i}$ have several terms with negative powers of $\lambda$, which constitute spurious kinematic singularities in the $\lambda\rightarrow0$ limit. These were introduced by the projectors that were used to extract the form factors from the quark loop amplitude, but they are explicitly cancelled in contributions from all other Wilson coefficients. In the case of the quark loop however there is implicit dependence on $\lambda$ coming from the massless triangle integral $F(Q_{1}^{2},Q_{2}^{2},Q_{3}^{2})$, which thus obscures the cancellation of these singularities. When $F$ is Taylor expanded around $\lambda=0$ it is possible to see that all negative powers of $\lambda$ cancel explicitly. Such expansion is necessary in the integration regions of the master formula in which two virtual photon momenta have a similar size and are much bigger than the third one, namely $Q_{1}\sim Q_{2}\gg Q_{3}\gg \Lambda_{QCD}$ and crossed versions. This regime is not quite the same as the $Q_{1}\sim Q_{2}\sim Q_{3}\gg \Lambda_{QCD}$ symmetric regime, but the OPE remains valid anyway as long as we remain in the perturbative QCD domain and the logarithms $\ln(Q_{i}/Q_{j})$ do not become too large and spoil the convergence of the perturbative series.

\section{Computation of the quark loop amplitude in our work}
We follow an alternative approach with respect to~\cite{Bijnens2020}.
Instead of projecting the quark loop amplitude onto the form factors of the master formula as a first step, we compute the amplitude in its tensor form. At intermediate stages of the computation we have to deal with tensor loop integrals, which we are able to write in terms of scalar ones by means of a kinematic--singularity--free tensor decomposition method first presented in~\cite{DavydychevDecomposition}. Once the tensor decomposition is performed we finally project on to the $\hat{\Pi}_{i}$ form factors of the master formula. In this way we are able to verify that there are no quark loop contributions neglected by the projection procedure, which is an implicit check of the generality of the tensor structures of the HLbL tensor found in~\cite{Colangelo2015,Colangelo2017} that we discussed in section~\ref{subsubsec:tensorDecompositionPi}. Finally, we compute the scalar integrals found in the tensor decomposition by means of their Mellin--Barnes representation~\cite{DavydychevMellinBarnes}. The series representation of Mellin--Barnes integrals provide a full systematic expansion of the chiral corrections to the massless part of the quark loop. Finally, we perform a numeric evaluation of the master formula~\eqref{eq:masterFormula} considering the quark loop contribution to the form factors $\hat{\Pi}_{i}$ and we discuss the results.

Our whole computation of the quark loop amplitude was done using the software \textit{Mathematica} and we also made extensive use of version 9.3.1 of \textit{FeynCalc} package~\cite{FeynCalc1,FeynCalc2,FeynCalc3} to compute Dirac traces and for intermediate steps involving tensors.

\subsection{First stages of the quark loop computation}

The first step in our computation was to perform the differentiation and take the limit with respect to $q_{4}^{\nu_{4}}$, whose effect is to duplicate the propagator that they act upon:\footnote{This formula is different to the one cited in~\cite{Bijnens2020} due to different quark propagator conventions.}
\begin{equation}\label{eq:propagatorDifferentiation}
    \lim_{q_{4}\rightarrow0}\frac{\partial}{\partial q_{\nu_{4}}} S(p+q_{4}) =i\lim_{q_{4}\rightarrow0} S(p+q_{4})\gamma^{\nu_{4}}S(p+q_{4}) =iS(p)\gamma^{\nu_{4}}S(p)\;.
\end{equation}
It is convenient to perform this differentiation and limit before computing the trace and the loop integral because by doing so we reduce the number of different propagators and external momenta from four to three.

After the Dirac trace was computed, several denominator simplifications were performed to reduce the complexity of the structure of the remaining tensor loop integrals. This lead to the appearance of integrals with only two different types of propagators in addition to the obvious ones with three. From these, the one with the most complex tensor structure was a fifth rank tensor with five propagators (but only three of them different from each other).

\subsubsection{Tensor loop integrals decomposition}

In general, computation of tensor loop integrals involves decomposing them in a linear combination of their external momenta and the metric tensor in which coefficients are given in terms of scalar loop integrals. A standard procedure to achieve this is the Passarino--Veltman decomposition~\cite{PassarinoVeltmanDecomposition,PassarinoVeltmanDecompositionCorrection}.  Scalar coefficients of this decomposition are obtained by contracting the tensor integral with each element of the tensor basis in which it is being decomposed. This yields a system of equations involving scalar integrals and form factors scalar products of the external momenta of the integral. One downside is that the form factors of the Passarino--Veltman decomposition always contain negative powers of the determinant of the Gram matrix of tensors used as a basis. These spurious kinematic singularities may be difficult to handle when the integrals of the~\eqref{eq:masterFormula} master formula are performed. Moreover, there already are unavoidable $\lambda^{-n}$ factors from the BTT projectors, so it is very inconvenient to introduce more singularities.

Since the Passarino-Veltman decomposition is technically inconvenient for our computation, we preferred to use an approach proposed by Davydychev in~\cite{DavydychevDecomposition} for tensor decomposition into scalar integrals which does not introduce kinematic singularities in the coefficients, at the cost of shifting the (space-time) dimension of the scalar integrals. Let us describe this decomposition procedure before continuing with the discussion of the quark loop computation. First, we need to introduce suitable notation. Tensor and scalar loop integrals are represented as:
\begin{equation}
    I^{(N)}_{\mu_{1}...\mu_{M}}(d;\nu_{1},...,\nu_{N}) \equiv \int \frac{d^{d}p}{(2\pi)^{d}} \frac{p_{\mu_{1}}...p_{\mu_{M}}}{D^{\nu_{1}}_{1}...D^{\nu_{N}}_{N}}\;,\hspace{0.75cm} I^{(N)}(d;\nu_{1},...,\nu_{N}) \equiv \int \frac{d^{d}p}{(2\pi)^{d}}\frac{1}{D^{\nu_{1}}_{1}...D^{\nu_{N}}_{N}}\;.
\end{equation}
where $D_{i}=(q_{i}+p)-m_{i}^{2} +i\epsilon$ represents the usual scalar (possibly massive) propagator, $\nu_{i}$ is the power of propagator $D_{i}$ in the integral, $q_{i}$ is an arbitrary external momentum and the Feynman prescription is implemented by $\epsilon\rightarrow0^{+}$. 

With this convention, the decomposition formula for tensor loop integrals in terms of scalar ones with shifted dimensions can be written as\footnote{Note that there is a difference in the equation we cite here and the one written in~\cite{DavydychevDecomposition} with respect to the factor of $4\pi$ due to the difference in the normalization convention for loop integrals.}~\cite{DavydychevDecomposition}:
\begin{equation}\label{eq:davydychevDecomposition}
\begin{split}
    I^{(N)}_{\mu_{1}...\mu_{M}}(d;\nu_{1},...,\nu_{N}) &=\sum_{\substack{\lambda,\kappa_{1},...,\kappa_{N}\\
    2\lambda +\sum_{i}\kappa_{i}=M}} \Big(-\frac{1}{2}\Big)^{\lambda}\{[g]^{\lambda}[q_{1}]^{\kappa_{1}}...[q_{N}]^{\kappa_{N}}\}_{\mu_{1}...\mu_{M}}\\
    &\hspace{-2cm}\times (\nu_{1})_{\kappa_{1}}...(\nu_{N})_{\kappa_{N}}(4\pi)^{M-\lambda}I^{(N)}(d+2(M-\lambda);\nu_{1}+\kappa_{1},...,\nu_{N}+\kappa_{N})\;,
\end{split}
\end{equation}
where $(\nu)_{\kappa}\equiv\Gamma(\nu+\kappa)/\Gamma(\nu)$ is the Pochhammer symbol. The structure between brackets represents the symmetrized tensor structure in which $g^{\mu_{1}\mu_{2}}$ appears $\lambda$ times, and each $q_{i}^{\mu_{j}}$ appears $\kappa_{i}$ times. Consequently, the restriction $2\lambda +\sum_{i}\kappa_{i}=M$ ensures that the tensor rank of the integral is conserved. The sum extends to all non--negative values of $\lambda,\kappa_{1},...,\kappa_{N}$. The proof of this formula rests mainly on the Schwinger representation of scalar loop integrals and recurrence formulas obtained by differentiation of such integrals with respect to each external momentum $q_{i}$. Finally, the result is generalized by induction. The proof of~\eqref{eq:davydychevDecomposition} is described in great detail in~\cite{DavydychevDecomposition} and we will not repeat it here. Nevertheless, there are some features of the formula which are worth to be motivated. First, note that the number of times that a tensor element $q_{i}^{\mu_{j}}$ appears in the decomposition is related to the power with which its associated denominator $D_{i}$ appears. This is in fact reminiscent of the external momentum derivatives which where used to obtain the formula. For example, the starting point of the proof of the formula for the vector integral $I_{\mu}^{(N)}(d;\nu_{1},...,\nu_{N})$ is the following differential identity:
\begin{equation}\label{eq:scalarIntegralDerivative}
    \frac{1}{2\nu_{1}}\frac{\partial}{\partial q_{1}^{\mu}}I^{(N)}(d;\nu_{1},...,\nu_{N}) =-I^{(N)}_{\mu}(d;\nu_{1}+1,...,\nu_{N}) -q_{1\mu}I^{(N)}(d;\nu_{1}+1,...,\nu_{N})\;.
\end{equation}
The difference in the powers of the $\nu_{1}$ in the derivative term and the two terms to the right is solved by using the Schwinger representation for scalar integrals, namely: 
\begin{align}\label{eq:SchwingerRepresentation}
    I^{(N)}(d;\nu_{1},...,\nu_{N}) &=\pi^{d/2}\,i^{1-d}\,\Gamma\Bigg(\sum_{i}\nu_{i}-\frac{d}{2}\Bigg)\Bigg[\prod_{i}\Gamma(\nu_{i})\Bigg]^{-1}\\\nonumber
    &\hspace{-1cm}\times\int^{1}_{0}...\int^{1}_{0}\prod \beta^{\nu_{i}-1}d\beta_{i}\,\delta\Big(\sum_{i}\beta_{i}-1\Big)\Bigg(\mathop{\sum\sum}\limits_{j\;<\;l}\beta_{j}\beta_{l}(p_{j}-p_{l})^{2}-\sum_{i}\beta_{i}m_{i}^{2}\Bigg)^{d/2-\sum_{i}\nu_{i}}\;,
\end{align}
which is valid for $\textit{Re}\{\nu_{i}\}>0$. There one can see how a shift in the sum of powers of denominators $\sum\nu_{i}$ may be offset by a twofold shift in the scalar integral's dimension. In the case of the metric tensor, its appearance is related to a reduction of the shift in the dimension of the scalar integral. This is due to the fact that metric tensors enter this decomposition from terms in which an external momentum derivative acts on its corresponding momentum, not on the scalar integral that is multiplying it, therefore it requires no additional offset and its dimensional shift is not increased. An explicit example of this situation can bee seen when taking a second derivative of the vector integral $I_{\mu}^{(N)}$ in~\eqref{eq:scalarIntegralDerivative}.

Finally, it is very important to note that although the dimension of the scalar integrals is increased in~\eqref{eq:davydychevDecomposition} with respect to the original tensor one, its superficial degree of divergence is not. A general tensor integral $I^{(N)}_{\mu_{1}...\mu_{M}}(d;\nu_{1},...,\nu_{N})$ has superficial degree of divergence equal to $d+M-2\sum_{i}\nu_{i}$ while for the scalar integrals in which it is decomposed it is equal to $d+M-2\sum_{i}\nu_{i} -\sum_{i}\kappa_{i}$. This result is a standard feature of decomposition algorithms for \emph{tensor} loop integrals, but for scalar integrals decomposition algorithms it is not always true.

We applied~\eqref{eq:davydychevDecomposition} to the tensor integrals appearing in our computation of the quark loop amplitude, thus its tensor structure was explicitly written in terms of the external momenta and the metric. As such, it was then possible to compare this structure to the $\partial_{q_{4}}^{\nu_{4}}\hat{T}_{i}^{\mu_{1}\mu_{2}\mu_{3}\mu_{4}}$ tensor basis that is used for the~\eqref{eq:masterFormula} master formula. To do this we extracted the quark loop contributions to the form factors $\hat{\Pi}_{i}$ with the help of the projectors of~\cite{Bijnens2020}. We found that all form factors received non--zero contributions from the quark loop. Furthermore, when we subtracted such contributions from the amplitude itself the result was equal to zero, which means that the quark loop amplitude contains no spurious parts that do not contribute to $a_{\mu}$. This implies that the first principles arguments presented in section~\ref{subsubsec:tensorDecompositionPi} to justify the decomposition of the HLbL tensor completely characterize the tensor structure of the quark amplitude, at least with respect to the soft derivative part of the decomposition.

It is worth noting that the tensor basis used in~\cite{Bijnens2020} is the one proposed in~\cite{Colangelo2017}, not the one from~\cite{Colangelo2015}. The choice of any of these two sets is of course irrelevant for $a_{\mu}$ and in this work we use the latter, because we are interested in using the projectors of~\cite{Bijnens2020}. 

\subsubsection{Computation of scalar integrals with shifted dimensions: first approach}

\begin{figure}
    \centering
    \begin{tabular}{ccc}
        \begin{fmffile}{tadpole}
        \setlength{\unitlength}{0.6cm}\small
        \begin{fmfgraph*}(5,5) 
            \fmfbottom{i1}
            \fmftop{o1}
            \fmf{plain}{i1,v1}
            \fmf{plain,right,tension=0.3}{v1,v2,v1}
            \fmf{phantom,tension=2}{v2,o1}
        \end{fmfgraph*}
        \end{fmffile} & 
        \begin{fmffile}{selfEnergy}
        \setlength{\unitlength}{0.6cm}\small
        \begin{fmfgraph*}(5,5) 
            \fmfleft{i1}
            \fmfright{o1}
            \fmf{plain}{i1,v1}
            \fmf{plain,right,tension=0.3}{v1,v2,v1}
            \fmf{plain}{v2,o1}
        \end{fmfgraph*}
        \end{fmffile} & 
        \begin{fmffile}{triangle}
        \setlength{\unitlength}{0.6cm}\small
        \begin{fmfgraph*}(5,5) 
            \fmfleft{i1,i2}
            \fmfright{o1,o2,o3}         
            \fmf{plain}{v1,v2,v3,v4,v5}
            \fmf{plain}{v5,v6,v7,v8,v9}
            \fmf{plain}{v9,v10,v11,v12,v1}
            \fmf{plain}{i1,v1}
            \fmf{plain}{v9,i2}
            \fmf{plain}{v5,o2}
            \end{fmfgraph*}
        \end{fmffile}
    \end{tabular}
    
    \caption{Three different basic topologies of one--loop diagrams with up to three external lines.}
    \label{fig:scalarIntegrals}
\end{figure}
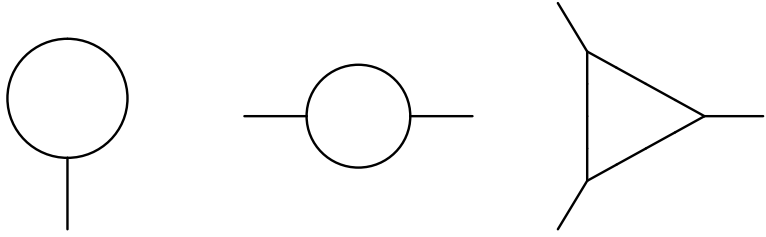

After tensor decomposing loop integrals and applying projectors on the quark loop amplitude, the form factors $\hat{\Pi}_{i}$ are given in terms of scalar integrals with shifted dimensions coming from~\eqref{eq:davydychevDecomposition}. It is necessary to compute them in order to perform the $|Q_{1}|$, $|Q_{2}|$ and $\tau$ integrals of the master formula. These scalar loops appear with two and three different propagators in the quark loop amplitude (see figure~\ref{fig:scalarIntegrals}).

Two--point integrals in four dimensions can be put in a relatively simple closed form (see section 5.1 in~\cite{BookPassarinoBardin}), but we are interested in dimension--shifted integrals. Besides this case is simple enough to be explained in detail and it gives one a natural introduction to the framework of hypergeometric functions and their corresponding Mellin--Barnes representation, which plays a central role for the triangle loop.

The tadpole formula~\eqref{eq:scalarLoopIntegralIdentity} can be used to compute self--energy integrals with the help of a well--known integral representation for products of propagators in terms of so-called ``Feynman parameters'', namely:
\begin{equation}\label{eq:feynmanParameters}
    \frac{1}{D_{1}^{\nu_{1}}...D_{N}^{\nu_{N}}}= \int^{1}_{0}\delta\Big(\sum_{1}^{N}x_{i}-1\Big)\prod_{i}dx_{i}\,\frac{\prod_{i}x_{i}^{\nu_{i}-1}}{\{x_{1}D_{1}+...+x_{N}D_{N}\}^{\sum_{i}\nu_{i}}}\frac{\Gamma(\sum_{i}\nu_{i})}{\prod_{i}\Gamma(\nu_{i})}\;.
\end{equation}
This formula applies for a set of arbitrary complex numbers $D_{i}$ and $\nu_{i}$, as long as $\textit{Re}\{\nu_{i}\}>0$ for every $i=1,...,N$. When this formula is used for the propagators in a self--energy integral one can turn the term $x_{1}D_{1}+...x_{N}D_{N}$ into the shape of a single new propagator whose external momentum and masses are a linear combination of the original ones. For the self--energy case with equal masses one has:
\begin{equation}\label{eq:selfEnergyIntermediate_1}
    I^{(2)}(d;\nu_{1},\nu_{2})
    =\frac{\Gamma(\nu_{1}+\nu_{2})}{\Gamma(\nu_{1})\Gamma(\nu_{2})}\int_{0}^{1} dx_{1}\,x_{1}^{\nu_{1}-1}x_{2}^{\nu_{2}-1}\int\frac{d^{d}p}{(2\pi)^{d}}\frac{1}{\{p^{2} -\Delta(x_{1})\}^{\nu_{1}+\nu_{2}}}\;,
\end{equation}
where $\Delta(x_{1})=-x_{1}(1-x_1)q_{1}^{2} +m^{2}$ and we have performed an integration--variable shift in the last step. For brevity it is always implicitly assumed that $x_{2}=1-x_{1}$ due to the delta function in~\eqref{eq:feynmanParameters}. Note that $\Delta$ becomes negative only for sufficiently large time--like external momentum, so we do not have to worry about the Feynman prescription in our kinematic regime. At this point we can use the expression~\eqref{eq:scalarLoopIntegralIdentity} for the tadpole scalar integral to obtain:
\begin{equation}\label{eq:selfEnergyIntermediate_2}
\begin{split}
    I^{(2)}(d;\nu_{1},\nu_{2}) &=\frac{i^{1-d}}{(4\pi)^{d/2}}\frac{\Gamma(\nu_{1}+\nu_{2}-\frac{d}{2})}{\Gamma(\nu_{1})\Gamma(\nu_{2})}\int_{0}^{1} dx_{1}\,x_{1}^{\nu_{1}-1}x_{2}^{\nu_{2}-1}\Big(\frac{1}{x_{1}x_2q_{1}^{2} -m^{2}}\Big)^{\nu_{1}+\nu_{2}-\frac{d}{2}}\;.
\end{split}
\end{equation}
To compute this integral it is necessary to introduce yet another computational trick, the so-called Mellin--Barnes representation
\begin{align}\nonumber
    \frac{1}{(A_{1}+...+A_{N})^{\beta}} &=\frac{1}{\Gamma(\beta)}\frac{1}{A_{N}^{\beta}}\int^{r+i\infty}_{r-i\infty}...\int^{r+i\infty}_{r-i\infty}\prod_{j}^{N-1}\frac{ds_{j}}{(2\pi i)}\,\Gamma(-s_{j})\Big(\frac{A_{j}}{A_{N}}\Big)^{s_{j}} \times\Gamma\Big(\beta+\sum_{i}s_{i}\Big)\\\label{eq:MBrepresentation}
    &\equiv\frac{1}{\Gamma(\beta)}\frac{1}{A_{N}^{\beta}}\int_{\prod_{j}s_{j}}\prod_{j}^{N-1}\Gamma(-s_{j})\Big(\frac{A_{j}}{A_{N}}\Big)^{s_{j}} \times\Gamma\Big(\beta+\sum_{i}s_{i}\Big)\;,
\end{align}
where $r$ and $\beta$ are real numbers and $r$ is chosen so that the path of integration separates the poles of the $\Gamma(-s_{i})$ functions to the left of the poles of the $\Gamma\Big(\beta+\sum_{i}s_{i}\Big)$ function. At its core, this seemingly very complex representation is in fact just a Taylor expansion in a geometric series. If we consider the case of a massive scalar propagator, it requires two different Taylor expansions in the infrared and ultraviolet regimes: 
\begin{equation}\label{eq:MBexample}
\begin{split}
    \frac{1}{(p^{2}-m^{2})^{\nu}} &= \frac{1}{p^{2\nu}}\frac{1}{\Gamma(\nu)}\sum_{n=0}\frac{1}{n!}\Big(\frac{m^{2}}{p^{2}}\Big)^{n}\Gamma(\nu+n) \hspace{1.75cm} \text{for} \hspace{0.5cm} \Big|\frac{m^{2}}{p^{2}}\Big|<1\;,\\
    &=\frac{1}{(-m^{2})^{\nu}}\frac{1}{\Gamma(\nu)}\sum_{n=0}\frac{1}{n!}\Big(\frac{p^{2}}{m^{2}}\Big)^{n}\Gamma(\nu+n) \hspace{1cm} \text{for}\hspace{0.5cm} \Big|\frac{p^{2}}{m^{2}}\Big|<1\;.
\end{split}
\end{equation}
The advantage of formula~\eqref{eq:MBrepresentation} is that it contains both series implicitly. The Mellin--Barnes representation for that case is:
\begin{equation}
    \frac{1}{(p^{2}-m^{2})^{\nu}} =\frac{1}{m^{2\nu}}\frac{1}{\Gamma(\nu)}\int_{s}\Big(-\frac{p^{2}}{m^{2}}\Big)^{s}\Gamma(-s)\Gamma(\nu+s)\;.
\end{equation}
To compute this integral we can use the standard framework of residues in complex variable by tracing a semi--circular integration contour with an infinite radius which is closed by a line that lies along the imaginary axis (see figure~\ref{fig:selfEnergyGeneralPoles}). The direction (positive or negative real) to which the semicircle points is restricted by the requirement that its contribution vanishes. Thus by Jordan's lemma one concludes that for $|k^{2}|<m^{2}$ the semicircle must point to the right (positive real) and for $m^{2}<|k^{2}|$ it must point to the left. In each case the contour of integration encloses the poles of $\Gamma(-s)$ or $\Gamma(\nu+s)$, respectively. Noting that the Euler Gamma function has simple poles at zero and all negative integers, then the sum of residues from Cauchy's theorem gives the correct geometric series of~\eqref{eq:MBexample} in the corresponding regimes. The prescription that the poles of these two functions are divided by the imaginary integration interval ensures that one series representation does not mix with the other.

In general, formula~\eqref{eq:MBrepresentation} is very useful to compute non--trivial Feynman parameters integrals; it is just a matter of carefully choosing the $A_{i}$'s so that each contains only terms with the same power of Feynman parameters. In this way the integral over the $x_{i}$ becomes straightforward. Inserting this identity into~\eqref{eq:selfEnergyIntermediate_2} one obtains:
\begin{equation}\label{eq:MBselfEnergy}
\begin{split}
     I^{(2)}(d;\nu_{1},\nu_{2}) &=\frac{i^{1-d}}{(4\pi)^{d/2}}\frac{(-m^{2})^{\frac{d}{2}-\nu_{1}-\nu_{2}}}{\Gamma(\nu_{1})\Gamma(\nu_{2})}\int_{s_{1}}\frac{\Gamma(s_{1}+\nu_{1})\Gamma(s_{1}+\nu_{2})}{\Gamma(2s_{1}+\nu_{1}+\nu_{2})}\Big(-\frac{q_{1}^{2}}{m^{2}}\Big)^{s_{1}}\\
     &\hspace{1cm}\times\Gamma(-s_{1})\Gamma\Big(s_{1}+\nu_{1}+\nu_{2}-\frac{d}{2}\Big)\;,
\end{split}
\end{equation}
where we have used the following formula: 
\begin{equation}
    \int_{0}^{1}...\int_{0}^{1} \prod_{i}dx_{i}\, x_{i}^{\rho_{i}-1} \delta\Big(\sum_{i}x_{i}-1\Big)=\frac{\prod_{i}\Gamma(\rho_{i})}{\Gamma(\sum_{i}\rho_{i})}
\end{equation}
to perform the Feynman parameters integral. This is in fact the simplest non--trivial case of a general formula for $N$--point scalar integrals with equal masses, which was first published in~\cite{DavydychevMellinBarnes}. For the triangle scalar integral with equal masses the general formula yields:
\begin{align}\nonumber
    I^{(3)}(d;\nu_{1},\nu_{2},\nu_{3}) &=\frac{i^{1-d}}{(4\pi)^{d/2}}\frac{(-m^{2})^{\frac{d}{2}-\sum_{i}\nu_{i}}}{\Gamma(\nu_{1})\Gamma(\nu_{2})\Gamma(\nu_{3})}\int_{\substack{\,s_{12}\\\,s_{13}\\\,s_{23}}}\Bigg(-\frac{q^{2}_{12}}{m^{2}}\Bigg)^{s_{12}}\Bigg(-\frac{q^{2}_{13}}{m^{2}}\Bigg)^{s_{13}}\Bigg(-\frac{q^{2}_{23}}{m^{2}}\Bigg)^{s_{23}}\\\label{eq:MBtriangle}
    &\hspace{1cm}\times \Gamma(-s_{12})\Gamma(-s_{13})\Gamma(-s_{23})\\\nonumber
    &\hspace{1cm}\times\Gamma\Big(\nu_{1} +s_{12}+s_{13}\Big)\Gamma\Big(\nu_{2} +s_{12} +s_{23}\Big)\Gamma\Big(\nu_{3} +s_{13}+s_{23}\Big)\\\nonumber
    &\hspace{1cm}\times\Gamma\Big(\sum_{i}\nu_{i}-\frac{d}{2}+s_{12}+s_{13}+s_{23}\Big)\Big[\Gamma\Big(\sum_{i}\nu_{i}+2s_{12}+2s_{13}+2s_{23}\Big)\Big]^{-1}\;.
\end{align}
Now the task is to compute the integrals in~\eqref{eq:MBselfEnergy} and~\eqref{eq:MBtriangle}. This is a complex task even for the self--energy case for both practical and conceptual reasons. From a practical perspective, the appearance of multiple Gamma functions, each with its own set of poles and zeros, renders the pole structure of the integrand unusually complex. Furthermore, the triangle integral has a triple nested integral and the poles of the Gamma functions are intertwined, which introduces a conceptual difficulty: the standard complex variable residues framework that is enough for the self--energy case cannot be naively expanded in general by iteration to consider multiple complex variable integrals. We stop the presentation of our computation of the quark loop to present the tools necessary to face this issue.

\subsection{Mellin--Barnes integrals, multivariate residues and hypergeometric functions}

The Davydychev tensor decomposition which has the benefit of not introducing additional  kinematic singularities has come at the cost of introducing scalar integrals in shifted dimensions. In~\eqref{eq:MBselfEnergy} and~\eqref{eq:MBtriangle} we have arrived at a representation for the emerging scalar integrals in terms of Mellin--Barnes representation. Analytical expressions to these are often given in terms of hypergeometric--like\footnote{The presence of logarithms and polygamma functions breaks some of the properties required on a power series to be considered a hypergeometric one.} series in one or more variables, therefore they can give us a systematic expansion of the quark masses effects on the loop. We will present a general framework of computation for nested Mellin--Barnes integrals.

\subsubsection{General properties of Mellin--Barnes integrals}

In particular, we have to deal with $P$--fold Mellin--Barnes integrals of the form: 
\begin{equation}\label{eq:generalMBintegral}
\begin{split}
    J(\{\boldsymbol{e}_{j}\},\boldsymbol{f};\{g_{j}\},h;u_{1},...,u_{P})&=\int^{+i\infty}_{-i\infty}...\int^{+i\infty}_{-i\infty}\prod_{i}^{P}\Big\{\frac{ds_{i}}{2\pi i} (-u_{i})^{s_{i}}\Big\}\frac{\prod_{j=1}^{k}\Gamma(\boldsymbol{e}_{j}\cdot \boldsymbol{s} +g_{j})}{\Gamma(\boldsymbol{f}\cdot \boldsymbol{s} +h)}\;,
\end{split}
\end{equation}
where $\boldsymbol{s}$ is a $P$--dimensional complex vector containing the integration variables, $\boldsymbol{e}_{j}$ and $\boldsymbol{f}$ are $P$--dimensional real vectors, $g_{j}$ and $h$ are real numbers, and $u_{i}$ is a complex number. Looking at~\eqref{eq:MBselfEnergy} and~\eqref{eq:MBtriangle}, we have $\boldsymbol{f}=(2,...,2)^{T}$ and $h=\sum_{i}\nu_{i}$, and that is a general feature of scalar loops. The vectors $\boldsymbol{e}_{j}$ and the numbers $g_{j}$ do not have a general form, but can be easily read from the integrand in each case. The integral paths are shifted from the origin by a finite real quantity $\gamma_{i}$ to prevent them from splitting the poles of a Gamma function in the numerator into subsets or passing through one of them.\footnote{If  one is computing the integral in dimensional regularization, the former purpose might not be compatible with the limit $\epsilon\rightarrow0$. We do not consider this situation here as it is not relevant for this work. Instead we refer the reader to the comprehensive study done in~\cite{straightContoursFriot}.} In general, the Gamma functions in both the numerator and denominator of the integrand may also appear with powers higher than one and there may be multiple gamma functions in the denominator, but we will not consider such cases as they do not happen in scalar loops.

There are two quantities upon which some important features of the integral in~\eqref{eq:generalMBintegral} depend: 
\begin{equation}
\begin{split}
    \boldsymbol{\Delta} &\equiv \sum_{i}\boldsymbol{e}_{i} -\boldsymbol{f} \hspace{1cm}
    \alpha \equiv \text{Min}_{||\hat{\boldsymbol{y}}||=1}\Big\{\sum_{i}|\boldsymbol{e}_{i}\cdot \hat{\boldsymbol{y}}| -|\boldsymbol{f}\cdot \hat{\boldsymbol{y}}|\Big\}\;,
\end{split}
\end{equation}
where $|.|$ symbolizes complex norm $||.||$ represents Euclidean vector norm. In particular, for all integrals of the type~\eqref{eq:davydychevDecomposition} we have $\boldsymbol{\Delta}=0$.  The asymptotic behaviour of the integrand is of course key for Mellin--Barnes integrals and these two quantities characterize it. First, let us see the meaning of $\alpha$. For this, let us consider the asymptotic behaviour of the integrand in~\eqref{eq:generalMBintegral} when the imaginary part of $s_{i}$ gets big:
\begin{equation}\label{eq:StirlingFormulaBigImaginary}
    \Gamma(r+i\tau) \longrightarrow \sqrt{2\pi}|\tau|^{r-1/2}e^{-\pi|\tau|/2}\hspace{2cm} \text{for}\hspace{0.5cm} |\tau|\rightarrow\infty\;.
\end{equation}
Then evaluating the complex norm of the integrand of~\eqref{eq:generalMBintegral} in the asymptotic regime $s_{i}=\lim_{|R_{i}|\rightarrow\infty}\gamma_{i}-x_{i}+iR_{i}$, where $x_{i}$ and $R_{i}$ are real numbers, and $\gamma_{i}$ represents the real shift to the integration paths, we obtain:
\begin{equation}
\begin{split}
    \Bigg|\prod_{i}^{P}\Big\{(-u_{i})^{s_{i}}\Big\}\frac{\prod_{j=1}^{k}\Gamma(\boldsymbol{e}_{j}\cdot \boldsymbol{s} +g_{j})}{\Gamma(\boldsymbol{f}\cdot \boldsymbol{s} +h)}\Bigg| \hspace{0.5cm}\longrightarrow&\hspace{0.5cm} \prod_{i}^{P}\Big\{|u_{i}|^{\gamma_{i}}\Big\}\frac{\prod_{j=1}^{k}|\boldsymbol{e}_{j}\cdot\boldsymbol{R}|^{\sum_{i}\boldsymbol{e}_{i}\cdot(\boldsymbol{\gamma}-\boldsymbol{x})+g_{j}-1/2}}{|\boldsymbol{f}\cdot\boldsymbol{R}|^{\boldsymbol{f}\cdot(\boldsymbol{\gamma}-\boldsymbol{x})+h-1/2}}\\
    &\hspace{0.5cm}\times\exp\Big\{-\Big(\textit{arg}\{u_{i}\}+\pi\Big)R_{i}-\Big(\sum_{j}|\boldsymbol{e}_{j}\cdot\boldsymbol{R}|-|\boldsymbol{f}\cdot\boldsymbol{R}|\Big)\frac{\pi}{2}\Big\}\;.
\end{split}
\end{equation}
The first line on the right hand side is a polynomial in $R_{i}$, while the other two are exponential. Thus we see that the integral in~\eqref{eq:generalMBintegral} is absolutely convergent for
\begin{equation}\label{eq:intermediateConvergenceInequality}
    -\textit{arg}\{-\boldsymbol{u}\}\cdot\boldsymbol{R} <\Big(\sum_{j}|\boldsymbol{e}_{j}\cdot\boldsymbol{R}|-|\boldsymbol{f}\cdot\boldsymbol{R}|\Big)\frac{\pi}{2}\;,
\end{equation}
where $\textit{arg}\{u_{i}\}$ is the argument of the complex variable $u_{i}$ and the components of $\textit{arg}\{-\boldsymbol{u}\}$ are equal to $\textit{arg}\{u_{i}\} +\pi$. Since the inequality~\eqref{eq:intermediateConvergenceInequality} is homogeneous in $\boldsymbol{R}$, then it can be simplified as 
\begin{equation}
    \text{Max}_{||\hat{\boldsymbol{y}}||=1}|\textit{arg}\{-\boldsymbol{u}\}\cdot\hat{\boldsymbol{y}}| <\alpha\frac{\pi}{2}\;.
\end{equation}
Finally, using the well--known Cauchy--Schwartz\footnote{Also known as Cauchy--Bunyakovsky or Cauchy--Bunyakovsky--Schwartz inequality.} inequality one concludes that
\begin{equation}\label{eq:MBconvergenceCondition}
    \text{Max}_{||\hat{\boldsymbol{y}}||=1}|\textit{arg}\{-\boldsymbol{u}\}\cdot\hat{\boldsymbol{y}}| =||\textit{arg}\{-\boldsymbol{u}\}|| \implies ||\textit{arg}\{-\boldsymbol{u}\}|| <\alpha\frac{\pi}{2}\;.
\end{equation}
Therefore, one sees that $\alpha$ characterizes the convergence regions of the Mellin--Barnes integral in~\eqref{eq:generalMBintegral}. For scalar loops one has $\alpha>\sum_{j}|(\hat{\boldsymbol{y}})_{j}| -|\sum_{j}(\hat{\boldsymbol{y}})_{j}|>0$, hence there is always a non--trivial region of convergence.

While $\alpha$ is related to the convergence of the integral as a function of $\boldsymbol{u}$, that is, the asymptotic behaviour of the integrand in imaginary directions, $\boldsymbol{\Delta}$ does the same with respect to the real part of the integration variables $\boldsymbol{s}$. This is key to know the direction to which the contours of integration can be closed. To justify this interpretation we follow a procedure analogous to that of $\alpha$ although this time the Stirling formula is specialized to the case of a big real part: 
\begin{equation}
    |\Gamma(r+i\tau)| \longrightarrow \sqrt{2\pi}|r|^{r-1/2}e^{-r}\;.
\end{equation}
With such formula we study the integrand in the limit $s_{i}=\lim_{|x_{i}|\rightarrow\infty}\gamma_{i}-x_{i}+iR_{i}$: 
\begin{equation}
\begin{split}
    \Bigg|\prod_{i}^{P}\Big\{(-u_{i})^{s_{i}}\Big\}\frac{\prod_{j=1}^{k}\Gamma(\boldsymbol{e}_{j}\cdot \boldsymbol{s} +g_{j})}{\Gamma(\boldsymbol{f}\cdot \boldsymbol{s} +h)}\Bigg| \hspace{0.5cm}\longrightarrow& \exp\Big\{-\Big(\sum_{j}\boldsymbol{e}_{j} -\boldsymbol{f}\Big)\cdot (\boldsymbol{\gamma}-\boldsymbol{x})\Big\} \Bigg|\prod_{i}^{P}\Big\{|u_{i}|^{-x_{i}}\Big\}\frac{\prod_{j=1}^{k}|\boldsymbol{e}_{j}\cdot \boldsymbol{x}|^{\boldsymbol{e}_{j}\cdot \boldsymbol{x} -1/2}}{|\boldsymbol{f}\cdot \boldsymbol{x}|^{\boldsymbol{f}\cdot \boldsymbol{x}-1/2}}\Bigg|\;,
\end{split}
\end{equation}
where $\boldsymbol{x}$ characterizes the direction to which the contour of integration closes and thus we see that for $\boldsymbol{\Delta}\not=0$ there are preferred directions in the complex plane. Instead, when $\boldsymbol{\Delta}=0$ there are many (infinitely many, as we will discuss later) regions where the integrand decreases depending on the values of $|u_{i}|$ and as such there are multiple series representations which, if $\alpha>0$, are analytic continuations of one another~\cite{twoDimensionsTsikh,BookMarichev}.

Let us introduce useful definitions to shed more light on the meaning of $\Delta$, which is crucial for the computation of Mellin--Barnes integrals. We have found that the exponential increase or decrease of the Mellin--Barnes integrand in infinite real directions of the complex space $\mathbb{C}^{P}$ depends on a scalar product with $\boldsymbol{\Delta}$. More specifically, we conclude that the integrand increases exponentially for any $\boldsymbol{s}\in\mathbb{C}^{P}$ with a large real part such that $\boldsymbol{\Delta}\cdot\textit{Re}\{\boldsymbol{s}\}>\boldsymbol{\Delta}\cdot\boldsymbol{\gamma}$ and the converse statement is valid if $\boldsymbol{\Delta}\cdot\textit{Re}\{\boldsymbol{s}\}<\boldsymbol{\Delta}\cdot\boldsymbol{\gamma}$. We will later see that one can compute Mellin--Barnes integrals by closing a multivariable infinite ``contour'' in the region in which the integrand vanishes asymptotically, as one can expect from a naive multivariate generalization of Jordan's lemma. Consequently, we now introduce definition that will come in handy below. Let $l_{\boldsymbol{\Delta}}$ be a hyperplane in the subspace $\mathbb{R}^{P}$ with normal vector $\boldsymbol{\Delta}$ whose points are defined by the condition $\boldsymbol{\Delta}\cdot\textit{Re}\{\boldsymbol{s}\}=\boldsymbol{\Delta}\cdot\boldsymbol{\gamma}$. Note that $l_{\boldsymbol{\Delta}}$ constitutes a critical region of the asymptotic behaviour of the Mellin--Barnes integrand. Let $\pi_{\boldsymbol{\Delta}}$ represent the ``half'' of $\mathbb{R}^{P}$ for which $\boldsymbol{\Delta}\cdot\textit{Re}\{\boldsymbol{s}\}<\boldsymbol{\Delta}\cdot\boldsymbol{\gamma}$, which is the region of exponential decrease of the integrand. $\pi_{\boldsymbol{\Delta}}$ can be regarded as the real projection of a section $\Pi_{\boldsymbol{\Delta}}$ of $\mathbb{C}^{P}$. Since $\boldsymbol{\Delta}$ is a real vector, then such section can be defined as a direct product: $\Pi_{\boldsymbol{\Delta}}\equiv\pi_{\boldsymbol{\Delta}}+i\mathbb{R}^{P}$. \footnote{For $P=1$ $l_{\boldsymbol{\Delta}}$ and $\pi_{\boldsymbol{\Delta}}$ are a point and a line, for $P=2$ they are a line and a plane, and for $P=3$ they are a plane and a 3D cube, respectively.} The points of $\Pi_{\boldsymbol{\Delta}}$ are characterized by the condition $\textit{Re}\{\boldsymbol{\Delta}\cdot\boldsymbol{s}\}<\boldsymbol{\Delta}\cdot\boldsymbol{\gamma}$, therefore, as we just discussed, it should be expected for the integrand poles that belong to $\Pi_{\boldsymbol{\Delta}}$ to play a major role in the computation of Mellin--Barnes integrals. 

\subsubsection{Multivariate generalization of Jordan's lemma for Mellin--Barnes integrals}

Let us now consider the actual computation of Mellin--Barnes integrals. In univariate residues we have the well--known Jordan's lemma:
\begin{equation}\label{eq:uniJordanLemma}
    \frac{1}{2\pi i}\int^{+\infty}_{-\infty}dx\,f(x)e^{i\lambda x} =\sum_{a\in S}\textit{Res}_{a}f(z)\;,
\end{equation}
where $\lambda>0$ and $S$ is the set of poles of $f(z)$ in the upper half of the complex plane. This formula is valid if $\lim_{|z|\rightarrow\infty}|f(z)|=0$ for $z$ in the upper half of the complex plane.\footnote{Note that Jordan's lemma is usually taken to be the result regarding the vanishing of the integral of a complex variable function along an infinite semicircle, of which~\eqref{eq:uniJordanLemma} is a famous application, but here we adhere to the convention of~\cite{BookShortTsikh}.} If $\lambda<0$ then the upper and lower halves of the complex plane change roles. This formula is only valid for one dimensional Mellin--Barnes integrals, that is, the self--energy ones. It can in principle be applied also for multiple integrals as long as the location of the poles remains univariate. An example of such situation would be a two--fold Mellin--Barnes integrals where such that in the numerator there are two gamma functions $\Gamma(z_{1})\Gamma(z_{2})$ and a counter example would be $\Gamma(z_{1})\Gamma(z_{2}+z_{1})$. In the latter case the poles become entangled and it is necessary to use multivariate residues machinery. It is evident that we face such situations with~\eqref{eq:MBselfEnergy} and~\eqref{eq:MBtriangle}.

It is possible to compute integrals~\eqref{eq:generalMBintegral} in the general multivariate case with a formula analogous to~\eqref{eq:uniJordanLemma}. Such formula is of course more abstract, so, before presenting the result, let us first point at certain features of the univariate formula that should be translated into the multivariate case. The basic idea behind~\eqref{eq:uniJordanLemma} is to use the straight path of integration of the original integral as a part of a larger closed contour. The integral along such contour can be computed with residues. The region to which the contour is closed is chosen such that contribution from the part of the contour that is additional to the original straight path vanishes. Since the original integration path is infinite and the contour is closed, then the additional parts are infinite too and must be placed in a region where the integrand vanishes, at least asymptotically. Such region is ultimately determined by $\lambda$ in the univariate case and by $\boldsymbol{\Delta}$ for the multivariate ones of~\eqref{eq:generalMBintegral}. Therefore one would expect the relevant poles of the multivariate case to be the ones in $\Pi_{\boldsymbol{\Delta}}$, just as the relevant ones for~\eqref{eq:uniJordanLemma} are in the upper half of $\mathbb{C}$ for $\lambda>0$.

Now we need to introduce the definition of multivariate poles and residues. These are slightly different from the univariate case. Let us consider the following general function: 
\begin{equation}\label{eq:multivariatePolarFunction}
    f(\boldsymbol{z}) =\frac{\eta(\boldsymbol{z})}{\phi_{1}(\boldsymbol{z})...\phi_{n}(\boldsymbol{z})}\;,
\end{equation}
where $\boldsymbol{z}=(z_{1},...,z_{n})\in\mathbb{C}^{n}$. A naive univariate generalization would tell us that $f(\boldsymbol{z})$ has poles in any $\boldsymbol{z}_{0}$ such that $\phi_{j}(\boldsymbol{z}_{0})=0$ for at least one $j\in\{1,...,n\}$, as long as $\eta(\boldsymbol{z}_{0})\not=0$. Instead, the correct definition states that $f(\boldsymbol{z})$ has poles in any $\boldsymbol{z}_{0}$ such that $\boldsymbol{\phi}(\boldsymbol{z}_{0})= (\phi_{1}(\boldsymbol{z}_{0}),...,\phi_{n}(\boldsymbol{z}_{0}))=0$, as long as $\eta(\boldsymbol{z}_{0})\not=0$. This definition is not as odd as it may seem: if we could arrange a variable change such that each $\phi_{j}$ becomes univariate then we would disentangle the multivariate poles and the closed integral of $f(\boldsymbol{z})$ would become a product of univariate integrals. For some integration contours such product would be equal to zero if not all $\phi_{j}$ had zeros at the same point. 

There is one more rather peculiar feature of the definition of poles that we have just given: it leaves space for ambiguities with respect to the way in which singular factors $\phi_{j}$ are grouped together. For example, let us consider the following function $f(z_{1},z_{2})$:
\begin{equation}
    f(z_{1},z_{2}) =\frac{\eta(z_{1},z_{2})}{z_{1}(z_{1}-z_{2}+1)(z_{1}+z_{2})}\;.
\end{equation}
There is no obvious way to define the singular functions. Three of the possibilities are:
\begin{align}\nonumber
    \phi_{1}&=z_{1}(z_{1}-z_{2}+1)\;, &\hspace{1cm} &\phi_{2}=(z_{1}+z_{2})\;,\\
    \phi_{1}&=(z_{1}-z_{2}+1)\;, &\hspace{1cm} &\phi_{2}=z_{1}(z_{1}+z_{2})\;,\\\nonumber
    \phi_{1}&=z_{1}\;, &\hspace{1cm} &\phi_{2}=(z_{1}-z_{2}+1)(z_{1}+z_{2})\;.
\end{align}
Each of these three combinations has different poles and they may have even different residues in the poles that they share.\footnote{An explicit computation of an example of the latter case is given in~\cite{multivariateLarsen}.} Furthermore, even if there were only two singular factors, the order in which they are defined introduces a sign ambiguity, as we will see later. Hence any residue formula must clearly specify the singular functions with respect to which its poles are defined. Each set of singular points defined by the condition $\phi_{j}(\boldsymbol{z})=0$ is called a divisor and we represent them with $F_{j}$. Consequently, the set $F_{1}\cap F_{2}\cap...\cap F_{n}$ contains the poles of $f(\boldsymbol{z})$ with respect to a certain set of divisors $\{F_{j}\}$.

Now let us consider the residues of $f(\boldsymbol{z})$ in this poles: 
\begin{equation}\label{eq:grothendieckResidues}
    \textit{Res}_{\{F_{1},...,F_{n}\},\boldsymbol{z}_{0}} f(\boldsymbol{z})=\frac{1}{(2\pi i)^{n}}\oint_{C_{\epsilon}}\frac{\eta(\boldsymbol{z})dz_{1}...dz_{n}}{\phi_{1}(\boldsymbol{z})...\phi_{n}(\boldsymbol{z})}\;,
\end{equation}
where $C_{\epsilon}\{\boldsymbol{z}\in\mathbb{C}^{P}|\,|\phi_{i}(\boldsymbol{z})|=\epsilon_{i}\}$ is called a cycle and $\epsilon_{i}$ has infinitesimal positive value. The orientation of the integration path $C_{\epsilon}$ is defined such that the change in the argument of every $\phi_{j}$ is always possible, which is analogous to the usual clockwise orientation although this time it refers to the functions $\phi_{j}$ rather than the integration variables $z_{j}$. Note that due to the definition of the orientation of $C_{\epsilon}$, one sees that residues are skew--symmetric with respect to the permutations of $\phi_{j}$. Equation~\eqref{eq:grothendieckResidues} defines local Grothendieck residues, which are a multivariate generalization of the univariate ones and are commonly used in the context of algebraic geometry~\cite{bookGriffiths}.

Now we are able to state the the formal mathematical generalization of~\eqref{eq:uniJordanLemma} for multiple variables, which is called ``multidimensional abstract Jordan lemma''~\cite{BookShortTsikh}. It asserts that for a complex variable function $f(\boldsymbol{z})$:
\begin{equation}\label{eq:multiJordanLemma}
    \frac{1}{(2\pi i)^{n}}\int_{\sigma} f(\boldsymbol{z}) dz_{1}...dz_{n}=\sum_{a\in\Pi}\textit{Res}_{a}f(\boldsymbol{z})\;.
\end{equation}
Let $\Pi$ be a polyhedron and $\sigma$ be the ``skeleton'' of $\Pi$, that is, the structure formed by the vertices and edges of $\Pi$. The residues in $\Pi$ are defined in terms of divisors $\{F_{j}\}$ such that each of them does not intersect one specific face of the polyhedron, that is, the polyhedron has $n$ faces $\sigma_{n}$ and the set of divisors verifies the condition $F_{j}\cap\sigma_{j}=\varnothing$ for each $j=1,...,n$. This is referred to as ``compatibility'' between divisors and the polyhedron.

In general, $\Pi$ may be bounded or not, however, we want to identify the edges in $\sigma$ with the infinite straight integration paths of~\eqref{eq:generalMBintegral}, so we are interested in the unbounded case. In this context there is an additional condition for the validity of~\eqref{eq:multiJordanLemma} which is essentially a multivariate generalization of the asymptotic behaviour condition on $f(z)$ when there is an infinite set of poles, which we omitted when discussing~\eqref{eq:uniJordanLemma} and we omit for this case, too, because it is not crucial for our analysis~\cite{BookShortTsikh,twoDimensionsTsikh}.

Applying~\eqref{eq:multiJordanLemma} to integrals of the type showed in~\eqref{eq:generalMBintegral} one obtains the following result~\cite{twoDimensionsTsikh,calabiYau_Tsikh}:
\begin{equation}\label{eq:JordanLemmaMB}
\begin{split}
    J(\{\boldsymbol{e}_{j}\},\boldsymbol{f};\{g_{j}\},h;u_{1},...,u_{P}) &=\sum_{a\in\Pi_{\boldsymbol{\Delta}}}\textit{Res}_{a} J\;.
\end{split}
\end{equation}
In addition, $\textit{Res}_{a} J$ represents the residue of the integrand in its pole $a$. The compatibility condition for the divisors and the polyhedron is of course still required for~\eqref{eq:JordanLemmaMB} to be valid. For $\boldsymbol{\Delta}=0$ one sees that there is no preferred region of the $\mathbb{C}^{P}$ space, hence such integrals are usually called ``degenerate''. In fact, in such cases formula~\eqref{eq:JordanLemmaMB} remains valid for any $\Pi_{\boldsymbol{\Delta}}$.

The analogy of this result with the standard one--dimensional Jordan lemma is more apparent in the one--dimensional case of~\eqref{eq:generalMBintegral}. In there, $l_{\boldsymbol{\Delta}}$ is just $\gamma+i\mathbb{R}$. Hence when $\Delta>0$ the sum of residues from the poles enclosed in the negative real half of the complex plane constitute a series representation convergent for any value of $u$, while the sum of residues from the other half forms a divergent asymptotic expansion~\cite{asymptoticExpansionBarnes}. For $\Delta<0$ the roles of these two halves of the complex plane are inverted, while for $\Delta=0$ one obtains two different series for each half that converge in non--overlapping complementary regions of the $u$ complex plane. If $\alpha>0$, then they are an analytical continuation of each other. In this way one can see the analogy of $\boldsymbol{\Delta}$ with the role of the time coordinate and its sign in Fourier transforms. Regarding the compatibility between the divisor and the polyhedron of integration, note that the face of $\Pi_{\boldsymbol{\Delta}}$ is just the integration path $\gamma+i\mathbb{R}$, therefore one sees that such prescription is just the multidimensional generalization of the requirement for~\eqref{eq:uniJordanLemma} that no poles lie on the integration path.
 
Now that we have presented the multivariate generalization of Jordan's lemma, we need to show how to compute Grothendieck residues of the integrand of~\eqref{eq:generalMBintegral} with respect to the poles and divisors that fulfill the requirements of~\eqref{eq:JordanLemmaMB}. 

Let us first start with poles. The ones that we are interested in exist at points where $P$ gamma functions become singular, that is, the intersection of $P$ singular hyperplanes of the gamma functions in the numerator. For example, in the case of the three--point function~\eqref{eq:MBtriangle}, we must have an intersection of three two--dimensional planes. Each gamma function in the numerator of the integrand generates a family $L^{j}$ with countably infinite singular hyperplanes $L_{n}^{j}$ defined as $L_{n}^{j}=\{\boldsymbol{s}\in\mathbb{C}^{P}|\,\boldsymbol{e}_{j}\cdot\boldsymbol{s} +g_{j}=-n\}$ for every $n\in\mathbb{N}$. One sees that each $\boldsymbol{e}_{j}$ is the normal vector of the family of singular hyperplanes of a given gamma function. They give us information about intersection of singular planes an therefore they are key to identify poles of the integrand. If a set of $P$ vectors $\{\boldsymbol{e}_{j_{1}},...,\boldsymbol{e}_{j_{P}}\}$ is linearly independent, then for any $n_{i}\in\mathbb{N}$ the set $L^{j_{1}}_{n_{1}}\cap...\cap L^{j_{P}}_{n_{P}}$ always has only one element $\boldsymbol{z}_{0}\in\mathbb{C}^{P}$, which constitutes a pole of the Mellin--Barnes integrand. Moreover, if each singular plane $L^{j_{i}}$ belongs to a different divisor $F_{j_{i}}$, then $\boldsymbol{z}_{0}$ is a relevant pole for~\eqref{eq:JordanLemmaMB}. With this definition of poles, the formula~\eqref{eq:JordanLemmaMB} requires us to:

\begin{itemize}
    \item Group the singular planes of the gamma functions in the numerator of~\eqref{eq:generalMBintegral} in $P$ divisors $F_{j}$ that satisfy the compatibility condition with respect to the faces of $\Pi_{\Delta}$.
    \item Study all possible $P$ combinations of gamma functions in the numerator of~\eqref{eq:generalMBintegral} such that each gamma functions belongs to a different divisor $F_{j}$.
    \item Determine which of these combinations have isolated intersection points, that is, poles.
    \item Discard all poles that do not belong to $\Pi_{\boldsymbol{\Delta}}$.
    \item Compute the residues of the integrand of~\eqref{eq:generalMBintegral} for all relevant poles.
\end{itemize}

In addition, there are situations in which things are more complicated. It is possible, and in fact it happens for the three--point function, that more than $P$ singular hyperplanes coincide at certain points. These cases are the multivariate versions of higher multiplicity poles and they are called ``resonant'' or ``logarithmic'' due to the logarithms that appear in the resulting series because of the derivatives of the terms $(-u)^{s}$ that are involved. Later in the section we present a useful tool to deal with such cases.

There is another subtlety that we have not addressed. The half space $\Pi_{\boldsymbol{\Delta}}$ plays a key role with in the computation, but it seems to be ill--defined for $\boldsymbol{\Delta}$, which is actually true for all the integrals that we need. The solution to this issue is very simple: one may define $\Pi_{\boldsymbol{\Delta}}$ arbitrarily. However, not that the key point for~\eqref{eq:multiJordanLemma} and~\eqref{eq:JordanLemmaMB} is that one computes an integral along the skeleton of a polyhedron in terms of the poles that lie within the polyhedron. Hence the polyhedrons $\Pi$ that one chooses for the computation must have $\boldsymbol{\gamma}$ as one of its vertices and $\boldsymbol{\gamma}+i\mathbb{R}^{P}$ as one of its edges. For a given $\boldsymbol{\gamma}$ there are still infinitely many options to define $\Pi_{\boldsymbol{\Delta}}$. Nevertheless, there are still only a finite number of series representations for~\eqref{eq:generalMBintegral} that, since $\alpha>0$, are analytic continuations of each other for different values of $|u_{i}|$. Once the residues have been computed and the corresponding series representation has been obtained, one can identify the convergence region of the series obtained by applying Horn's theorem~\cite{BookBateman,BookSrivastava,HornTheorem}. It is even possible to determine the convergence region of a series before performing the full computation~\cite{algorithmFriot} in order to compute only the series representation that converges for the kinematic regime that in one is interested in. We expect to shed more light on these issues with examples later in this subsection.

Now that we have studied the poles that we need to compute~\eqref{eq:generalMBintegral}, we have only left to consider how to compute the residues on the right hand side of~\eqref{eq:JordanLemmaMB}. As happens in the single variable case, the formal definition~\eqref{eq:grothendieckResidues} usually is not the most appropriate tool.

Let us begin with the simple case in which there is a straightforward connection between univariate residues and multivariate ones. For this, let us consider again the general function $f(\boldsymbol{z})$ of~\eqref{eq:multivariatePolarFunction}. If the Jacobian determinant evaluated at the pole $\boldsymbol{z}_{0}$:
\begin{equation}\label{eq:jacobianDeterminant}
    \textit{det}\Big(\frac{\partial \phi_{j}}{\partial z_{i}}\Big)\Big|_{\boldsymbol{z}=\boldsymbol{z}_{0}}
\end{equation}
is not equal to zero then one can perform the variable change $w_{i}\equiv \phi_{i}$, which disentangles the poles and hence allows for the multivariate integral to become a product of univariate integrals. The latter can be evaluated by the usual methods. These are called ``simple poles''~\cite{BookLongTsikh}. As we mentioned previously, this is usually not the case for~\eqref{eq:generalMBintegral}.

When the Jacobian~\eqref{eq:jacobianDeterminant} is zero, then one has to use another formula called the ``Transformation law'' for multivariate residues (see page 20 of~\cite{BookLongTsikh}), which is valid for residues of any function $f(\boldsymbol{z})$ irrespective of the value of its Jacobian determinant. For a function $f(\boldsymbol{z})$ with an isolated pole at $\boldsymbol{z}=\boldsymbol{z}_{0}$ one has:
\begin{equation}\label{eq:transformationLaw}
    \textit{Res}_{\boldsymbol{z}_{0}}\frac{\eta(\boldsymbol{z})}{\phi_{1}(\boldsymbol{z})...\phi_{n}(\boldsymbol{z})} = \textit{Res}_{\boldsymbol{z}_{0}}\frac{\eta(\boldsymbol{z})\textit{det}\hat{A}}{\rho_{1}(\boldsymbol{z})...\rho_{n}(\boldsymbol{z})}\;,
\end{equation}
such that:
\begin{equation}
    \rho_{i}(\boldsymbol{z}) =\sum_{j}a_{ij}(\boldsymbol{z})\phi_{j}(\boldsymbol{z}) \hspace{1cm}\longrightarrow\hspace{1cm} \boldsymbol{\rho}(\boldsymbol{z}) =\hat{A}(\boldsymbol{z})\boldsymbol{\phi}(\boldsymbol{z})\;,
\end{equation}
where the coefficients $a_{ij}(\boldsymbol{z})$ are holomorphic functions that form the matrix $\hat{A}$ and $\boldsymbol{\rho}=(\rho_{1},...,\rho_{n})$. The holomorphy condition for these matrix elements is important to ensure that they do not cancel zeros in any $\phi_{j}$. Another requisite for~\eqref{eq:transformationLaw} to hold is that all the poles of $\boldsymbol{\rho}$ and $\boldsymbol{\phi}$ are isolated.\footnote{In the mathematical literature this result is often presented in terms of ideals noted as $\langle\phi_{1},...,\phi_{n}\rangle$ and $\langle\rho_{1},...,\rho_{n}\rangle$. The condition of isolation for the poles is equivalent to the assertion that these two are \emph{zero dimensional} ideals.} The transformation law is useful to compute multivariate residues as long as one is able to find a set $\{\rho_{j}\}$ such that each element is an univariate function, because then one may factorize the integrals and use the standard univariate machinery for residue computation. From this formula it is also easy to see that even a change in the order of the denominators $\phi_{j}$ introduces a minus sign from $\textit{A}$, which illustrates the importance of properly taking into account the orientation of the cycles in multivariate integrals.

\subsubsection{Example of the computation of scalar integrals with Mellin--Barnes integrals with one variable.}

In this subsection we compute a self--energy scalar integral, that is, a single variable Mellin--Barnes integral.

The general shape of scalar self--energy integrals is~\eqref{eq:MBselfEnergy}. Since that formula is symmetric with respect to the interchange $\mathbb{N}\setminus\{0\}\ni\nu_{1}\leftrightarrow\nu_{2}$, we can choose without loss of generality $\nu_{1}<\nu_{2}$ such that $\nu_{1}\equiv\nu\geq1$ and $\nu_{2}\equiv\nu+n$ where $n\in\mathbb{N}$ is a natural number.\footnote{There is actually loss of generality, since Mellin--Barnes representations are also valid for integrals with real propagator powers. Nevertheless, for the quark loop computation no such terms appear.} For this example we focus on the special case $n=1$ with space--time dimension $d=6$ for brevity and we will use it to provide insight about the general case. With that notation, the self--energy integral becomes:
\begin{equation}\label{eq:MBselfEnergyTransformed_1}
\begin{split}
     I^{(2)}(6;\nu,\nu+1) &=\frac{i^{-5}}{(4\pi)^{3}}\frac{(-m^{2})^{2-2\nu}}{\Gamma(\nu)\Gamma(\nu+1)}\int_{s}\Big(-\frac{q^{2}}{m^{2}}\Big)^{s}\frac{\Gamma(s+\nu)\Gamma(s+\nu+1)}{\Gamma(2s+2\nu+1)}\Gamma(-s)\Gamma\Big(s+2\nu-2\Big)\;.
\end{split}
\end{equation}
For this integral we can use the standard complex calculus tools in one variable. As mentioned previously, we always have $\Delta=0$ and $\alpha>0$, therefore the region of the complex plane where we can close the integration contour to obtain a convergent series representation is defined by the asymptotic behaviour of $(|q^{2}|/m^{2})^{s}$. For the quark loop we are interested in the high energy regime, therefore we have $|q^{2}|\gg m^{2}$. For such kinematic regime, the integrand of~\eqref{eq:MBselfEnergyTransformed_1} is decreasing in the negative real half of the complex plane. 

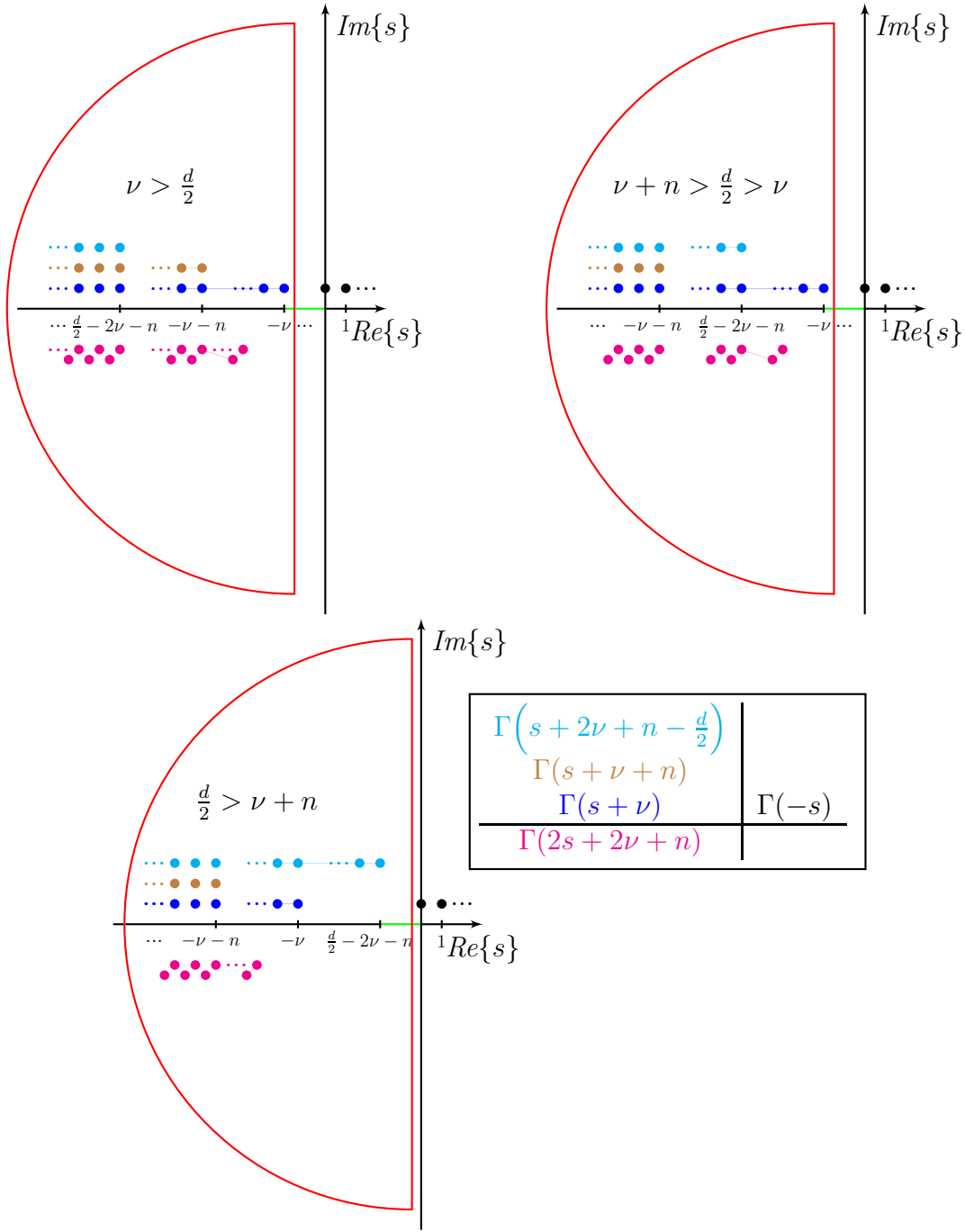
\begin{figure}
    \centering
        \begin{tikzpicture}[thick,x=0.30cm,y=0.30cm]
            \def\xr{12} \def\yr{12}
            \def\TickSize{0.2}
    
            \draw (-3-\xr,0) -- (-2,0); 
            \draw[green] (-2,0) -- (0,0); 
            \draw[-latex'] (0,0) -- (3,0) node[below] {$\textit{Re}\{s\}$};
            \draw[-latex'] (0,-3-\yr) -- (0,3+\yr) node[below right] {$\textit{Im}\{s\}$};
            \path (-1,-\TickSize-0.2) node[below,scale=0.75] {$...$} -- (-1-\xr,-\TickSize-0.2) node[below,scale=0.75] {$...$};
    
            \draw (-2,+\TickSize) -- (-2,-\TickSize) node[below,scale=0.65] at (-2-0.25,-\TickSize) {$-\nu$};
            \draw (-6,+\TickSize) -- (-6,-\TickSize) node[below,scale=0.65] at (-6-0.25,-\TickSize) {$-\nu-n$};
            \draw (-10,+\TickSize) -- (-10,-\TickSize) node[below,scale=0.65] at (-10-0.25,-\TickSize+0.2) {$\frac{d}{2}-2\nu-n$};

            \foreach \n [evaluate=\n using int(\n)] in {1,...,1}{
                \draw (\n,+\TickSize) -- (\n,-\TickSize) node[below, scale=0.65] {$1$} ;
            }

            \draw[color=white,-stealth'] (-2+0.5,\xr+2) arc[start angle=90, end angle=135, radius=\xr+2];
            \draw[color=white,-stealth'] (-2+0.5,\xr+2) arc[start angle=90, end angle=225, radius=\xr+2];
            \draw[color=white,-stealth'] (-2+0.5,-\xr-2) -- (-2+0.5,-\xr/2-2/2);
            \draw[color=white,-stealth'] (-2+0.5,-\xr-2) -- (-2+0.5,\xr/2+2/2);
            \draw[red] (-2+0.5,\xr+2) arc[start angle=90, end angle=270, radius=\xr+2] -- cycle;

            \foreach \n [evaluate=\n using int(\n)] in {0,...,1}{
                \fill (\n,1) circle[radius=2pt];
            }
            \path (2,1) node {$...$};
            \foreach \n [evaluate=\n using int(\n)] in {-10,-11,...,-12}{
                \fill[color=blue] (\n,1) circle[radius=2pt];}
            \path[color=blue] (-4,1) node {$...$} -- (-8,1) node {$...$} -- (-\xr-1,1) node {$...$};
            \fill[color=blue] (-2,1) circle[radius=2pt] -- (-3,1) circle[radius=2pt] -- (-6,1) circle[radius=2pt] -- (-7,1) circle[radius=2pt];
            \foreach \n [evaluate=\n using int(\n)] in {-10,-11,...,-12}{
                \fill[color=brown] (\n,2) circle[radius=2pt];}
            \fill[color=brown] (-6,2) circle[radius=2pt] -- (-7,2) circle[radius=2pt];
            \path[color=brown] (-\xr-1,2) node {$...$} -- (-8,2) node {$...$};
            \foreach \n [evaluate=\n using int(\n)] in {-10,-11,...,-12}{
                \fill[color=cyan] (\n,3) circle[radius=2pt];}
            \path[color=cyan] (-\xr-1,3) node {$...$};
            \foreach \n [evaluate=\n using int(\n)] in {-10,-11,...,-12}{
                \fill[color=magenta] (\n,-2) circle[radius=2pt];
                \fill[color=magenta] (\n-0.5,-2.5) circle[radius=2pt];}
            \fill[color=magenta] (-4,-2) circle[radius=2pt] -- (-4-0.5,-2.5) circle[radius=2pt] -- (-6,-2) circle[radius=2pt] -- (-6-0.5,-2.5) circle[radius=2pt] -- (-7,-2) circle[radius=2pt] -- (-7-0.5,-2.5) circle[radius=2pt];
            \path[color=magenta] (-\xr-1,-2) node {$...$} -- (-5,-2) node {$...$} -- (-8,-2) node {$...$};
    
            \path (-8,6) node{$\nu>\frac{d}{2}$};
        \end{tikzpicture}%
        \hspace{1.5cm}        
        \begin{tikzpicture}[thick,x=0.30cm,y=0.30cm]
            \def\xr{12} \def\yr{12}
            \def\TickSize{0.2}
    
            \draw (-3-\xr,0) -- (-2,0); 
            \draw[green] (-2,0) -- (0,0); 
            \draw[-latex'] (0,0) -- (3,0) node[below] {$\textit{Re}\{s\}$};
            \draw[-latex'] (0,-3-\yr) -- (0,3+\yr) node[below right] {$\textit{Im}\{s\}$};
            \path (-1,-\TickSize-0.2) node[below,scale=0.75] {$...$} -- (-1-\xr,-\TickSize-0.2) node[below,scale=0.75] {$...$};
    
            \draw (-2,+\TickSize) -- (-2,-\TickSize) node[below,scale=0.65] at (-2-0.25,-\TickSize) {$-\nu$};
            \draw (-6,+\TickSize) -- (-6,-\TickSize) node[below,scale=0.65] at (-6-0.0,-\TickSize+0.2) {$\frac{d}{2}-2\nu-n$};
            \draw (-10,+\TickSize) -- (-10,-\TickSize) node[below,scale=0.65] at (-10-0.35,-\TickSize) {$-\nu-n$};

            \foreach \n [evaluate=\n using int(\n)] in {1,...,1}{
                \draw (\n,+\TickSize) -- (\n,-\TickSize) node[below, scale=0.65] {$1$} ;
            }

            \draw[color=white,-stealth'] (-2+0.5,\xr+2) arc[start angle=90, end angle=135, radius=\xr+2];
            \draw[color=white,-stealth'] (-2+0.5,\xr+2) arc[start angle=90, end angle=225, radius=\xr+2];
            \draw[color=white,-stealth'] (-2+0.5,-\xr-2) -- (-2+0.5,-\xr/2-2/2);
            \draw[color=white,-stealth'] (-2+0.5,-\xr-2) -- (-2+0.5,\xr/2+2/2);
            \draw[red] (-2+0.5,\xr+2) arc[start angle=90, end angle=270, radius=\xr+2] -- cycle;

            \foreach \n [evaluate=\n using int(\n)] in {0,...,1}{
                \fill (\n,1) circle[radius=2pt];
            }
            \path (2,1) node {$...$};
            \foreach \n [evaluate=\n using int(\n)] in {-10,-11,...,-12}{
                \fill[color=blue] (\n,1) circle[radius=2pt];
            }
            \path[color=blue] (-4,1) node {$...$} -- (-8,1) node {$...$} -- (-\xr-1,1) node {$...$};
            \fill[color=blue] (-2,1) circle[radius=2pt] -- (-3,1) circle[radius=2pt] -- (-6,1) circle[radius=2pt] -- (-7,1) circle[radius=2pt];
            \foreach \n [evaluate=\n using int(\n)] in {-10,-11,...,-12}{
                \fill[color=brown] (\n,2) circle[radius=2pt];
            }
            \fill[color=cyan] (-6,3) circle[radius=2pt] -- (-7,3) circle[radius=2pt];
            \path[color=cyan] (-\xr-1,3) node {$...$} -- (-8,3) node {$...$};
            \foreach \n [evaluate=\n using int(\n)] in {-10,-11,...,-12}{
                \fill[color=cyan] (\n,3) circle[radius=2pt];
            }
            \path[color=brown] (-\xr-1,2) node {$...$};
            \foreach \n [evaluate=\n using int(\n)] in {-10,-11,...,-12}{
                \fill[color=magenta] (\n,-2) circle[radius=2pt];
                \fill[color=magenta] (\n-0.5,-2.5) circle[radius=2pt];
            }
            \fill[color=magenta] (-4,-2) circle[radius=2pt] -- (-4-0.5,-2.5) circle[radius=2pt] -- (-6,-2) circle[radius=2pt] -- (-6-0.5,-2.5) circle[radius=2pt] -- (-7,-2) circle[radius=2pt] -- (-7-0.5,-2.5) circle[radius=2pt];
    
            \path (-8,6) node{$\nu+n>\frac{d}{2}>\nu$};
        \end{tikzpicture}
        \hspace{1.5cm}
        \begin{tikzpicture}[thick,x=0.30cm,y=0.30cm]
            \def\xr{12} \def\yr{12}
            \def\TickSize{0.2}
    
            \draw (-3-\xr,0) -- (-2,0); 
            \draw[green] (-2,0) -- (0,0); 
            \draw[-latex'] (0,0) -- (3,0) node[below] {$\textit{Re}\{s\}$};
            \draw[-latex'] (0,-3-\yr) -- (0,3+\yr) node[below right] {$\textit{Im}\{s\}$};
            \path (-1-\xr,-\TickSize-0.2) node[below,scale=0.75] {$...$};
    
            \draw (-2,+\TickSize) -- (-2,-\TickSize) node[below,scale=0.65] at (-2-0.5,-\TickSize+0.2) {$\frac{d}{2}-2\nu-n$};
            \draw (-6,+\TickSize) -- (-6,-\TickSize) node[below,scale=0.65] at (-6-0.25,-\TickSize) {$-\nu$};
            \draw (-10,+\TickSize) -- (-10,-\TickSize) node[below,scale=0.65] at (-10-0.25,-\TickSize) {$-\nu-n$};

            \foreach \n [evaluate=\n using int(\n)] in {1,...,1}{
                \draw (\n,+\TickSize) -- (\n,-\TickSize) node[below, scale=0.65] {$1$} ;
            }

            \draw[color=white,-stealth'] (-0.45,\xr+2) arc[start angle=90, end angle=135, radius=\xr+2];
            \draw[color=white,-stealth'] (-0.45,\xr+2) arc[start angle=90, end angle=225, radius=\xr+2];
            \draw[color=white,-stealth'] (-0.45,-\xr-2) -- (-0.45,-\xr/2-2/2);
            \draw[color=white,-stealth'] (-0.45,-\xr-2) -- (-0.45,\xr/2+2/2);
            \draw[red] (-0.45,\xr+2) arc[start angle=90, end angle=270, radius=\xr+2] -- cycle;

            \foreach \n [evaluate=\n using int(\n)] in {0,...,1}{
                \fill (\n,1) circle[radius=2pt];
            }
            \path (2,1) node {$...$};
            \foreach \n [evaluate=\n using int(\n)] in {-10,-11,...,-12}{
                \fill[color=blue] (\n,1) circle[radius=2pt];
            }
            \path[color=cyan] (-4,3) node {$...$} -- (-8,3) node {$...$} -- (-\xr-1,3) node {$...$};
            \fill[color=cyan] (-2,3) circle[radius=2pt] -- (-3,3) circle[radius=2pt] -- (-6,3) circle[radius=2pt] -- (-7,3) circle[radius=2pt];
            \foreach \n [evaluate=\n using int(\n)] in {-10,-11,...,-12}{
                \fill[color=brown] (\n,2) circle[radius=2pt];
            }
            \fill[color=blue] (-6,1) circle[radius=2pt] -- (-7,1) circle[radius=2pt];
            \path[color=blue] (-\xr-1,1) node {$...$} -- (-8,1) node {$...$};
            \foreach \n [evaluate=\n using int(\n)] in {-10,-11,...,-12}{
                \fill[color=cyan] (\n,3) circle[radius=2pt];
            }
            \path[color=brown] (-\xr-1,2) node {$...$};
            \path[color=blue] (-\xr-1,1) node {$...$};
            \foreach \n [evaluate=\n using int(\n)] in {-10,-11,...,-12}{
                \fill[color=magenta] (\n,-2) circle[radius=2pt];
                \fill[color=magenta] (\n-0.5,-2.5) circle[radius=2pt];
            }
            \fill[color=magenta] (-8,-2) circle[radius=2pt] -- (-8-0.5,-2.5) circle[radius=2pt];
            \path[color=magenta] (-\xr-1,-2) -- (-9,-2) node {$...$};
    
            \path (-8,6) node{$\frac{d}{2}>\nu+n$};
            \path (12,7) node[draw,align=left] {
            \begin{tabular}{c|c}
                \textcolor{cyan}{$\Gamma\Big(s+2\nu+n-\frac{d}{2}\Big)$} & \\ 
                \textcolor{brown}{$\Gamma(s+\nu+n)$} & \\
                \textcolor{blue}{$\Gamma(s+\nu)$} & \textcolor{black}{$\Gamma(-s)$}\\\hline
                \textcolor{magenta}{$\Gamma(2s+2\nu+n)$} & \\
            \end{tabular}
        };
        \end{tikzpicture}
    \caption{Graphical representation of the polar structure of a residue computation of the scalar self--energy integral~\eqref{eq:MBselfEnergy} for positive propagator powers $\nu$ and $\nu+n$ with $n\in\mathbb{N}$ for three illustrative cases. The green interval in the real axis represents the possible values that $\gamma$ can take. Dots represent singularities and their color shows the gamma function to which they belong, as shown in the legend box. From these figures one sees that no third order poles can possibly arise from the self--energy scalar integral with positive propagator powers.}
    \label{fig:selfEnergyGeneralPoles}
\end{figure}

There are three gamma functions in the numerator that have poles in the negative part of the real axis. Using the multidimensional terminology we previously introduced, we can say that there are three families of singular hyperplanes in the negative half of the real plane: 
\begin{align}\nonumber
    &S_{1}\equiv\{s\in\mathbb{C}|\,s=-p-\nu\hspace{0.5cm} \text{for $p\in\mathbb{N}$ }\}\;, \hspace{0.75cm}S_{2}\equiv\{s\in\mathbb{C}|\,s=-p-\nu-1\hspace{0.5cm} \text{for $p\in\mathbb{N}$ }\}\;,\\\label{eq:examplePoleSets}
    &S_{3}\equiv\Big\{s\in\mathbb{C}\Big|\,s=-p+2-2\nu\hspace{0.5cm} \text{for $p\in\mathbb{N}$ }\Big\}\;.
\end{align}
For each of these three sets, the rightmost pole is obtained for $p=0$, therefore we have $s=-\nu$ for $S_{1}$, $s=-\nu-1$ for $S_{2}$ and $s=2-2\nu$ for $S_{3}$. Consequently, the rightmost pole of all three sets is $-\nu$ if $2-\nu<0$ or $2-2\nu$ otherwise. From these two cases, we specialize this computation for the case $\nu>2$, since it has a wider range of use. There is also a set of poles on the positive real axis, and its leftmost pole is $s=0$. Therefore $\gamma$ may be anywhere within the interval $(-\nu,0)$ in order not to split any of these four sets of poles (see figure~\ref{fig:selfEnergyGeneralPoles} for the general case). In multivariate residues language, this means that the polyhedron $\Pi$ that interests us is defined by $\{s\in\mathbb{C}|\,\textit{Re}\{s\}<\gamma\}$ and the integration path $\gamma+i\mathbb{R}$ is both its edge and its face. In the single variable case one still does not see much freedom to choose $\Pi$ even though $\Delta=0$, because there exist only two polyhedrons for which $\gamma+i\mathbb{R}$ is an edge, namely, the positive and negative real halves.

To expose removable singularities caused by the presence of a gamma function in the denominator we use the duplication formula:
\begin{equation}
    \Gamma(2s) =\frac{2^{2s-1}}{\sqrt{\pi}}\Gamma(s)\Gamma\Big(s+\frac{1}{2}\Big)\;,
\end{equation}
which, when applied on to the gamma functions in the integral~\eqref{eq:MBselfEnergyTransformed_1}, yields 
\begin{equation}
    \sqrt{4\pi}\Big(-\frac{q^{2}}{4m^{2}}\Big)^{s}\frac{\Gamma(s+\nu)\Gamma(s+\nu+1)}{\Gamma(s+\nu+\frac{1}{2})\Gamma(s+\nu+1)}\Gamma(-s)\Gamma(s+2\nu-2)\;.
\end{equation}
We have not performed an obvious cancellation of gamma functions to emphasize that in the general $n$ case it is necessary to define if $n$ is odd or even in order to know which of the two $\Gamma$ in the denominator is removing singularities. Applying basic recurrence formulas to shed light on the actual multiplicity of the remaining poles one obtains a the following expression for the integrand of~\eqref{eq:MBselfEnergyTransformed_1}:
\begin{equation}
    \sqrt{4\pi}\Big(-\frac{q^{2}}{4m^{2}}\Big)^{s}\frac{\Gamma(-s)}{\Gamma(s+\nu+\frac{1}{2})}\frac{\Gamma^{2}(s+2\nu-2)}{\prod_{j=0}^{\nu-3}(s+\nu+j)}\;.
\end{equation}
We have explicitly extracted the finite set of poles that only belong to the $\Gamma(s+\nu)$ in order the separate them from the rest which are shared with $\Gamma(s+2\nu-2)$ and whose multiplicity equal to two is now evident by the power of the gamma function. Let us first consider the residues of the finite set of simple poles:
\begin{equation}\label{eq:simplePoles}
    \sqrt{4\pi}\Big(-\frac{4m^{2}}{q^{2}}\Big)^{\nu}\sum_{k=0}^{\nu-3}\Big(\frac{4m^{2}}{q^{2}}\Big)^{k}\frac{\Gamma(\nu+k)}{\Gamma(-k+\frac{1}{2})}\frac{\Gamma(-k+\nu-2)}{\Gamma(k+1)}\;.
\end{equation}
where $k\in\{0,...,\nu-3\}$ and we have represented the integrand of~\eqref{eq:MBselfEnergyTransformed_1} by $I^{(2)}$ for brevity.

Second order poles are defined by $s=2-2\nu-l$ for every $l\in\mathbb{N}$. It is therefore convenient to place them at the origin by the following change of variables: $s\rightarrow s+2-2\nu-l$. Then, the singularity is exposed by using the generalized reflection formula for the gamma function: 
\begin{equation}\label{eq:reflectionFormula}
    \Gamma(s-n) =(-1)^{n}\frac{\Gamma(1-s)\Gamma(s+1)}{s\Gamma(n+1-s)}\hspace{1cm} \text{For $n\in\mathbb{Z}$}
\end{equation}
and thus the integrand becomes: 
\begin{equation}
    \sqrt{4\pi}\Big(-\frac{q^{2}}{4m^{2}}\Big)^{s+2-2\nu-l}\frac{\Gamma(-s-2+2\nu+l)}{\Gamma(s-\nu-l+\frac{5}{2})}\frac{\Gamma^{2}(1-s)\Gamma^{2}(s+1)}{s^{2}\Gamma^{2}(l+1-s)\prod_{j=0}^{\nu-3}(s+2-\nu-l+j)}\;.
\end{equation}
Using the residues formula for higher multiplicities we find the full contribution from second order poles to be:
\begin{align}\label{eq:secondOrderPoles}
    &-\sqrt{4\pi}\Big(-\frac{16m^{4}}{q^{4}}\Big)^{\nu-1}\sum_{l=0}^{\infty}\Big(-\frac{4m^{2}}{q^{2}}\Big)^{l}\frac{\Gamma(-2+2\nu+l)}{\Gamma(-\nu-l+\frac{5}{2})}\frac{1}{\Gamma(l+1)\Gamma(l+\nu-1)}\\\nonumber
    &\hspace{0cm}\times\Bigg(\ln\Big\{-\frac{q^{2}}{4m^{2}}\Big\} -\psi^{0}(-2+2\nu+l) -\psi^{(0)}(-\nu-l+\frac{5}{2})-3\psi^{(0)}(l+1) +\psi^{0}(l+\nu-1)\Bigg)\;.
\end{align}
where $\Gamma'(z)$ is the first derivative of the gamma function and $\psi^{0}(z)\equiv\frac{d}{dz}\ln\Gamma(z)=\Gamma'(z)/\Gamma(z)$ is the digamma function. To simplify some terms we have used the fact that for $p\in\mathbb{N}\setminus\{0\}$ and $H_n=\sum_k^n 1/k$ an harmonic number we have $\psi^{(0)}(p)=H_{p-1} -\gamma_E$.

In conclusion we find that the scalar self--energy integral~\eqref{eq:MBselfEnergyTransformed_1}, which belongs to a special family of the general type in~\eqref{eq:MBselfEnergy} with $\nu_{1}\equiv\nu\geq2$, $\nu_{2}\equiv\nu+1$ and $d=6$, is equal to the sum of the expressions in~\eqref{eq:simplePoles} and~\eqref{eq:secondOrderPoles} multiplied by the factor of the Mellin--Barnes integral in~\eqref{eq:MBselfEnergy}. 

Note that this result is quite different from the one cited in equation (18) of~\cite{BoosDavydychev}. In such case there appear three hypergeometric series and no logarithms or digamma functions. The reason for such discrepancy is that the result obtained in~\cite{BoosDavydychev} is valid for ``generic'' values of $\nu_{1}$, $\nu_{2}$ and $d$, that is, they consider the case in which none of them produces higher multiplicity poles of removable singularities, which does happen in our example. Consequently, closing the integration contour to the left half of the complex plane only implies summing over three independent sets of simple poles from three gamma functions, thus three hypergeometric series appear.

Several important general properties of the scalar integrals that appear in the quark loop with shifted dimensions are present in this result. First of all, we can see in~\eqref{eq:secondOrderPoles} the infrared divergent logarithms $\ln\{q^{2}/m^{2}\}$ that we renormalized in a previous section. In our computation these came due to the presence of higher multiplicity poles in the integrand. In general, self--energy integrals have three gamma functions in the numerator with poles in the negative real half of the complex plane (see equation~\eqref{eq:MBselfEnergy}), so one might think that poles of order $3$ may arise. This would be problematic, since it would introduce further infrared divergent terms like $\ln^{2}\{q^{2}/m^{2}\}$, which we have not taken into account in the renormalization procedure. The absence of such poles is however not a lucky coincidence of the example that we have just considered, but rather a general property of self--energy scalar integrals. It is easy to see that the general version of the pole sets in~\eqref{eq:examplePoleSets} for arbitrary $d$ and $n$ is:
\begin{align}\nonumber
    &S_{1}\equiv\{s\in\mathbb{C}|\,s=-p-\nu\hspace{0.5cm} \text{for $p\in\mathbb{N}$ }\}\;,\hspace{0.75cm}S_{2}\equiv\{s\in\mathbb{C}|\,s=-p-\nu-n\hspace{0.5cm} \text{for $p\in\mathbb{N}$ }\}\;,\\\label{eq:threeSingularSets}
    &S_{3}\equiv\Big\{s\in\mathbb{C}\Big|\,s=\frac{d}{2}-p-2\nu-n\hspace{0.5cm} \text{for $p\in\mathbb{N}$ }\Big\}\;.
\end{align}
 If $d$ is an odd integer, then the intersection of $S_{1}$ or $S_{2}$ with $S_{3}$ is empty and no third order poles appear. Now let us consider the case when $d$ is an even integer. Since all of them extend infinitely along the negative integers, the intersection of these sets has an infinite number of elements and, in fact, it is equal to whichever of the three sets starts further to the left in the real axis. Just as we did in the example, for the general case the gamma function in the denominator can be decomposed by means of the duplication formula to get a divisor proportional to $\Gamma(s+\nu+\frac{n+1}{2})\Gamma(s+\nu+\frac{n}{2})$. Since $\nu$ and $n$ are always positive integers in the quark loop computation, then one these two gamma function will always remove singularities from the gamma functions in the numerator. Which one of them does it depends on whether $n$ is even or odd and we consider the latter case. In brief we argue that it is a worst--case scenario. The factor $1/\Gamma(s+\nu+\frac{n+1}{2})$ has a set of zeros $\zeta=\{s\in\mathbb{C}|\,s=-p-\nu-\frac{n+1}{2}\hspace{0.5cm} \text{for $p\in\mathbb{N}$ }\}$. Note that even though the gamma function in the denominator lowers the order of an infinite number of singularities, it is not enough to lower the order of all singularities: singularities which lie to the right of $\zeta$ do not have their order lowered. Consequently, if this set of zeros starts further enough to the left\footnote{We refer to right or left in the sense of the negative and positive directions, respectively, of the real axis.} such that it does not lower the multiplicity of one or more of the poles in $S_{1}\cap S_{2} \cap S_{3}$, then there will appear new infrared singularities of the type $\ln^{2}\{q^{2}/m^{2}\}$. To determine if or when this is possible, we need to show that the rightmost zero in $\zeta$ is always further to the right than the rightmost pole of at least one set $S_{i}$. Looking at figure~\ref{fig:selfEnergyGeneralPoles} it is very easy to understand that such is the case for $S_{2}$ as long as $n$ is a natural number. Hence, we prove that only simple logarithmic infrared divergences are introduced by self--energy scalar integrals $I^{(2)}(d;\nu_{1},\nu_{2})$ with any positive integers $d$, $\nu_{1}$ and $\nu_{2}$ in the high energy regime. It is obvious that for the low energy not even logarithmic divergences can appear.

As a conclusion of our analysis of our self--energy example, we consider the convergence properties of the series representation found and in particular of~\eqref{eq:secondOrderPoles}, which has an infinite number of terms and a rather complex structure that requires careful study. The standard tools for the analysis of convergence properties of hypergeometric--like functions of a single variable are d'Alembert's ratio test and Raabe's test~\cite{BookBateman}. For this example we choose the former. Let us define $a_l$ as the $l$--th term of the series in~\eqref{eq:secondOrderPoles}. Hence we find:
\begin{equation}
\begin{split}
    \Big|\frac{a_{l+1}}{a_{l}}\Big| &=\Big(-\frac{4m^{2}}{q^{2}}\Big)\frac{(-2+2\nu+l)(-\nu-l+\frac{3}{2})}{(l+1)(l+\nu-1)}\\
    &\hspace{-1cm}\times\Bigg(\frac{\ln\Big\{-\frac{q^{2}}{4m^{2}}\Big\} -\psi^{(0)}(-2+2\nu+l+1) -\psi^{(0)}(-\nu-l-1+\frac{5}{2})-3\psi^{(0)}(l+2) +\psi^{(0)}(l+\nu)}{\ln\Big\{-\frac{q^{2}}{4m^{2}}\Big\} -\psi^{(0)}(-2+2\nu+l) -\psi^{(0)}(-\nu-l+\frac{5}{2})-3\psi^{(0)}(l+1) +\psi^{(0)}(l+\nu-1)}\Bigg)\;.
\end{split}
\end{equation}
Taking into account the asymptotic behaviour of the digamma function: $\psi^{(0)}(z)\rightarrow\ln z +\frac{1}{2z} +O(\frac{1}{z^{2}})$, one concludes that $\lim_{l\rightarrow\infty}|a_{l+1}/a_{l}|=4m^2/q^2$.
Therefore, we see that the series representation that we found converges absolutely for $q^{2}>4m^{2}$, that is, above the threshold for particle--antiparticle production.

\subsection{Final stages of the quark loop computation and analysis}\label{subsec:finalStagesQuarkLoop}

At this point we have all the necessary tools to compute self--energy and tadpole integrals (\eqref{eq:MBselfEnergy} and~\eqref{eq:MBtriangle}). Nevertheless, in the quark loop expression there appear more than one hundred different scalar integrals of these two types, hence automation is required. For this we have used a \textit{Mathematica} package called \textit{MBConicHulls}\footnote{This package requires \textit{Mathematica} 12 or a more recent version.}~\cite{packageFriot}, which calls upon functions of another package called \textit{MultivariateResidues}~\cite{multivariateLarsen} that has to be installed as a dependency. In~\cite{packageFriot} and~\cite{algorithmFriot} the authors describe how the computation of Mellin--Barnes integrals with multivariate residues, which we have just reviewed, can be organized in a very compact algorithm that uses very intuitive geometric concepts and allows to understand the practical implications of the rather abstract results of multivariate complex calculus.

A typical Mellin--Barnes integral representing an scalar triangle loop has 16 different series representations, and each of them contains up to six different subseries. Consequently, the assessment of the convergence regions of the series representations found by the \textit{MBConicHulls} package requires automation as well. We have developed a program that evaluates the asymptotic behaviour of a given triple series and finds its region of convergence by comparing it with the behaviour of other series whose convergence conditions are already known. The concept of the program is based on Horn's theorem for the convergence of hypergeometric series of up to three variables~\cite{HornTheorem}, which is a rather natural extension of D'Alembert's ratio test to the multivariate case. Let us consider the triple series
\begin{equation}
    \sum_{n_1,n_2,n_3}^\infty C(n_1,n_2,n_3) x^{n_1}y^{n_2}z^{n_3}\;.
\end{equation}
It is considered hypergeometric as long as the coefficients 
\begin{equation}
\begin{split}
    f(n_1,n_2,n_3) &=\frac{C(n_1+1,n_2,n_3)}{C(n_1,n_2,n_3)},\hspace{0.5cm}
    g(n_1,n_2,n_3) =\frac{C(n_1,n_2+1,n_3)}{C(n_1,n_2,n_3)},\hspace{0.5cm}
    h(n_1,n_2,n_3) =\frac{C(n_1,n_2,n_3+1)}{C(n_1,n_2,n_3)}
\end{split}
\end{equation}
are rational functions of $n_1$, $n_2$ and $n_3$. If so, then the convergence region of the integral is given by the intersection of the following five sets:
\begin{equation}
\begin{split}
    C&=\Big\{(|x|,|y|,|z|)\;\Big|\;|x|<\rho(1,0,0)\land|y|<\sigma(1,0,0)\land |z|<\tau(1,0,0)\Big\}\\
    X&=\Big\{(|x|,|y|,|z|)\;\Big|\;\forall(n_2,n_3)\in\mathbb{R}^{2}_{+}\,:\,|x|<\rho(0,n_2,n_3)\lor|y|<\sigma(0,n_2,n_3)\lor |z|<\tau(0,n_2,n_3)\Big\}\\
    Y&=\Big\{(|x|,|y|,|z|)\;\Big|\;\forall(n_1,n_3)\in\mathbb{R}^{2}_{+}\,:\,|x|<\rho(n_1,0,n_3)\lor|y|<\sigma(n_1,0,n_3)\lor |z|<\tau(n_1,0,n_3)\Big\}\\
    Z&=\Big\{(|x|,|y|,|z|)\;\Big|\;\forall(n_1,n_2)\in\mathbb{R}^{2}_{+}\,:\,|x|<\rho(n_1,n_2,0)\lor|y|<\sigma(n_1,n_2,0)\lor |z|<\tau(n_1,n_2,0)\Big\}\\
    E&=\Big\{(|x|,|y|,|z|)\;\Big|\;\forall(n_1,n_2,n_3)\in\mathbb{R}^{3}_{+}\,:\,|x|<\rho(n_1,n_2,n_3)\lor|y|<\sigma(n_1,n_2,n_3)\lor |z|<\tau(n_1,n_2,n_3)\Big\}\;,
\end{split}
\end{equation}
where $\mathbb{R}_{+}$ represents the set of positive reals and $\rho$, $\sigma$ and $\tau$ capture the asymptotic behaviour of $f$, $g$ and $h$: 
\begin{align}\nonumber
    \rho(n_1,n_2,n_3) &= \Big|\lim_{u\rightarrow\infty}f(un_1,un_2,un_3)\Big|^{-1},\hspace{0.5cm}
    \sigma(n_1,n_2,n_3) = \Big|\lim_{u\rightarrow\infty}g(un_1,un_2,un_3)\Big|^{-1},\\
    \tau(n_1,n_2,n_3) &= \Big|\lim_{u\rightarrow\infty}h(un_1,un_2,un_3)\Big|^{-1}\;.
\end{align}
The program that we have developed computes $\rho$, $\sigma$ and $\tau$ for each subseries that form a series representation of a Mellin--Barnes integral and identifies its region of convergence by comparing them to the $\rho$, $\sigma$ and $\tau$ of series whose convergence conditions are known. Care had to be taken for the program not be misled by redefinitions of the arguments of the series or the presence of logarithms. We have also taken into account the result found in~\cite{algorithmFriot} which extends the use of Horn's theorem to series that are not hypergeometric by the definition given previously, because they include polygamma functions. Finally, the program chooses the appropriate series representation according to the kinematic regime indicated beforehand.

The convergence region of some triple series representations of triangle loops in shifted dimensions could not be found in the mathematical literature due to them being quite non--standard. In such cases the approach presented in~\cite{Phan2019}, alternate to~\cite{DavydychevMellinBarnes}, was followed. That paper refers to scalar triangle loop integrals in arbitrary space--time dimension with three different masses, but unit propagator powers: 
\begin{equation}
\begin{split}
    J_{3}^{(d)} &=\int\frac{d^dp}{(2\pi)^d}\frac{1}{(p+p_1+p_2)^2-m_3^2}\frac{1}{(p+p_1)^2-m_2^2}\frac{1}{p^2-m_1^2}\;.
\end{split}
\end{equation}

The first step of the computation is to use Feynman parameters in the  standard way, as we described for the self--energy loop when introducing formula~\eqref{eq:MBselfEnergy}. An appropriate change of variables renders one of the two Feynman parameter integrals straightforward to perform. After using~\eqref{eq:MBrepresentation} on the integrand, the remaining Feynman parameter integral has the one--variable Gaussian hypergeometric function $_2F_1$ as its solution. Finally, the Mellin--Barnes integral of $_2F_1$ yields the double Appell hypergeometric function $F_1$. The key point of this result is that $F_1$ belongs to the well--known family of Gaussian hypergeometric functions and as such its convergence and analytical continuation properties are well--known~\cite{BookSrivastava,BookSlater}. We quote here the result for arbitrary space--time dimension $d$ valid in the high energy regime: 
\begin{equation}\label{eq:scalarTriangleArbitraryDimension}
\begin{split}
    J_3^{(d)} &=\frac{i\Gamma\Big(\frac{4-d}{2}\Big)}{(4\pi)^{\frac{d}{2}}\lambda^{1/2}(p_{1}^{2},p_{2}^{2},p_{3}^{2})}\Bigg\{J^{(d)}_{123} -(M_3-i\epsilon)^{\frac{d-4}{2}}J^{(d=4)}_{123} +(1,2,3)\leftrightarrow(2,3,1)\\
    &\hspace{4cm}+(1,2,3)\leftrightarrow(3,1,2)\Bigg\}\;,
\end{split}
\end{equation}
where $p_3=-p_1-p_2$ and: 
\begin{equation}
\begin{split}
    J_{ijk}^{(d)} &=\frac{x_{ij}}{(x_{k}-x_{ij})}(M_{ij}-i\epsilon)^{\frac{d-4}{2}}F_1\Big(\frac{1}{2};1,\frac{4-d}{2};\frac{3}{2};\frac{x_{ij}^{2}}{(x_k-x_{ij})^{2}},-\frac{p_i^2x^{2}_{ij}}{M_{ij}-i\epsilon}\Big)\\
    &-\frac{x_{ij}^{2}}{2(x_k-x_{ij})^2}(M_{ij}-i\epsilon)^{\frac{d-4}{2}}F_1\Big(1;1,\frac{4-d}{2};2;\frac{x^2_{ij}}{(x_k-x_{ij})^2},-\frac{p_i^2x^2_{ij}}{M_{ij}-i\epsilon}\Big)\\
    &-\Big\{x_{ij}\rightarrow 1-x_{ij}\;\;;\;\;x_{k}\rightarrow 1-x_{k}\Big\}
\end{split}
\end{equation}
\begin{equation}
    x_{ij} =\frac{p_i^2+m_i^2-m_j^2}{2p_i^2}\;.
\end{equation}
$M_3$ and $M_{ij}$ for $i,j=1,2,3$ are defined in terms of Cayley and Gramm determinants for the triangle loop. Their definition and properties are given in appendix~\ref{AppChapter3}. The definition of $x_k$ is rather lengthy and not very relevant, so it is written in the appendix as well. The Appell function $F_1$ has the convergent series representation: 
\begin{equation}
    F_{1}(a;b,b';c;x,y) =\sum_{n_1,n_2=0}\frac{(a)_{n_1+n_2}(b)_{n_1}(b')_{n_2}}{(c)_{n_1+n_2}}\frac{x^{n_1}}{n_1!}\frac{y^{n_2}}{n_2!}
\end{equation}
for $|x|<1$ and $|y|<1$. In the high energy regime we have $\Big|\frac{x_{ij}^2}{(x_k-x_{ij})^2}\Big|=\Big|\frac{m_i^2-M_{ij}}{M_3-M_{ij}}\Big|<1$ and $\Big|\frac{p_i^{2}x_{ij}^2}{M_{ij}}\Big|=\Big|1-\frac{m_i^2}{M_{ij}}\Big|<1$, therefore this representation is valid for our quark loop computation. From this result, triangle loops with arbitrary propagator powers can be computed from $J_3^{(d)}$ via derivatives with respect to the masses:
\begin{equation}
    I^{(N)}(d;\nu_1,...,\nu_N) =\prod_i\Bigg(\frac{1}{\Gamma(\nu_i)}\Big(\frac{\partial}{\partial m_i^2}\Big)^{\nu_i-1}\Bigg)J_3^{(d)}\Bigg|_{m_{i}=m}\;.
\end{equation}
We are interested in integrals with $d\in\{4,6,8,10,12\}$. It is not difficult to note that the gamma function pole at $d=4$ in $J_3^{(d)}$ is of course a spurious singularity. On the other hand for $d\geq6$ there are actual ultraviolet singularities but, it is possible to check in a lengthy but straightforward way that singular terms vanish and dependence on the renormalization scale disappears when propagator powers get high enough in $I^{(N)}$. This is relevant for us, because of the loop integrals we find are ultraviolet finite.

After the Mellin--Barnes representation of the scalar integrals with shifted dimensions are computed by the methods described previously\footnote{The script that performs the steps that we have described throughout this section can be found in~\href{https://github.com/DanielMelo2000/QuarkLoopCode}{this} repository.}, the last step is the computation of the integrals over $|Q_{1}|$, $|Q_{2}|$ and $\tau$ that remain in the master formula.

First, let us remember that even though we followed a kinematic--singularity--free tensor loop decomposition, there are spurious kinematic singularities which have been introduced by the negative powers of $\lambda$ that are present in the projectors with which one extracts the HLbL form factors from the quark loop. As discussed at the beginning of this section, these singularities cancel explicitly for contributions of all Wilson coefficients, except the quark loop. This is expected, since the spurious nature of the singularities implies that they must disappear in tree--level contributions. For self--energy and triangle loop integrals in shifted dimensions we do not arrive in general to closed analytical expression, but rather a series representation. Therefore, spurious kinematic zeros inside these terms do not necessarily show explicitly to cancel singularities. This introduces numerical instability in the region of the master formula's angular integral when $\tau\equiv\hat{Q}_{1}\cdot\hat{Q}_{2}\rightarrow\pm1$, because that is when $\lambda$ is equal to or approaches zero: 
\begin{equation}
    \lambda(q_{1}^{2},q_{2}^{2},q_{3}^{2}) =(q_{3}^{2} -q_{1}^{2} -q_{2}^{2})^{2} -4q_{1}^{2}q_{2}^{2} = 4Q_{1}^{2}Q_{2}^{2}\Big(\tau^{2}-1\Big)\;,
\end{equation}
where we have switched back to the euclidean versions of the virtual photon momenta $q_{i}\cdot q_{j}\rightarrow -Q_{i}\cdot Q_{j}$. When computing the contribution to the master formula's integral from these regions, it is convenient to expand the integral's series representations around $\tau=\pm1$ to avoid numerical instability. The fact that we traced the $\lambda=0$ singularity to a value in $\tau$ has useful practical implications. Indeed self--energy and triangle scalar integrals are computed in a single and triple series representation, respectively, where the expansion variable are the differences between external momenta. The external momenta that can appear in quark loop scalar integrals are $Q_{1}$ and $Q_{3}$ or $Q_{2}$ and $Q_{3}$, depending on the permutation one is considering. However, only $Q_{3}$ depends on $\tau$. Therefore any integral that does not depend on $Q_{3}$ does not require special treatments.

To perform the integrals on the euclidean norm of the virtual photons' momenta, it is important to keep in mind that the quark loop was obtained from an OPE in perturbative QCD. Therefore its range of validity starts above $\Lambda_{\text{QCD}}$, the perturbative threshold. $\Lambda_{\text{QCD}}$ is usually taken to be close to the proton's mass, which is about $\sim 940$~MeV. In principle, this means that one can compute the quark loop contribution to $a_{\mu}$ starting from $|Q_{1}|=|Q_{2}| =1~\text{GeV}\equiv Q_{min}$, however, taking into account that the OPE framework discussed in previous sections introduces an implicit counting parameter $\Lambda_{\text{QCD}}/|Q|$, one would expect the error coming from neglected higher non-perturbative effects to be large right above $\Lambda_{\text{QCD}}$. The relation between the size of such error and the values of $Q_{min}$ was studied in~\cite{Bijnens2020}. To that end, they computed the quark loop contribution as a function of $Q_{min}$ in the interval $[1~\text{GeV},4~\text{GeV}]$ and the contributions from the non--perturbative condensates of the previous sections were considered as well. Their results showed that massless quark loop contributions fall like $1/Q_{min}^{2}$ and, in general, contributions from elements of the OPE with dimension $d$ behave like $1/Q_{min}^{d}$. This is expected from the asymptotic behaviour of the integral kernels $T_{i}$ of the master formula~\eqref{eq:masterFormula}, which for $|Q_{i}|\rightarrow\infty$ behave like $m_{\mu}^{2}/Q_{i}^{2}$, except for $T_{1}$, which falls like $m_{\mu}^{4}/Q_{i}^{4}$. Since mass effects become small for large momenta $|Q_{i}|$ and the massless quark loop contribution does not introduce an energy scale, then it must fall like $1/Q_{min}$. Mass corrections to the quark loop are suppressed by $m^{2}_{f}/Q_{i}^{2}$ with respect to the massless part. In addition, contributions from other OPE elements $S_{i,\mu\nu}$ of dimension $d$ are comparatively suppressed as well by a factor $(\Lambda_{QCD}/|Q_{i}|)^{d-2}$, thus the asymptotic behaviour of their contributions is explained.\footnote{Asymptotic freedom also plays a role in this result. As we mentioned in the OPE sections, the correction to the naive dimensional counting of the OPE is given by the anomalous dimension of each OPE element, but QCD's asymptotic freedom ensures that, at high enough energy, corrections are small.}

For the value of the quarks' masses $m_{f}$ and the renormalization scale $\mu$ we followed the simplified choice of~\cite{Bijnens2020}, which was:\footnote{In~\cite{WP} constituent masses are used, because they are more appropriate when comparing with low--energy results.}
\begin{align}
    m_{u}=m_{d}= 5~\text{MeV}\hspace{1cm}m_{s}= 100~\text{MeV}\hspace{1cm} \mu=Q_{min}\;.
\end{align}    
Note that no running of the masses is performed. This is justified by the very small size of mass corrections to the quark loop. With this values we computed the quark loop contribution for $Q_{min}=1$ or $2$~GeV. As discussed previously in this section, we obtained a systematic expansion of the quark loop in terms of the quarks masses. This allowed us to study the mass corrections to the massless quark loop contribution and found them to be very small, even at $m^{2}_{f}$ order. Furthermore, we found the result for $Q_{min}=1$~GeV to be about four times bigger than the $Q_{min}=2$~GeV case. 

In~\cite{Bijnens2020} the massless quark loop was found to be the largest contribution to $a_{\mu}$ by two orders of magnitude and the leading mass corrections were even smaller than the non--perturbative di--quark magnetic susceptibility ($S_{2,\mu\nu}$ in the OPE) by two further orders of magnitude. The complete results of~\cite{Bijnens2020} are summarized in table~\ref{tab:resultsBijnens}, where one can see that the dominance of quark loop contributions with respect to the other. Note however that those results do not show the complete picture, because the quark loop contribution, which is the leading perturbative contribution to $a_{\mu}^{\text{HLbL}}$, does not really involve strong interactions and hence it does not depend on $\alpha_{s}$. Nevertheless, in~\cite{Bijnens2021} the next--to--leading--order (NLO) gluonic correction to the massless part of the quark loop was computed (see figure~\ref{fig:quarkLoopNLO}) taking into account the running of $\alpha_{s}(\mu)$ and its contribution to $a_{\mu}^{\text{HLbL}}$ was found to be about 10 \% of the leading order and negative. 

\begin{table}
    \centering
    \begin{tabular}{|c|r|c|r|r|}\hline
        \multirow{2}{*}{OPE element} & \multicolumn{1}{c|}{Magnetic} & \multirow{2}{*}{Mass order} & \multicolumn{2}{c|}{Contribution to $a_{\mu}$ from $Q_{min}$} \\\cline{4-5}
         & \multicolumn{1}{c|}{susceptibility} & & \multicolumn{1}{c|}{$1$~GeV} & \multicolumn{1}{c|}{$2$~GeV}\\\hline\hline
        \multirow{2}{*}{$S_{1,\mu\nu}$} & \multicolumn{1}{c|}{\multirow{2}{*}{$1$}} & $m^{0}$ & $1.73\cdot10^{-10}$ & $4.35\cdot10^{-11}$\\
         &  & $m^{2}$ & $-5.7\cdot 10^{-14}$ & $-3.6\cdot 10^{-15}$\\\hline
        \multirow{2}{*}{$S_{2,\mu\nu}$} & \multicolumn{1}{c|}{\multirow{2}{*}{$-4\cdot10^{-2}$~GeV}} & $m^{1}$ & $-1.2\cdot10^{-12}$ & $-7.3\cdot10^{-14}$\\
        &  & $m^{3}$ & $6.4\cdot 10^{-15}$ & $1.0\cdot 10^{-16}$\\\hline
        $S_{3,\mu\nu}$ & $3.5\cdot 10^{-3}$~GeV$^3$ & \multirow{7}{*}{$m^{0}$}  & $-3.0\cdot10^{-14}$ & $-4.7\cdot10^{-16}$\\
        $S_{4,\mu\nu}$ & $3.5\cdot10^{-3}$~GeV$^3$ &  & $3.3\cdot10^{-14}$ & $5.3\cdot10^{-16}$\\
        $S_{5,\mu\nu}$ & $-1.6\cdot10^{-2}$~GeV$^3$ &  & $-1.8\cdot10^{-13}$ & $-2.8\cdot10^{-15}$\\
        $S_{6,\mu\nu}$ & $2\cdot10^{-2}$~GeV$^4$ &  & $1.3\cdot10^{-13}$ & $2.0\cdot10^{-15}$\\
        $S_{7,\mu\nu}$ & $3.3\cdot10^{-3}$~GeV$^4$ &  & $9.2\cdot10^{-13}$ & $1.5\cdot10^{-14}$\\
        $S_{8,1,\mu\nu}$ & $-1.5\cdot10^{-4}$~GeV$^4$ &  & $3.0\cdot10^{-13}$ & $4.7\cdot10^{-15}$\\
        $S_{8,2,\mu\nu}$ & $ -1.4\cdot10^{-4}$~GeV$^4$ &  & $-1.3\cdot10^{-13}$ & $-2.0\cdot10^{-15}$\\\hline
    \end{tabular}
    \caption{Results published in~\cite{Bijnens2020} about the contribution of the quark loop and the rest of OPE elements $S_{i,\mu\nu}$ to $a_{\mu}$ as function of the cutoff $Q_{min}$ from which the master formula integral is performed.}
    \label{tab:resultsBijnens}
\end{table}

\begin{figure}[b]
    \centering
    \begin{fmffile}{quarkLoopNLO}
        \setlength{\unitlength}{0.6cm}\small
        \begin{fmfgraph*}(5,5) 
            \fmfleft{i1,i2}
            \fmfright{o1,o2,o3}         
            \fmf{phantom}{v1,v2,v3,v4,v5}
            \fmf{phantom}{v5,v6,v7,v8,v9}
            \fmf{phantom}{v9,v10,v11,v12,v1}
            \fmf{photon}{i1,v1}
            \fmf{photon}{v9,i2}
            \fmf{photon}{v5,o2}
            \fmffreeze
            \fmf{gluon}{v11,v7}
            \fmf{photon}{v3,o1}
            \fmf{fermion}{v1,v3}
            \fmf{plain}{v3,v5}
            \fmf{fermion}{v5,v7}
            \fmf{plain}{v7,v9}
            \fmf{fermion}{v9,v11}
            \fmf{plain}{v11,v1}
            \fmfset{arrow_len}{2.5mm}
            \fmfset{curly_len}{2mm}
            \fmfv{decor.shape=circle,decor.size=1mm}{o1}
            \fmffreeze
            \fmfshift{(-0.3w,0.2h)}{o1}
        \end{fmfgraph*}
    \end{fmffile}
    \caption{Representative diagram of the NLO contribution to the Wilson coefficient of $S_{1,\mu\nu}$ in the OPE of $\Pi^{\mu_{1}\mu_{2}\mu_{3}}$. The black dot represents creation/annihilation of a line by the background fields in the vacuum. This diagram represents the first QCD correction to the quark loop.}
    \label{fig:quarkLoopNLO}
\end{figure}
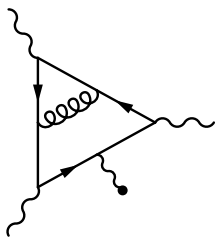

\begin{figure}[b]
    \centering
    \begin{fmffile}{quarkLoopMixedVirtualities}
        \setlength{\unitlength}{0.6cm}\small
        \begin{fmfgraph*}(5,5) 
            \fmfleft{i}
            \fmftop{iup}
            \fmfright{o}
            \fmfbottom{odown}
            \fmf{photon}{i,v1}
            \fmf{photon}{v3,o}
            \fmf{phantom}{i,v1,v2,v3,o}
            \fmf{phantom}{i,v1,v4,v3,o}
            \fmffreeze
            \fmfshift{(0.0w,0.2h)}{v2}
            \fmfshift{(0.0w,-0.2h)}{v4}
            \fmf{fermion,left=0.3,tension=1}{v1,v2}
            \fmf{fermion,left=0.3,tension=1}{v2,v3}
            \fmf{fermion,left=0.3,tension=1}{v3,v4}
            \fmf{fermion,left=0.3,tension=1}{v4,v1}
            \fmf{photon}{iup,v2}
            \fmf{photon}{odown,v4}
            \fmfset{arrow_len}{2.5mm}
            \fmfv{decor.shape=circle,decor.size=1mm}{odown,iup}
            \fmfshift{(0.0w,-0.1h)}{iup}
            \fmfshift{(0.0w,0.1h)}{odown}
        \end{fmfgraph*}
    \end{fmffile}
    \caption{Representative diagram of the fully perturbative contribution to the OPE of $\Pi^{\mu_{1}\mu_{2}}$ in the mixed virtualities regime. A black dot represents creation/annihilation of a line by the background fields in the vacuum. Depending on the value of $q_{3}$, one of these photons may be interact perturbatively with vacuum.}
    \label{fig:quarkLoopMixed}
\end{figure}
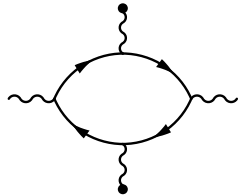

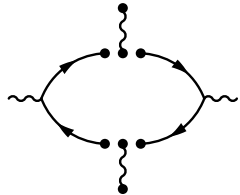
\begin{figure}[b]
    \centering
    \begin{fmffile}{cutQuarkMixedVirtualities}
        \setlength{\unitlength}{0.6cm}\small
        \begin{fmfgraph*}(5,5) 
            \fmfleft{i}
            \fmftop{iup}
            \fmfright{o}
            \fmfbottom{odown}
            \fmf{photon}{i,v1}
            \fmf{photon}{v5,o}
            \fmf{phantom,tension=2}{i,v1,v2,v3,v4,v5,o}
            \fmf{phantom,tension=2}{i,v1,v6,v7,v8,v5,o}
            \fmffreeze
            \fmfshift{(0.0w,0.2h)}{v3}
            \fmfshift{(0.1w,0.2h)}{v2}
            \fmfshift{(-0.1w,0.2h)}{v4}
            \fmfshift{(-0.1w,-0.2h)}{v8}
            \fmfshift{(0.1w,-0.2h)}{v6}
            \fmfshift{(0.0w,-0.2h)}{v7}
            \fmf{fermion,left=0.3,tension=0.5}{v1,v2}
            \fmf{fermion,left=0.3,tension=0.5}{v4,v5}
            \fmf{fermion,left=0.3,tension=0.5}{v5,v8}
            \fmf{fermion,left=0.3,tension=0.5}{v6,v1}
            \fmf{photon}{iup,v3}
            \fmf{photon}{odown,v7}
            \fmfset{arrow_len}{2.5mm}
            \fmfv{decor.shape=circle,decor.size=1mm}{odown,iup,v2,v3,v4,v6,v7,v8}
            \fmfshift{(0.0w,-0.1h)}{iup}
            \fmfshift{(0.0w,0.1h)}{odown}
        \end{fmfgraph*}
    \end{fmffile}
    \caption{Representative diagram of the one--cut--quark contributions to the OPE of $\Pi^{\mu_{1}\mu_{2}}$ in the mixed virtualities regime. A black dot represents creation/annihilation of a line by the background fields in the vacuum. Depending on the value of $q_{3}$, one of these photons may be interact perturbatively with vacuum.}
    \label{fig:quarkCutMixed}
\end{figure}

At the beginning of section~\ref{sec:OPEfirst} we mentioned two different roles that SDC play in the computation of $a_{\mu}$. First, one can obtain the high--energy asymptotic behaviour of a Green function to learn the asymptotic behaviour of an hadronic form factor, in order to fill the gap of missing or scarce experimental data in the high energy parts of dispersive integrals or the master formula. Secondly, one can similarly use this approach directly to the HLbL tensor in the high energy regions of the master integral, which is the purpose of the quark loop computation that we did. However, in the latter case the interplay between low and high energy contributions is not clear cut, because low energy contributions are computed for the full $|Q_{1}|$ and $|Q_{2}|$ intervals of the master integral, not only up to $1$ or $2$~GeV. Thus one sees that some overlapping of contributions is present and there is a risk of double counting. Therefore, for the high energy contribution to be successfully accounted for it is necessary to understand how much of it has already been taken into account by low--energy computations. 

One way to answer that question is to determine how much the low--energy contributions' asymptotic behaviour resembles the results of the high energy framework. It was argued in~\cite{BijnensSDC2003,KnechtNyffelerSDC} that it is impossible to fulfill all QCD SDC with a finite number of resonances. To obtain an estimate of the missing high energy contributions caused by such mismatch, one can use a top--down approach: to constrain hadronic contributions to fulfill SDC and study how much the result differs from when they are constrained by experimental data, that is, by their low--energy behaviour. For example, the mixed virtualities regime $Q_{1}\sim Q_{2}\equiv Q\gg Q_{3}\gg\Lambda_{QCD}$ of the HLbL tensor, first studied in~\cite{MixedVirtualities}, imposes the following constraint: 
\begin{equation}
    \lim_{Q,Q_{3}\rightarrow\infty}Q^{2}Q_{3}^{2}\overline{\Pi}_{1} =-\frac{2}{3\pi^{2}}
\end{equation}
and a similar one for crossed condition for $\overline{\Pi}_{2}$. In addition, as we already argued, the symmetric regime $Q_{1}\sim Q_{2}\sim Q_{3}\equiv Q\gg \Lambda_{QCD}$, via the massless quark loop, imposes the following asymptotic behaviour: 
\begin{equation}
    \lim_{Q\rightarrow\infty}Q^{4}\overline{\Pi}_{1} =-\frac{4}{9\pi^{2}}\;.
\end{equation}
The proposals to ensure that the transition form factors match the mixed virtualities behaviour have ranged from ignoring their momentum dependence~\cite{MixedVirtualities} to summing an infinite tower of axial and vector resonances in holographic QCD~\cite{holographicQCD,holographicQCD_2}. In~\cite{ColangeloSDC,ColangeloSDC_2}, a hybrid approach is followed: pseudo scalar pole contributions are computed in a large--$N_{c}$ Regge model such that they satisfy SDC, but those results are only used in the low--energy region of integration of the master formula. The integral over the remaining part is computed with the quark loop expression, taking advantage of the asymptotic behaviour of the massless quark loop contribution to $\overline{\Pi}_{1}$, which fulfills the mixed--virtualities SDC as well. This reduces model--dependence with respect to the first two approaches mentioned and allows to clearly separate the effect of SDC on low-- and high--energy contributions to lower double counting risks. Nevertheless, such risks still remain with respect to axial vector contributions, lies in a transition region between the perturbative and non--perturbative domain of QCD and is still a significant source of uncertainty for $a_{\mu}^{\text{HLbL}}$. Compared to the data--driven computation, there is an increase in the contribution from pseudo scalar poles: 
\begin{equation}\label{eq:LSDCresult}
    \Delta a_{\mu}^{LSDC} =\Big[8.7(5.5)_{\text{PS--poles}} +4.6(9)_{\text{pQCD}}\Big]\times10^{-11} = 13(6)\times10^{-11}\;,
\end{equation}
where the superindex $LSDC$ illustrates the fact that we are only considering the constraints regarding the asymptotic behaviour of the ``longitudinal'' part of the HLbL tensor in the mixed virtualities regime. The ``transversal'' form factors are $\overline{\Pi}_{3-12}$. They are related to the contribution from axial vectors and obey a different SDC in the mixed virtualities regime. This result is in very good agreement with the holographic QCD one~\cite{holographicQCD}. In contrast, it hints at an overestimation from the approach proposed in~\cite{MixedVirtualities}. When $\Delta a_{\mu}^{LSDC}$ is computed fully with the large--$N_c$ Regge model, the result is very close to~\eqref{eq:LSDCresult}. In the end, the net increase of the HLbL contribution due to SDC is estimated to be $\Delta a_{\mu}^{SDC}=15(10)\cdot10^{-11}$~\cite{WP}. A part of the uncertainty of~\eqref{eq:LSDCresult} is estimated by varying the matching scale between the Regge model and the quark loop, and is then added to each element's model or theoretical uncertainty. Therefore higher order corrections to the quark loop can decrease the uncertainty of $\Delta a_{\mu}^{LSDC}$. In~\cite{ColangeloSDC_3}, the SDC contribution was reassessed taking into consideration the perturbative corrections to the quark loop and the result was: 
\begin{equation}\label{eq:LSDCresult_2}
    \Delta a_{\mu}^{LSDC} =\Big[8.7(5.3)_{\text{PS--poles}} +4.2(1)_{\text{pQCD}}\Big]\times10^{-11} = 13(5)\times10^{-11}\;,
\end{equation}
which reduces the uncertainty of the previous result. It is worth mentioning that the negative $O(\alpha_{s})$ correction to the massless quark loop improves the agreement between the Regge sum of pseudoscalars and the perturbative result. However, further study regarding the matching procedure is still needed.

Recently, two works~\cite{backgrounOPEmixedVirtualities_1,backgrounOPEmixedVirtualities_2} were published regarding the extension of the background OPE framework to the mixed virtualities regimes of the HLbL tensor. To illustrate the broader range of use of the background field framework that we used in this work, we briefly review their results. The analogue of $\Pi^{\mu_{1}\mu_{2}\mu_{3}}$ is now: 
\begin{equation}
\begin{split}
    \Pi^{\mu_{1}\mu_{2}} &=\frac{i}{e^{2}}\int\frac{d^{4}q_{4}}{(2\pi)^{4}}\int d^{4}x\int d^{4}y\;e^{-i(q_{1}x+q_{2}y)}\langle0|T\,J^{\mu_{1}}(x)J^{\mu_{2}}(y)|\gamma^{*}(q_{3})\gamma(q_{4})\rangle\\
    &=-\epsilon_{\mu_{3}}(q_{3})\epsilon_{\mu_{4}}(q_{4})\Pi^{\mu_{1}\mu_{2}\mu_{3}\mu_{4}}(q_{1},q_{2},q_{3})\;.
\end{split}
\end{equation}
When the OPE is performed up to operators with mass dimension $D=4$, the quoted result is: 
\begin{equation}
\begin{split}
    \Pi^{\mu_{1}\mu_{2}} &= -\frac{1}{4}\langle F_{\nu_{3}\mu_{3}}F_{\nu_{4}\mu_{4}}\rangle\frac{\partial}{\partial q_{3\nu_{3}}}\frac{\partial}{\partial q_{4\nu_{4}}}\Pi^{\mu_{1}\mu_{2}\mu_{3}\mu_{4}}_{\text{quark loop}}\Big|_{q_{3}=q_{4}=0}\\
    &-\frac{e_{f}^{2}}{e^{2}}\langle\overline{\psi}(0)\Big(\gamma^{\mu_{1}}S^{0}(-\hat{q})\gamma^{\mu_{2}} -\gamma^{\mu_{2}}S^{0}(-\hat{q})\gamma^{\mu_{1}}\Big)\psi(0)\rangle\\
    &-\frac{ie_{f}^{2}}{e^{2}\hat{q}^{2}}\Big(g^{\mu_{1}\delta}g^{\mu_{2}}_{\beta} +g^{\mu_{2}\delta}g^{\mu_{1}}_{\beta} -g^{\mu_{1}\mu_{2}}g^{\delta}_{\beta}\Big)\Big(g_{\alpha\delta}-2\frac{\hat{q}_{\alpha}\hat{q}_{\delta}}{\hat{q}^{2}}\Big)\langle\overline{\psi}(0)\Big[\overrightarrow{D}^{\alpha} -\overleftarrow{D}^{\alpha}\Big]\gamma^{\beta}\psi(0)\rangle\;,
\end{split}
\end{equation}
where $\hat{q}\equiv (q_{1}-q_{2})/2$ and the matrix element $\langle...\rangle$ now includes the virtual photon $\gamma^{*}(q_{3})$ and the real soft one $\gamma(q_{4})$. The term $\Pi_{\text{quark loop}}$ is proportional to the quark loop amplitude discussed in this work. The origin of the first term is quite clear: it comes from matrix elements with four contracted quark fluctuations in which the resulting two fermion propagators have a total of two soft photon insertions between the two (see figure~\ref{fig:quarkLoopMixed}), hence the two derivatives and field strength tensors. The second and third terms come instead from terms with two background quark fields and two fluctuations (see figure~\ref{fig:quarkCutMixed}), where the background fields are Taylor expanded as usual up to order $O(x_{1}-x_{2})$. The appearance of $\hat{q}$ comes from the fact that only $x_{1}-x_{2}$ is close to zero in the mixed virtualities regime, in contrast the symmetric regime, where the three currents' coordinates are close. It is worth mentioning that in this case quark operators start at dimension $D=3$ and therefore they are, in principle, the leading term of the OPE instead of the perturbative quark loop.


\section{Conclusions}

We have reviewed the basic framework for the dispersive computation of the HLbL contribution to the anomalous magnetic moment of the muon, $a_{\mu}^{\text{HLbL}}$, focusing on the kinematic singularity free tensor decomposition that allowed it. Unlike hadronic models used previously, dispersive estimates have allowed for the computation of unambiguous contributions, at least for pseudoscalar poles, which has improved the uncertainty estimation. We have also discussed the current consensus of low--energy contributions to $a_{\mu}^{\text{HLbL}}$ and the corresponding role of SDC as a means of uncertainty assessment and high energy contribution computation.

The main focus of this work is on the HLbL scattering amplitude in the symmetric high energy regime, in which it can be represented via and OPE in which the soft photon is regarded as a background field to avoid infrared-divergent Wilson coefficients. To present a through discussion we introduced  the background field method and stressed how renormalization and operator mixing are included in a very natural way within that framework. The same applies for the derivation of the Wilson coefficients, in which perturbative and non--perturbative contributions are systematically separated and do not require much decision--making from the user. From the OPE of $\Pi^{\mu_{1}\mu_{2}\mu_{3}\mu_{4}}$, we found that the quark loop is leading contribution after infrared divergent logarithms have been subtracted by renormalization, in agreement with the literature.

Finally, we presented the computation of the quark loop with its full tensor structure, that is, without projecting the form factors of the HLbL tensor out of it, in contrast with previous computations. This allowed us to check the generality of the basis elements of the kinematic--singularity--free tensor decomposition of the HLbL tensor. We concluded that these elements do span the tensor structures of the quark amplitude, thus obtaining an explicit check that we have not found in the literature. To compute the full quark loop amplitude it was necessary to use a decomposition algorithm for tensor loop integrals and we used one that did not introduce further spurious kinematic singularities, at the cost of shifting the dimensions of the integrals. Using a Mellin--Barnes representation for the resulting scalar integrals allowed us to keep full mass dependence and  obtain a complete series representation of the required integrals that contains all quark mass effects at any order in the high energy regime. We also presented the fundamentals of single and multiple complex variable residues computation necessary to provide a reasonably thorough mathematical foundation for the procedure.  The aforementioned computations were implemented by a \textit{Mathematica} script that used \textit{FeynCalc}. To highlight the importance of the quark loop computation result, we described the use of the quark loop computation as a SDC to the low--energy contributions to $a_{\mu}^{\text{HLbL}}$ and its effects in the critical task of lowering the uncertainty of the anomalous magnetic moment of the muon value in the Standard Model.

As we mentioned in the previous section, the $O(\alpha_{s})$ (two--loop) correction to the massless quark loop has been performed and it has yielded a $\sim10\%$ correction. This suggests that a three--loop correction would probably have a size comparable to that of some non--perturbative contributions, whose magnetic susceptibilities $X_{i}$ have not been rigorously computed yet. Consequently, a more detailed study of these from lattice groups is necessary.

The extension of the background OPE formalism to the mixed virtualities high energy regime of the HLbL amplitude has recently been done for the first time, which provides an alternative to systematically expand upon the knowledge that we had on this regime~\cite{MixedVirtualities}. The case when the lowest virtual momentum is within the perturbative regime of QCD has already been computed at leading order, but results for perturbative corrections have not been published yet. The complementary case in which the lowest virtual momentum is below the perturbative threshold presents added complexities due to the lack of knowledge on the resulting non--perturbative matrix elements. This creates an opportunity for numerical studies from lattice. Ultimately, improvements on SDC coming from the mixed virtualities regime can help to lower the uncertainty from the lower--energy contributions to $a_{\mu}^{\text{HLbL}}$ or contribute to a better estimation of axial vectors contributions, which continue to be a rather large source of uncertainty.

\vspace{6pt} 


\appendix
\section{Triangle scalar loop integrals in arbitrary dimensions} 

\label{AppChapter3} 

In equation~\eqref{eq:scalarTriangleArbitraryDimension} we quoted the result of~\cite{Phan2019} for scalar triangle loop integrals in arbitrary space--time dimensions with unit propagator powers, $J_3^{(d)}$. In this appendix we complete the definition of relevant quantities that we used and present the results in a way that clearly shows the appearance of logarithms and ultraviolet singularities.

First, let us define the Cayley determinant $S_3$:
\begin{equation}
\begin{split}
    S_3 &=
    \begin{vmatrix}
        2m_1^2 & -p_1^2 +m_1^2 +m_2^2 & -p_3^2 +m_1^2 +m_3^2\\
        -p_1^2 +m_1^2 +m_2^2 & 2m_2^2 & -p_2^2 +m_2^2 +m_3^2\\
        -p_3^2 +m_1^2 +m_3^2 & -p_2^2 +m_2^2 +m_3^2 & 2m_3^2
    \end{vmatrix}\;.
\end{split}
\end{equation}
In the same fashion we define the Cayley determinants $S_{ij}$ of self--energy integrals, which are obtained by suppressing one of the three propagators in the triangle:
\begin{equation}
    S_{ij} = 
    \begin{vmatrix}
        2m_i^2 & -p_i^2 +m_i^2 +m_j^2\\
        -p_i^2 +m_i^2 +m_j^2 & 2m_j^2
    \end{vmatrix}
    =-\lambda(p_i^2,m_i^2,m_j^2)\;.
\end{equation}

Similarly for the Gram determinants we have: 
\begin{equation}
\begin{split}
    G_3 &= -8
    \begin{vmatrix}
        p_1^2 & p_1\cdot p_2\\
        p_1\cdot p_2 & p_2^2
    \end{vmatrix}=2\lambda(p_1^2,p_2^2,p_3^2)\;,\\
    G_{12} &=-4p_1^2\;,\hspace{1cm} G_{13} =-4p_3^2 \;,\hspace{1cm} G_{23} =-4p_2^2\;,
\end{split}
\end{equation}
where $p_3=-p_1-p_2$. Finally we have $M_3=S_3/G_3$ and $M_{ij}=S_{ij}/S_3$.

Now let us define $x_k$: 
\begin{equation}
    x_1 =1-\frac{D-E\beta +2(C-B\beta)}{2(1-\beta)(C-B\beta)} \;,\hspace{1cm} x_2=1+\frac{D-E\beta}{2(C-B\beta)} \;,\hspace{1cm} x_3=-\frac{D-E\beta}{2\beta(C-B\beta)}\;,
\end{equation}
where
\begin{equation}
\begin{split}
    A&=p_1^2\;,\hspace{1cm} B=p_3^2\;,\hspace{1cm} C=-p_1\cdot p_3\;, \hspace{1cm} D=-(p_1^2+m_1^2-m_2^2)\;,\\
    E&=-(p_3^2+m_1^2-m_3^2)\;, \hspace{1cm} F=m_1^2\;,\hspace{1cm} \beta=\frac{C+\sqrt{C^2-AB}}{B}\;.
\end{split}
\end{equation}
$x_{ij}$ and $x_{k}$ fulfill the following relevant identities:
\begin{equation}
    p_i^2x_{ij}^{2}=m_i^2-M_{ij}\;, \hspace{1cm} p_i^2(x_k-x_{ij})^{2} =M_3-M_{ij}\;.
\end{equation}

Finally, we will present the formulas for the case $d=4+2k-2\epsilon$ with $k\in\mathbb{N}$, which are relevant for our computation. Keeping full $\epsilon$ dependence we have:
\begin{equation}
\begin{split}
    J^{(4+2k-2\epsilon)}_3\times&\frac{(4\pi)^{2+k}\lambda^{1/2}(p_1^2,p_2^2,p_3^2)}{i(4\pi)^{\epsilon}}\\
    &=\Gamma(-k+\epsilon)\Big(M_3^k\Big(\frac{\mu^{2}}{M_3}\Big)^{\epsilon}-M_{ij}^k\Big(\frac{\mu^{2}}{M_{ij}}\Big)^{\epsilon}\Big)\sum_{\substack{n_1=1}}\frac{1}{n_1}\Bigg(-\frac{x_{ij}}{x_k-x_{ij}}\Bigg)^{n_1}\\
    &\hspace{0cm}-M_{ij}^k\Big(\frac{\mu^{2}}{M_{ij}}\Big)^{\epsilon}\Bigg[\sum^{n_2=k}_{\substack{n_1=1\\n_2=1}}\frac{\Gamma(-k+n_2+\epsilon\Big)}{(n_1+2n_2)n_2!}\Bigg(-\frac{x_{ij}}{x_k-x_{ij}}\Bigg)^{n_1}\Bigg(-\frac{p_i^2x^{2}_{ij}}{M_{ij}}\Bigg)^{n_2}\\
    &\hspace{1cm}+\sum_{\substack{n_1=1\\n_2=k+1}}\frac{\Gamma(-k+n_2\Big)}{(n_1+2n_2)n_2!}\Bigg(-\frac{x_{ij}}{x_k-x_{ij}}\Bigg)^{n_1}\Bigg(-\frac{p_i^2x^{2}_{ij}}{M_{ij}}\Bigg)^{n_2}\Bigg]\\
    &\hspace{0cm}-\Big\{x_{ij}\rightarrow 1-x_{ij}\;\;;\;\;x_{k}\rightarrow 1-x_{k}\Big\}\;.
\end{split}
\end{equation}
Note that we neglected the infinitesimal term $i\epsilon$ that gives the Feynman prescription, because it is not relevant in the deep space--like region that we are interested in. Nevertheless, it can be easily reinstated by replacing $M_{ij}\rightarrow M_{ij}-i\epsilon$ and $M_{3}\rightarrow M_{3}-i\epsilon$. Taking the limit $\epsilon\rightarrow0$ we have:
\begin{align}\nonumber
    J^{(4+2k)}_3&\times\frac{(4\pi)^{2+k}\lambda_-^{1/2}}{i}\\\nonumber
    &=\frac{(-1)^k}{k!}\Bigg\{M_3^k\Big(\frac{1}{\hat{\epsilon}}+\ln\Big\{\frac{\mu^2}{M_{3}}\Big\}\Big)-M_{ij}^k\Big(\frac{1}{\hat{\epsilon}}+\ln\Big\{\frac{\mu^2}{M_{ij}}\Big\}\Big)\Bigg\}\sum_{\substack{n_1=1}}\frac{1}{n_1}\Bigg(-\frac{x_{ij}}{x_k-x_{ij}}\Bigg)^{n_1}\\\nonumber
    &+\frac{(-1)^k}{k!}\sum_{j=1}^k\frac{1}{j}\Big(M_3^k-M_{ij}^k\Big)\sum_{\substack{n_1=1}}\frac{1}{n_1}\Bigg(-\frac{x_{ij}}{x_k-x_{ij}}\Bigg)^{n_1}\\\nonumber
    &-M_{ij}^k\Big(\frac{1}{\hat{\epsilon}}+\ln\Big\{\frac{\mu^2}{M_{ij}}\Big\}\Big)\Bigg[\sum^{n_2=k}_{\substack{n_1=1\\n_2=1}}\frac{(-1)^{k-n_2}}{(n_1+2n_2)(k-n_2)!n_2!}\Bigg(-\frac{x_{ij}}{x_k-x_{ij}}\Bigg)^{n_1}\Bigg(-\frac{p_i^2x^{2}_{ij}}{M_{ij}}\Bigg)^{n_2}\\\nonumber
    &-M_{ij}^k\Bigg[\sum^{n_2=k}_{\substack{n_1=1\\n_2=1}}\frac{1}{(n_1+2n_2)n_2!}\Bigg(-\frac{x_{ij}}{x_k-x_{ij}}\Bigg)^{n_1}\Bigg(-\frac{p_i^2x^{2}_{ij}}{M_{ij}}\Bigg)^{n_2}\frac{(-1)^{k-n_2}}{(k-n_2)!}\sum_{j=1}^{k-n_2}\frac{1}{j}\\\nonumber
    &\hspace{1cm}+\sum_{\substack{n_1=1\\n_2=k+1}}\frac{\Gamma(-k+n_2\Big)}{(n_1+2n_2)n_2!}\Bigg(-\frac{x_{ij}}{x_k-x_{ij}}\Bigg)^{n_1}\Bigg(-\frac{p_i^2x^{2}_{ij}}{M_{ij}}\Bigg)^{n_2}\Bigg]\\
    &-\Big\{x_{ij}\rightarrow 1-x_{ij}\;\;;\;\;x_{k}\rightarrow 1-x_{k}\Big\}\;,
\end{align}
where a sum over the three permutations $(i,j,k)\rightarrow(1,2,3)\rightarrow(2,3,1)\rightarrow(3,1,2)$ is implied.  $\mu$ is the renormalization scale. The singular terms $1/\hat{\epsilon}$ vanish and dependence on $\mu$ disappears when powers of the propagators get high enough via derivatives with respect to the masses.

\appto{\bibsetup}{\raggedright}
\printbibliography[heading=bibintoc]

@book{Barut,
     author = {A.O.Barut} ,
     title = {The theory of the scattering matrix },
     publisher = {The MacMillan company} ,
     year = {1967}
 }

@article{Nima,
    Author = {N. Arkani-Hamed, T.C. Huang, Y.T. Huang},
    title = {Scattering amplitudes for all masses and spins},
    journal = {Journal of High energy Physics, 11 070} ,
    year = {2021}
}

@book{BookStrichartz,
  title = {A guide to distribution theory and Fourier transform},
  author = {R. Strichartz},
  year = {2003},
  publisher = {World Scientific Publishing Company}}

@book{BookPassarinoBardin,
  title = {The Standard Model in the Making: Precision Study of the Elecroweak Interactions},
  author = {Dima Bardin and Giampiero Passarino},
  year = {1999},
  publisher = {Clarendon Press, Oxford}}

@book{BookPeskin,
  title = {An introduction to quantum field theory},
  author = {M. Peskin and D. Schroeder},
  year = {1995},
  publisher = {Addison--Wesley Publishing Company}}

@book{BookWeinberg,
  title = {The Quantum Theory of Fields},
  author = {Steven Weinberg},
  year = {1996},
  publisher = {Cambridge University Press}}

@book{BookMarichev,
  title = {Methods for Computing Integrals of Special Functions},
  author = {Oleg Igorevich Marichev},
  year = {1978},
  publisher = {},
  address = {Minsk}}

@book{BookLongTsikh,
  title = {Translations of Mathematical Monographs: Multidimensional Residues and Their Applications},
  author = {A. K. Tsikh},
  volume = {103},
  year = {1992},
  publisher = {American Mathematical Society}}

@book{BookShortTsikh,
  title = {Aspects of Mathematics: Contributions to Complex Analysis and Analytic 
 Geometry},
  editor = {Henri Skoda and Jean-Marie Trepreau},
  author = {},
  volume = {E26},
  pages = {233--241},
  year = {1994},
  publisher = {Springer}}

@book{BookBateman,
  title = {Higher Trascendental Functions -- Volume 1},
  author = {Harry Bateman and Arthur Erdélyi and Wilhelm Magnus and Fritz Oberhettinger and Francesco G. Tricomi},
  year = {1953},
  publisher = {McGraw--Hill}}

@book{HLbL_LO9,
    author = {Jegerlehner, Fred},
    year = {2017},
    publisher = {Springer},
    title = {STMP -- The Anomalous Magnetic Moment of the Muon},
    volume = {274}}

@book{BookGriffiths,
    author = {P. Griffiths and J. Harris},
    year = {1978},
    publisher = {John Wiley and Sons},
    title = {Principles of algebraic geometry}}

@book{BookSpearman,
    author = {Martin, A. D. and Spearman, T. D.},
    year = {1970},
    publisher = {North Holland Publishing Company},
    title = {Elementary Particle Theory}
}

@book{BookSrivastava,
  title = {Multiple Gaussian Hypergeometric Series},
  author = {H. M. Srivastava and Per W. Karlsson},
  year = {1985},
  publisher = {John Wiley and Sons}}

@book{BookSlater,
  title = {Generalized Hypergeometric Functions},
  author = {Lucy Joan Slater},
  year = {1966},
  publisher = {Cambridge University Press}}

@article{Aldins1970,
    author = {Aldins, Janis and Brodsky, Stanley J. and Kinoshita, Toichiro},
    year = {1970},
    Journal = {Physical Review D},
    volume = {1},
    number = {8}}

@article{Mandelstam1958,
    author = {S. Mandelstam},
    year = {1958},
    Journal = {Physical Review},
    volume = {112},
    number = {4}}

@article{WP,
    Author = {T. Aoyama and others},
    Journal = {Physics Reports},
    Number = {887},
    Pages = {1--166},
    Year = {2020}}

@article{Knecht2002,
    Author = {Knecht, Marc and Nyffeler Andreas},
    Journal = {Physical Review D},
    Number = {073034},
    Volume = {65},
    Year = {2002}}

@article{kloe,
    Author = {Anastasi, A. and others},
    Journal = {J. High Energ. Phys.},
    Number = {2018},
    Volume = {173},
    Pages = {},
    Year = {2018},
    URL = {https://doi.org/10.1007/JHEP03(2018)173}
}

@article{eMagneticMomentMeasurement_1,
    Author = {D. Hanneke and S. Hoogerheide and
    G. Gabrielse},
    Journal = {Physical Review A},
    Number = {052122},
    Volume = {83},
    Pages = {},
    Year = {2011}}

@article{eMagneticMomentMeasurement_2,
    Author = {D. Hanneke and S. Fogwell and
    G. Gabrielse},
    Journal = {Physical Review A},
    Number = {120801},
    Volume = {100},
    Pages = {},
    Year = {2008}}

@article{eMagneticMomentComputation,
    Author = {T. Aoyama and T. Kinoshita and M.Nio},
    Journal = {Physical Review D},
    Number = {036001},
    Volume = {97},
    Pages = {},
    Year = {2018}}

@article{BNL,
    Author = {G. W. Bennet and others},
    Journal = {Physical Review D},
    Number = {072003},
    Volume = {73},
    Pages = {},
    Year = {2006}}

@inproceedings{Colangelo2023discussion,
    author = "Colangelo, Gilberto and Hoferichter, Martin and Stoffer, Peter",
    title = "{Puzzles in the hadronic contributions to the muon anomalous magnetic moment}",
    booktitle = "{21st Conference on Flavor Physics and CP Violation}",
    eprint = "2308.04217",
    archivePrefix = "arXiv",
    primaryClass = "hep-ph",
    reportNumber = "PSI-PR-23-28, ZH-TH 42/23",
    month = "8",
    year = "2023"
}

@article{Fermilab1,
    Author = {B.Abi and others},
    Journal = {Physical Review Letters},
    Number = {141801},
    Volume = {126},
    Pages = {},
    Year = {2021}}

@article{Fermilab2,
    Author = {D. P. Aguillard and others},
    Journal = {Physical Review Letters},
    Number = {161802},
    Volume = {131},
    Pages = {},
    Year = {2023}}

@article{BESIII,
title = {},
journal = {Physics Letters B},
volume = {753},
pages = {629-638},
year = {2016},
issn = {0370-2693},
related = {BESIIIerratum},
relatedstring = {Corrected in:},
doi = {https://doi.org/10.1016/j.physletb.2015.11.043},
url = {https://www.sciencedirect.com/science/article/pii/S0370269315008990},
author = {M. Ablikim and M.N. Achasov and X.C. Ai and O. Albayrak and M. Albrecht and D.J. Ambrose and A. Amoroso and F.F. An and Q. An and J.Z. Bai and R. {Baldini Ferroli} and Y. Ban and D.W. Bennett and J.V. Bennett and M. Bertani and D. Bettoni and J.M. Bian and F. Bianchi and E. Boger and I. Boyko and R.A. Briere and H. Cai and X. Cai and O. Cakir and A. Calcaterra and G.F. Cao and S.A. Cetin and J.F. Chang and G. Chelkov and G. Chen and H.S. Chen and H.Y. Chen and J.C. Chen and M.L. Chen and S.J. Chen and X. Chen and X.R. Chen and Y.B. Chen and H.P. Cheng and X.K. Chu and G. Cibinetto and H.L. Dai and J.P. Dai and A. Dbeyssi and D. Dedovich and Z.Y. Deng and A. Denig and I. Denysenko and M. Destefanis and F. {De Mori} and Y. Ding and C. Dong and J. Dong and L.Y. Dong and M.Y. Dong and S.X. Du and P.F. Duan and E.E. Eren and J.Z. Fan and J. Fang and S.S. Fang and X. Fang and Y. Fang and L. Fava and F. Feldbauer and G. Felici and C.Q. Feng and E. Fioravanti and M. Fritsch and C.D. Fu and Q. Gao and X.Y. Gao and Y. Gao and Z. Gao and I. Garzia and K. Goetzen and W.X. Gong and W. Gradl and M. Greco and M.H. Gu and Y.T. Gu and Y.H. Guan and A.Q. Guo and L.B. Guo and Y. Guo and Y.P. Guo and Z. Haddadi and A. Hafner and S. Han and X.Q. Hao and F.A. Harris and K.L. He and X.Q. He and T. Held and Y.K. Heng and Z.L. Hou and C. Hu and H.M. Hu and J.F. Hu and T. Hu and Y. Hu and G.M. Huang and G.S. Huang and J.S. Huang and X.T. Huang and Y. Huang and T. Hussain and Q. Ji and Q.P. Ji and X.B. Ji and X.L. Ji and L.W. Jiang and X.S. Jiang and X.Y. Jiang and J.B. Jiao and Z. Jiao and D.P. Jin and S. Jin and T. Johansson and A. Julin and N. Kalantar-Nayestanaki and X.L. Kang and X.S. Kang and M. Kavatsyuk and B.C. Ke and P. Kiese and R. Kliemt and B. Kloss and O.B. Kolcu and B. Kopf and M. Kornicer and W. Kühn and A. Kupsc and J.S. Lange and M. Lara and P. Larin and C. Leng and C. Li and Cheng Li and D.M. Li and F. Li and F.Y. Li and G. Li and H.B. Li and J.C. Li and Jin Li and K. Li and K. Li and Lei Li and P.R. Li and T. Li and W.D. Li and W.G. Li and X.L. Li and X.M. Li and X.N. Li and X.Q. Li and Z.B. Li and H. Liang and Y.F. Liang and Y.T. Liang and G.R. Liao and D.X. Lin and B.J. Liu and C.X. Liu and F.H. Liu and Fang Liu and Feng Liu and H.B. Liu and H.H. Liu and H.H. Liu and H.M. Liu and J. Liu and J.B. Liu and J.P. Liu and J.Y. Liu and K. Liu and K.Y. Liu and L.D. Liu and P.L. Liu and Q. Liu and S.B. Liu and X. Liu and Y.B. Liu and Z.A. Liu and Zhiqing Liu and H. Loehner and X.C. Lou and H.J. Lu and J.G. Lu and Y. Lu and Y.P. Lu and C.L. Luo and M.X. Luo and T. Luo and X.L. Luo and X.R. Lyu and F.C. Ma and H.L. Ma and L.L. Ma and Q.M. Ma and T. Ma and X.N. Ma and X.Y. Ma and F.E. Maas and M. Maggiora and Y.J. Mao and Z.P. Mao and S. Marcello and J.G. Messchendorp and J. Min and R.E. Mitchell and X.H. Mo and Y.J. Mo and C. {Morales Morales} and K. Moriya and N.Yu. Muchnoi and H. Muramatsu and Y. Nefedov and F. Nerling and I.B. Nikolaev and Z. Ning and S. Nisar and S.L. Niu and X.Y. Niu and S.L. Olsen and Q. Ouyang and S. Pacetti and P. Patteri and M. Pelizaeus and H.P. Peng and K. Peters and J. Pettersson and J.L. Ping and R.G. Ping and R. Poling and V. Prasad and M. Qi and S. Qian and C.F. Qiao and L.Q. Qin and N. Qin and X.S. Qin and Z.H. Qin and J.F. Qiu and K.H. Rashid and C.F. Redmer and M. Ripka and G. Rong and Ch. Rosner and X.D. Ruan and V. Santoro and A. Sarantsev and M. Savrié and K. Schoenning and S. Schumann and W. Shan and M. Shao and C.P. Shen and P.X. Shen and X.Y. Shen and H.Y. Sheng and W.M. Song and M.R. Shepherd and X.Y. Song and S. Sosio and S. Spataro and G.X. Sun and J.F. Sun and S.S. Sun and Y.J. Sun and Y.Z. Sun and Z.J. Sun and Z.T. Sun and C.J. Tang and X. Tang and I. Tapan and E.H. Thorndike and M. Tiemens and M. Ullrich and I. Uman and G.S. Varner and B. Wang and D. Wang and D.Y. Wang and K. Wang and L.L. Wang and L.S. Wang and M. Wang and P. Wang and P.L. Wang and S.G. Wang and W. Wang and X.F. Wang and Y.D. Wang and Y.F. Wang and Y.Q. Wang and Z. Wang and Z.G. Wang and Z.H. Wang and Z.Y. Wang and T. Weber and D.H. Wei and J.B. Wei and P. Weidenkaff and S.P. Wen and U. Wiedner and M. Wolke and L.H. Wu and Z. Wu and L.G. Xia and Y. Xia and D. Xiao and H. Xiao and Z.J. Xiao and Y.G. Xie and Q.L. Xiu and G.F. Xu and L. Xu and Q.J. Xu and X.P. Xu and L. Yan and W.B. Yan and W.C. Yan and Y.H. Yan and H.J. Yang and H.X. Yang and L. Yang and Y. Yang and Y.X. Yang and M. Ye and M.H. Ye and J.H. Yin and B.X. Yu and C.X. Yu and J.S. Yu and C.Z. Yuan and W.L. Yuan and Y. Yuan and A. Yuncu and A.A. Zafar and A. Zallo and Y. Zeng and B.X. Zhang and B.Y. Zhang and C. Zhang and C.C. Zhang and D.H. Zhang and H.H. Zhang and H.Y. Zhang and J.J. Zhang and J.L. Zhang and J.Q. Zhang and J.W. Zhang and J.Y. Zhang and J.Z. Zhang and K. Zhang and L. Zhang and X.Y. Zhang and Y. Zhang and Y.N. Zhang and Y.H. Zhang and Y.T. Zhang and Yu Zhang and Z.H. Zhang and Z.P. Zhang and Z.Y. Zhang and G. Zhao and J.W. Zhao and J.Y. Zhao and J.Z. Zhao and Lei Zhao and Ling Zhao and M.G. Zhao and Q. Zhao and Q.W. Zhao and S.J. Zhao and T.C. Zhao and Y.B. Zhao and Z.G. Zhao and A. Zhemchugov and B. Zheng and J.P. Zheng and W.J. Zheng and Y.H. Zheng and B. Zhong and L. Zhou and X. Zhou and X.K. Zhou and X.R. Zhou and X.Y. Zhou and K. Zhu and K.J. Zhu and S. Zhu and S.H. Zhu and X.L. Zhu and Y.C. Zhu and Y.S. Zhu and Z.A. Zhu and J. Zhuang and L. Zotti and B.S. Zou and J.H. Zou}
}

@article{windowColangelo2022,
title = {Data-driven evaluations of Euclidean windows to scrutinize hadronic vacuum polarization},
journal = {Physics Letters B},
volume = {833},
pages = {137313},
year = {2022},
issn = {0370-2693},
doi = {https://doi.org/10.1016/j.physletb.2022.137313},
url = {https://www.sciencedirect.com/science/article/pii/S0370269322004476},
author = {G. Colangelo and A.X. El-Khadra and M. Hoferichter and A. Keshavarzi and C. Lehner and P. Stoffer and T. Teubner}
}

@article{SND20,
    Journal = { J. High Energ. Phys.},
volume = {2021},
number = {113},
pages = {},
year = {2021},
issn = {},
doi = {https://doi.org/10.1007/JHEP01(2021)113},
url = {},
author = {Achasov, M.N. and Baykov, A.A. and others}}

@article{DirectScanChinese,
    Journal = {Nuclear Instruments and Methods},
volume = {659},
number = {1},
pages = {21-29},
year = {2011},
issn = {0168-9002},
doi = {https://doi.org/10.1016/j.nima.2011.08.050},
url = {https://www.sciencedirect.com/science/article/pii/S0168900211017104},
author = {E.V. Abakumova and M.N. Achasov and V.E. Blinov and X. Cai and H.Y. Dong and C.D. Fu and F.A. Harris and V.V. Kaminsky and A.A. Krasnov and Q. Liu and X.H. Mo and N.Yu. Muchnoi and I.B. Nikolaev and Q. Qin and H.M. Qu and S.L. Olsen and E.E. Pyata and A.G. Shamov and C.P. Shen and K.Yu. Todyshev and G.S. Varner and Y.F. Wang and Q. Xiao and J.Q. Xu and J.Y. Zhang and T.B. Zhang and Y.H. Zhang and A.A. Zhukov}}

@article{cmd2,
journal = {Physics Letters B},
volume = {648},
number = {1},
pages = {28-38},
year = {2007},
issn = {0370-2693},
doi = {https://doi.org/10.1016/j.physletb.2007.01.073},
url = {https://www.sciencedirect.com/science/article/pii/S0370269307001931},
author = {R.R. Akhmetshin and V.M. Aulchenko and V.Sh. Banzarov and L.M. Barkov and N.S. Bashtovoy and A.E. Bondar and D.V. Bondarev and A.V. Bragin and S.K. Dhawan and S.I. Eidelman and D.A. Epifanov and G.V. Fedotovich and N.I. Gabyshev and D.A. Gorbachev and A.A. Grebenuk and D.N. Grigoriev and V.W. Hughes and F.V. Ignatov and S.V. Karpov and V.F. Kazanin and B.I. Khazin and I.A. Koop and P.P. Krokovny and A.S. Kuzmin and I.B. Logashenko and P.A. Lukin and A.P. Lysenko and K.Yu. Mikhailov and J.P. Miller and A.I. Milshtein and I.N. Nesterenko and M.A. Nikulin and V.S. Okhapkin and A.V. Otboev and E.A. Perevedentsev and A.S. Popov and S.I. Redin and B.L. Roberts and N.I. Root and A.A. Ruban and N.M. Ryskulov and A.G. Shamov and Yu.M. Shatunov and B.A. Shwartz and A.L. Sibidanov and V.A. Sidorov and A.N. Skrinsky and V.P. Smakhtin and I.G. Snopkov and E.P. Solodov and J.A. Thompson and Yu.V. Yudin and A.S. Zaitsev and S.G. Zverev}
}

@article{DirectScanRussians,
    Author = {V.L. Ivanov and G.V. Fedotovich and R.R. Akhmetshin and A.N. Amirkhanov and others},
    Journal = {},
    Number = {arXiv:2008.05548},
    Pages = {},
    Year = {2020}}

@article{cmd3,
    Author = {F. V. Ignatov and others},
    Journal = {},
    Number = {},
    Volume = {},
    Pages = {},
    note = {arXiv:2302.08834 [hep-ex]},
    Year = {2023}}

@article{RadiativeReturn_1,
    Author = {F. Ambrosino and others},
    Journal = {Physical Letters B},
    Number = {285},
    Volume = {670},
    Pages = {},
    Year = {2009}}

@article{Tau,
    Author = {R. Alemany and M. Davier and A. Hoecker},
    Journal = {European Physical Journal},
    Number = {123},
    Volume = {C2},
    Pages = {},
    Year = {1998}}

@article{RadiativeReturn_2,
    Author = {B. Aubert and others},
    Journal = {Physical Review Letters},
    Number = {231801},
    Volume = {103},
    Pages = {},
    Year = {2009}}

@article{Muone_1,
    Author = {C. M. Carloni Calame and M. Passera and L. Trentadue and G. Venanzoni},
    Journal = {Physical Letters B},
    Number = {325},
    Pages = {746},
    Year = {2015}}

@article{Muone_2,
    Author = {G. Abbiendi and others},
    Journal = {European Physical Journal C},
    Number = {139},
    Pages = {77},
    Year = {2017}}

@article{Muone_3,
    Author = {P. Banerjee and C. M. Carloni Calame and M. Chiesa and S. Di Vita and T. Engel and others},
    Journal = {European Physical Journal C},
    Number = {591},
    Volume = {80},
    Pages = {},
    Year = {2020}}

@article{BijnensSDC2003,
    Author = {Johan Bijnens and Elvira Gamiz and Edisher Lipartia and Joaquim Prades},
    Journal = {JHEP},
    Number = {0304},
    Volume = {2003},
    URL = {https://doi.org/10.1088/1126-6708/2003/04/055},
    Note = {arXiv:hep-ph/0304222},
    Year = {2003}}

@article{KnechtNyffelerSDC,
    Author = {M. Knecht and A. Nyffeler},
    Journal = {The European Physical Journal C},
    Number = {},
    Volume = {21},
    URL = {https://doi.org/10.1007/s100520100755},
    Note = {arXiv:hep-ph/0106034},
    Pages ={659–-678},
    Year = {2001}}

@article{PionPole_1,
    Author = {M. Hoferichter and B. Hoid an B. Kubis and S. Leupold and S. P. Schneider},
    Journal = {Physical Review Letters},
    Number = {112002},
    Volume = {121},
    Pages = {},
    Year = {2018}}

@article{Colangelo2017,
    Author = {G. Colangelo and M. Hoferichter and M. Procura and others},
    Journal = {JHEP},
    Number = {161},
    Volume = {},
    Pages = {},
    Year = {2017}}

@article{ColangeloSWaves,
    Author = {G. Colangelo and M. Hoferichter and M. Procura and P. Stoffer},
    Journal = {Physical Review Letters},
    Number = {232001},
    Volume ={118},
    Pages = {},
    Year = {2017}}

@article{HadronicModel1,
    Author = {M. Hayakawa and T. Kinoshita and A. I. Sanda},
    Journal = {Physical Review Letters},
    Number = {790},
    Pages = {},
    Volume = {75},
    Year = {1995}}

@article{HadronicModel2,
    Author = {J. Bijnens and E. de Rafael and H. Zheng},
    Journal = {Zeitschrift für Physik C Particles and Fields},
    Title = {Low-Energy Behaviour of Two-Point Functions of Quark Currents},
    Number = {},
    Volume = {62},
    Pages = {437--454},
    Year = {1994}}

@article{Colangelo2014,
    Author = {G. Colangelo and M.Hoferichter and M. Procura and others},
    Journal = {JHEP},
    Number = {},
    Volume = {91},
    Pages = {},
    Year = {2014}}

@article{Colangelo2015,
    Author = {G. Colangelo and M. Hoferichter and M. Procura and others},
    Journal = {JHEP},
    Number = {},
    Volume = {74},
    Pages = {},
    Year = {2015}}

@article{Bijnens2019,
    Author = {J. Bijnens and N. Hermansson-Truedsson and A. Rodriguez-Sanchez},
    Journal = {Physical Letters B},
    Number = {134994},
    Volume = {798},
    Pages = {},
    Note = {arXiv:1908.03331 [hep-ph]},
    Year = {2019}}

@article{Brodsky1967Projector,
    Author = {Stanley J. Brodsky and J. D. Sullivan},
    Journal = {Physical Review},
    Number = {5},
    Volume = {156},
    Pages = {1644--1647},
    Year = {1967}}

@article{KarplusNeuman1950,
    Author = {Robert Karplus and Maurice Neuman},
    Journal = {Physical Review},
    Number = {3},
    Volume = {80},
    Pages = {380--385},
    Year = {1950}}

@article{BardeenTung,
    Author = {W. Bardeen and W. Tung},
    Journal = {Physical Review},
    Number = {5},
    Volume = {173},
    Pages = {1423--1433},
    Year = {1968}}

@article{low1958_SoftPhotonZeros,
    Author = {F. E. Low},
    Journal = {Physical Review},
    Number = {4},
    Volume = {110},
    Pages = {974--977},
    Year = {1958}}

@article{Bjorken1964_GegenbauerPolynomials,
    Author = {James D. Bjorken},
    Journal = {Journal of Mathematical Physics},
    Number = {2},
    Volume = {5},
    Pages = {192--198},
    Year = {1964}}

@article{Bijnens2020,
    Author = {J. Bijnens and N. Hermansson-Truedsson and L. Laub and A. Rodríguez-Sánchez},
    Journal = {Nuclear and Particle Physics Proceedings},
    Volume = {312-317},
    Pages = {180--184},
    Year = {2020}}

@article{AlternateDispersive1,
    Author = {V. Pascalutsa and V. Pauk and M. Vanderhaeghen},
    Journal = {Physical Review D},
    Volume = {85},
    Number = {116001},
    Note = {1204.0740},
    Year = {2012}}

@article{AlternateDispersive2,
    Author = {V. Pauk and M. Vanderhaeghen},
    Journal = {Physical Review D},
    Volume = {90},
    Number = {11},
    Year = {2014}}

@article{AlternateDispersive3,
    Author = {J. Green and O. Gryniuk and G. von Hippel and H. B. Meyer and V. Pascalutsa},
    Journal = {Physical Review Letters},
    Volume = {115},
    Number = {222003},
    Note = {1507.01577},
    Year = {2015}}

@article{AlternateDispersive4,
    Author = {I. Danilkin and M. Vanderhaeghen},
    Journal = {Physical Review D},
    Volume = {95},
    Number = {014019},
    Note = {1611.04646},
    Year = {2017}}

@article{AlternateDispersive5,
    Author = {F. Hagelstein and V. Pascalutsa},
    Journal = {Physical Review Letters},
    Volume = {120},
    Number = {072002},
    Note = {1710.04571},
    Year = {2018}}

@article{protonMass,
    Author = {(Particle Data Group) R. L. Workman and others },
    Journal = {Progress of Theoretical and Experimental Physics},
    Volume = {2022},
    Number = {083C01},
    Year = {2022}}

@article{dispersiveTFF_pion,
    Author = {M. Hoferichter and B. Kubis and S. Leupold and F. Niecknig and S. P. Schneider},
    Journal = {European Physical Journal},
    Volume = {C74},
    Number = {3180},
    Year = {2014}}

@article{CA_pion2,
    Author = {P. Masjuan and S. Peris},
    Journal = {Physical Review Letters},
    Volume = {B686},
    Number = {307},
    Note = {arXiv:0903.0294 [hep-ph]},
    Year = {2010}}

@article{dispersiveTFF_eta,
    Author = {C. Hanhart and A. Kups´c and U.-G. Meißner and F. Stollenwerk and A. Wirzba},
    Journal = {European Physical Journal},
    Volume = {C73},
    Number = {2668},
    Note = {arXiv:1307.5654 [hep-ph], Erratum: Eur. Phys. J. C75, 242 (2015)},
    Year = {2013}}

@article{dispersiveTFF_eta2,
    Author = {C.-W. Xiao and T. Dato and C. Hanhart and B. Kubis and U.-G. Meißner and A. Wirzba},
    Journal = {European Physical Journal},
    Volume = {C81},
    Number = {1002},
    Note = {arXiv:1509.02194 [hep-ph]},
    Year = {2021}}

@article{dispersiveTFF_eta3,
    Author = {Simon Holz and Christoph Hanhart and Martin Hoferichter and Bastian Kubis},
    Journal = {European Physical Journal},
    Volume = {C82},
    Number = {434},
    Note = {arXiv:2202.05846 [hep-ph]},
    Year = {2022}}

@article{pionEta1,
    Author = {O. Deineka and I. Danilkin and M. Vanderhaeghen},
    Journal = {European Physical Journal Web Conference},
    Volume = {199},
    Number = {02005},
    Note = {arXiv:1808.04117 [hep-ph]},
    Year = {2019}}

@article{pionEta2,
    Author = {I. Danilkin and O. Deineka and M. Vanderhaeghen},
    Journal = {Physical Review Letters},
    Volume = {D96},
    Number = {114018},
    Note = {arXiv:1709.08595 [hep-ph]},
    Year = {2017}}

@article{AxialFormFactor1,
    Author = {R. N. Cahn},
    Journal = {Physical Review Letters},
    Volume = {D35},
    Number = {3342},
    Note = {},
    Year = {1987}}

@article{AxialFormFactor2,
    Author = {R. N. Cahn},
    Journal = {Physical Review Letters},
    Volume = {D37},
    Number = {833},
    Note = {},
    Year = {1988}}

@article{AxialFormFactor3,
    Author = {[L3 Collaboration] P. Achard and others},
    Journal = {Physical Review Letters},
    Volume = {B526},
    Number = {269},
    Note = {},
    Year = {2002}}

@article{AxialFormFactor4,
    Author = {[L3 Collaboration] P. Achard and others},
    Journal = {JHEP},
    Volume = {0703},
    Number = {018},
    Note = {},
    Year = {2007}}

@article{AxialResult1,
    Author = {P. Roig and P. Sanchez-Puertas},
    Journal = {Physical Review D},
    Volume = {101},
    Number = {074019},
    Note = {arXiv:1910.02881 [hep-ph]},
    Year = {2020}}

@article{FourDimensionsRelations,
    Author = {Gernot Eichmann and Christian S. Fischer and Walter Heupel and Richard Williams},
    Note = {arXiv:1411.7876v2 [hep-ph]},
    Year = {2014}}

@article{BMW2021,
    Author = {Sz. Borsanyi and Z. Fodor and J. N. Guenther and C. Hoelbling and S. D. Katz and L. Lellouch and T. Lippert and K. Miura and L. Parato and K. K. Szabo and F. Stokes and B. C. Toth and Cs. Torok and L. Varnhorst},
    Journal = {Nature},
    Volume = {593},
    Number = {},
    Pages = {51--55},
    Note = {},
    Year = {2021}}

@article{Mainz2021,
    Author = {E.-H. Chao and R. J. Hudspith and A. G´erardin and J. R. Green and H. B. Meyer and K. Ottnad},
    Journal = {European Physical Journal C},
    Volume = {81},
    Number = {651},
    Note = {arXiv:2104.02632 [hep-lat]},
    Year = {2021}}

@article{HLbLLattice,
    Author = {T. Blum and N. Christ and M. Hayakawa and T. Izubuchi and L. Jin and C. Jung and C. Lehner},
    Journal = {Physical Review Letters},
    Volume = {124},
    Number = {132002},
    Note = {arXiv:1911.08123 [hep-lat]},
    Year = {2020}}

@article{BL1,
    Author = {G. P. Lepage and S. J. Brodsky},
    Journal = {Physics Letters B},
    Volume = {87},
    Number = {359},
    Year = {1979}}

@article{BL2,
    Author = {G. P. Lepage and S. J. Brodsky},
    Journal = {Physical Review D},
    Volume = {22},
    Number = {2157},
    Year = {1980}}

@article{TFFAsymptotic1,
    Author = {V. A. Nesterenko and A. V. Radyushkin},
    Journal = {Soviet Journal of Nuclear Physics},
    Volume = {38},
    Number = {284},
    Year = {1983}}

@article{TFFAsymptotic2,
    Author = {V. A. Novikov and M. A. Shifman and A. I. Vainshtein and M. B. Voloshin and V. I. Zakharov},
    Journal = {Nuclear Physics B},
    Volume = {237},
    Number = {3},
    Pages = {525--550},
    Year = {1984}}

@article{TFFAsymptotic3,
    Author = {A. S. Gorsky},
    Journal = {Soviet Journal of Nuclear Physics},
    Volume = {46},
    Number = {537},
    Year = {1987}}

@article{TFFAsymptotic4,
    Author = {A. V. Manohar},
    Journal = {Physics Letters B},
    Volume = {244},
    Number = {101},
    Year = {1990}}

@article{pionTFFAsymptotic1,
    Author = {M. Hoferichter and B.L. Hoid and B. Kubis and S. Leupold and S. P. Schneider},
    Journal = {JHEP},
    Volume = {10},
    Number = {141},
    Note = {arXiv:1808.04823 [hep-ph]},
    Year = {2018}}

@article{pionTFFAsymptotic2,
    Author = {M. Hoferichter and B.-L. Hoid and B. Kubis and S. Leupold and S. P. Schneider},
    Journal = {Physical Review Letters},
    Volume = {121},
    Number = {112002},
    Note = {arXiv:1805.01471 [hep-ph]},
    Year = {2018}}

@article{OPE1,
    Author = {Kenneth Wilson},
    Journal = {Physical Review},
    Volume = {179},
    Number = {1499},
    Note = {},
    Year = {1969}}

@article{OPEBackgroundViejo,
    Author = {B. L. Ioffe and A. V. Smilga},
    Journal = {Nuclear Physics B},
    Volume = {232},
    Number = {},
    Pages = {109--142},
    URL = {doi:10.1016/0550-3213(84)90364-X},
    Year = {1984}}

@article{OPEBackgroundObsoleto_1,
    Author = {M. A. Shifman and A. I Vainshtein and V. I. Zakharov},
    Journal = {Nuclear Physics B},
    Volume = {147},
    Number = {},
    Pages = {385--518},
    Note = {},
    Year = {1979}}

@article{OPEBackgroundObsoleto_2,
    Author = {M. A. Shifman and A. I. Vainshtein and M. Voloshin and V. I. Zakharov},
    Journal = {Physics Letters B},
    Volume = {77},
    Number = {1},
    Pages = {80--83},
    URL = {https://doi.org/10.1016/0370-2693(78)90206-X},
    Year = {1978}}

@article{OPEBackgroundReciente,
    Author = {A. Czarnecki and W. J. Marciano and A. Vainshtein},
    Journal = {Physical Review D},
    Volume = {67},
    Number = {073006},
    Note = {arXiv:hep-ph/0212229},
    related = {OPEBackgroundReciente_Erratum},
    relatedstring = {Corrected in:},
    Year = {2003}}

@article{OPEBackgroundReview,
    Author = {V. A. Novikov and M. A. Shifman and A. I. Vainshtein and V. I. Zakharov},
    Journal = {Fortschritte der Physik},
    Volume = {32},
    Number = {11},
    Note = {},
    pages = {585--622},
    Year = {1984}}

@article{Bijnens2021,
    Author = {J. Bijnens and N. Hermansson-Truedsson and A. Rodríguez-Sánchez},
    Journal = {Journal of High Energy Physics},
    Volume = {04},
    Number = {240},
    Note = {arXiv:2101.09169v2 [hep-ph]},
    URL = {https://doi.org/10.1007/JHEP04%282021%29240},
    Year = {2021}}

@article{Fock1937,
    Author = {V. Fock},
    Journal = {Physikalische Zeitschrift der Sowjetunion},
    Volume = {12},
    Number = {},
    Pages = {404--425},
    Note = {},
    Year = {1937}}

@article{GaugeInvariantExpansion,
    Author = {M. A. Shifman},
    Journal = {Nuclear Physics},
    Volume = {B173},
    Number = {1},
    Pages = {13--31},
    Note = {},
    Year = {1980}}

@article{FockSchwingerPropagator,
    Author = {W. Kummer and J. Weiser},
    Journal = {Zeitschrift für Physik C Particles and Fields},
    Volume = {31},
    Number = {},
    Pages = {105--110},
    URL = {https://doi.org/10.1007/BF01559599},
    Year = {1986}}

@article{FockSchwingerPropagatorInversion,
    Author = {E. V. Shuryak and A. I. Vainshtein},
    Journal = {Nuclear Physics B},
    Volume = {201},
    Number = {1},
    Pages = {141--158},
    URL = {https://doi.org/10.1016/0550-3213(82)90377-7},
    Year = {1982}}

@article{ZuberCompositeOperatorRenormalization1,
    Author = {H. Kluberg--Stern and J. B. Zuber},
    Journal = {Physical Review D},
    Volume = {12},
    Number = {2},
    Pages = {467--481},
    URL = {https://doi.org/10.1103/PhysRevD.12.467},
    Year = {1975}}

@article{ZuberCompositeOperatorRenormalization2,
    Author = {H. Kluberg--Stern and J. B. Zuber},
    Journal = {Physical Review D},
    Volume = {12},
    Number = {2},
    Pages = {482--488},
    URL = {https://doi.org/10.1103/PhysRevD.12.482},
    Year = {1975}}

@article{ZuberCompositeOperatorRenormalization3,
    Author = {H. Kluberg--Stern and J. B. Zuber},
    Journal = {Physical Review D},
    Volume = {12},
    Number = {10},
    Pages = {3159--3180},
    URL = {https://doi.org/10.1103/PhysRevD.12.3159},
    Year = {1975}}

@article{PassarinoVeltmanDecomposition,
    Author = {Giampiero Passarino and Martinus Veltman},
    Journal = {Nuclear Physics B},
    Volume = {160},
    Number = {1},
    Pages = {151--207},
    URL = {https://doi.org/10.1016/0550-3213(79)90234-7},
    Year = {1979}}

@article{PassarinoVeltmanDecompositionCorrection,
    Author = {R. Keith Ellis and Zoltan Kunszt and Kirill Melnikov and Giulia Zanderighi},
    Journal = {Physics Reports},
    Volume = {518},
    Number = {4--5},
    Pages = {141--250},
    URL = {https://doi.org/10.1016/j.physrep.2012.01.008},
    Year = {2012}}

@article{MixedVirtualities,
    Author = {K. Melnikov and A. Vainshtein},
    Journal = {Physical Review D},
    Volume = {70},
    Number = {113006},
    Pages = {},
    Note = {arXiv:hep-ph/0312226},
    URL = {https://doi.org/10.1103/PhysRevD.70.113006},
    Year = {2004}}

@article{holographicQCD_2,
  title = {Axial vector transition form factors in holographic QCD and their contribution to the anomalous magnetic moment of the muon},
  author = {Leutgeb, Josef and Rebhan, Anton},
  journal = {Phys. Rev. D},
  volume = {101},
  issue = {11},
  pages = {114015},
  numpages = {16},
  year = {2020},
  month = {Jun},
  publisher = {American Physical Society},
  doi = {10.1103/PhysRevD.101.114015},
  url = {https://link.aps.org/doi/10.1103/PhysRevD.101.114015}
}

@article{holographicQCD_3,
  title = {Hadronic light-by-light contribution to the muon $g\ensuremath{-}2$ from holographic QCD with massive pions},
  author = {Leutgeb, Josef and Rebhan, Anton},
  journal = {Phys. Rev. D},
  volume = {104},
  issue = {9},
  pages = {094017},
  numpages = {21},
  year = {2021},
  month = {Nov},
  publisher = {American Physical Society},
  doi = {10.1103/PhysRevD.104.094017},
  url = {https://link.aps.org/doi/10.1103/PhysRevD.104.094017}
}

@article{holographicQCD_4,
  title = {Hadronic light-by-light contribution to the muon $g\ensuremath{-}2$ from holographic QCD with solved $U(1{)}_{A}$ problem},
  author = {Leutgeb, Josef and Mager, Jonas and Rebhan, Anton},
  journal = {Phys. Rev. D},
  volume = {107},
  issue = {5},
  pages = {054021},
  numpages = {13},
  year = {2023},
  month = {Mar},
  publisher = {American Physical Society},
  doi = {10.1103/PhysRevD.107.054021},
  url = {https://link.aps.org/doi/10.1103/PhysRevD.107.054021}
}

@article{holographicQCD,
    Author = {Luigi Cappiello and Oscar Cata and Giancarlo D'Ambrosio and David Greynat and Abhishek Iyer},
    Journal = {Physical Review D},
    Volume = {102},
    Number = {016009},
    Pages = {},
    Note = {arXiv:1912.02779 [hep-ph]},
    URL = {https://doi.org/10.1103/PhysRevD.102.016009},
    Year = {2020}}

@article{ColangeloSDC,
    Author = {G. Colangelo and F. Hagelstein and M. Hoferichter and L. Laub and P. Stoffer},
    Journal = {Physical Review D},
    Volume = {101},
    Number = {051501},
    Pages = {},
    Note = {arXiv:1910.11881 [hep-ph]},
    URL = {https://doi.org/10.1103/PhysRevD.101.051501},
    Year = {2020}}

@article{ColangeloSDC_2,
    Author = {Gilberto Colangelo and Franziska Hagelstein and Martin Hoferichter and Laetitia Laub and Peter Stoffer},
    Journal = {JHEP},
    Volume = {2020},
    Number = {101},
    Pages = {},
    Note = {arXiv:1910.13432 [hep-ph]},
    URL = {https://doi.org/10.1007/JHEP03%282020%29101},
    Year = {2020}}

@article{ColangeloSDC_3,
    Author = {Gilberto Colangelo and Franziska Hagelstein and Martin Hoferichter and Laetitia Laub and Peter Stoffer},
    Journal = {The European Physical Journal C},
    Volume = {81},
    Number = {702},
    Pages = {},
    Note = {arXiv:2106.13222 [hep-ph]},
    URL = {https://doi.org/10.1140/epjc/s10052-021-09513-x},
    Year = {2021}}

@article{backgrounOPEmixedVirtualities_1,
    Author = {Johan Bijnens and Nils Hermansson-Truedsson and Antonio Rodríguez-Sánchez},
    Journal = {},
    Volume = {},
    Number = {},
    Pages = {},
    Note = {arXiv:2211.17183 [hep-ph]},
    URL = {https://doi.org/10.48550/arXiv.2211.17183},
    Year = {2022}}

@article{backgrounOPEmixedVirtualities_2,
    Author = {Johan Bijnens and Nils Hermansson-Truedsson and Antonio Rodríguez-Sánchez},
    Journal = {EPJ Web of Conferences},
    Volume = {274},
    Number = {06010},
    Pages = {},
    Note = {arXiv:2211.04068 [hep-ph]},
    URL = {https://doi.org/10.1051/epjconf/202227406010},
    Year = {2022}}

@article{HLbL_LO4,
    Author = {M. Hoferichter and B.-L. Hoid and B. Kubis and S. Leupold and S. P. Schneider},
    Journal = {JHEP},
    Volume = {10},
    Number = {141},
    Note = {arXiv:1808.04823 [hep-ph]},
    Year = {2018}}

@article{HLbL_LO7,
    Author = {V. Pauk and M. Vanderhaeghen},
    Journal = {European Physical Journal},
    Volume = {C74},
    Number = {3008},
    Note = {arXiv:1401.0832 [hep-ph]},
    Year = {2014}}

@article{HLbL_LO8,
    Author = {I. Danilkin and M. Vanderhaeghen},
    Journal = {Physical Review D},
    Volume = {95},
    Number = {014019},
    Note = {arXiv:1611.04646 [hep-ph]},
    Year = {2017}}

@article{HLbL_L010,
    Author = {M. Knecht and S. Narison and A. Rabemananjara and D. Rabetiarivony},
    Journal = {Physical Review Letters},
    Volume = {B787},
    Number = {111},
    Note = {arXiv:1808.03848 [hep-ph]},
    Year = {2018}}

@book{bookCA,
  title = {Essentials of Padé Approximants},
  author = {G. A. Baker},
  year = {1975},
  address = {New York},
  edition = {First},
  publisher = {Academic Press}}

@article{Tarrach,
    Author = {R. Tarrach},
    Journal = {Nuov. Cim. A},
    Volume = {28},
    Number = {},
    URL = {https://doi.org/10.1007/BF02894857},
    Pages = {409–-422},
    Year = {1975}}

@article{SugawaraKanazawa,
    Author = {M. Sugawara and A. Kanazawa},
    Journal = {Physical Review},
    Volume = {123},
    Number = {1895},
    URL = {https://doi.org/10.1103/PhysRev.123.1895},
    Year = {1961}}

@article{CA_pion,
    Author = {P. Masjuan and P. Sanchez-Puertas},
    Journal = {Physical Review Letters},
    Volume = {D95},
    Number = {054026},
    Note = {arXiv:1701.05829 [hep-ph]},
    Year = {2017}}

@article{LatticeReview2021,
    Author = {Y. Aoki and T. Blum and G. Colangelo and S. Collins and M. Della Morte and P. Dimopoulos and S. Dürr and X. Feng and H. Fukaya and M. Golterman and others},
    Journal = {The European Physical Journal C},
    Volume = {82},
    Number = {869},
    Note = {arXiv:2111.09849v2 [hep-lat]},
    Year = {2022}}

@article{DavydychevDecomposition,
    Author = {A. I. Davydychev},
    Journal = {Physics Letters B},
    Volume = {263},
    Number = {1},
    Pages ={107--111},
    URL = {https://doi.org/10.1016/0370-2693(91)91715-8},
    Year = {1991}}

@article{DavydychevMellinBarnes,
    Author = {A. I. Davydychev},
    Journal = {Journal of Mathematical Physics},
    Volume = {32},
    Number = {1052},
    URL = {doi: 10.1063/1.529383},
    Year = {1991}}

@article{FeynCalc1,
    Author = {V. Shtabovenko and R. Mertig and F. Orellana},
    Journal = {Computer Physics Communications},
    Volume = {256},
    Number = {107478},
    Note = {arXiv:2001.04407},
    Year = {2020}}

@article{FeynCalc2,
    Author = {V. Shtabovenko and R. Mertig and F. Orellana},
    Journal = {Computer Physics Communications},
    Volume = {207},
    Number = {},
    Pages = {432--444},
    Note = {arXiv:1601.01167},
    Year = {2016}}

@article{FeynCalc3,
    Author = {R. Mertig and M. Böhm and A. Denner},
    Journal = {Computer Physics Communications},
    Volume = {64},
    Number = {3},
    Pages = {345--359},
    URL = {https://doi.org/10.1016/0010-4655(91)90130-D},
    Year = {1991}}

@article{BoosDavydychev,
    Author = {E. E. Boos and A. I. Davydychev},
    Journal = {Theoretical and Mathematical Physics},
    Volume = {89},
    Number = {},
    Pages = {1052–-1064},
    URL = {https://doi.org/10.1007/BF01016805},
    Year = {1991}}

@article{twoDimensionsTsikh,
    Author = {O. N. Zhdanov and A. K. Tsikh},
    Journal = {Siberian Mathematical Journal},
    Number = {2},
    Volume = {39},
    Pages = {245--260},
    URL = {https://doi.org/10.1007/BF02677509},
    Year = {1998}}

@article{calabiYau_Tsikh,
    Author = {M. Passare and A .K. Tsikh and A. A. Cheshel},
    Journal = {Theoretical and Mathematical Physics},
    Number = {},
    Volume = {109},
    Pages = {1544–-1555},
    URL = {https://doi.org/10.1007/BF02073871},
    Year = {1996}}

@article{packageFriot,
    Author = {B. Ananthanarayan and Sumit Banik and Samuel Friot and Shayan Ghosh},
    Journal = {Physical Review Letters},
    Number = {151601},
    Volume = {127},
    Pages = {},
    Note = {arXiv:2012.15108 [hep-th]},
    URL = {https://doi.org/10.1103/PhysRevLett.127.151601},
    Year = {2021}}

@article{algorithmFriot,
    Author = {Samuel Friot and David Greynat},
    Journal = {Journal of Mathematical Physics},
    Number = {023508},
    Volume = {53},
    Pages = {},
    Note = {arXiv:1107.0328 [math-ph]},
    URL = {https://doi.org/10.1063/1.3679686},
    Year = {2012}}

@article{straightContoursFriot,
    Author = {Sumit Banik and Samuel Friot},
    Journal = {},
    Number = {},
    Volume = {},
    Pages = {},
    Note = {arXiv:2212.11839 [hep-ph]},
    URL = {https://doi.org/10.48550/arXiv.2212.11839},
    Year = {2022}}

@article{multivariateLarsen,
    Author = {Kasper J. Larsen and Robbert Rietkerk},
    Journal = {Computer Physics Communications},
    Number = {},
    Volume = {222},
    Pages = {250--262},
    Note = {arXiv:1701.01040 [hep-th]},
    URL = {https://doi.org/10.1016/j.cpc.2017.08.025},
    Year = {2018}}

@article{asymptoticExpansionBarnes,
    Author = {E. W. Barnes},
    Journal = {Proceedings of the London Mathematical Society},
    Number = {},
    Volume = {s2-5},
    Pages = {59--116},
    Note = {},
    URL = {},
    Year = {1907}}

@article{babar2013,
  title = {Precise measurement of the ${e}^{\mathbf{+}}{e}^{\mathbf{\ensuremath{-}}}\ensuremath{\rightarrow}{\ensuremath{\pi}}^{\mathbf{+}}{\ensuremath{\pi}}^{\mathbf{\ensuremath{-}}}\mathbf{(}\ensuremath{\gamma}\mathbf{)}$ cross section with the initial-state radiation method at BABAR},
  author = {Lees, J. P. and Poireau, V. and Tisserand, V. and Garra Tico, J. and Grauges, E. and Palano, A. and Eigen, G. and Stugu, B. and Brown, D. N. and Kerth, L. T. and Kolomensky, Yu. G. and Lynch, G. and Koch, H. and Schroeder, T. and Asgeirsson, D. J. and Hearty, C. and Mattison, T. S. and McKenna, J. A. and Khan, A. and Blinov, V. E. and Buzykaev, A. R. and Druzhinin, V. P. and Golubev, V. B. and Kravchenko, E. A. and Onuchin, A. P. and Serednyakov, S. I. and Skovpen, Yu. I. and Solodov, E. P. and Todyshev, K. Yu. and Yushkov, A. N. and Bondioli, M. and Kirkby, D. and Lankford, A. J. and Mandelkern, M. and Atmacan, H. and Gary, J. W. and Liu, F. and Long, O. and Vitug, G. M. and Campagnari, C. and Hong, T. M. and Kovalskyi, D. and Richman, J. D. and West, C. A. and Eisner, A. M. and Kroseberg, J. and Lockman, W. S. and Martinez, A. J. and Schumm, B. A. and Seiden, A. and Chao, D. S. and Cheng, C. H. and Echenard, B. and Flood, K. T. and Hitlin, D. G. and Ongmongkolkul, P. and Porter, F. C. and Rakitin, A. Y. and Andreassen, R. and Huard, Z. and Meadows, B. T. and Sokoloff, M. D. and Sun, L. and Bloom, P. C. and Ford, W. T. and Gaz, A. and Nauenberg, U. and Smith, J. G. and Wagner, S. R. and Ayad, R. and Toki, W. H. and Spaan, B. and Schubert, K. R. and Schwierz, R. and Bernard, D. and Verderi, M. and Clark, P. J. and Playfer, S. and Bettoni, D. and Bozzi, C. and Calabrese, R. and Cibinetto, G. and Fioravanti, E. and Garzia, I. and Luppi, E. and Munerato, M. and Negrini, M. and Piemontese, L. and Santoro, V. and Baldini-Ferroli, R. and Calcaterra, A. and de Sangro, R. and Finocchiaro, G. and Patteri, P. and Peruzzi, I. M. and Piccolo, M. and Rama, M. and Zallo, A. and Contri, R. and Guido, E. and Lo Vetere, M. and Monge, M. R. and Passaggio, S. and Patrignani, C. and Robutti, E. and Bhuyan, B. and Prasad, V. and Lee, C. L. and Morii, M. and Edwards, A. J. and Adametz, A. and Uwer, U. and Lacker, H. M. and Lueck, T. and Dauncey, P. D. and Behera, P. K. and Mallik, U. and Chen, C. and Cochran, J. and Meyer, W. T. and Prell, S. and Rubin, A. E. and Gritsan, A. V. and Guo, Z. J. and Arnaud, N. and Davier, M. and Derkach, D. and Grosdidier, G. and Le Diberder, F. and Lutz, A. M. and Malaescu, B. and Roudeau, P. and Schune, M. H. and Stocchi, A. and Wang, L. L. and Wormser, G. and Lange, D. J. and Wright, D. M. and Chavez, C. A. and Coleman, J. P. and Fry, J. R. and Gabathuler, E. and Hutchcroft, D. E. and Payne, D. J. and Touramanis, C. and Bevan, A. J. and Di Lodovico, F. and Sacco, R. and Sigamani, M. and Cowan, G. and Brown, D. N. and Davis, C. L. and Denig, A. G. and Fritsch, M. and Gradl, W. and Griessinger, K. and Hafner, A. and Prencipe, E. and Barlow, R. J. and Jackson, G. and Lafferty, G. D. and Behn, E. and Cenci, R. and Hamilton, B. and Jawahery, A. and Roberts, D. A. and Dallapiccola, C. and Cowan, R. and Dujmic, D. and Sciolla, G. and Cheaib, R. and Lindemann, D. and Patel, P. M. and Robertson, S. H. and Biassoni, P. and Neri, N. and Palombo, F. and Stracka, S. and Cremaldi, L. and Godang, R. and Kroeger, R. and Sonnek, P. and Summers, D. J. and Nguyen, X. and Simard, M. and Taras, P. and De Nardo, G. and Monorchio, D. and Onorato, G. and Sciacca, C. and Martinelli, M. and Raven, G. and Jessop, C. P. and LoSecco, J. M. and Wang, W. F. and Honscheid, K. and Kass, R. and Brau, J. and Frey, R. and Sinev, N. B. and Strom, D. and Torrence, E. and Feltresi, E. and Gagliardi, N. and Margoni, M. and Morandin, M. and Posocco, M. and Rotondo, M. and Simi, G. and Simonetto, F. and Stroili, R. and Akar, S. and Ben-Haim, E. and Bomben, M. and Bonneaud, G. R. and Briand, H. and Calderini, G. and Chauveau, J. and Hamon, O. and Leruste, Ph. and Marchiori, G. and Ocariz, J. and Sitt, S. and Biasini, M. and Manoni, E. and Pacetti, S. and Rossi, A. and Angelini, C. and Batignani, G. and Bettarini, S. and Carpinelli, M. and Casarosa, G. and Cervelli, A. and Forti, F. and Giorgi, M. A. and Lusiani, A. and Oberhof, B. and Paoloni, E. and Perez, A. and Rizzo, G. and Walsh, J. J. and Lopes Pegna, D. and Olsen, J. and Smith, A. J. S. and Telnov, A. V. and Anulli, F. and Faccini, R. and Ferrarotto, F. and Ferroni, F. and Gaspero, M. and Li Gioi, L. and Mazzoni, M. A. and Piredda, G. and B\"unger, C. and Gr\"unberg, O. and Hartmann, T. and Leddig, T. and Schr\"oder, H. and Voss, C. and Waldi, R. and Adye, T. and Olaiya, E. O. and Wilson, F. F. and Emery, S. and Hamel de Monchenault, G. and Vasseur, G. and Y\`eche, Ch. and Aston, D. and Bard, D. J. and Bartoldus, R. and Benitez, J. F. and Cartaro, C. and Convery, M. R. and Dorfan, J. and Dubois-Felsmann, G. P. and Dunwoodie, W. and Ebert, M. and Field, R. C. and Franco Sevilla, M. and Fulsom, B. G. and Gabareen, A. M. and Graham, M. T. and Grenier, P. and Hast, C. and Innes, W. R. and Kelsey, M. H. and Kim, P. and Kocian, M. L. and Leith, D. W. G. S. and Lewis, P. and Lindquist, B. and Luitz, S. and Luth, V. and Lynch, H. L. and MacFarlane, D. B. and Muller, D. R. and Neal, H. and Nelson, S. and Perl, M. and Pulliam, T. and Ratcliff, B. N. and Roodman, A. and Salnikov, A. A. and Schindler, R. H. and Snyder, A. and Su, D. and Sullivan, M. K. and Va'vra, J. and Wagner, A. P. and Wisniewski, W. J. and Wittgen, M. and Wright, D. H. and Wulsin, H. W. and Young, C. C. and Ziegler, V. and Park, W. and Purohit, M. V. and White, R. M. and Wilson, J. R. and Randle-Conde, A. and Sekula, S. J. and Bellis, M. and Burchat, P. R. and Miyashita, T. S. and Alam, M. S. and Ernst, J. A. and Gorodeisky, R. and Guttman, N. and Peimer, D. R. and Soffer, A. and Lund, P. and Spanier, S. M. and Ritchie, J. L. and Ruland, A. M. and Schwitters, R. F. and Wray, B. C. and Izen, J. M. and Lou, X. C. and Bianchi, F. and Gamba, D. and Lanceri, L. and Vitale, L. and Martinez-Vidal, F. and Oyanguren, A. and Ahmed, H. and Albert, J. and Banerjee, Sw. and Bernlochner, F. U. and Choi, H. H. F. and King, G. J. and Kowalewski, R. and Lewczuk, M. J. and Nugent, I. M. and Roney, J. M. and Sobie, R. J. and Tasneem, N. and Gershon, T. J. and Harrison, P. F. and Latham, T. E. and Puccio, E. M. T. and Band, H. R. and Dasu, S. and Pan, Y. and Prepost, R. and Wu, S. L.},
  collaboration = {BABAR Collaboration},
  journal = {Phys. Rev. D},
  volume = {86},
  issue = {3},
  pages = {032013},
  numpages = {49},
  year = {2012},
  month = {Aug},
  publisher = {American Physical Society},
  doi = {10.1103/PhysRevD.86.032013},
  url = {https://link.aps.org/doi/10.1103/PhysRevD.86.032013}
}

@article{recentG2Review,
    Author = {G. Colangelo and others},
    Journal = {},
    Number = {},
    Volume = {},
    Pages = {},
    Note = {arXiv:2203.15810 [hep-ph]},
    URL = {https://doi.org/10.48550/arXiv.2203.15810},
    Year = {2022}}

@article{HVPrc,
    Author = {Anian Altherr and Lucius Bushnaq and Isabel Campos and Marco Catillo and Alessandro Cotellucci and Madeleine Dale and Patrick Fritzsch and Roman Gruber and Javad Komijani and Jens Lücke and Marina Krstić Marinković and Sofie Martins and Agostino Patella and Nazario Tantalo and Paola Tavella},
    Journal = {},
    Number = {},
    Volume = {},
    Pages = {},
    Note = {arXiv:2301.04385 [hep-lat]},
    URL = {},
    Year = {2023}}

@article{etmcHVP,
  title = {Lattice calculation of the short and intermediate time-distance hadronic vacuum polarization contributions to the muon magnetic moment using twisted-mass fermions},
  author = {Alexandrou, C. and Bacchio, S. and Dimopoulos, P. and Finkenrath, J. and Frezzotti, R. and Gagliardi, G. and Garofalo, M. and Hadjiyiannakou, K. and Kostrzewa, B. and Jansen, K. and Lubicz, V. and Petschlies, M. and Sanfilippo, F. and Simula, S. and Urbach, C. and Wenger, U.},
  collaboration = {Extended Twisted Mass Collaboration},
  journal = {Phys. Rev. D},
  volume = {107},
  issue = {7},
  pages = {074506},
  numpages = {42},
  year = {2023},
  month = {Apr},
  publisher = {American Physical Society},
  doi = {10.1103/PhysRevD.107.074506},
  url = {https://link.aps.org/doi/10.1103/PhysRevD.107.074506}
}

@article{MainzHVP,
  title = {Window observable for the hadronic vacuum polarization contribution to the muon $g\ensuremath{-}2$ from lattice QCD},
  author = {C\`e, M. and G\'erardin, A. and von Hippel, G. and Hudspith, R. J. and Kuberski, S. and Meyer, H. B. and Miura, K. and Mohler, D. and Ottnad, K. and Paul, S. and Risch, A. and San Jos\'e, T. and Wittig, H.},
  journal = {Phys. Rev. D},
  volume = {106},
  issue = {11},
  pages = {114502},
  numpages = {33},
  year = {2022},
  month = {Dec},
  publisher = {American Physical Society},
  doi = {10.1103/PhysRevD.106.114502},
  url = {https://link.aps.org/doi/10.1103/PhysRevD.106.114502}
}

@article{hpqcdHVP,
  title = {Light-quark connected intermediate-window contributions to the muon $g\ensuremath{-}2$ hadronic vacuum polarization from lattice QCD},
  author = {Bazavov, Alexei and Davies, Christine and DeTar, Carleton and El-Khadra, Aida X. and G\'amiz, Elvira and Gottlieb, Steven and Jay, William I. and Jeong, Hwancheol and Kronfeld, Andreas S. and Lahert, Shaun and Lepage, G. Peter and Lynch, Michael and Lytle, Andrew T. and Mackenzie, Paul B. and McNeile, Craig and Neil, Ethan T. and Peterson, Curtis T. and Ray, Gaurav and Simone, James N. and Van de Water, Ruth S. and Vaquero, Alejandro},
  collaboration = {Fermilab Lattice, HPQCD, and MILC Collaborations},
  journal = {Phys. Rev. D},
  volume = {107},
  issue = {11},
  pages = {114514},
  numpages = {27},
  year = {2023},
  month = {Jun},
  publisher = {American Physical Society},
  doi = {10.1103/PhysRevD.107.114514},
  url = {https://link.aps.org/doi/10.1103/PhysRevD.107.114514}
}

@article{ukqcdHVP,
    Author = {T. Blum and P. A. Boyle and M. Bruno and D. Giusti and V. Gülpers and R. C. Hill and T. Izubuchi and Y.-C. Jang and L. Jin and C. Jung and A. Jüttner and C. Kelly and C. Lehner and N. Matsumoto and R. D. Mawhinney and A. S. Meyer and J. T. Tsang},
    Journal = {},
    Number = {},
    Volume = {},
    Pages = {},
    Note = {arXiv:2301.08696 [hep-lat]},
    URL = {},
    Year = {2023}}

@article{MainzHLbL2023,
    Author = {Nils Asmussen and En-Hung Chao and Antoine Gérardin and Jeremy R. Green and Renwick J. Hudspith and Harvey B. Meyer and Andreas Nyffeler},
    Journal = {JHEP},
    Number = {40},
    Volume = {2023},
    Pages = {},
    Note = {},
    doi = {https://doi.org/10.1007/JHEP04(2023)040},
    URL = {},
    Year = {2023}}

@article{BMWhlbl2023,
    Author = {Antoine Gérardin and Willem E. A. Verplanke and Gen Wang and Zoltan Fodor and Jana N. Guenther and Laurent Lellouch and Kalman K. Szabo and Lukas Varnhorst},
    Journal = {},
    Number = {},
    Volume = {},
    Pages = {},
    Note = {arXiv:2305.04570 [hep-lat]},
    URL = {},
    Year = {2023}}

@article{ETMChlbl2022,
    Author = {Constantia Alexandrou and Simone Bacchio and Sebastian Burri and Jacob Finkenrath and Andrew Gasbarro and Kyriakos Hadjiyiannakou and Karl Jansen and Gurtej Kanwar and Bartosz Kostrzewa and Konstantin Ottnad and Marcus Petschlies and Ferenc Pittler and Carsten Urbach and Urs Wenger},
    Journal = {},
    Number = {},
    Volume = {},
    Pages = {},
    Note = {arXiv:2212.06704 [hep-lat]},
    URL = {},
    Year = {2022}}

@article{ETMChlbl2023,
    Author = {C. Alexandrou and S. Bacchio and G. Bergner and S. Burri and J. Finkenrath and A. Gasbarro and K. Hadjiyiannakou and K. Jansen and G. Kanwar and B. Kostrzewa and G. Koutsou and K. Ottnad and M. Petschlies and F. Pittler and F. Steffens and C. Urbach and U. Wenger},
    Journal = {},
    Number = {},
    Volume = {},
    Pages = {},
    Note = {	arXiv:2308.12458 [hep-lat]},
    URL = {},
    Year = {2023}}

@article{ukqcdHLbL,
    Author = {Thomas Blum and Norman Christ and Masashi Hayakawa and Taku Izubuchi and Luchang Jin and Chulwoo Jung and Christoph Lehner and Cheng Tu},
    Journal = {},
    Number = {},
    Volume = {},
    Pages = {},
    Note = {	arXiv:2304.04423 [hep-lat]},
    URL = {},
    Year = {2023}}

@article{Zanke2021,
author={Zanke, Marvin
and Hoferichter, Martin
and Kubis, Bastian},
journal={Journal of High Energy Physics},
year={2021},
month={Jul},
day={16},
volume={2021},
number={7},
pages={106},
issn={1029-8479},
doi={10.1007/JHEP07(2021)106},
url={https://doi.org/10.1007/JHEP07(2021)106}
}

@article{newScalar2021,
title = {A dispersive estimate of scalar contributions to hadronic light-by-light scattering},
journal = {Physics Letters B},
volume = {820},
pages = {136502},
year = {2021},
issn = {0370-2693},
doi = {https://doi.org/10.1016/j.physletb.2021.136502},
url = {https://www.sciencedirect.com/science/article/pii/S0370269321004421},
author = {Igor Danilkin and Martin Hoferichter and Peter Stoffer}
}

@article{Phan2019,
    Author = {Khiem Hong Phan and Dzung Tri Tran},
    Journal = {Progress of Theoretical and Experimental Physics},
    Number = {6},
    Volume = {2019},
    Pages = {},
    Note = {arXiv:1904.07430 [hep-ph]},
    URL = {https://doi.org/10.1093/ptep/ptz050},
    Year = {2019}}

@article{HornTheorem,
    Author = {J. Horn},
    Journal = {Mathematische Annalen},
    Number = {},
    Volume = {34},
    Pages = {544–-600},
    Note = {},
    URL = {https://doi.org/10.1007/BF01443681},
    Year = {1889}}

\end{document}